\documentclass[11pt,twoside,a4paper,cmspaper,final,collab]{cms-tdr}
\def\svnVersion{741766d-D}\def\svnDate{2023/08/21}\def\cmsCernNoTag{CERN-EP-2023-065}\def\cmsCernDate{\today}\def\cmsMessage{Published in the Journal of High Energy Physics as \href{http://dx.doi.org/10.1007/JHEP08(2023)040}{\doi{10.1007/JHEP08(2023)040}.}\\This version contains a correction to the acknowledgments.}
\begin{document}\cmsNoteHeader{HIG-21-009}

\newcommand{\ggH}{\ensuremath{\Pg\Pg\PH}\xspace}
\newcommand{\VBF}{\ensuremath{\mathrm{VBF}}\xspace}
\newcommand{\WH}{\ensuremath{\PW\PH}\xspace}
\newcommand{\ZH}{\ensuremath{\PZ\PH}\xspace}
\newcommand{\VH}{\ensuremath{\PV\PH}\xspace}
\newcommand{\ttH}{\ensuremath{\ttbar\PH}\xspace}
\newcommand{\ZZ}{\ensuremath{\PZ\PZ}\xspace}
\newcommand{\qqZZ}{\ensuremath{\PQq\PAQq\to\PZ\PZ}\xspace}
\newcommand{\ggZZ}{\ensuremath{\Pg\Pg\to\PZ\PZ}\xspace}
\newcommand{\HZZfl}{\ensuremath{\PH\to\PZ\PZ\to4\ell}\xspace}
\newcommand{\HZZ}{\ensuremath{\PH\to\PZ\PZ}\xspace}
\newcommand{\mH}{\ensuremath{m_{\PH}}\xspace}
\newcommand{\mlplm}{\ensuremath{m_{\ell^{+}\ell^{-}}}\xspace}
\newcommand{\mllll}{\ensuremath{m_{4\ell}}}
\newcommand{\Dalt}{\ensuremath{{\mathcal D}_{\text{alt}}}\xspace}
\newcommand{\Dint}{\ensuremath{{\mathcal D}^{\text{dec}}_{\text{int}}}\xspace}
\newcommand{\Dzm}{\ensuremath{{\mathcal D}^{\text{dec}}_{\text{0-}}}\xspace}
\newcommand{\Dzhp}{\ensuremath{{\mathcal D}^{\text{dec}}_{\text{0h+}}}\xspace}
\newcommand{\DLone}{\ensuremath{{\mathcal D}^{\text{dec}}_{\Lambda\text{1}}} }
\newcommand{\DLoneZg}{\ensuremath{{\mathcal D}_{\Lambda\text{1}}^{\PZ\gamma, \text{dec}}}\xspace}
\newcommand{\DCP}{\ensuremath{{\mathcal D}^{\text{dec}}_{\text{CP}}}\xspace}
\newcommand{\Zll}{\ensuremath{\PZ\to\Pell^{+}\Pell^{-}}\xspace}
\newcommand{\usedLumiABC}{138\fbinv\xspace}
\newcommand{\JHUGEN}{\textsc{JHUGen}\xspace}

\providecommand{\cmsTable}[1]{\resizebox{\textwidth}{!}{#1}}
\newlength\cmsTabSkip\setlength{\cmsTabSkip}{1ex}

\providecommand{\cmsLeft}{left\xspace}
\providecommand{\cmsRight}{right\xspace}

\cmsNoteHeader{HIG-21-009}
\title{Measurements of inclusive and differential cross sections for the Higgs boson production and decay to four-leptons in proton-proton collisions at \texorpdfstring{$\sqrt{s} = 13\TeV$}{sqrt(s) = 13 TeV}}

\date{\today}

\abstract{
	Measurements of the inclusive and differential fiducial cross sections for the Higgs boson production in the $\PH\to\PZ\PZ\to 4\ell$ ($\ell=\Pe,\Pgm$) decay channel are presented.
	The results are obtained from the analysis of proton-proton collision data recorded by the CMS experiment at the CERN LHC at a center-of-mass energy of 13\TeV, corresponding to an
	integrated luminosity of $\usedLumiABC$. The measured inclusive fiducial cross section is $2.73\pm0.26\unit{fb}$, in agreement with the standard model expectation of $2.86\pm0.1\unit{fb}$.
	Differential cross sections are measured as a function of several kinematic observables sensitive to the Higgs boson production and decay to four leptons.
	A set of double-differential measurements is also performed, yielding a comprehensive characterization of the four leptons final state.
	Constraints on the Higgs boson trilinear coupling and on the bottom and charm quark coupling modifiers are derived from its transverse momentum distribution.
	All results are consistent with theoretical predictions from the standard model.
}

\hypersetup{%
	pdfauthor={CMS Collaboration},%
	pdftitle={Inclusive and differential cross section measurements in the Higgs boson to four-lepton decay channel in proton-proton collisions at \texorpdfstring{$\sqrt{s} = 13\TeV$}{sqrt(s) = 13 TeV}},%
	pdfsubject={CMS},%
	pdfkeywords={CMS, Higgs}} 

\maketitle 

\section{Introduction}
\label{sec:intro}
The discovery of the Higgs (\PH) boson in 2012 by the ATLAS and CMS Collaborations~\cite{Aad:2012tfa,Chatrchyan:2012ufa,Chatrchyan:2013lba} is a major confirmation of the correctness of the theoretical approach involving the electroweak (EW) symmetry breaking mechanism~\cite{Englert:1964et,Higgs:1964ia,Higgs:1964pj,Guralnik:1964eu,Higgs:1966ev,Kibble:1967sv}.
Subsequent measurements of the properties of this particle~\cite{ATLASPropertiesRun1,CMSPropertiesRun1,ATLASCMSMassRun1,ATLASCMSPropertiesRun1}, including its mass, quantum numbers, and couplings, further confirmed the consistency of these measurements with the standard model (SM) predictions.

The \PH boson decay into four charged leptons ($\HZZfl$, $\ell=\Pe,\Pgm$), with its fully reconstructible final state and large signal-to-background ratio, has been one of the pillars for the characterization of  this particle since its discovery. Several properties of the \PH boson were measured in this decay channel at the CERN LHC, based on the Run 1 data set at center-of-mass energies of 7 and 8\TeV and on the Run 2 data set at 13\TeV. These include the determination of its mass, spin and  parity~\cite{ATLASH4lLegacyRun1,CMSH4lLegacyRun1,CMSH4lSpinParity,CMSH4lAnomalousCouplings,CMSH4l2016,ATLASH4l2016},
width~\cite{CMSH4lWidth,CMSH4lLifetime,ATLASH4lWidth,ATLASH4lWidth2016}, inclusive and differential fiducial cross sections~\cite{ATLASH4lFiducial8TeV,CMSH4lFiducial8TeV, CMSH4l2016,ATLASH4lFiducial2016, ATLASH4lLegacyRun2, ATLASH4lFiducialRun2}, and tensor structure for interactions with a pair of gauge bosons~\cite{CMSH4lAnomalousCouplings,CMSH4lLifetime, CMSH4lAnomalousCouplings2016,ATLASH4l2016,CMSHVVAnomalousCouplings2016,CMSHIG19009}.
The most precise value of the \PH boson mass to date, measured by the CMS Collaboration, is $\text{m}_\PH=125.38\pm0.14\GeV$, obtained from the combination of the $\HZZfl$ and $\PH\to\PGg\PGg$ decay channels from the analysis of the Run 1 and 2016 Run 2 data sets~\cite{CMS:2020xrn}. From the analysis of  the full Run 2 data set, the CMS Collaboration reported the first evidence for the off-shell \PH boson production in events with a final state of two $\PZ$ bosons decaying into either four charged leptons, or two charged leptons and two neutrinos, with a measured value of the \PH boson width of $\Gamma_\PH=3.2^{+2.4}_{-1.7}\MeVns$~\cite{CMS:2022ley}.

The \PH boson production is often experimentally characterized via the so called simplified template cross section (STXS) framework, which defines mutually exclusive phase space regions designed to maximize the experimental sensitivity to physics beyond the SM (BSM) effects and reduce, at the same time, the theoretical model dependence in the measurements~\cite{deFlorian:2016spz}.
The ATLAS and CMS Collaborations published results of cross section measurements in the STXS framework using the full Run 2 data set in the $\HZZfl$~\cite{ATLASH4lLegacyRun2,CMSHIG19001} and other decay channels~\cite{ATLAS:2022tnm,CMS:2021kom,ATLAS:2022ooq,CMS:2022uhn,ATLAS:2020fcp,ATLAS:2022yrq,CMS:2022kdi}. The ATLAS and CMS Collaborations recently published results obtained from the combination of all the decay channels focusing on measurements of simplified template cross sections (STXS)~\cite{ATLAS:2022vkf} and \PH boson couplings~\cite{CMS:2022dwd}, respectively.

Fiducial cross section measurements constitute a complementary approach for the characterization of the \PH boson production and decay that provide a set of less model-dependent results by unfolding detector effects from the data, thus allowing a direct comparison with state-of-the-art theoretical predictions.
The ATLAS and CMS Collaborations published fiducial cross section measurements in the $\PH\to\PGg\PGg$~\cite{ATLAS:2022fnp,CMS:2022hyj}, $\PH\to\PW\PW$~\cite{ATLAS:2023hyd,ATLAS:2023pwa,CMS:2020dvg}, and
$\HZZfl$~\cite{ATLASH4lFiducialRun2,CMSHIG19001} decay channels using the full Run 2 data set.
The CMS Collaboration also published results in the $\PH\to\PGt\PGt$~\cite{CMS:2021gxc} decay channel, while the ATLAS Collaboration presented results from the combination of the $\PH\to\PGg\PGg$ and $\HZZfl$ decay channels~\cite{ATLAS:2022qef}.

This paper presents measurements of inclusive and differential cross sections for the \PH boson production in the $\HZZfl$ decay channel using data from proton-proton ($\Pp\Pp$) collisions recorded with the CMS detector at the LHC in 2016--2018 and corresponding to an integrated luminosity of $\usedLumiABC$. 
To reduce the model dependence, all the measurements are performed within a fiducial phase space region defined to closely reproduce the experimental acceptance and reconstruction-level selection criteria.

Differential cross sections are measured for several kinematic observables sensitive to the \PH boson production and its decay into four leptons, providing a complete characterization of this channel and coverage of the entire fiducial phase space. This includes the measurement of six double-differential cross sections.
Fiducial cross sections are also measured in bins of matrix element kinematic discriminants sensitive to possible anomalous couplings of the \PH boson to vector bosons.
This provides a valuable test of the SM predictions and may reveal possible BSM physics.

The analysis builds upon the methods used in previous measurements of \PH boson properties in the four-lepton decay channel~\cite{CMSH4lFiducial8TeV, CMSHIG19001},  featuring the latest CMS Run 2 calibrations and a reduction by $\sim40$\% for the leading systematic uncertainty in lepton reconstruction and selection efficiencies.

The measurement of the fiducial cross section in bins of transverse momentum $\pt$ of the \PH boson ($\pt^\PH$) is also used to set constraints on the \PH boson trilinear self-coupling and on the coupling modifiers of the \PH boson to \Pb and \PQc quarks.
The CMS Collaboration recently reported constraints on the \PH boson self-coupling  from the combination of several decay channels using the full Run 2 data set~\cite{CMS:2022dwd}. 
These constraints are obtained from the interpretation of the STXS results, while in this paper an alternative and complementary approach using differential cross section measurements is explored.
The ATLAS Collaboration set limits on the coupling modifiers of the \PH to \Pb and \PQc quarks from the combination of the $\HZZfl$ and $\PH\to\PGg\PGg$ decay channels using the 2016--2018 Run 2 data set~\cite{ATLAS:2022qef}.
The CMS Collaboration reported similar results using the 2016 Run 2 data set and combining the $\HZZfl$ and $\PH\to\PGg\PGg$ decay channels~\cite{CMS:2018gwt}.
The constraints derived from the analysis presented in this paper supersede the ones reported for the $\HZZfl$ channel in Ref.~\cite{CMS:2018gwt}.

This paper is organized as follows. 
The CMS detector is briefly described in Section~\ref{sec:detector}.
The data set used is presented in Section~\ref{sec:samples}, along with a description of the simulated signal and background samples.
The event reconstruction techniques and the selection criteria used to identify \PH boson candidates are outlined in Section~\ref{sec:reconstruction}.
The definition of the restricted phase space region where the differential cross sections are measured is given in Section~\ref{sec:fiducialvolume}.
A complete description of all the kinematic observables, with a particular emphasis on matrix element discriminants, is presented in Section~\ref{sec:observables}.
The background modeling is presented in Section~\ref{sec:bkgd}.
The signal modeling and the statistical procedure adopted in the extraction of the inclusive and differential cross sections are presented in Section~\ref{sec:measurement}. 
The systematic uncertainties that affect the measurement are described in Section~\ref{sec:systematics}.
The results of the analysis and their comparison to the SM expectations  are outlined in Section~\ref{sec:results}.
In Section~\ref{sec:interpretations} the measurement of the fiducial cross section in differential bins of  $\pt^\PH$ are used to set constraints to the trilinear self-coupling of the \PH boson and to its couplings with charm and bottom quarks.
A summary highlighting the main findings of the analysis is given in Section~\ref{sec:summary}.

\section{The CMS detector}
\label{sec:detector}
The central feature of the CMS apparatus is a superconducting solenoid of 6\unit{m} internal diameter, providing a magnetic field of 3.8\unit{T}. Within the solenoid volume are a silicon pixel and strip tracker, a lead tungstate crystal electromagnetic calorimeter (ECAL), and a brass and scintillator hadron calorimeter (HCAL), each composed of a barrel and two endcap sections. Forward calorimeters extend the pseudorapidity ($\eta$) coverage provided by the barrel and endcap detectors. Muons are detected in gas-ionization chambers embedded in the steel flux-return yoke outside the solenoid. The electromagnetic calorimeter consists of 75\,848 lead tungstate crystals, which provide coverage in pseudorapidity $\abs{\eta} < 1.48 $ in a barrel region (EB) and $1.48 < \abs{\eta} < 3.0$ in two endcap regions (EE). Preshower detectors consisting of two planes of silicon sensors interleaved with a total of three radiation lengths of lead are located in front of each EE detector. The hadron forward (HF) calorimeter uses steel as an absorber and quartz fibers as the sensitive material. The two halves of the HF are located 11.2\unit{m} from the interaction region, one on each end, and together they provide coverage in the range $3.0 < \abs{\eta} < 5.2$. They also serve as luminosity monitors. 

Events of interest are selected using a two-tiered trigger system. The first level (L1), composed of custom hardware processors, uses information from the calorimeters and muon detectors to select events at a rate of approximately 100\unit{kHz} within a fixed latency of 4\mus~\cite{Sirunyan:2020zal}. The second level, known as the high-level trigger (HLT), consists of a farm of processors running a version of the full event reconstruction software optimized for fast processing, and reduces the event rate to around 1\unit{kHz} before data storage~\cite{Khachatryan:2016bia}. 

The primary vertex (PV) is taken to be the vertex corresponding to the hardest scattering in the event, evaluated using tracking information alone, as described in Section 9.4.1 of Ref.~\cite{CMS-TDR-15-02}.

The electron momentum is estimated by combining the energy measurement in the ECAL with the momentum measurement in the tracker. The momentum resolution for electrons with transverse momentum $\pt \approx 45\GeV$ from $\PZ \to \Pe \Pe$ decays ranges 1.6--5\%. It is generally better in the EB than in EE, and also depends on the bremsstrahlung energy emitted by the electron as it traverses the material in front of the ECAL~\cite{CMS:2020uim,CMS-DP-2020-021}. 

Muons are reconstructed with detection planes made using three technologies: drift tubes, cathode strip chambers, and resistive-plate chambers. Matching muons to tracks measured in the silicon tracker results in a relative \pt resolution, for muons with \pt up to 100\GeV, of 1\% in the barrel and 3\% in the endcaps. The \pt resolution in the barrel is better than 7\% for muons with \pt up to 1\TeV~\cite{Sirunyan:2018}. 

A more detailed description of the CMS detector, together with the definition of the coordinate system used and the relevant kinematic variables, can be found in Ref.~\cite{Chatrchyan:2008zzk}. 

\section{Data and simulated samples}
\label{sec:samples}
This analysis is based on proton-proton collisions recorded by the CMS experiment at the LHC in 2016, 2017, and 2018, corresponding to integrated luminosities of 36.3, 41.5, and 59.8\fbinv, respectively~\cite{CMS-LUM-17-003,CMS-PAS-LUM-17-004,CMS-PAS-LUM-18-002}.

Candidate events are selected offline from leptons passing loose identification and isolation requirements~\cite{CMS:2020uim,Sirunyan:2018}, following the online selection based on dielectron, dimuon, and electron-muon HLT algorithms. The various lepton \pt thresholds used in the online selection for each data-taking period are reported in Table~\ref{tab:HLT}.
Additional triggers that require three leptons with lower \pt thresholds and no isolation criteria, as well as single-electron and single-muon triggers, are used to increase the efficiency.
Events selected with the single-lepton triggers are used to measure the trigger efficiency by means of the ``tag-and-probe'' technique~\cite{CMS:2011aa}, following the strategy of Ref.~\cite{CMSHIG19001}. A ``tag'' lepton is defined as a lepton that matches geometrically a candidate from the single-lepton triggers, whereas the other leptons are used as ``probes'' and are combined together to form any of the triggers. The overall trigger efficiency measured in data is larger than 99\% and is in agreement with that estimated from simulated samples.

\begin{table}[htb]
	\centering
	\topcaption{
		Thresholds applied on the \pt of the leading/subleading leptons in each data-taking period for the main dielectron (\Pe/\Pe), dimuon (\Pgm/\Pgm), and electron-muon (\Pe/\Pgm, \Pgm/\Pe) HLT algorithms.
		\label{tab:HLT}
	}
	\begin{tabular}{cccc}
		& \Pe/\Pe ({\GeVns}) & \Pgm/\Pgm ({\GeVns})& \Pe/\Pgm, \Pgm/\Pe ({\GeVns})\\
		\hline
		2016 & 17/12 & 17/8 & 17/8, 8/23 \\
		2017 & 23/12 & 17/8 & 23/8, 12/23 \\
		2018 & 23/12 & 17/8 & 23/8, 12/23
	\end{tabular}
\end{table}

{\tolerance=800
	Signal samples are simulated at next-to-leading order (NLO) in perturbative quantum chromodynamics (pQCD) using the \POWHEG~2.0~\cite{Nason:2004rx,Frixione:2007vw,Alioli:2010xd} generator for the five main production mechanisms of the SM \PH boson: gluon fusion ($\ggH$)~\cite{Alioli:2008tz}, vector boson fusion ($\VBF$)~\cite{Nason:2009ai},
	associated production with a vector boson ($\VH$, where \PV = \PW, \PZ)~\cite{Luisoni2013},
	and associated production with a pair of top quarks ($\ttH$)~\cite{Hartanto:2015uka}. Events produced via the $\ggH$ mechanism are simulated at NLO with \POWHEG~2.0 and reweighted to match the predictions at next-to-next-to-leading order in the strong coupling, including matching to a parton shower (\textsc{NNLOPS})~\cite{Hamilton:2013fea} as a function of the $\pt^\PH$ and of the number of jets in the event. The $\Pg\Pg\to\ZH$ contribution to the $\ZH$ production mode is simulated at leading order (LO) using \JHUGEN~7.3.0~\cite{Gao:2010qx, Bolognesi:2012mm,Anderson:2013afp,Gritsan:2016hjl,Gritsan:2020pib}.
	The $\ggH$ production mechanism is simulated at NLO also with \MGvATNLO using the \textsc{HC\_NLO\_X0\_UFO-heft} model in the 5 flavor scheme~\cite{Wiesemann:2014ioa}. The \PH boson is produced in association with 0, 1, or 2 jets in the final states, merged with the FxFx scheme. The top quark mass is set to $173.0\GeV$ in the simulation, but finite top mass effects in loops are filtered out.
    The \PH boson production in association with $\PQb$ quarks is not considered in this analysis as its impact on the unfolded distributions is expected to be negligible with respect to all the other production modes.
	The decay of the \PH boson to four leptons is modeled with \JHUGEN~7.0.2.
	The simulation of the various production and decay modes is based on the theoretical predictions from Refs.~\cite{Anastasiou:2015ema,Anastasiou2016,Ciccolini:2007jr,Ciccolini:2007ec,Bolzoni:2010xr,Bolzoni:2011cu,Brein:2003wg,Ciccolini:2003jy,Beenakker:2001rj,Beenakker:2002nc,Dawson:2002tg,Dawson:2003zu,Yu:2014cka,Frixione:2014qaa,Demartin:2015uha,Demartin:2016axk,Denner:2011mq,Djouadi:1997yw,hdecay2,Bredenstein:2006rh,Bredenstein:2006ha,Boselli:2015aha,Actis:2008ts}, which are summarized in Ref.~\cite{deFlorian:2016spz}.\par}

{\tolerance=5000
	The main background processes originate from $\PZ\PZ$ production from quark-antiquark annihilation and gluon fusion. The former is simulated at NLO in pQCD with \POWHEG~2.0~\cite{Melia:2011tj}, while the latter is generated at LO with \MCFM~7.0.1~\cite{MCFM,Campbell:2011bn,Campbell:2013una,Campbell:2015vwa}.
	The reducible background contribution arising from the production of $\PZ$ bosons with associated jets ($\PZ+$jets) is estimated with the data-driven technique already used in Ref.~\cite{CMSHIG19001} and described in Section~\ref{sec:redbkgd}. \par}

An additional sample of Drell--Yan plus jets (DY+jets) events is produced with \linebreak \MGvATNLO~2.4.2 for validation studies and for the training of the boosted decision tree (BDT) used for the identification and isolation requirements on electrons, as described in Section~\ref{sec:reconstruction}.
All other simulated samples are used to model the signal shape, estimate backgrounds, optimize the analysis strategy, and evaluate the systematic uncertainties.

All Monte Carlo (MC) generators are interfaced with \PYTHIA to simulate the parton showering and hadronization effects. Version 8.230~\cite{Sjostrand:2014zea} is used for the three data-taking years with the CUETP8M1 tune~\cite{Khachatryan:2015pea} for 2016 and the CP5 tune~\cite{Sirunyan:2019dfx} for 2017 and 2018. Parton distribution functions (PDFs) are taken from the NNPDF3.0 set~\cite{Ball:2014uwa} for the three data taking periods.

The response of the CMS detector is modeled using the \GEANTfour~\cite{Agostinelli:2002hh,GEANT} package.
The simulated events are reconstructed with the same algorithms used for data and the distribution of the number of pileup events per bunch crossing is reweighted to match that observed in the data.

\section{Event reconstruction and selection}
\label{sec:reconstruction}
The particle-flow (PF) algorithm~\cite{CMS-PRF-14-001} aims to reconstruct and identify each individual particle in an event, with an optimized combination of information from the various elements of the CMS detector. The energy of photons is obtained from the ECAL measurement. The energy of electrons is determined from a combination of the electron momentum at the PV, as determined by the tracker, the energy of the corresponding ECAL cluster, and the energy sum of all bremsstrahlung photons spatially compatible with originating from the electron track. The momentum of muons is obtained from the combined information of the tracker and the muon chambers. The energy of charged hadrons is determined from a combination of their momentum measured in the tracker and the matching ECAL and HCAL energy deposits, corrected for the response of the calorimeters to hadronic showers. Finally, the energy of neutral hadrons is obtained from the corresponding corrected ECAL and HCAL energy deposits.

{\tolerance=1200 
	The information from the ECAL and the tracker is combined to reconstruct electrons~\cite{CMS:2020uim} with $\pt^{\Pe} > 7\GeV$ within the geometrical acceptance of the detector, defined by the pseudorapidity region $\abs{\eta^{\Pe}} < 2.5$.
	The identification of electrons is performed with a BDT algorithm sensitive to the presence of bremsstrahlung along the electron trajectory, the geometrical and momentum-energy matching with the corresponding cluster in the ECAL, the features of the electromagnetic shower in the ECAL, and observables that discriminate against electrons originating from photon conversions.
	The isolation sums for electrons, defined similarly as for muons, are included in the BDT discriminant.
	This choice is proven to enhance the suppression of nonprompt electrons originating from hadron decays and from overlap of neutral and charged hadrons within jets~\cite{CMS:2020uim} and has a better performance than a cutoff-based approach using the relative isolation.
	The BDT for the electron identification and isolation is implemented using the \textsc{xgboost} library~\cite{Chen:2016btl}.
	The training is performed on a dedicated sample of DY+jets
	simulated events.
	Electron samples are divided into six mutually exclusive categories defined by two $\pt$ ranges ($7<\pt^{\Pe}<10\GeV$ and $\pt^{\Pe}>10\GeV$) and three $\eta$ selections corresponding to the central barrel ($\abs{\eta^{\Pe}}<0.8$), outer barrel ($0.8<\abs{\eta^{\Pe}}<1.479$), and endcaps ($1.479<\abs{\eta^{\Pe}}<2.5$).
	The BDT is trained separately for the three data-taking periods and the selection requirements are defined to achieve the same signal efficiency for the three data taking periods (97\% for $p_T^{\Pe}>10\GeV$; 80\% for $p_T^{\Pe}<10\GeV$ in the barrel; 74\% for $p_T^{\Pe}<10\GeV$ in the endcap). 
	\par}

The information from the silicon tracker and the muon system~\cite{Sirunyan:2018} is combined to reconstruct muons with $\pt^{\Pgm} > 5\GeV$ and $\abs{\eta^{\Pgm}} < 2.4$.
The matching between inner and outer tracks is performed starting either from the tracks in the silicon trackers or from those reconstructed in the muon system.
Cases where inner tracks are matched to segments in only one or two muon detector layers are also considered,  to cope with very-low-\pt muons that do not traverse the entire detector.
Muon objects are selected from the muon track candidates by applying loose requirements on the track in the muon system and the inner tracker, taking into account  also their compatibility with small energy deposits in the ECAL and HCAL.

A requirement on the relative isolation, ${\mathcal I}^{\Pgm}<0.35$, is introduced to discriminate between muons from \PZ boson decays and those originating from hadron decays within jets, where ${\mathcal I}^{\Pgm}$ is defined as:

\begin{linenomath}
	\begin{equation}
		\label{eqn:pfiso}
		{\mathcal I}^{\Pgm} \equiv \Big( \sum \pt^\text{charged} + \max\big[ 0, \sum \pt^\text{neutral} +
		\sum \pt^{\Pgg} - \pt^{\Pgm,\mathrm{PU}} \big] \Big) / \pt^{\Pgm} ,
	\end{equation}
\end{linenomath}

{\tolerance=800 
and where $\sum \pt^\text{charged}$ is the scalar sum of the transverse momenta of charged hadrons originating from the PV, whereas $\sum \pt^\text{neutral}$ and $\sum \pt^{\Pgg}$ are the scalar sums for neutral hadrons and photons, respectively.
The isolation requirement is defined using a cone of radius $\Delta R=0.3$ around the muon direction at the PV, with the angular distance between two particles $i$ and $j$ defined as $\Delta R(i,j) = \sqrt{\smash[b]{(\Delta\eta_{i,j})^{2} + (\Delta\phi_{i,j})^{2}}}$.
The quantity $\pt^{\Pgm,\text{PU}}$ in Eq.~(\ref{eqn:pfiso}) is defined from the $\pt$ sum of all the charged hadrons $i$ not originating from the PV as $\pt^{\Pgm,\text{PU}}\equiv0.5\sum_{i}\ensuremath{p_{\mathrm{T},i}^{\Pgm,\mathrm{PU}}}\xspace$, where the factor of 0.5 corrects for using only the charged particles in the isolation cone~\cite{PUmitigationCMS}. The $\pt^{\Pgm,\text{PU}}$ contribution is subtracted in the definition of ${\mathcal I}^{\Pgm}$ to correct for energy deposits arising from pileup interactions.
\par}

Final-state radiation (FSR) photons arising from Z boson decays are recovered as follows. The PF photon candidates with $\abs{\eta^\Pgg}<2.4$ are considered as FSR objects if they have $\pt^{\Pgg} > 2\GeV$ and a relative isolation ${\mathcal I}^{\Pgg} <1.8$, where ${\mathcal I}^{\Pgg}$ is defined similarly as for muons in Eq.~(\ref{eqn:pfiso}).
These FSR candidates are associated with the closest lepton in the event and are not retained if $\Delta R(\Pgg,\ell)/(\pt^{\Pgg})^2>0.012\GeV^{-2}$ and $\Delta R(\Pgg,\ell) > 0.5$.
For each lepton, the FSR candidate with the lowest value of $\Delta R(\Pgg,\ell)/(\pt^{\Pgg})^2$, if any, is selected.
The photon candidates identified from the FSR recovery algorithm are excluded from the computation of the muon isolation.

Nonprompt leptons from decays of hadrons or photon conversions are suppressed based on the impact parameter significance. This variable is defined as the ratio of the 3-dimensional impact parameter, computed with respect to the position of the PV, to its uncertainty, and leptons are rejected if the value of this quantity is greater than 4.

The leptonic decays of known dilepton resonances are used to calibrate the momentum scale and resolution of electrons and muons in bins of $\pt^{\ell}$ and $\eta^{\ell}$, as described in Refs.~\cite{CMS:2020uim,Sirunyan:2018}.
Efficiencies for the lepton reconstruction and selection are measured in several bins of $\pt^{\ell}$ and $\eta^{\ell}$ by means of the tag-and-probe technique using samples of $\PZ$ boson events both in data and simulation. Simulated yields are corrected by the measured efficiency ratio between data and simulation.

For each event, hadronic jets are clustered from reconstructed particles using the infrared- and collinear-safe anti-\kt algorithm~\cite{Cacciari:2008gp} with a distance parameter of 0.4~\cite{Cacciari:2011ma}. The jet momentum is computed from the vectorial sum of all particle momenta in the jet, and is found in simulation to be, on average, within 5 to 10\% from the true momentum over the whole \pt spectrum and detector acceptance. Additional pp interactions within the same or nearby bunch crossings can contribute with additional tracks and calorimetric energy deposits, increasing the apparent jet momentum. To mitigate this effect, tracks identified as originating from pileup vertices are discarded and an offset correction is applied to correct for remaining contributions. Jet energy corrections are derived from simulation studies so that the average measured energy of jets becomes identical to that of particle-level jets. In situ measurements of the momentum balance in dijet, $\text{photon} + \text{jet}$, $\PZ + \text{jet}$, and multijet events are used to account for any residual differences in the jet energy scale between data and simulation~\cite{Khachatryan:2016kdb}. Additional selection criteria are applied to remove jets potentially dominated by instrumental effects or reconstruction failures. Only jets with $\pt^{\text{jet}}>30\GeV$,  $\abs{\eta^{\text{jet}}} < 4.7$, and a distance parameter of $\Delta R(\ell/\cPgg,\text{jet})>0.4$ from all selected leptons and FSR photons, are considered.
Jets not satisfying the tight identification criteria and the criteria corresponding to the tight working point of the pileup jet identification algorithm described in Ref.~\cite{PUmitigationCMS} are also discarded.

The PF objects mentioned above serve as input to the event selection, which targets events containing at least four well-identified and isolated leptons originating from the PV and possibly accompanied by a FSR photon. The FSR photons are included in the invariant mass computations.
The event selection, which closely follows that employed in Ref.~\cite{CMSHIG19001}, is detailed below.

The $\PZ$ boson candidates are formed from pairs of same-flavor and opposite-charge leptons ($\Pep\Pem$, $\PGmp\PGmm$) with an invariant mass within $12 < \mlplm  < 120\GeV$. 
Two such pairs are required to create $\PZ\PZ$ candidates, where the $\PZ$ boson candidate with invariant mass closest to the world-average $\PZ$ boson mass~\cite{Zyla:2020zbs} is referred to as $\PZ_1$, whereas $\PZ_2$ denotes the other $\PZ$ boson candidate.
Three mutually exclusive subchannels are defined from the flavors of the four leptons in the event: $4\Pe$, $4\Pgm$, and $2\Pe 2\Pgm$.

The $\PZ\PZ$ candidates must fulfill additional requirements designed to improve the sensitivity to \PH boson decays.
The $\PZ_1$ candidates are required to have an invariant mass larger than 40\GeV.
All lepton pairs $(\ell_i, \ell_j)$ must be separated by an angular distance of  $\Delta R(\ell_i, \ell_j) > 0.02$.
Events must contain two leptons with $\pt > 10\GeV$ and at least one with $\pt > 20\GeV$.	
In the $4\Pe$ and $4\Pgm$ channels, where the same four leptons can be used to build an alternative $\PZ_a \PZ_b$ candidate, candidates with $m_{\PZ_b}<12\GeV$ are not considered if $\PZ_a$ is closer to the world-average $\PZ$ boson mass than $\PZ_1$ and the event is rejected. This suppresses events with an on-shell $\PZ$ accompanied by a low-mass dilepton resonance (\eg, $\cPJgy$ or $\Upsilon$).
The invariant mass of the four possible opposite-charge lepton pairs (irrespective of flavor), computed without FSR photons, must satisfy $m_{\ell^{+}\ell'^{-}} > 4\GeV$ in order to further suppress events with leptons originating from hadron decays in jet fragmentation or from leptonic decays of low-mass resonances.
The $\PZ\PZ$ candidates are retained if the invariant mass of the four-lepton system $\mllll$ is larger than 70\GeV.

In events where more than one $\PZ\PZ$ candidate satisfies the selection requirements above, the one with the largest scalar sum of transverse momenta of the two leptons defining the  $\PZ_2$ is retained.

Finally, only events with $105<\mllll<160\GeV$ are considered for the statistical analysis. 

\section{Fiducial phase space definition}
\label{sec:fiducialvolume}
Cross sections are measured in a fiducial phase space defined to match closely the experimental acceptance of the reconstruction-level selections. 
The fiducial phase space is defined at generator-level, following the strategy adopted in previous $\HZZfl$ analyses~\cite{CMSH4l2016,CMSHIG19001}. It relies on requirements on the lepton kinematics and isolation, and on the event topology, in order to minimize the model dependence of the results.

The definition of the fiducial phase space is summarized in Table~\ref{tab:FidDef}. The events are retained if the leading (subleading) lepton has $\pt>20\, (10)\GeV$. Additional electrons (muons) that may be present in the event are required to have  $\pt>7\,(5)\GeV$ and $\abs{\eta}<2.5\, (2.4)$. Lepton isolation is ensured by requiring the scalar sum of the \pt of all stable particles, \ie, those particles not decaying in the detector volume, within a cone of radius $\Delta R=0.3$ to be less than 0.35 times the \pt of the lepton.
Neutrinos, FSR photons, and leptons (electrons and muons) are not included in the computation of the isolation sum to enhance the model independence of the measurements, following the findings of Ref.~\cite{CMSH4lFiducial8TeV}.
Events passing these requirements are retained if they have at least two same-flavor, opposite-sign lepton pairs.
The pair with invariant mass closest to the world-average $\PZ$ boson mass~\cite{Zyla:2020zbs} is labeled as $\PZ_{1}$ and it must have $40<m_{\PZ_{1}}<120\GeV$.
The second $\PZ$ boson candidate is referred to as $\PZ_{2}$ and it must have $12<m_{\PZ_{2}}<120\GeV$.
Each lepton pair $\ell_i, \ell_j$ must be separated by $\Delta R(\ell_i, \ell_j) > 0.02$, while any opposite-sign lepton pair must satisfy  $m_{\ell^{+}\ell'^{-}} > 4\GeV$, reflecting the selection criteria used at reconstruction level.

Leptons at the fiducial level are considered as $\textit{dressed}$, \ie, FSR photons are collected within a cone of radius 0.3. Jets do not enter in the definition of the fiducial phase space, but they are used when dealing with jet observables. Jets at the fiducial level are built with the anti-\kt clustering algorithm with a distance parameter of 0.4 out of stable particles, excluding neutrinos. Jets are retained if they satisfy $\pt^{\text{jet}}>30\GeV$ and $\abs{\eta^{\text{jet}}} < 4.7$, similarly to the condition used at reconstruction level. Only jets with no leptons inside a cone of radius 0.4 are kept.

\begin{table*}[!htb]
\centering
\topcaption{
	Summary of the requirements used in the definition of the fiducial phase space for the $\HZZfl$ cross section measurements.
	\label{tab:FidDef}
}
\cmsTable
{
	\begin{tabular}{lc}
		\multicolumn{2}{c}{Requirements for the $\HZZfl$ fiducial phase space} \\
		\hline
		\multicolumn{2}{c}{Lepton kinematics and isolation} \\
		\hline
		\vspace{-0.4cm} & \\
		Leading lepton $\pt$ & $\pt > 20\GeV$ \\
		\vspace{-0.4cm} & \\
		Sub-leading lepton $\pt$ & $\pt > 10\GeV$ \\
		\vspace{-0.4cm} & \\
		Additional electrons (muons) $\pt$ & $\pt > 7 (5)\GeV$ \\
		\vspace{-0.4cm} & \\
		Pseudorapidity of electrons (muons) & $\abs{\eta} <$ 2.5 (2.4) \\
		\vspace{-0.4cm} & \\
		Sum of scalar $\pt$ of all stable particles within $\Delta R < 0.3$ from lepton & $<0.35 \pt$ \\
		\hline
		\multicolumn{2}{c}{Event topology} \\
		\hline
		\multicolumn{2}{l}{Existence of at least two same-flavor OS lepton pairs, where leptons satisfy criteria above} \\
		Inv. mass of the $\PZ_1$ candidate & $40 < m_{\PZ_{1}} < 120\GeV$ \\
		\vspace{-0.4cm} & \\
		Inv. mass of the $\PZ_2$ candidate & $12 < m_{\PZ_{2}} < 120\GeV$ \\
		\vspace{-0.4cm} & \\
		Distance between selected four leptons & $\Delta R(\ell_{i},\ell_{j})>0.02$ for any $i\neq j$  \\
		\vspace{-0.4cm} & \\
		Inv. mass of any opposite sign lepton pair & $m_{\ell^{+}\ell'^{-}}>4\GeV$ \\
		\vspace{-0.4cm} & \\
		Inv. mass of the selected four leptons & $105 < \mllll < 160\GeV$  \\
\end{tabular}}
\normalsize
\end{table*}

\section{Observables}
\label{sec:observables}
\begin{figure}[tbp]
\centering
{\includegraphics[width=0.57\columnwidth]{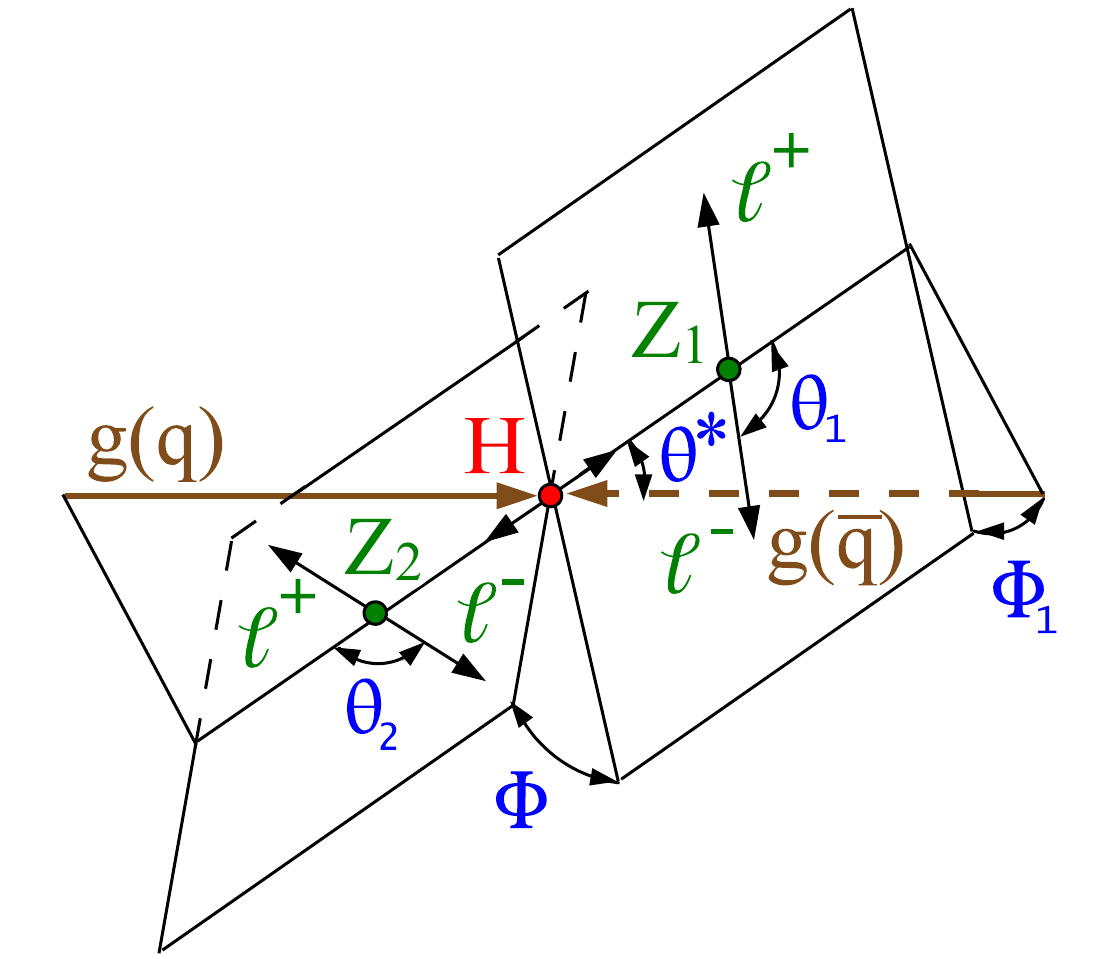}}
\caption{Schematic representation of the $\Pg\Pg/\PQq\PAQq\to \PH\to ZZ\to 4\ell$ process. The five angles depicted in blue are considered in the differential analysis, as detailed in the text. }
\label{fig:decayHZZ4l}
\end{figure}

Fiducial cross sections are measured in bins of several kinematic observables sensitive to the H boson production and decay $\Pp\Pp\to\HZZfl$, of which a schematic representation is given in Fig.~\ref{fig:decayHZZ4l}. 

The decay of the \PH boson to four leptons is fully described by the invariant mass of the two $\PZ$ boson candidates, three angles describing the $\PZ$  boson decays ($\Phi$, $\theta_1$, $\theta_2$), and two angles connecting production to decay ($\Phi_1$, $\theta^*$). 
The angle $\theta^*$ is defined in the \PH rest frame as the angle between the beam axis and the direction of the $\PZ_1$ candidate.
$\Phi$ and $\Phi_1$ are the azimuthal angles between the three planes constructed from the \PH decay products and the decay products of the two $\PZ$ bosons in the \PH rest frame. 
The $\theta_1$ and $\theta_2$ angles are defined in the $\PZ_1$ and $\PZ_2$ rest frames, respectively, as the angles between the $\PZ$ boson direction in the \PH boson rest frame and the direction of the negative decay lepton. 
The set comprising these seven observables is hereafter referred to as $\vec{\Omega}^{\PH\to\Z\Z\to 4\ell}(\theta^*,\theta_1,\theta_2,\Phi,\Phi_1,m_{Z1},m_{Z2}|m_{4\ell})$ and can be used to build matrix element discriminants sensitive to the $4\ell$ decay, as detailed in Section~\ref{sec:discriminants}.
Fiducial cross sections are measured in differential bins of these observables, except for the $\theta$ angles for which their cosine is used.
The distributions as a function of $\pt^\PH$ and pseudorapidity of the reconstructed \PH boson are also measured.

Fiducial cross sections are also measured in differential bins of the number of associated jets ($N_\text{jets}$) and $\pt$ of the leading ($\pt^{j_1}$) and subleading ($\pt^{j_2}$) jet in the event.
For events with two or more jets the properties of the dijet system constituted by the two leading jets are assessed by measuring differential cross sections in bins of its invariant mass ($m_{jj}$), of the difference in pseudorapidity ($\Delta\eta_{jj}$), and of the difference in azimuthal angle ($\Delta\phi_{jj}$) between the two jets. The angle $\Delta\phi_{jj}$ is defined to be invariant under the exchange of the two jets as follow:
\begin{equation}
	\Delta\phi_{jj} = \frac{(\hat{j}_{T1} \times \hat{j}_{T2}) \cdot \hat{z}}{\abs{(\hat{j}_{T1} \times \hat{j}_{T2}) \cdot \hat{z}}} \cdot \frac{(\vec{j}_{1} - \vec{j}_{2}) \cdot \hat{z}}{\abs{(\vec{j}_{1} - \vec{j}_{2}) \cdot \hat{z}}} \cdot \cos^{-1}(\hat{j}_{T1}\cdot\hat{j}_{T2})
\end{equation}
where the vectors $\vec{j}_{1,2}$ represent the direction of the leading and subleading jet in the laboratory frame, and the unit vectors $\hat{j}_{T1,2}$ the corresponding transverse component. This definition is also independent of the choice of the positive $z$ axis direction, $\hat{z}$.

The rapidity-weighted jet vetoes $\mathcal{T}_{\text{C}}^{\text{max}}$ and $\mathcal{T}_{\text{B}}^{\text{max}}$ are also studied. These are defined, following Ref.~\cite{Gangal:2014qda}, as:
\begin{linenomath*}
\begin{equation}
	\label{eqn:tauC}
	\mathcal{T}_{\text{C}}^{\text{max}} = \max_{j}\left(\frac{\sqrt{E^2_j-p_{z,j}^2}}{2\cosh\left(y_j-y_\PH\right)}\right),
\end{equation}
\end{linenomath*}
\begin{linenomath*}
\begin{equation}
	\label{eqn:tauB}
	\mathcal{T}_{\text{B}}^{\text{max}} = \max_{j}\left(m^j_{\text{T}}e^{-\mid y_j-y_\PH\mid}\right),
\end{equation}
\end{linenomath*}
where $y_j$ and $m^j_{\text{T}}$ are the rapidity and transverse mass of the jet, defined from its mass $m$ and momentum $p$ as $m^j_{\text{T}}=\sqrt{m^2+p^2_\text{x}+p^2_\text{y}}$, while $y_\PH$ is the rapidity of the \PH boson.
The value of each observable is computed for each jet in the event and its maximum value is taken for each event.
Since their resummation structure is different from the canonical $\pt^{j}$, they give complementary information on the properties of jets in an event and can be used as a test of quantum chromodynamics (QCD) resummation. The 0-jet phase space can be redefined using these observables. The events with no jets are defined as the ones with $\mathcal{T}_{\text{C}}^{\text{max}}<15\GeV$ and $\mathcal{T}_{\text{B}}^{\text{max}}<30\GeV$, where the values of these cuts are chosen accordingly to the findings of Ref.~\cite{Gangal:2014qda}. In the following, these events will be defined as 0-jet$|{\mathcal{T}_{\text{C}}^{\text{max}}}$ and 0-jet$|{\mathcal{T}_{\text{B}}^{\text{max}}}$, respectively.

The properties of the \PH{}+jet(s) system are also studied by measuring differential cross sections in bins of  the transverse momentum and invariant mass of the \PH plus leading jet system for events with at least one jet, or of the \PH plus leading and subleading jet system for events with at least two jets.
The observables characteristic of the $\PH+j(j)$ system can be defined only in events with at least one (two) jets. In all other cases, an underflow bin is introduced to consider all events for which the observable is undefined. 

\subsection{Matrix element discriminants}
\label{sec:discriminants}
The \JHUGEN and \MCFM generators are used to compute the matrix element probability $\mathcal{P}_i$ for  an event to arise from a physical process $i$, given the value of the reconstructed invariant mass of the four-lepton system $\mllll$.
These probabilities are defined as a function of $\vec\Omega^{\HZZfl}$ and retain the maximal information on the underlying physics content of each event.
Hence, the $\mathcal{P}_i(\vec\Omega^{\HZZfl})$ probabilities are used to construct likelihood-ratio-like matrix element discriminants sensitive to the difference between two physical processes $a$ and $b$, when considering two production mechanisms, or to be used to test a BSM hypothesis against the SM scenario.
These matrix element discriminants have been widely used in the context of  $\HZZfl$ analyses, from the measurement of the \PH boson properties~\cite{CMSHIG19001} to the constraints on possible anomalous couplings~\cite{CMSHIG19009}.
The general structure of these discriminants is an adaptation of the more classic likelihood ratio, properly rescaled to ensure that the discriminants are always bounded between 0 and 1.
Two types of kinematic discriminants can be built to test the compatibility between signal (``sig'')  and alternative (``alt'') hypotheses and their interference (``int''):
\begin{equation}
		\mathcal{D}_\mathrm{alt}\left(\vec{\Omega}\right) =  \frac{\mathcal{P}_\text{sig}\left(\vec{\Omega}\right) }
		{\mathcal{P}_\text{sig}\left(\vec{\Omega}\right) +\mathcal{P}_\mathrm{alt}\left(\vec{\Omega}\right) } ,
		\qquad
		\mathcal{D}_\mathrm{int}\left(\vec{\Omega}\right) = 
		\frac{\mathcal{P}_\mathrm{int}\left(\vec{\Omega}\right) }
		{2 \ \sqrt{{\mathcal{P}_\text{sig}\left(\vec{\Omega}\right) \ \mathcal{P}_\mathrm{alt}\left(\vec{\Omega}\right) }}} ,
		\label{eqn:mela}
\end{equation}

where $\mathcal{P}_{\text{sig}}$ and $\mathcal{P}_{\text{alt}}$ are the probabilities of an event under the two considered hypotheses, given their kinematic properties $\vec\Omega$, and $\mathcal{P}_{\text{int}}$ is the probability for the interference between the two model contributions (``sig'' and ``alt''). This definition of $\mathcal{D}_\mathrm{int}$ is bounded between $-1$ and $1$ for any value of $\mathcal{D}_\mathrm{alt}$.

A total of six matrix element discriminants sensitive to different values of possible anomalous couplings of the \PH boson to vector bosons are considered.
The general scattering amplitude describing the interaction between a spin-zero \PH boson and two spin-one gauge bosons $\PV_1$ and $\PV_2$ can be written, following the conventions of Ref.~\cite{CMSHIG19009}, as:
\begin{equation}
	\begin{aligned}
		A(\PH\PV_1\PV_2) =
		\frac{1}{v}
		\left[ a_{1}^{\PV\PV}
		+ \frac{\kappa_1^{\PV\PV}q_{\PV1}^2 + \kappa_2^{\PV\PV} q_{\PV2}^{2}}{\left(\Lambda_{1}^{\PV\PV} \right)^{2}} 
		+ \frac{\kappa_3^{\PV\PV}(q_{\PV1} + q_{\PV2})^{2}}{\left(\Lambda_{Q}^{\PV\PV} \right)^{2}} \right]
		m_{\PV1}^2 \epsilon_{\PV1}^* \epsilon_{\PV2}^*  
		\\
		+ \frac{1}{v}a_{2}^{\PV\PV}  f_{\PGm\Pgn}^{*(1)}f^{*(2),\PGm\Pgn}
		+ \frac{1}{v}a_{3}^{\PV\PV}   f^{*(1)}_{\PGm\Pgn} {\tilde f}^{*(2),\PGm\Pgn} ,
		\label{eq:formfact-fullampl-spin0}
	\end{aligned}
\end{equation}

where $v$ is the vacuum expectation value of the \PH potential, $f^{(i)\PGm\Pgn}=\epsilon_{\PV i}^{\PGm}q_{\PV i}^{\PGn}-\epsilon_{\PV i}^{\PGn}q_{\PV i}^{\PGm}$, $\tilde{f}_{\PGm\Pgn}^{(i)}=\frac{1}{2}\epsilon_{\PGm\Pgn\rho\sigma}f^{(i),\rho\sigma}$, and $\epsilon_{\PV i}$, $q_{\PV i}$, and $m_{\PV i}$ are the polarization vector, four-momentum, and pole mass of a gauge boson, respectively. The constants $\Lambda_1$ and $\Lambda_{Q}$ are the scales of BSM physics. In the above equation, the only leading tree-level contributions are $a_1^{\PZ\PZ}\neq 0$ and $a_1^{\PW\PW}\neq 0$. The rest of the ZZ and WW couplings are considered as anomalous contributions, which are either small contributions arising in the SM due to loop corrections or new BSM contributions. The SM value of those are not yet distinguishable from zero experimentally with the available data.

The $a_i$  and $\kappa_i$ terms correspond to the strengths of vector boson couplings, following the notation adopted in Ref.~\cite{CMSHIG19009}.
In particular, the  $a_3$ CP-odd term is expected to be null in the SM and is sensitive to possible BSM effects that would result in CP violation.
The $a_2$ term corresponds to the CP-even contribution to the $\PH\PV\PV$ coupling and is sensitive to possible BSM contributions from heavy \PH bosons.
The $\kappa_{1,2}/(\Lambda_{1})^2$ and $\kappa_{3}/(\Lambda_{Q})^2$ terms are sensitive to possible physics at a new energy scale represented by the denominator. The $\kappa_{3}/(\Lambda_{Q})^2$ coupling allows for scenarios that violate the gauge symmetries of the SM but is not considered in this analysis. Symmetries and gauge invariance force $\kappa_{1}^{ZZ}=\kappa_{2}^{ZZ}$, leading to the single coupling $\kappa_{1}^{ZZ}/(\Lambda_{1}^{ZZ})^2$ to investigate and denoted $\Lambda_{1}$ in what follows. Gauge invariance imposes $\kappa_{1}^{\PZ\gamma}=0$, making it impossible to measure the $\Lambda_{1}^{\PZ\gamma}$ coupling in any process involving an on-shell photon. However, the $\PH\to 4\ell$ channel contains events featuring an off-shell photon, \ie, $H\to\PZ\PGg*\to 4\ell$, that can be used to study the $\Lambda_{1}^{Z\gamma}$ coupling.
Table~\ref{tab:MEdiscriminants} details the set of kinematic discriminants considered and the couplings to which they are sensitive. 
The index ``dec'' indicates that only decay information is used to build these discriminants.

\begin{table}[tbp]
	\centering 
	\topcaption{Matrix element kinematic discriminants considered in the analysis. Some discriminants have a special label to identify the targeted Higgs boson property rather than the name of the coupling. \Dzm is sensitive to a CP-odd Higgs boson, \DCP is the observable sensitive to the CP-mixing, and \Dzhp is sensitive to heavy CP-even Higgs boson.}
	\renewcommand{\arraystretch}{1.5}
	\begin{tabular}{ ccccc|cc  }
		& \multicolumn{4}{c}{\Dalt}& \multicolumn{2}{c}{$\mathcal{D}_{\text{int}}$} \\
		\cline{2-7}
		& \multicolumn{6}{c}{Coupling} \\
		& a$_3$ & a$_2$ & $\Lambda_1$ & $\Lambda_1^{\PZ\gamma}$ & a$_3$ & a$_2$\\
		Discriminant & \Dzm & \Dzhp & \DLone & \DLoneZg & \DCP & \Dint \\
	\end{tabular}
	\label{tab:MEdiscriminants}
\end{table}

Differential cross sections are measured in bins of these six matrix element discriminants under the SM hypothesis.
The compatibility of the measurements with the SM predictions is assessed by  comparing the results with the discriminants built for alternative BSM scenarios, where HVV anomalous couplings are introduced by modifying the $a_i$ and $\kappa_i$ values in Eq.~(\ref{eq:formfact-fullampl-spin0}) with respect to their SM values.

Tables~\ref{tab:binBoundaries_production} and~\ref{tab:binBoundaries_decay} summarize the bin boundaries for all the observables considered in this analysis that target the \PH boson production and the $\HZZfl$ decay, respectively.

\begin{table}[tbp!]
	\centering 
	\topcaption{Bin boundaries for one-dimensional observables targeting the \PH boson production. The bin boundaries denoted with $\infty$ correspond to no upper limit applied on the observable value.}
	\cmsTable{
		\renewcommand{\arraystretch}{1.5}
		\begin{tabular}{llc}
			Observable & Definition & Bin boundaries \\
			\hline
			$\pt^\PH$ & Transverse momentum of the $4\ell$ system  & [0,10,20,30,45,60,80,120,200,$\infty$[ $\GeVns$ \\
			$\abs{y_\text{\PH}}$ & Rapidity of the $4\ell$ system  & [0,0.15,0.3,0.45,0.6,0.75,0.9,1.2,1.6,2.5]  \\
			$N_\text{jets}$&  Number of associated jets in the event& $=$0,$=$1,$=$2,$=$3,$\geq$4 \\
			$\pt^{\text{j}_1}$ & Transverse momentum of the leading jet& [0-jet,30,55,95,200,$\infty$[ $\GeVns$  \\
			$\pt^{\text{j}_2}$ & Transverse momentum of the subleading jet& [0/1-jet,30,40,65,90,$\infty$[ $\GeVns$  \\
			$\mathcal{T}_{\text{C}}^{\text{max}}$ &  Rapidity-weighted jet veto & [0-jet$|{\mathcal{T}_{\text{C}}^{\text{max}}}$,15,20,30,50,80,$\infty$[ $\GeVns$  \\
			$\mathcal{T}_{\text{B}}^{\text{max}}$ &  Rapidity-weighted jet veto & [0-jet$|{\mathcal{T}_{\text{B}}^{\text{max}}}$,30,45,75,150,$\infty$[ $\GeVns$  \\
			$m_\text{jj}$ & Invariant mass of the leading and subleading jets system & [0/1-jet,0,120,300,$\infty$[ $\GeVns$ \\
			$\abs{\Delta\eta_\text{jj}}$ & Difference in pseudorapidities of the leading and subleading jets & [0/1-jet,0.0,1.6,3.0,10.0] \\
			$\Delta\phi_\text{jj}$ & Azimuthal angle difference between the leading and subleading jets & [0/1-jet,$-\pi$, $-\pi/2$,  0, $\pi/2$, $\pi$] \\
			$\pt^{\PH \text{j}}$ & Transverse momentum of the $4\ell$ and leading jet system & [0-jet,0,30,50,110,$\infty$[ $\GeVns$ \\
			$m_{\PH \text{j}}$ & Invariant mass of the $4\ell$ and leading jet system & [0-jet,110,180,220,300,400,600,$\infty$[ $\GeVns$ \\
			$\pt^{\PH \text{jj}}$ & Transverse momentum of the $4\ell$, leading and subleading jets system & [0/1-jet,0,20,60,$\infty$[ $\GeVns$ \\
		\end{tabular}
	}
	\label{tab:binBoundaries_production}
\end{table}

\begin{table}[htbp!]
	\centering 
	\topcaption{Bin boundaries for one-dimensional observables targeting the $\HZZfl$ decay.}
	\cmsTable{
		\renewcommand{\arraystretch}{1.5}
		\begin{tabular}{llc}
			Observable & Definition & Bin boundaries \\
			\hline
			$\cos \theta^*$ &  Cosine of the decay angle of the leading lepton pair in the $4\ell$ rest frame  & [-1.0,-0.75,-0.50,-0.25,0.0,0.25,0.50,0.75,1.0]  \\
			$\cos \theta_1$, $\cos \theta_2$  & \bgroup\renewcommand{\arraystretch}{1} \begin{tabular}[t]{@{}l@{}} Cosine of the production angle,\\ relative to the $\PZ$ vector, of the antileptons from the two $\PZ$ bosons\end{tabular}\egroup & [-1.0,-0.75,-0.50,-0.25,0.0,0.25,0.50,0.75,1.0]  \\
			$\Phi$, $\Phi_1$ & Azimuthal angles between the decay planes & [$-\pi$, $-3\pi/4$, $-\pi/2$, $-\pi/4$,  0, $\pi/4$, $\pi/2$, $3\pi/4$, $\pi$]  \\
			$m_{\PZ_{1}}$ & Invariant mass of the two leading leptons & [40,65,75,85,92,120] $\GeVns$   \\
			$m_{\PZ_{2}}$ & Invariant mass of the two subleading leptons & [12,20,25,28,32,40,50,65] $\GeVns$  \\
			\Dzm & Matrix element discriminant targeting $a_3$ coupling & [0.0,0.4,0.5,0.6,0.7,0.8,0.9,1.0] \\
			\Dzhp & Matrix element discriminant targeting $a_2$ coupling & [0.0,0.35,0.4,0.45,0.55,0.65,0.75,1.0] \\
			\DLone & Matrix element discriminant targeting $k_1$ coupling  & [0.0,0.45,0.5,0.6,0.7,1.0] \\
			\DLoneZg & Matrix element discriminant targeting $k_2^{Z\gamma}$ coupling & [0.0,0.35,0.45,0.5,0.55,0.65,1.0] \\
			\DCP & Interference matrix element discriminant targeting $a_3$ coupling  & [-0.75,-0.25,-0.1,0.0,0.1,0.25,0.75] \\
			\Dint & Interference matrix element discriminant targeting $a_2$ coupling  & [0.0,0.7,0.8,0.9,0.95,1.0] \\
		\end{tabular}	\label{tab:binBoundaries_decay}
	}
\end{table}

\begin{table}[tbp!]
	\centering 
	\topcaption{Double-differential observables with their corresponding bin boundaries. The bin boundaries denoted with $\infty$ correspond to no upper limit applied on the observable value.}
	\cmsTable{
		\renewcommand{\arraystretch}{1.5}
		\begin{tabular}{lcccccccccccc}
			Observable & Bin 1 & Bin 2 & Bin 3 & Bin 4 & Bin 5 & Bin 6 & Bin 7  & Bin 8  & Bin 9  & Bin 10 & Bin 11  & Bin 12   \\
			\hline
			$m_{\PZ_{1}}(\text{GeV})$ & [40,85] & [40,70] & [70,120] & [85,120] & [85,120] & [85,120] & & & & & & \\
			$m_{\PZ_{2}}(\text{GeV})$ & [12,35] & [35,65] & [35,65] & [30,35] & [24,30] & [12,24] & & & & & & \\
			\hline
			$\abs{y_{\PH}}$ & [0,0.5] & [0,0.5] & [0,0.5] & \multicolumn{1}{c|}{[0,0.5]} & [0.5,1.0] & [0.5,1.0] & \multicolumn{1}{c|}{[0.5,1.0]} & [1.0,2.5] & [1.0,2.5] & [1.0,2.5] & & \\
			$\pt^{\PH}(\text{GeV})$ & [0,40] & [40,80] & [80,150] & \multicolumn{1}{c|}{[150,$\infty$[} & [0,45] & [45,120] & \multicolumn{1}{c|}{[120,$\infty$[}& [0,45] & [45,120] & [120,$\infty$[ & & \\
			\hline
			$N_\text{jets}$ & 0 & 0 & \multicolumn{1}{c|}{0} & 1 & 1 & 1 & \multicolumn{1}{c|}{1} & $>=2$ & $>=2$ & $>=2$ & $>=2$ & \\
			$\pt^{\PH}(\text{GeV})$& [0,15] & [15,30] & \multicolumn{1}{c|}{[30,$\infty$[} & [0,60] & [60,80] & [80,120] & \multicolumn{1}{c|}{[120,$\infty$[} & [0,100] & [100,170] & [170,250]& [250,$\infty$[ & \\
			\hline
			$\pt^{\text{j}_1}(\text{GeV})$ & \multirow{2}{*}{$N_{jets}<2$} & [30,60] & [60,350] & [60,350] &  & &  & & &  & & \\
			$\pt^{\text{j}_2}(\text{GeV})$& & [30,60] & [30,60] & [60,350]& & & &  & & & &  \\
			\hline
			$\pt^{\PH \text{j}}(\text{GeV})$  & \multirow{2}{*}{$N_{jets}<1$} & [0,30] & [0,45] & [30,350] & [45,350] & &  & & &  & & \\
			$\pt^{\PH}(\text{GeV})$& & [0,85] & [85,350] & [0,85] & [85,350] & & & & & & &  \\
			\hline
			$\mathcal{T}_{\text{C}}^{\text{max}}(\text{GeV})$  & 0-jet$|{\mathcal{T}_{\text{C}}^{\text{max}}}$ & 0-jet$|{\mathcal{T}_{\text{C}}^{\text{max}}}$ & 0-jet$|{\mathcal{T}_{\text{C}}^{\text{max}}}$ & 
			0-jet$|{\mathcal{T}_{\text{C}}^{\text{max}}}$ & 0-jet$|{\mathcal{T}_{\text{C}}^{\text{max}}}$ & \multicolumn{1}{c|}{0-jet$|{\mathcal{T}_{\text{C}}^{\text{max}}}$} & [15,25] & \multicolumn{1}{c|}{[15,25]} & [25,40] & \multicolumn{1}{c|}{[25,40]} & [40,$\infty$[ & [40,$\infty$[\\
			$\pt^{\PH}(\text{GeV})$ & [0,15] & [15,30] & [30,45] & [45,70] & [70,120] & \multicolumn{1}{c|}{[120,$\infty$[} & [0,120] & \multicolumn{1}{c|}{[120,$\infty$[} & [0,120] & \multicolumn{1}{c|}{[120,$\infty$[} & [0,200] & [200,$\infty$[ \\
		\end{tabular}	\label{tab:binBoundaries2D}
	}
\end{table}

To ensure a complete characterization of the $\HZZfl$ decay channel and to maximize the coverage of  different phase space regions, a set of double-differential measurements is also performed.
Cross sections are measured in bins of $m_{\PZ_{1}}$ vs. $m_{\PZ_{2}}$, $\abs{y_{\PH}}$ vs. $\pt^{\PH}$, number of associated jets vs. $\pt^{\PH}$, \pt of the leading  vs. subleading jet, $\pt^{\PH j}$ vs. $\pt^{\PH}$, and $\mathcal{T}_{\text{C}}^{\text{max}}$ vs. $\pt^{\PH}$.
The corresponding bin boundaries are listed in Table~\ref{tab:binBoundaries2D}.

\section{Background estimation}
\label{sec:bkgd}

\subsection{Irreducible backgrounds}
\label{sec:irrbkgd}
Irreducible $\PZ\PZ$ background contributions arising from $\Pq\Paq$ annihilation or gluon fusion are estimated from simulation. 
The former is simulated at NLO in pQCD with \POWHEG~2.0 and reweighted to NNLO using a $K$ factor computed as a function of $m_{\PZ\PZ}$ exploiting the NNLO computation of the $\qqZZ$ fully differential cross section~\cite{Grazzini2015407}.
The $K$ factor ranges 1.0--1.2 and is 1.1 at $m_{\PZ\PZ}=125\GeV$.
The NLO EW corrections are applied as a function of $m_{\PZ\PZ}$ according to the computation presented in Ref.~\cite{Bierweiler:2013dja}.

The soft collinear approximation has been shown to describe accurately the cross section and the interference term for the gluon fusion $\PZ\PZ$ production at NNLO in pQCD~\cite{Bonvini:1304.3053}.
Additional calculations demonstrate that the $K$ factors are very similar at NLO for signal and background~\cite{Melnikov:2015laa} and at NNLO for the signal and interference terms~\cite{Li:2015jva}.
Hence, the same $K$ factor is used for the signal and the background~\cite{Passarino:1312.2397v1}.
The \textsc{hnnlo}~v2 program~\cite{Catani:2007vq,Grazzini:2008tf,Grazzini:2013mca} is used to obtain the signal NNLO $K$ factor as a function of $m_{\PZ\PZ}$ from the ratio of the NNLO and LO $\Pg\Pg\to\PH\to2\ell2\ell^\prime$ cross sections for the predicted SM \PH boson decay width of 4.07\MeV~\cite{Zyla:2020zbs}. 
The NNLO/LO $K$ factor for {\ggZZ} varies from $\approx$2.0 to 2.6 and is 2.27 at $m_{\PZ\PZ}=125\GeV$, and a 10\% systematic uncertainty is used when it is applied to the background.

The irreducible background contributions are included as binned templates in the likelihood function separately for the three considered final states ($4\Pe$, $4\Pgm$, and $2\Pe 2\Pgm$).
The templates are normalized to the most accurate theoretical calculations for the $\qqbar\to\ZZ\to 4\ell$ and $\Pg\Pg\to\ZZ \to 4\ell$ cross sections~\cite{Grazzini2015407,Bierweiler:2013dja,Bonvini:1304.3053,Melnikov:2015laa,Li:2015jva,Passarino:1312.2397v1}.
A second method for the measurement of the inclusive fiducial cross section is presented in Section~\ref{sec:results},  where the normalization of these processes is treated as an unconstrained parameter in the fit to assess the constraint that can be derived from sidebands in data.

\subsection{Reducible background}
\label{sec:redbkgd}
The reducible background contribution to the \PH boson signal in the $4\ell$ channel mainly comes from the $\PZ\text{+jets}$,  $\ttbar\text{+jets}$,
$\PZ\gamma\text{+jets}$, $\PW\PW\text{+jets}$, and $\PW\PZ\text{+jets}$ production, hereafter collectively referred to as ``$\PZ\text{+X}$'' since the $\PZ\text{+jets}$ contribution is the dominant one.

The contribution from the reducible background is estimated with the technique explained in Ref.~\cite{CMSHIG19001}. The method is based on lepton misidentification rates, which are defined as the fraction of non-signal leptons that satisfy the selection criteria, computed in a control region in data that includes a Z boson and exactly one additional ``loose'' lepton ($\PZ+\ell$), \ie, leptons with \pt, $\eta$, and PV cuts  but without identification nor isolation cuts applied. The lepton misidentification rates are then applied to another control region, comprised of a $\PZ$ boson candidate and two opposite-sign or same-sign ``loose'' leptons  ($\PZ+\ell\ell$), to reweigh the number of events to the signal region.

The distributions as functions of $m_{4\ell}$ of the $\PZ\text{+jets}$ reducible background are derived for the three final states ($4\Pgm$, $4\Pe$, and $2\Pe 2\Pgm$) separately and are included as binned templates in the likelihood function.

\section{Measurement methodology}
\label{sec:measurement}

The $\Pp\Pp\to\HZZfl$ fiducial cross sections are extracted from a maximum likelihood fit of the signal and background expected distributions to the observed $4\ell$ mass distribution, $N_{\text{obs}}(\mllll)$, parametrized for each final state $f$, in each kinematic bin $i$ of a given observable, and year of data taking $y$ as:

\begin{linenomath}
	\begin{equation}
		\label{eqn:m4l}
		\begin{aligned}
			N_{\text{obs}}^{f,i,y}(m_{4\ell}) & = N_{\text{fid}}^{f,i,y}(m_{4\ell}) + N_{\text{nonfid}}^{f,i,y}(m_{4\ell}) + N_{\text{\text{nonres}}}^{f,i,y}(m_{4\ell}) + N_{\text{bkg}}^{f,i,y}(m_{4\ell}) \\ 
			& = \sum_j^{\text{genBin}}\epsilon_{i,j}^{f,y}\,(1+f_{\text{nonfid}}^{f,i,y})\,\sigma_{\text{fid}}^{f,j,y}\,\mathcal{L}\,\mathcal{P}_{\text{res}}^{f,y}(m_{4\ell}) \\
			& + N_{\text{nonres}}^{f,i,y}\,\mathcal{P}_{\text{nonres}}^{f,y}(m_{4\ell}) + N_{\text{bkg}}^{f,i,y}\,\mathcal{P}_{\text{bkg}}^{f,i,y}(m_{4\ell}).
		\end{aligned}
	\end{equation}
\end{linenomath}

The $N_{\text{fid}}^{f,i,y}(m_{4\ell})$ and $N_{\text{nonfid}}^{f,i,y}(m_{4\ell})$ terms represent the distributions of resonant events originating from within and outside the fiducial volume, respectively.
The  $N_{\text{\text{nonres}}}^{f,i,y}(m_{4\ell})$ and $N_{\text{bkg}}^{f,i,y}(m_{4\ell})$ terms represent the distributions of nonresonant and background events.

The \PH resonant signal distribution is parametrized with a double-sided Crystal Ball (DCB) function~\cite{Skwarnicki:1986xj} around $\mH=125\GeV$.
The corresponding probability density function, $\mathcal{P}_{\text{res}}(m_{4\ell})$, is scaled by the fiducial cross section, $\sigma_{\text{fid}}$, and the integrated luminosity $\mathcal{L}$.
The DCB function parameters are obtained from a simultaneous fit of the $\mllll$ distributions corresponding to the various mass points in the $\mH$ range of 105--160\GeV, which allows to express the dependency of the fitted parameters in $\mH$ directly in the fit, following the same strategy of Ref.~\cite{CMSHIG19001}.

A Landau distribution is introduced to empirically model the shape of the nonresonant signal contribution, $\mathcal{P}_{\text{nonres}}(m_{4\ell})$, for the $\WH$, $\ZH$, and $\ttH$ processes where one of the leptons from the \PH boson decay is either not selected or falls outside the acceptance.
The fraction of such events in the mass range considered is about 5, 22, and 17\% for $\WH$, $\ZH$, and $\ttH$, respectively.
These nonresonant events are treated as a background in the measurements.
The reducible and irreducible backgrounds are included in the fit as normalized binned templates, $\mathcal{P}_{\text{bkg}}$, of the mass distribution of these processes.

{\tolerance=800 
An additional contribution ($f_{\text{nonfid}}$) is introduced to take into account the presence of events not originating from the fiducial volume but satisfying the selections and is treated as background in the measurements.
This contribution is referred to as the ``nonfiducial signal'' and is estimated from simulation for each signal model.
The values of  $f_{\text{nonfid}}$ are found to be consistent across the different observables considered, for the same production mechanism.
Dedicated simulations have shown that the $m_{4\ell}$ distribution of these events is identical to that of the resonant fiducial signal. To minimize the model dependence of the measurement, the value of $f_{\text{nonfid}}$ is fixed to be a fraction of the fiducial signal component.
The values of this fraction are reported in Table~\ref{tab:summarySM} and range between $4\%$ for the $\VBF$ production mechanism and up to $18\%$ for the $\ttH$ mode.
The acceptance of the events originating from $\VH$ or $\ttH$ is lower than $\ggH$ and $\VBF$ events, reflecting the possible presence of leptons in the final states not produced by the H boson decay and resulting in larger values of $f_{\text{nonfid}}$ for these production mechanisms.
\par}

Generator-level observables used in the definition of the fiducial phase space are smeared by detector effects at reconstruction level.
The  $\epsilon_{i,j}^{f}$ response matrix is obtained from simulation, for each final state $f$, and is used to unfold the number of expected events in bin \textit{i} at the reconstruction level to the number of expected events of a given observable in bin \textit{j} at the fiducial level.
For the measurement of the inclusive fiducial cross section, $\epsilon_{i,j}^{f}$ corresponds to a single number, the efficiency, listed in the second column of Table~\ref{tab:summarySM} for the various production mechanisms.
The table also shows the acceptance $\mathcal{A}_{\text{fid}}$, defined as the fraction of signal events that fall within the fiducial phase space.

\begin{table*}[!h!tb]
\centering
\topcaption{
	Summary of the inputs to the maximum likelihood based unfolding. 
	The fraction of signal events within the fiducial phase space (acceptance $\mathcal{A}_{\text{fid}}$), the reconstruction efficiency ($\epsilon$) in the fiducial phase space,
	and the ratio of the number of reconstructed events outside the fiducial phase space to that of the ones inside the fiducial phase space ($f_{\text{nonfid}}$) are quoted for each production mechanism for $\mH=125.38\GeV$.
	The last column shows the value of $(1+f_{\text{nonfid}})\epsilon$, which regulates the signal yield for a given fiducial cross section.
	All values are shown with their statistical uncertainty.
	The values for the $\ggH$ production mode are obtained using the \POWHEG generator.
	\label{tab:summarySM}
}
\begin{tabular}{lcccc}
	Signal process & $\mathcal{A}_{\text{fid}}$ & $\epsilon$ & $f_{\mathrm{nonfid}}$  & $(1+f_{\text{nonfid}})\epsilon$ \\
	\hline
	$\ggH$ (\POWHEG) & 0.408 $\pm$ 0.001 & 0.619 $\pm$ 0.001 & 0.053 $\pm$ 0.001 & 0.652 $\pm$ 0.001 \\
	$\VBF$ & 0.448 $\pm$ 0.001 & 0.632 $\pm$ 0.002 & 0.043 $\pm$ 0.001 & 0.659 $\pm$ 0.002 \\
	$\WH$ & 0.332 $\pm$ 0.001 & 0.616 $\pm$ 0.002 & 0.077 $\pm$ 0.001 & 0.664 $\pm$ 0.002 \\
	$\ZH$  & 0.344 $\pm$ 0.002 & 0.626 $\pm$ 0.003 & 0.083 $\pm$ 0.002 & 0.678 $\pm$ 0.003 \\
	$\ttH$ & 0.320 $\pm$ 0.002 & 0.614 $\pm$ 0.003 & 0.179 $\pm$ 0.003 & 0.725 $\pm$ 0.005 \\
\end{tabular}
\end{table*}

Systematic uncertainties are included in the form of nuisance parameters and the fiducial cross section measurements are obtained using an asymptotic approach~\cite{LHC-HCG} with a test statistic based on the profile likelihood ratio~\cite{Cowan_2011}.
A maximum likelihood fit is performed simultaneously in all final states and bins of each observable, assuming $\mH=125.38\GeV$.
The branching fractions of the \PH boson to the different final states ($4\Pe,4\Pgm,2\Pe2\Pgm$) are unconstrained parameters in the fit to increase the model independence of the measurements, following the strategy adopted in Ref.~\cite{CMSHIG19001}.
A likelihood-based unfolding is performed to resolve the detector effects from the observed distributions to the fiducial phase space. 
This approach is the same as in
Refs.~\cite{CMSHggFiducial8TeV,CMSHIG19001} and allows to simultaneously unfold detector effects and perform the fit to extract the fiducial cross section.
The analysis strategy of Ref.~\cite{CMSHIG19001} is extended by measuring separately the fiducial cross sections in $4\Pe+4\Pgm$ and $2\Pe 2\Pgm$ final states for observables targeting the $\HZZfl$ decay.
This choice is driven by the different physics in the final states containing different- and same-flavor leptons arising from the destructive interference between the two alternative methods of constructing the $\HZZfl$ diagrams in the same-helicity states in the case of identical leptons.

\section{Systematic uncertainties}
\label{sec:systematics}
The integrated luminosities of the 2016, 2017, and 2018 data-taking periods are individually known with uncertainties in the range 1.2--2.5\%~\cite{CMS-LUM-17-003,CMS-PAS-LUM-17-004,CMS-PAS-LUM-18-002}, while the 2016--2018 integrated luminosity has an uncertainty of 1.6\%.
The partial correlation scheme considered for this systematic uncertainty is summarized in Table~\ref{tab:SystOverviewABC}.

Experimental systematic uncertainties in trigger and lepton reconstruction and selection efficiencies are estimated from data for different final states. 
These uncertainties are derived from a tag-and-probe technique using $\cPJgy$ and $\PZ$ decays into a pair of leptons and range  4.3--10.9\% in the 4\Pe channel and 0.6--1.9\% in the 4\Pgm channel, depending on the \pt region.
In this paper, a new way to estimate the systematic uncertainty in the electron efficiency measurements is introduced. Alternative variations in the tag-and-probe fit used to derive the systematic uncertainty are combined and the RMS of the values is used for the systematic uncertainty evaluation. This makes the evaluation of the systematic uncertainty for the low-\pt bins more solid, and leads to their reduction of a factor approximately $40$\% in the electron reconstruction and selection efficiency with respect to the values reported in Ref.~\cite{CMSHIG19001}.

{\tolerance=800 The systematic uncertainties in the lepton momentum scale and resolution are estimated from dedicated studies of the $\Zll$ mass distribution in data and simulation.
The momentum scale uncertainty is 0.06\% in the 4\Pe channel and 0.01\% in the 4\Pgm channel, while the resolution uncertainty is 10\% in the 4\Pe channel and 3\% in the 4\Pgm channel.
The effect of these uncertainties is evaluated by allowing the corresponding parameters of the DCB function used to model the resonant signal to remain unconstrained in the fit.\par} 

Jet-related observables are affected by systematic uncertainties in the estimation of the jet energy scale. These uncertainties affect the normalization of the processes and are modeled with a set of nuisance parameters representing the various sources and accounting for the partial correlation among the various final states and years. Their values depend on the kinematic bin and range 0.1--33\%, with an average value of 3\%.

Furthermore, experimental systematic uncertainties in the reducible background estimation are considered. These uncertainties originate from the evaluation of the lepton misidentification rates and vary between 23 and 43\%, depending on the final state. The impact of these uncertainties is negligible.

Table~\ref{tab:SystOverviewABC} summarizes the experimental systematic uncertainties considered in the analysis.

\begin{table}[tbp]
\centering
\topcaption{
	Summary of the input values of the experimental systematic uncertainties.
	\label{tab:SystOverviewABC}
}
\cmsTable{
\begin{tabular}{l c c c} 
	\hline
	\multicolumn{4}{c}{Common experimental uncertainties} \\
	\hline
	& 2016 & 2017 & 2018 \\
	\hline
	Luminosity uncorrelated & 1 \% & 2 \% & 1.5 \% \\ 
	Luminosity correlated 2016--2018 & 0.6 \% & 0.9 \% & 2 \% \\ 
	Luminosity correlated 2017--2018 & \NA & 0.6 \% & 0.2 \% \\ 
	Lepton id/reco efficiencies & 0.7--10 \% & 0.6--8.5 \% & 0.6--9.5 \% \\ 
	Jet energy scale & 0.1\%--27\% & 0.1\%--33\% & 0.1\%--33\% \\
	\hline
	\multicolumn{4}{c}{Background related uncertainties} \\
	\hline
	Reducible background (Z+X) & 25--43 \% & 23--36 \%  & 24--36 \% \\ 
	\hline
	\multicolumn{4}{c}{Signal related uncertainties} \\
	\hline
	Lepton energy scale & 0.06\% (\Pe)--0.01\% (\Pgm) & 0.06\% (\Pe)--0.01\% (\Pgm) & 0.06\% (\Pe)--0.01\% (\Pgm) \\ 
	Lepton energy resolution & 10\% (\Pe)--3\% (\Pgm) & 10\% (\Pe)--3\% (\Pgm)  & 10\% (\Pe)--3\% (\Pgm) \\ 
	\hline
\end{tabular}
}
\end{table}

Theoretical uncertainties in the renormalization and factorization scales, and in the choice of the PDF set affect both the signal and background rates.
The scale uncertainty is determined by varying renormalization and factorization scales between 0.5 and 2 times their nominal value, while keeping their ratio between 0.5 and 2.
The uncertainty in the PDF set is determined following the PDF4LHC recommendations taking the root mean square of the variation of the results when using different replicas of the default NNPDF set~\cite{Butterworth:2015oua}.
An additional 10\% uncertainty in the $K$ factor used for the $\ggZZ$ background prediction is applied.
A systematic uncertainty of 2\%~\cite{deFlorian:2016spz} in the branching fraction of $\HZZfl$ is considered and only affects the signal yield.
The theoretical uncertainties affecting the signal are not included in the fit but evaluated and indicated in Figs. 4--24. 

\section{Results}
\label{sec:results}
\begin{figure}[!htb]
\centering
\includegraphics[width=0.49\textwidth]{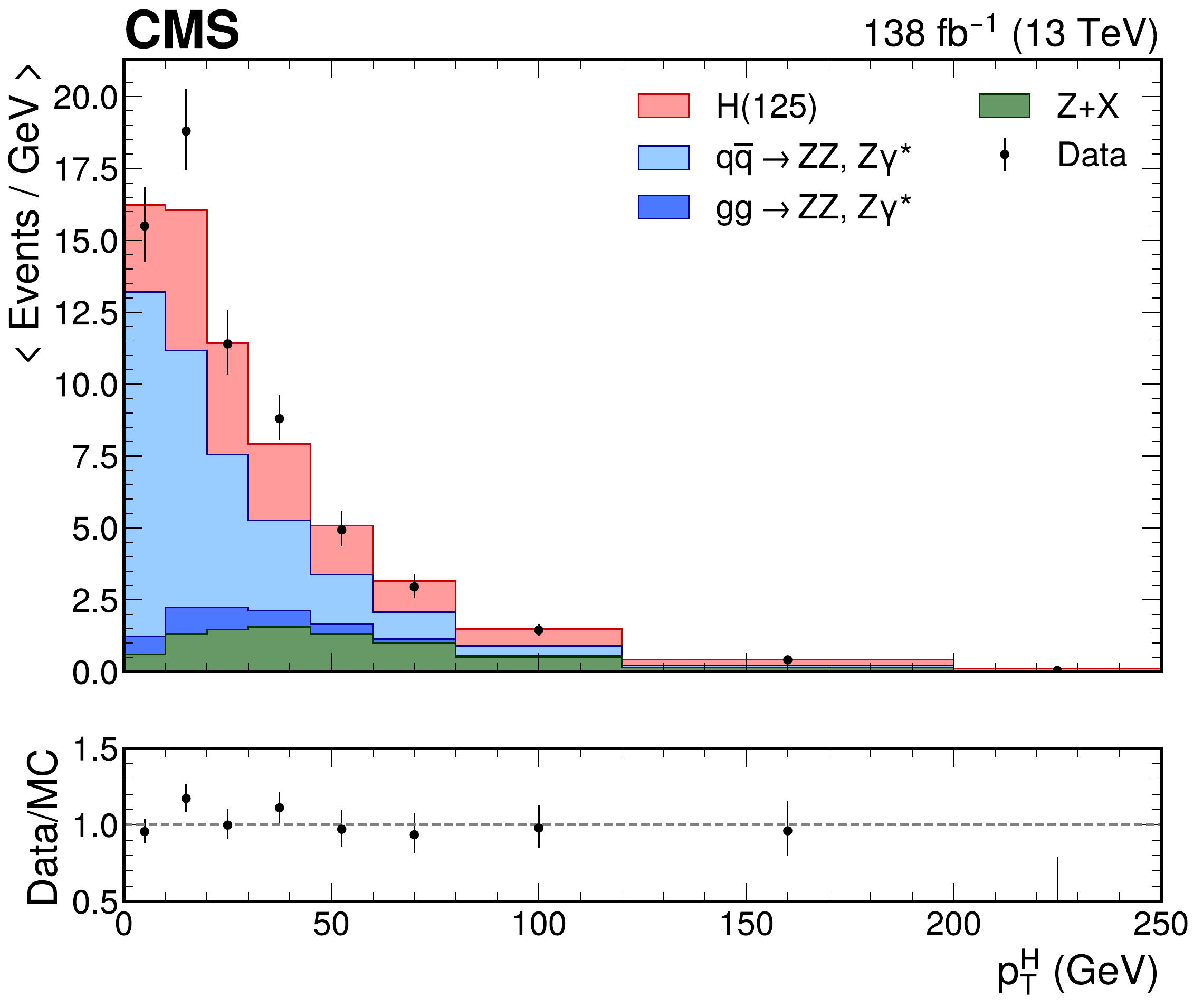}
\includegraphics[width=0.49\textwidth]{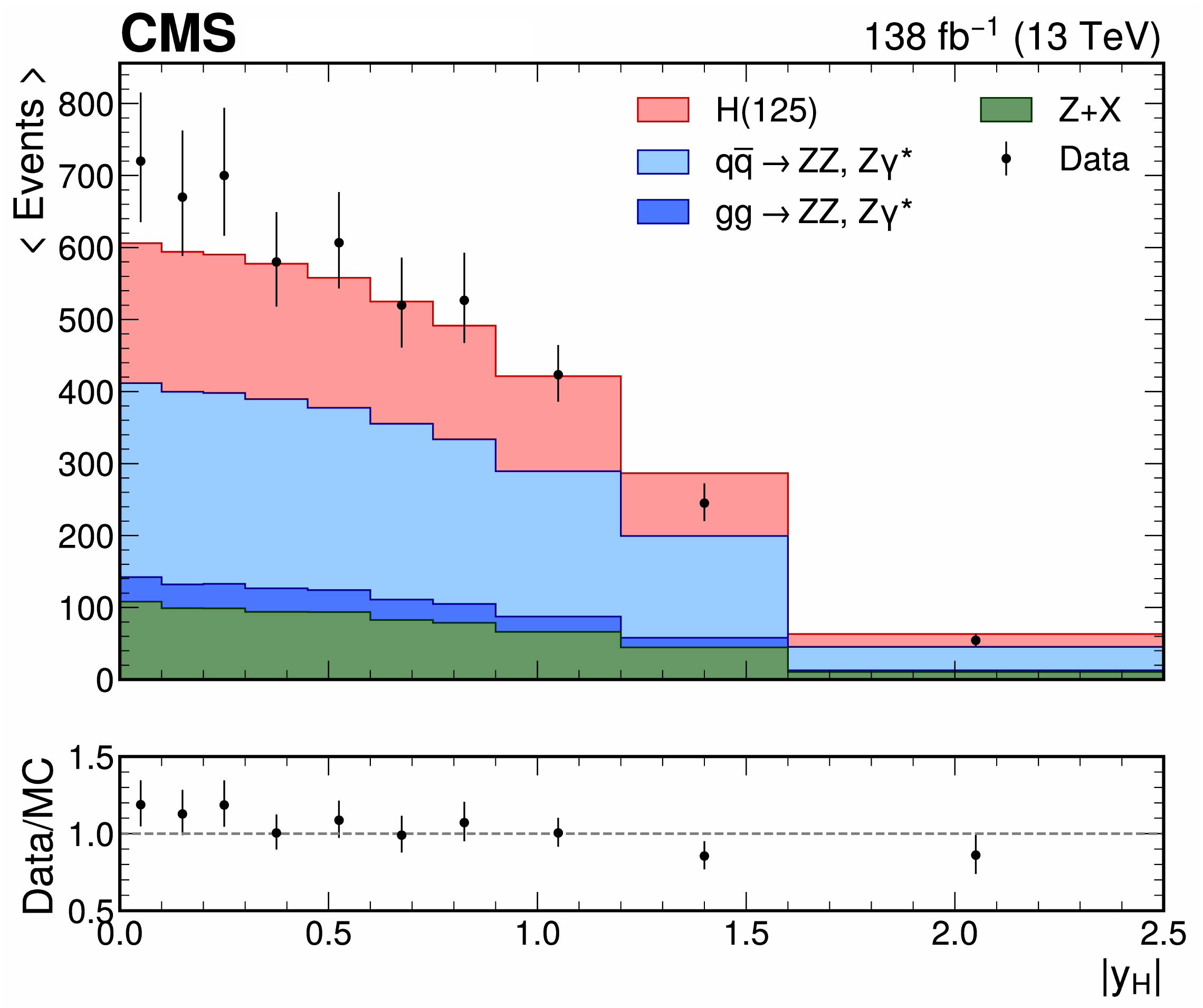}
\caption{
	Reconstructed transverse momentum (\cmsLeft) and rapidity (\cmsRight) of the four-lepton system for events with $105<\mllll<160\GeV$.
	Points with error bars represent the data, solid histograms the predictions from simulation.
	The $y$ axes of the top panels have been rescaled to display the number of events per bin, divided by the width of each bin.
	The lower panels show the ratio of the measured values to the expectations from the simulation.
	\label{fig:reco_pTYH}
}
\end{figure}

This section reports the measurements of the fiducial cross sections in differential bins of the kinematic observables introduced in Section~\ref{sec:observables}. 
All fiducial cross section measurements are in agreement with the SM predictions within uncertainties.
The compatibility of the results with the theoretical predictions is quantified by reporting the $p$-value for each observable, computed using the negative log-likelihood ratio as test statistic evaluated at the SM point.
The $p$-value is calculated using a $\chi^2$ probability density function with the number of bins used in the measurements taken as the number of degrees of freedom.
The observed $p$-values range from 0.05 to 0.99. 
The inclusive cross section is measured with an overall precision of $10\%$, with statistical and systematic contributions of $8\%$ and $6\%$, respectively.
All the differential measurements are limited by the statistical uncertainty.
The differential cross sections as functions of $\pt^{\PH}$ and $\abs{y_{\PH}}$ are measured with an average precision of $35\%$, while the most precise cross sections are measured with a precision of $20\%$.
The measurements are compared to the theoretical predictions from various generators.
The uncertainties in these predictions come from the uncertainty in the fiducial acceptance, the $\HZZfl$ branching fraction and variations of the PDF replicas, $\alpha_s$ value, and renormalization and factorization scales.
Figure~\ref{fig:reco_pTYH} depicts the distributions of these two observables comparing data to predictions from simulation.
With respect to Ref.~\cite{CMSHIG19001}, the data set used in this analysis benefits from an improved object calibration. This leads to a better precision in the final results and permits measurements of jet-related observables in a phase space region that extends up to $\abs{\eta}=4.7$. The inclusive fiducial result features a reduction of 15\% of the uncertainty, particularly evident in the 40\% reduction of the systematic component obtained using a root-mean-square approach to compute the uncertainties in the electrons selection efficiency~\cite{ATLAS:2019jvq}, which are the leading source of systematic uncertainty on the measurements performed in this analysis.
Tabulated results are provided in the HEPData record for this analysis~\cite{hepdata}.

The kinematic properties of the four-lepton system are studied by measuring differential cross sections in bins of angular observables sensitive to the $\PH\PV\PV$ decay.
These results are reported for the inclusive four-lepton final state and for the same-flavor and different-flavor final states separately to account for interference effects in the case of identical helicity states.
In all cases the results agree with the distributions predicted by the SM.
The largest deviations with respect to the expected values are observed in the central bins of  $\cos\theta_2$ and $\Phi$ and are compatible with statistical fluctuations in the observed data.
The $p$-values of these two measurements are found to be 0.23 and 0.24, respectively, thus corroborating the compatibility with the SM predictions.
For the first time, differential cross sections are measured in bins of kinematic discriminants sensitive to the presence of possible $\PH\PV\PV$ anomalous couplings.
These measurements are compared to the distributions of these discriminants computed under the SM hypothesis and under various anomalous coupling hypotheses. The former is always favored in the comparison with data.

A total of six double-differential measurements are also reported.

\subsection{Inclusive cross section}
\begin{figure}[!htb]
\centering
\includegraphics[width=0.6\textwidth]{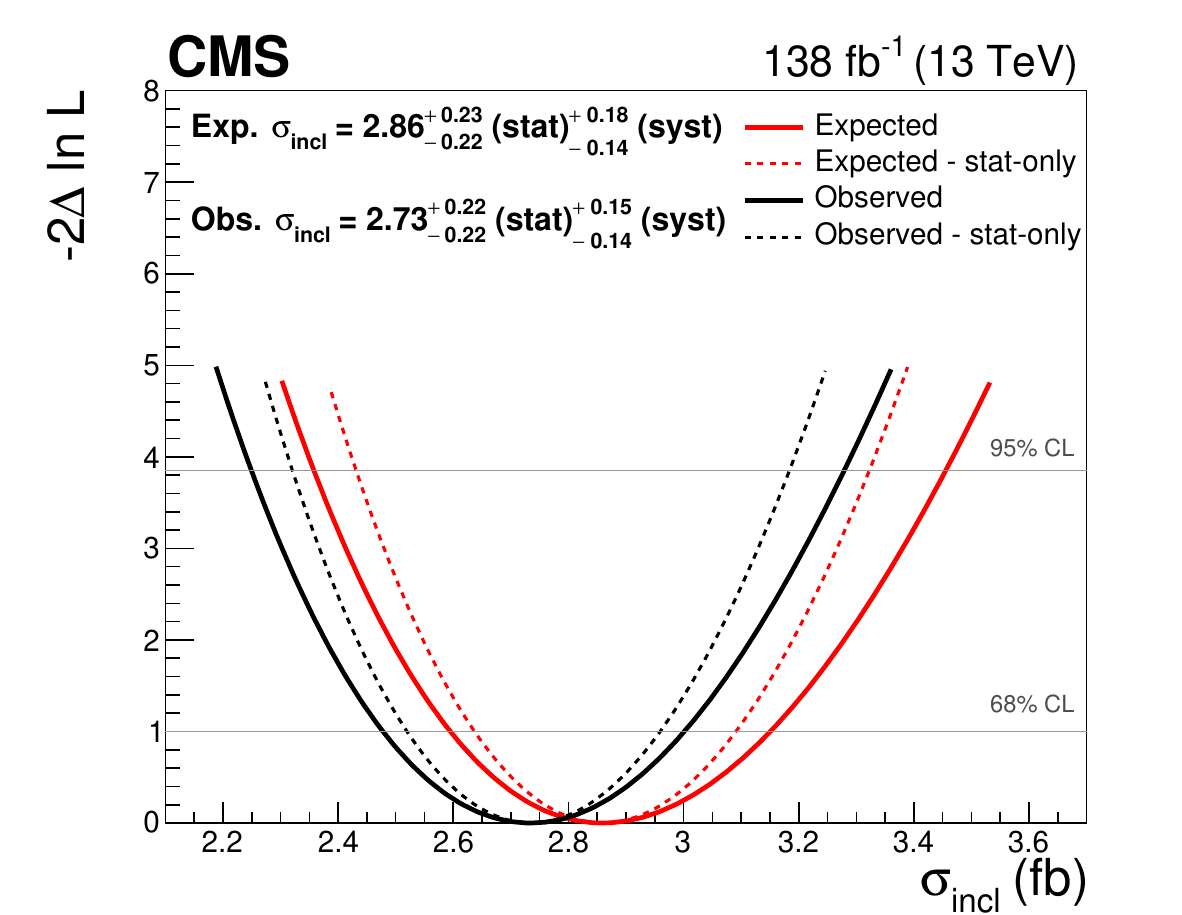}
\caption{
	Log-likelihood scan for the measured inclusive fiducial cross section. The scan is shown with (solid line) and without (dashed line) systematic uncertainties profiled in the fit.
	\label{fig:fiducial_LLScan}}
\end{figure}

\begin{figure}[!htb]
\centering
\includegraphics[width=0.48\textwidth]{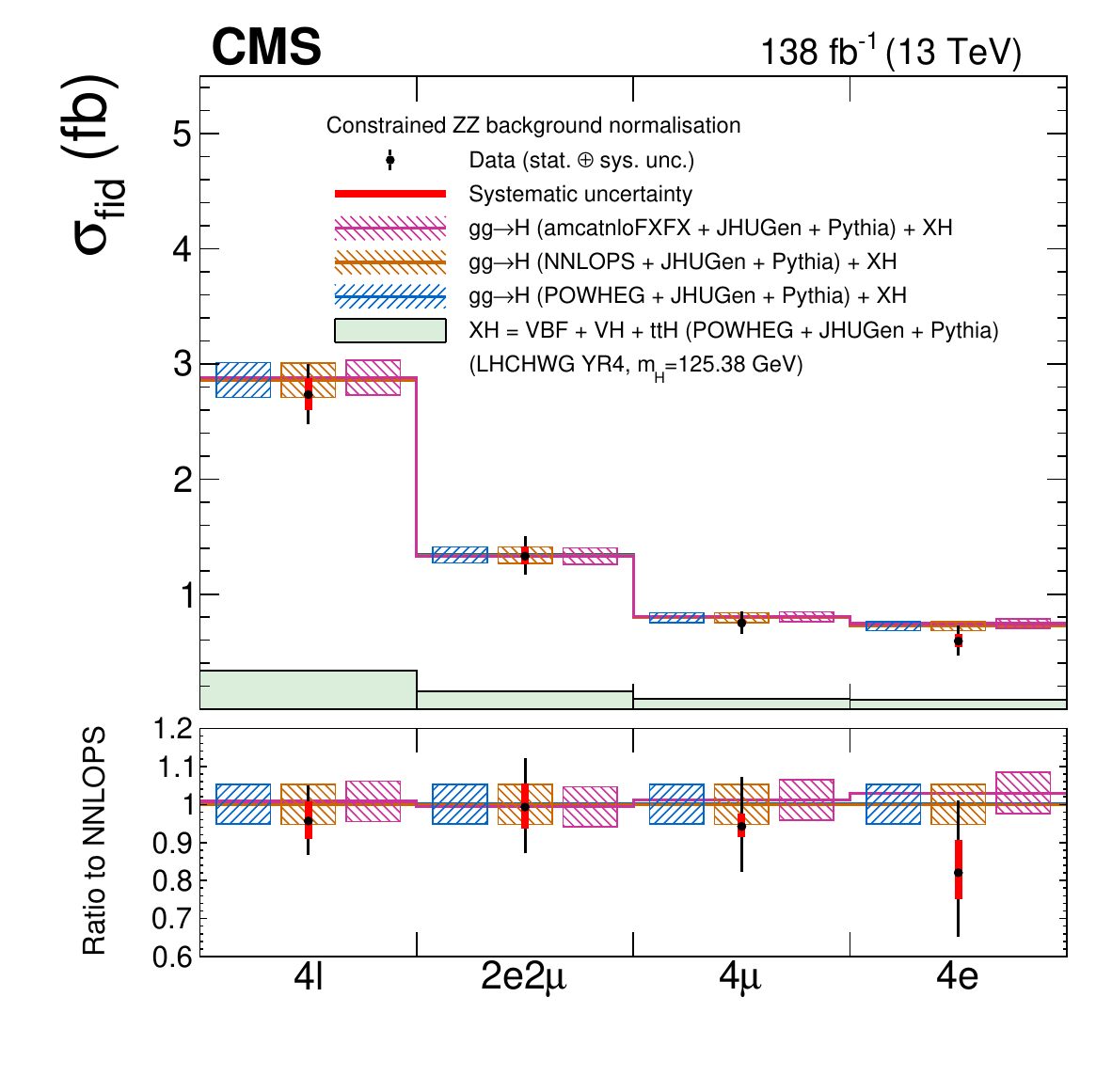}
\includegraphics[width=0.48\textwidth]{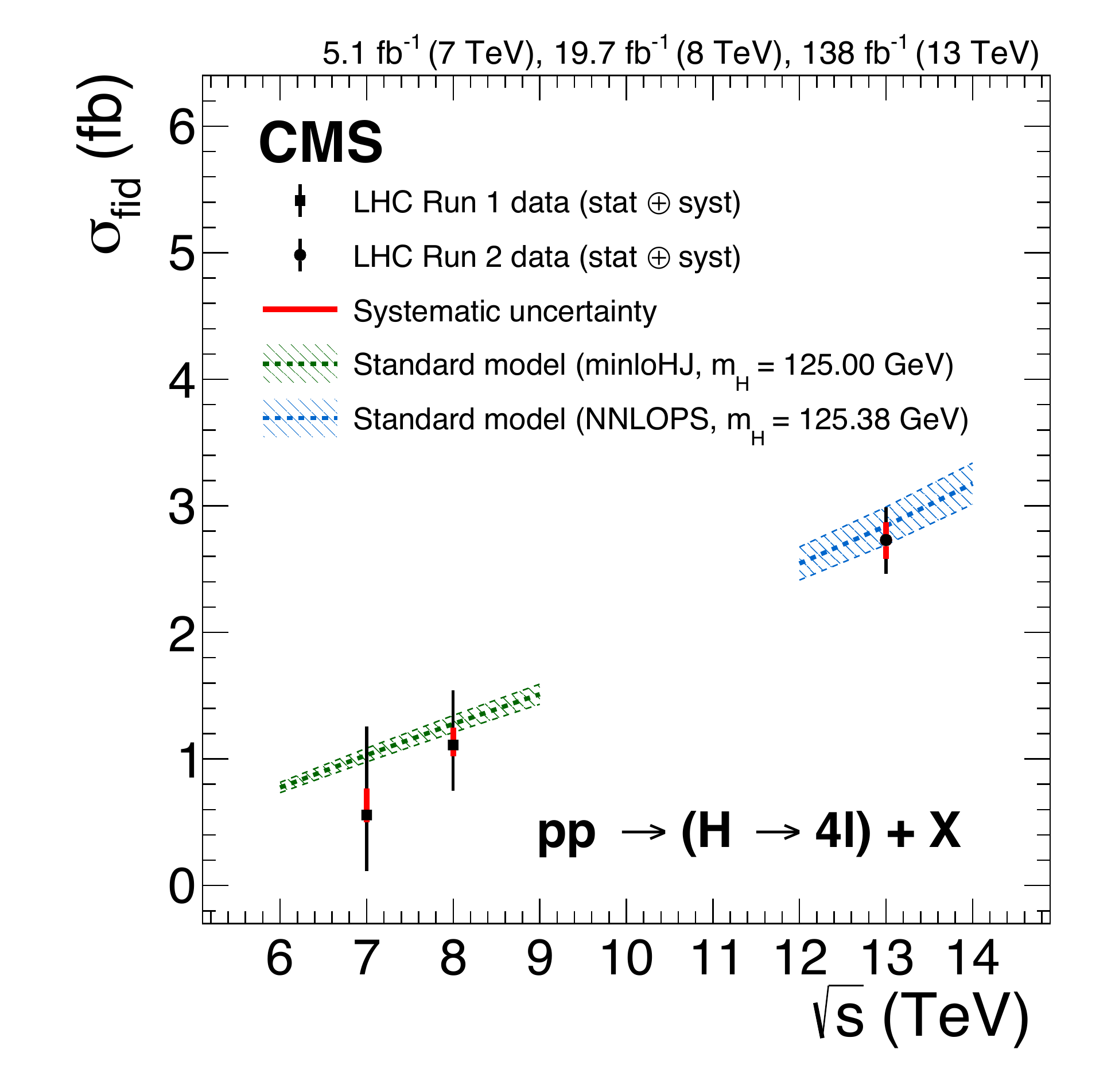}
\caption{
	Measured inclusive fiducial cross section for the various final states (\cmsLeft); and as a function of the center-of-mass energy $\sqrt{s}$ (\cmsRight).
	In the \cmsLeft panel the acceptance and theoretical uncertainties are calculated using \POWHEG (blue), \textsc{NNLOPS} (orange), and \MGvATNLO (pink).
	The subdominant component of the signal ($\VBF + \VH + \ttH$) is denoted as XH and is fixed to the SM prediction.
	In the \cmsRight panel the acceptance is calculated using \textsc{minloHJ} at  $\sqrt{s}=13\TeV$ and \textsc{NNLOPS}~\cite{Grazzini:2013mca,deFlorian:2012mx} at $\sqrt{s}=$~7 and 8\TeV.
	\label{fig:fiducial_inclusive}}
\end{figure}

\begin{figure}[!htb]
\centering
\includegraphics[width=0.48\textwidth]{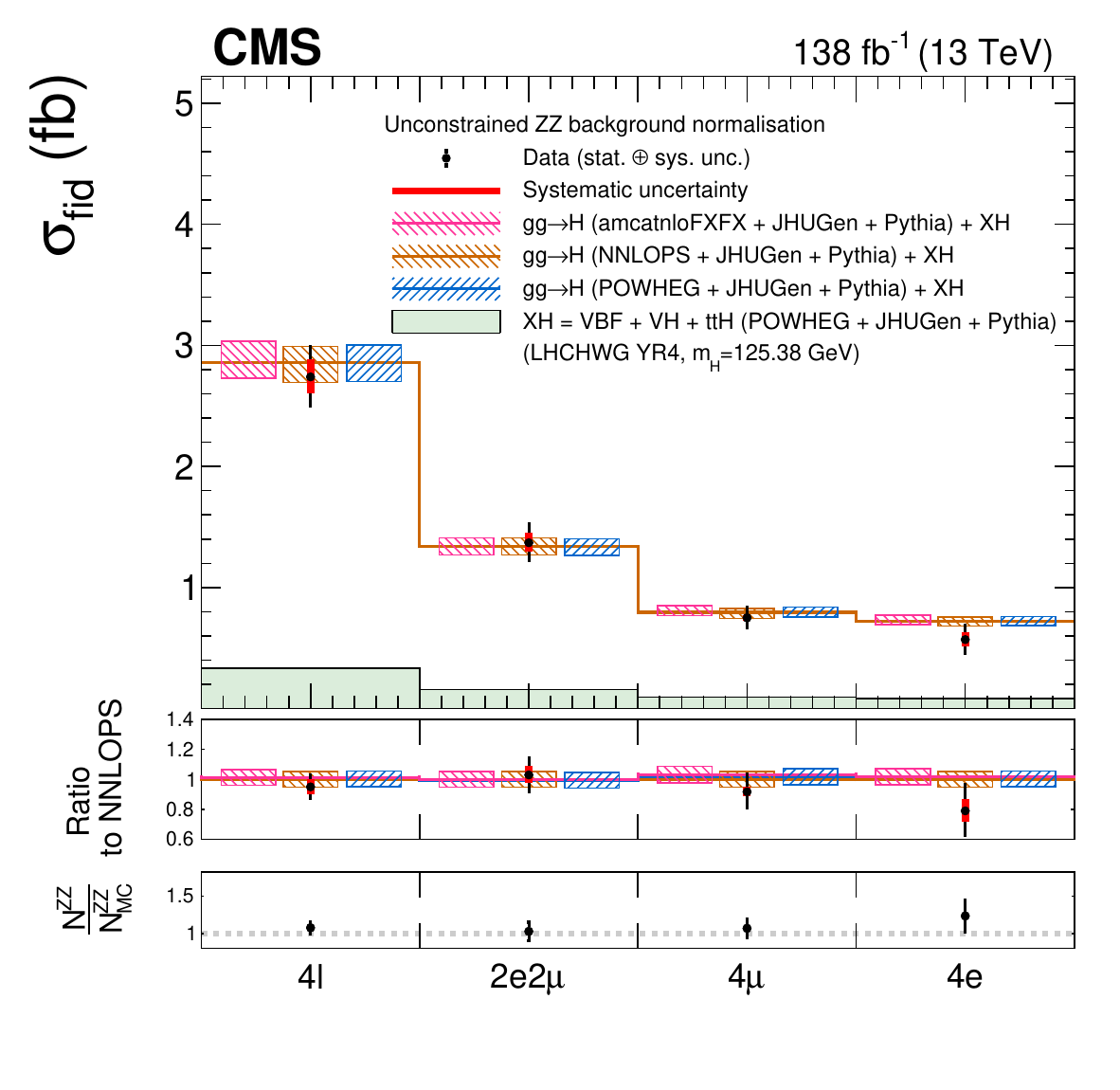}
\raisebox{0.08\height}{\includegraphics[width=0.5\textwidth]{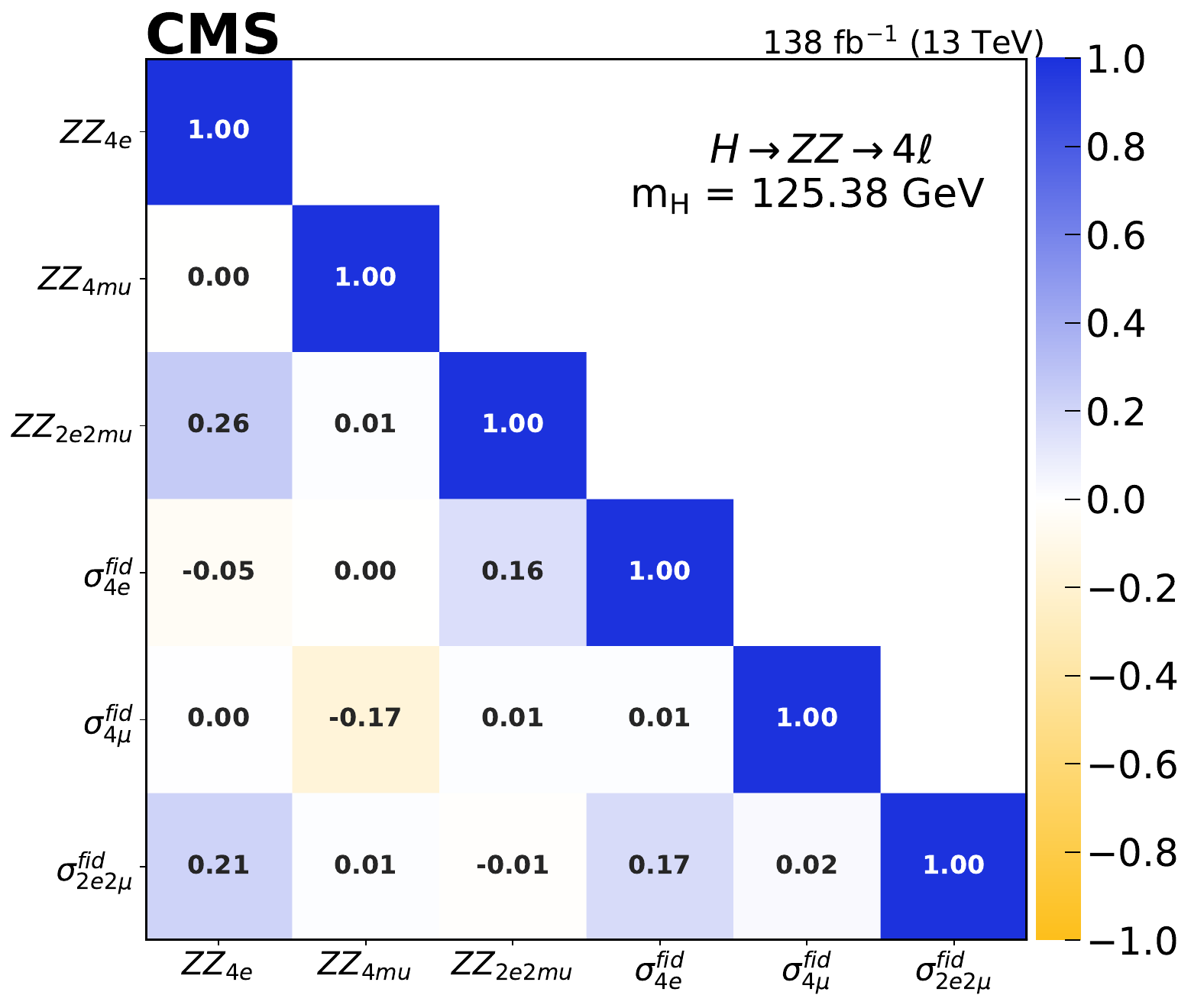}}
\caption{
	Inclusive fiducial cross section measured for the various final states with the irreducible backgrounds normalization $\PZ\PZ$ unconstrained in the fit (\cmsLeft) and the corresponding correlation matrix (\cmsRight).
	The acceptance and theoretical uncertainties in the differential bins are calculated using \POWHEG (blue), \textsc{NNLOPS} (orange), and \MGvATNLO (pink).
	The subdominant component of the signal ($\VBF + \VH + \ttH$) is denoted as XH and it is fixed to the SM prediction.
	The ratio of the measured cross section to the theoretical prediction obtained from each generator is shown in the central panel, while the lower panel shows the ratio between the values derived from the measured $\PZ\PZ$ normalization and the MC prediction.
	\label{fig:fiducial_zzfloating}}
\end{figure}

The measured inclusive fiducial cross section for the $\HZZfl$ process is 

\begin{equation}
\begin{split}
	\sigma^{\text{fid}} & = 2.73\pm0.22\stat\pm0.15\syst \unit{fb} \\
	& = 2.73\pm0.22\stat\pm 0.12~\text{(electrons)}\pm 0.05~\text{(lumi)}\pm 0.05~\text{(bkg)}\pm 0.03~\text{(muons)}\unit{fb} 
\end{split}
\end{equation}

for a \PH boson mass of $\mH=125.38\GeV$, in agreement with the SM expectation of $\sigma_{{\text{fid}}}^{\text{SM}}=2.86\pm0.15\unit{fb}$. Figure~\ref{fig:fiducial_LLScan} shows the corresponding log-likelihood scan.
The systematic component of the uncertainty is dominated by electron-related nuisance parameters (electrons), especially the electron selection efficiency that is the leading nuisance parameter in the four-lepton decay channel.
The muon-related nuisance parameters (muons) and the uncertainties on the luminosity measurement (lumi)  and on the background predictions (bkg) play a minor role on the overall systematic uncertainty on $\sigma^{\text{fid}}$.
The inclusive fiducial cross section measured in the three final states ($4\Pe,\,4\Pgm,\,$ and $2\Pe 2\Pgm$) is shown in the left panel of Fig.~\ref{fig:fiducial_inclusive}, while the right panel depicts the evolution of the $\HZZfl$ fiducial cross section as a function of the center-of-mass energy.
The results are compared with the cross sections predicted by the $\POWHEG$, $\MGvATNLO$, and \textsc{NNLOPS}  generators for the \PH boson production and parton showering, while the decay is always modeled by \JHUGEN.
The measurement of the inclusive fiducial cross section is repeated treating the normalization of the $\PZ\PZ$ irreducible background processes as an unconstrained parameter in the fit.
The results are presented in the left panel of Fig.~\ref{fig:fiducial_zzfloating} for the inclusive $\HZZfl$ measurement and the three final states considered. The correlation coefficient ($\rho$) between the inclusive fiducial cross section measurement and the $\PZ\PZ$ normalization in the $4\ell$ final state is found to be $\rho=-0.03$, while the correlations between the $\PZ\PZ$ normalization in each final state and the corresponding fiducial cross section measurements are shown in the right panel of Fig.~\ref{fig:fiducial_zzfloating}.
The positive correlations observed between the $\PZ\PZ$ background normalizations and the cross sections measured in the $4\Pe$ and $2\Pe2\Pgm$ final states are driven by the systematic uncertainties on the electrons reconstruction and identification, which are the leading nuisance parameters in the analysis.

The results are summarized in Table~\ref{tab:fiducial_xsec}.
The measured cross sections, obtained with the $\PZ\PZ$ normalization treated as an unconstrained parameter in the fit, are in agreement with the results obtained when the irreducible background normalization is constrained to the theoretical expectation.
The uncertainty in this parameter when extracted from sidebands in data (7.5\%) is larger than the theoretical uncertainty in its predictions (6.3\%).
For these reasons, in the following the $\PZ\PZ$ normalization is fixed to the SM prediction.

\begin{table*}[!htb]
\centering
\topcaption{
	Measured inclusive fiducial cross section and $\pm 1$ standard deviation uncertainties for the various final states at $\mH=125.38\GeV$.
	The upper row summarizes the results obtained when the irreducible background normalization is constrained to the SM expectation and theoretical uncertainty, while the lower section present the results from a fit with the $\PZ\PZ$ normalization treated as an unconstrained parameter. The first row presents the fiducial cross section, the middle row the ZZ background normalization extracted from the fit, and the bottom row the ZZ estimation from MC. The uncertainties on $N^{ZZ}_{MC}$ are the pre-fit uncertainties summing the statistical and systematic uncertainty.
	\label{tab:fiducial_xsec}
}
\renewcommand{\arraystretch}{1.5}
\begin{tabular}{ccccc}
	& $4\Pe$ & $4\Pgm$ & $2\Pe 2\Pgm$ & Inclusive \\
	\hline
	\multicolumn{5}{c}{Constrained ZZ background}\\
	$\sigma_{{\text{fid}}} $ & $0.59^{+0.13}_{-0.12} \unit{fb}$ & $0.75^{+0.10}_{-0.09} \unit{fb}$ & $1.33^{+0.17}_{-0.16} \unit{fb}$ & $2.73^{+0.22}_{-0.22}\stat^{+0.15}_{-0.14}\syst \unit{fb}$  \\
	\hline
	\multicolumn{5}{c}{Unconstrained ZZ background}\\
	$\sigma_{{\text{fid}}} $& $0.57^{+0.15}_{-0.12} \unit{fb}$ & $0.75^{+0.10}_{-0.09} \unit{fb}$ & $1.37^{+0.17}_{-0.16} \unit{fb}$ & $2.74^{+0.24}_{-0.23}\stat^{+0.14}_{-0.11}\syst \unit{fb}$  \\
	$N^{ZZ}$ & $92^{+16}_{-13}$ & $162^{+19}_{-18}$ & $193^{+23}_{-21}$ & $445^{+27}_{-26}\stat^{+21}_{-19}\syst$   \\
	$N^{ZZ}_{MC}$ & $74^{+7}_{-8}$ & $152^{+7}_{-8}$ & $188^{+13}_{-14}$ & $414^{+24}_{-28}$ \\
\end{tabular}
\end{table*}

\subsection{Differential cross sections: production}
\begin{center}
	\begin{figure}[!htbp]
		\centering
		\includegraphics[width=0.48\textwidth]{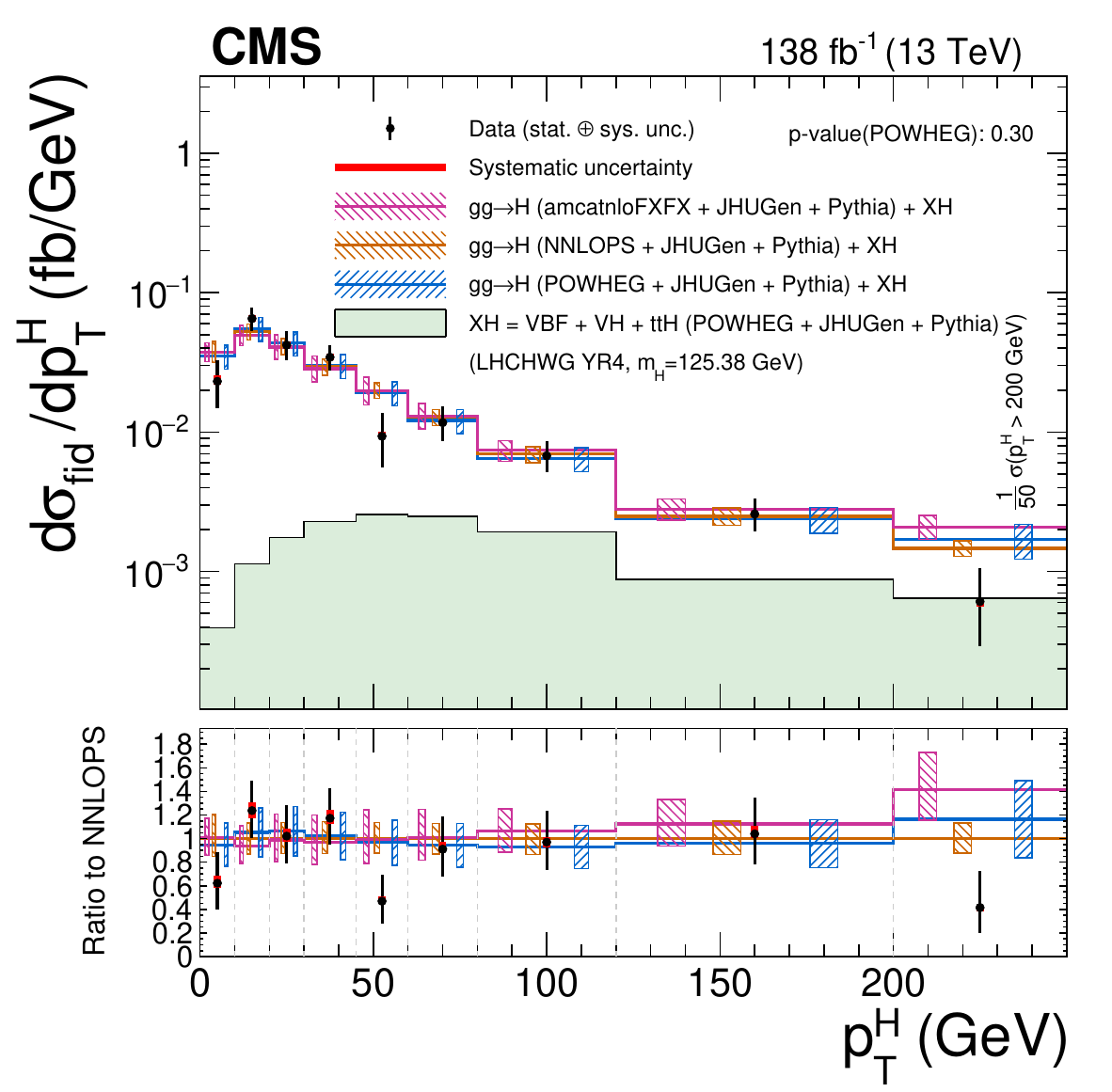}
		\includegraphics[width=0.48\textwidth]{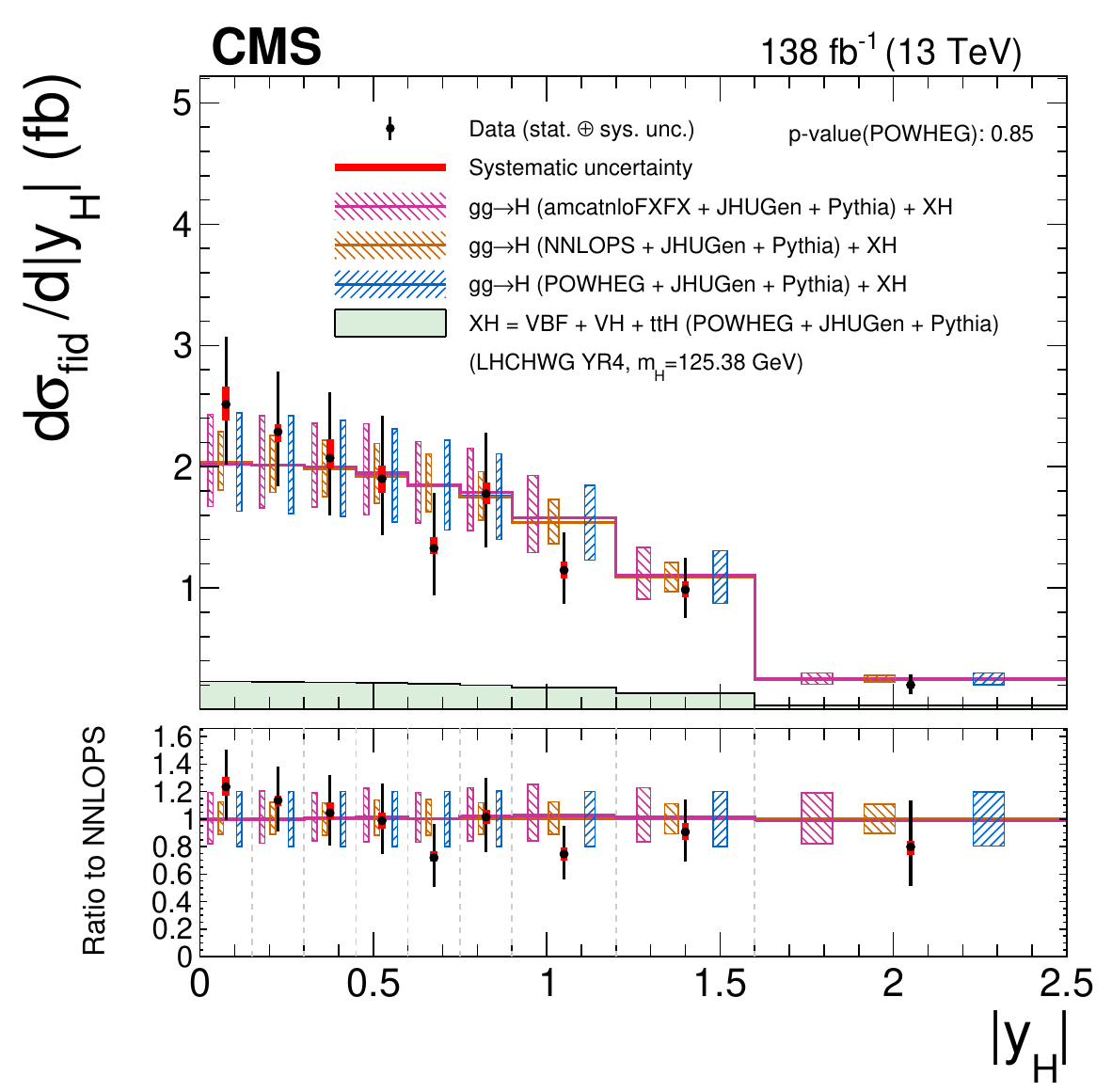}
		
		\caption{
			Differential cross sections as functions of the transverse momentum of the Higgs boson $\pt^{\PH}$ (\cmsLeft) and of the rapidity of the Higgs boson $\abs{y_{\PH}}$  (\cmsRight). 
			The fiducial cross section in the last bin (\cmsLeft) is measured for events with $\pt^{\PH}>200\GeV$ and normalized to a bin width of $50\GeV$.
			The acceptance and theoretical uncertainties in the differential bins are calculated using the $\ggH$ predictions from three different generators normalized to next-to-next-to-next-to-leading order ($\mathrm{N^3LO}$)~\cite{deFlorian:2016spz}.
			The subdominant component of the signal ($\VBF + \VH + \ttH$) is denoted as XH and is fixed to the SM prediction.
			The measured cross sections are compared with the  $\ggH$ predictions from \POWHEG (blue), \textsc{NNLOPS} (orange), and \MGvATNLO (pink).
			The hatched areas correspond to the systematic uncertainties in the theoretical predictions.
			Black points represent the measured fiducial cross sections in each bin, black error bars the total uncertainty in each measurement, red boxes the systematic uncertainties.
			The lower panels display the ratios of the measured cross sections and of the predictions from \POWHEG and \MGvATNLO to the \textsc{NNLOPS} theoretical predictions.
			\label{fig:fidPTH_YH}}
	\end{figure}
\end{center}

\begin{center}
	\begin{figure}[!htb]
		\centering
		\includegraphics[width=0.48\textwidth]{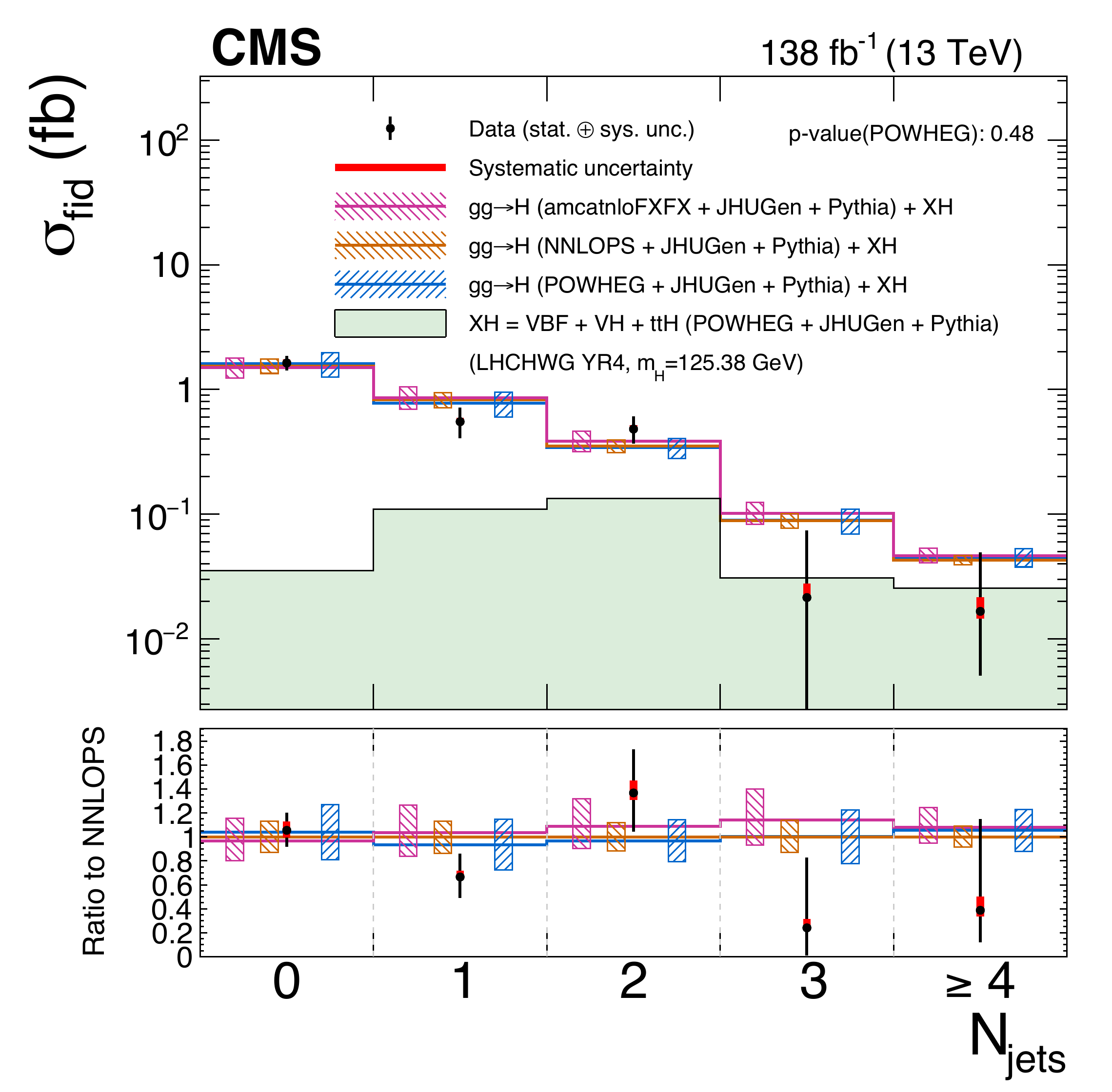}
		\includegraphics[width=0.48\textwidth]{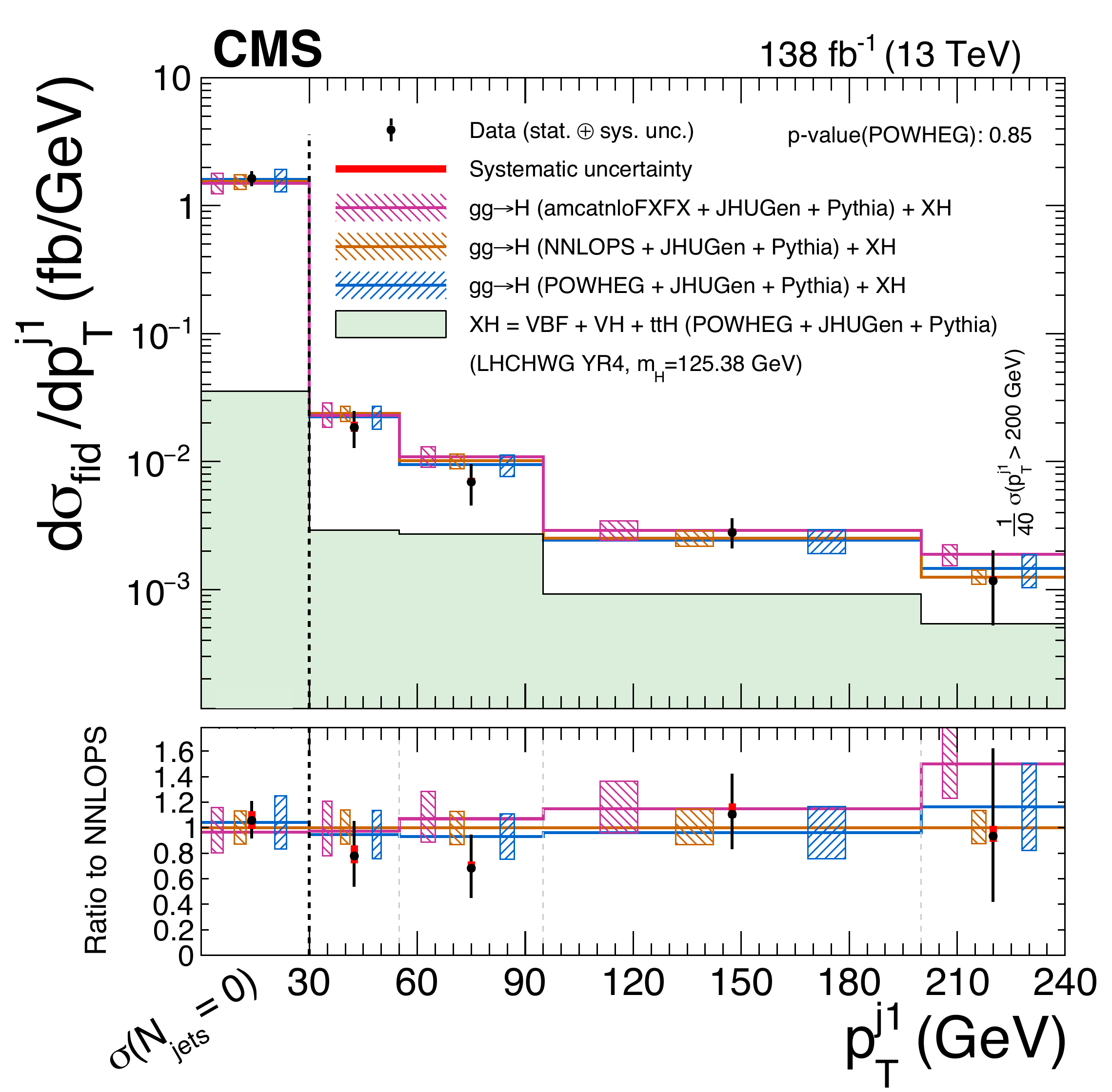}\\
		\includegraphics[width=0.48\textwidth]{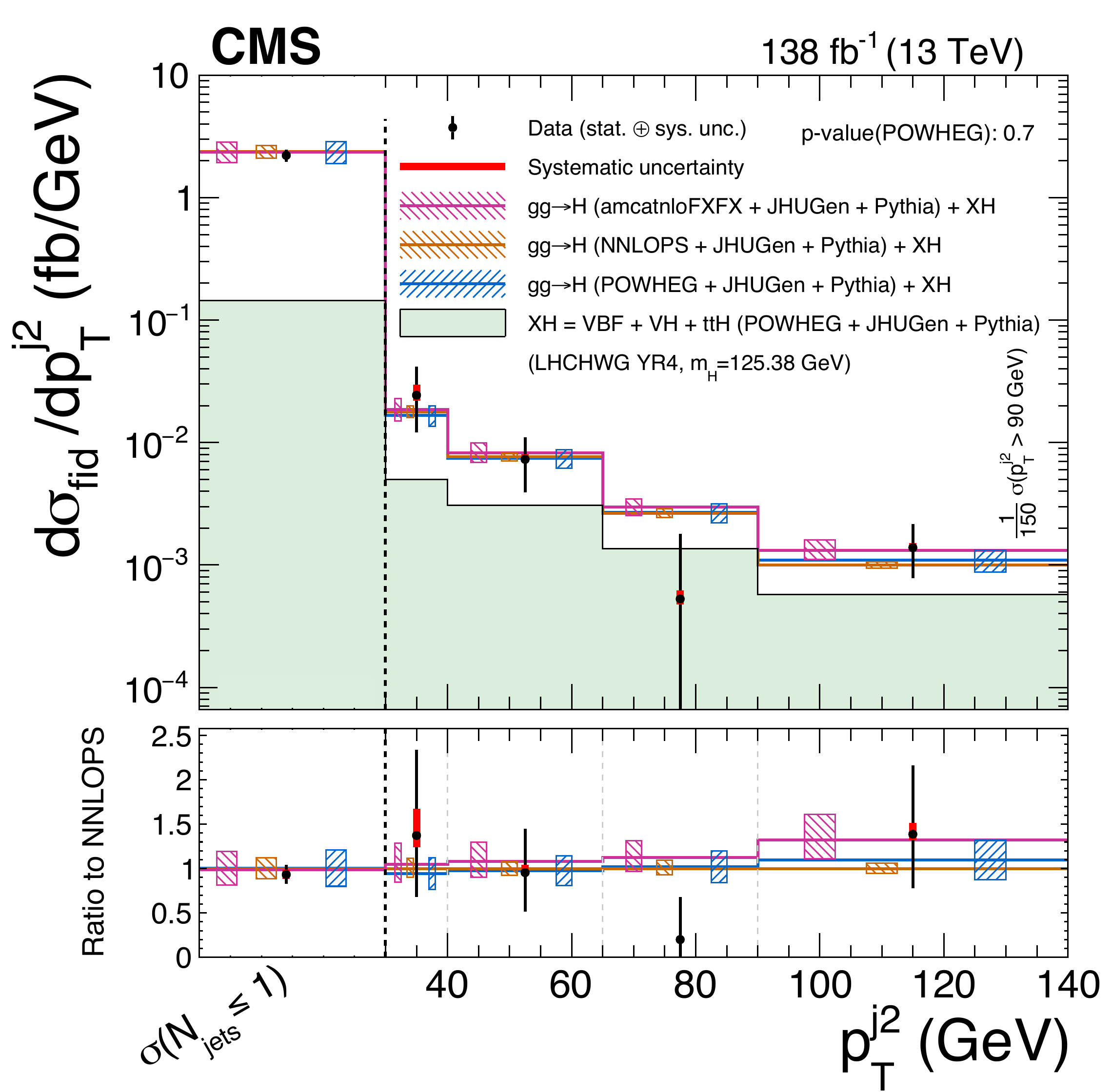}	
		\caption{
			Differential cross sections as functions of the number of jets in the event (upper \cmsLeft) and of the \pt of the leading (upper \cmsRight) and subleading (lower) jet.
			Upper right: the fiducial cross section in the last bin is measured for events with $\pt^{\text{j1}}>200\GeV$ and normalized to a bin width of $40\GeV$.
			The first bin comprises all events with less than one jet, for which $\pt^{\text{j1}}$  is undefined.
			Lower: the fiducial cross section in the last bin is measured for events with $\pt^{\text{j2}}>90\GeV$ and normalized to a bin width of $150\GeV$.
			The first bin comprises all events with less than two jet, for which $\pt^{\text{j2}}$  is undefined.
			The acceptance and theoretical uncertainties in the differential bins are calculated using the $\ggH$ predictions from three different generators normalized to next-to-next-to-next-to-leading order ($\mathrm{N^3LO}$)~\cite{deFlorian:2016spz}.
			The subdominant component of the signal ($\VBF + \VH + \ttH$) is denoted as XH and is fixed to the SM prediction.
			The measured cross sections are compared with the  $\ggH$ predictions from \POWHEG (blue), \textsc{NNLOPS} (orange), and \MGvATNLO (pink).
			The hatched areas correspond to the systematic uncertainties in the theoretical predictions.
			Black points represent the measured fiducial cross sections in each bin, black error bars the total uncertainty in each measurement, red boxes the systematic uncertainties.
			The lower panels display the ratios of the measured cross sections and of the predictions from \POWHEG and \MGvATNLO to the \textsc{NNLOPS} theoretical predictions.
			\label{fig:fidJets}}
	\end{figure}
\end{center}

\begin{center}
	\begin{figure}[!htb]
		\centering
		\includegraphics[width=0.48\textwidth]{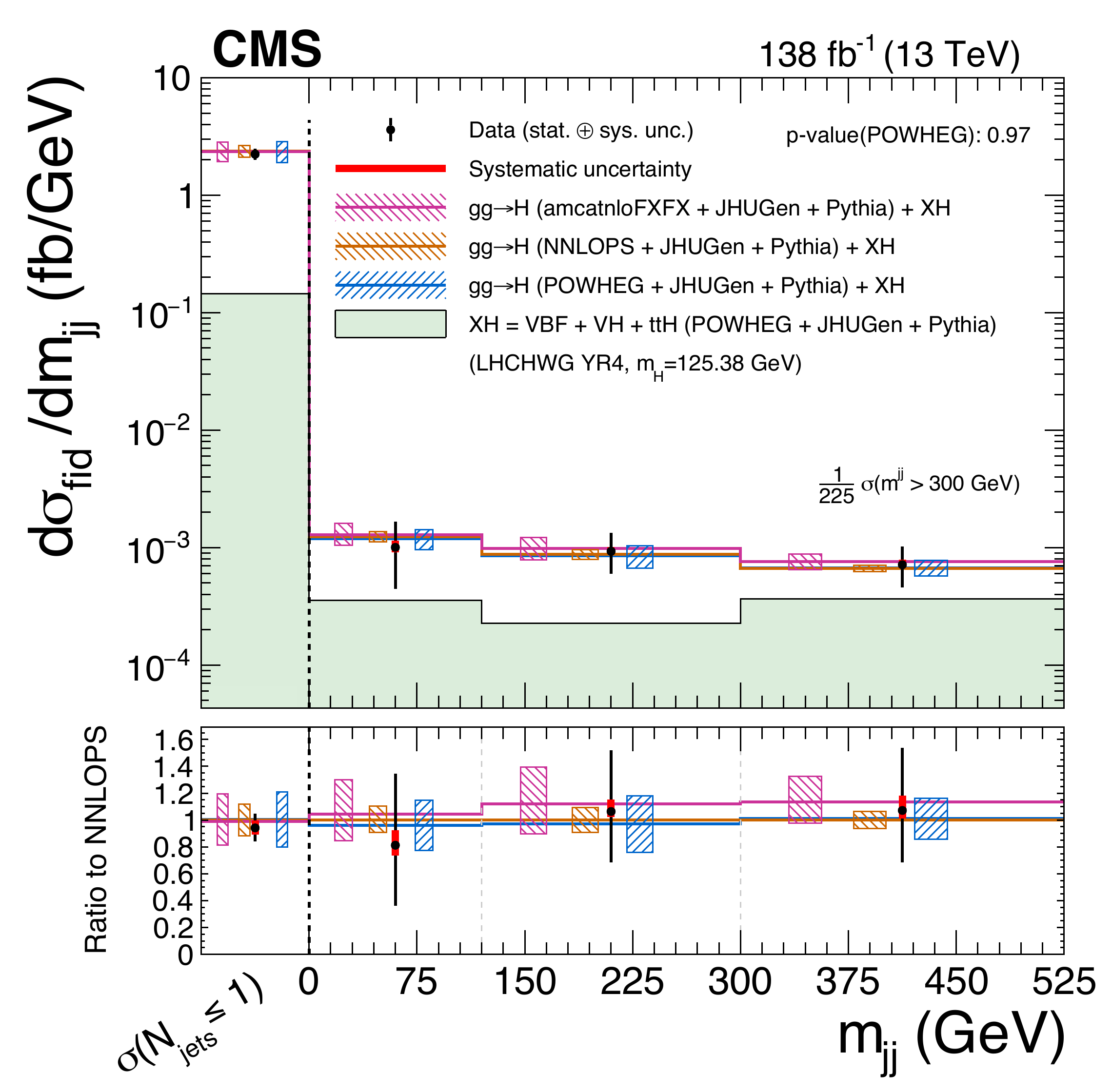}
		\includegraphics[width=0.48\textwidth]{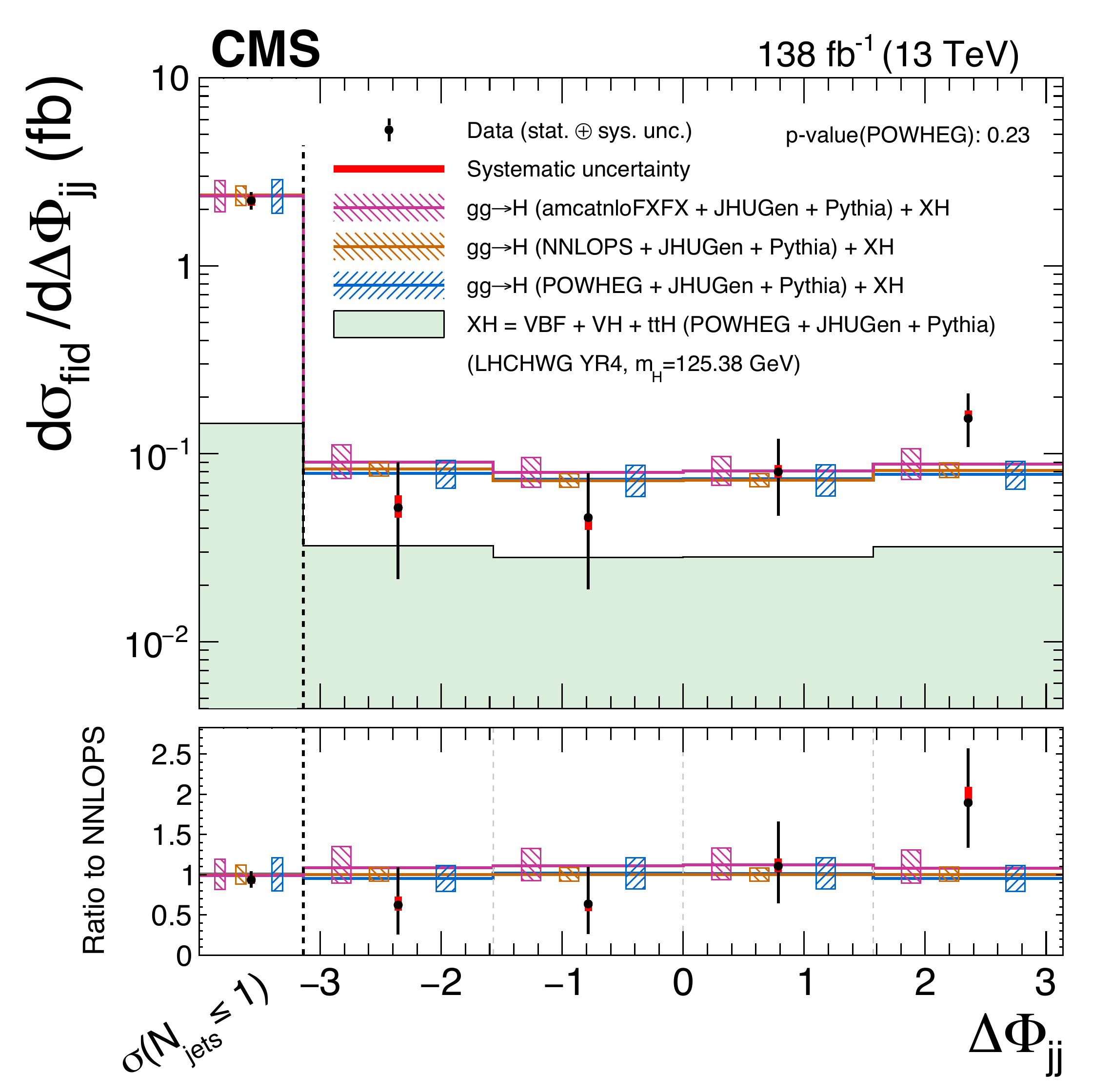}\\
		\includegraphics[width=0.48\textwidth]{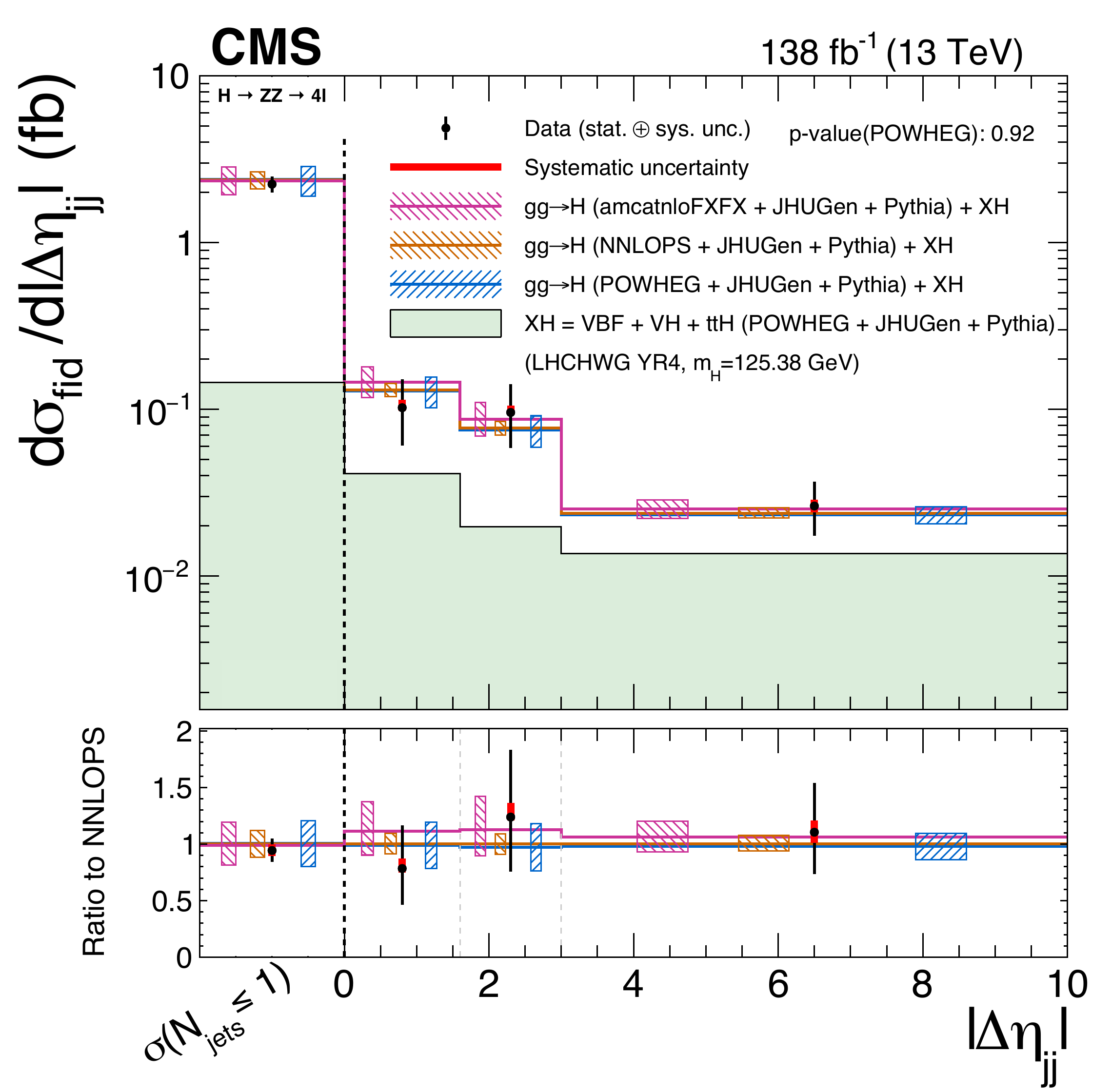}
		\caption{
			Differential cross sections as functions of the invariant mass $m_\text{jj}$ (upper \cmsLeft), the difference in azimuthal angle $\Delta\phi_{\text{jj}}$ (upper \cmsRight) the difference in pseudorapidity $\abs{\Delta\eta_\text{jj}}$ (lower) of the dijet system.
			Upper Left: the fiducial cross section in the last bin is measured for events with $m_\text{jj}>300\GeV$ and normalized to a bin width of $225\GeV$.
			The first bin comprises all events with less than two jets, for which $m_\text{jj}$  is undefined.
			Upper right: the first bin comprises all events with less than two jet, for which  $\abs{\Delta\phi_\text{jj}}$  is undefined.
			Lower: the first bin comprises all events with less than two jet, for which  $\abs{\Delta\eta_\text{jj}}$  is undefined.
			The acceptance and theoretical uncertainties in the differential bins are calculated using the $\ggH$ predictions from three different generators normalized to next-to-next-to-next-to-leading order ($\mathrm{N^3LO}$)~\cite{deFlorian:2016spz}.
			The subdominant component of the signal ($\VBF + \VH + \ttH$) is 		denoted as XH and is fixed to the SM prediction.
			The measured cross sections are compared with the  $\ggH$ predictions from \POWHEG (blue), \textsc{NNLOPS} (orange), and \MGvATNLO (pink).
			The hatched areas correspond to the systematic uncertainties in the theoretical predictions.
			Black points represent the measured fiducial cross sections in each bin, black error bars the total uncertainty in each measurement, red boxes the systematic uncertainties.
			The lower panels display the ratios of the measured cross sections and of the predictions from \POWHEG and \MGvATNLO to the \textsc{NNLOPS} theoretical predictions.
			\label{fig:fidDiJet}}
	\end{figure}
\end{center}

Fiducial cross sections are measured in differential bins of observables sensitive to the \PH boson production.
The results for the $\pt^\PH$ and $\abs{y_\text{\PH}}$ are shown in Fig.~\ref{fig:fidPTH_YH}.
Figure~\ref{fig:fidJets}
shows the measurements of the fiducial cross sections in bins of the number of associated jets and of the \pt of the leading and subleading jet in the event.
The fiducial cross section is also measured in bins of the invariant mass and difference in pseudorapidity of the dijet system, as shown in Fig.~\ref{fig:fidDiJet}. These measurements enhance the sensitivity to phase space regions where $\VBF$ and $\ttH$ production mechanisms dominate and where a larger jet multiplicity is expected.

\begin{center}
	\begin{figure}[!htb]
		\centering
		\includegraphics[width=0.48\textwidth]{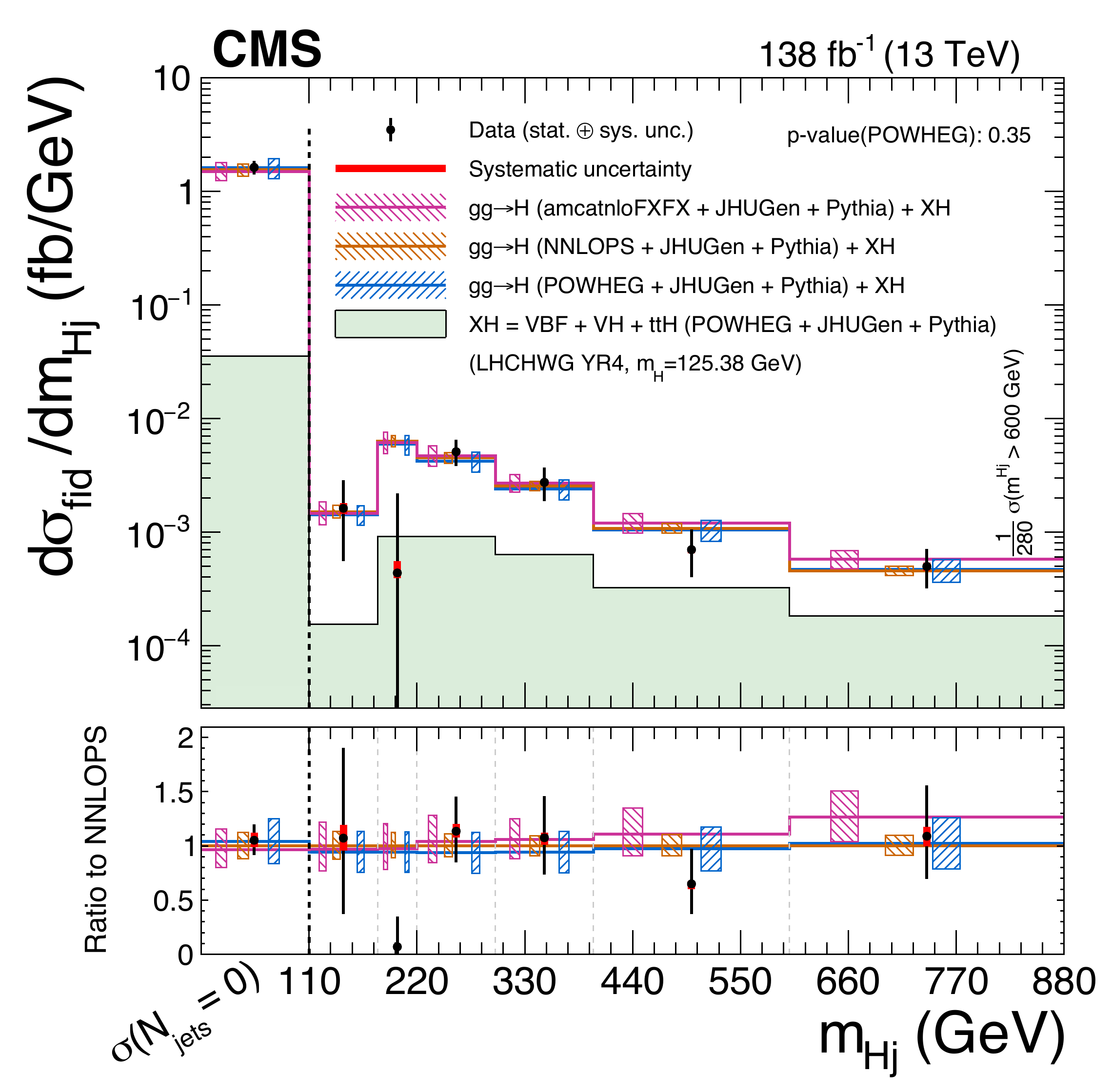}
		\includegraphics[width=0.48\textwidth]{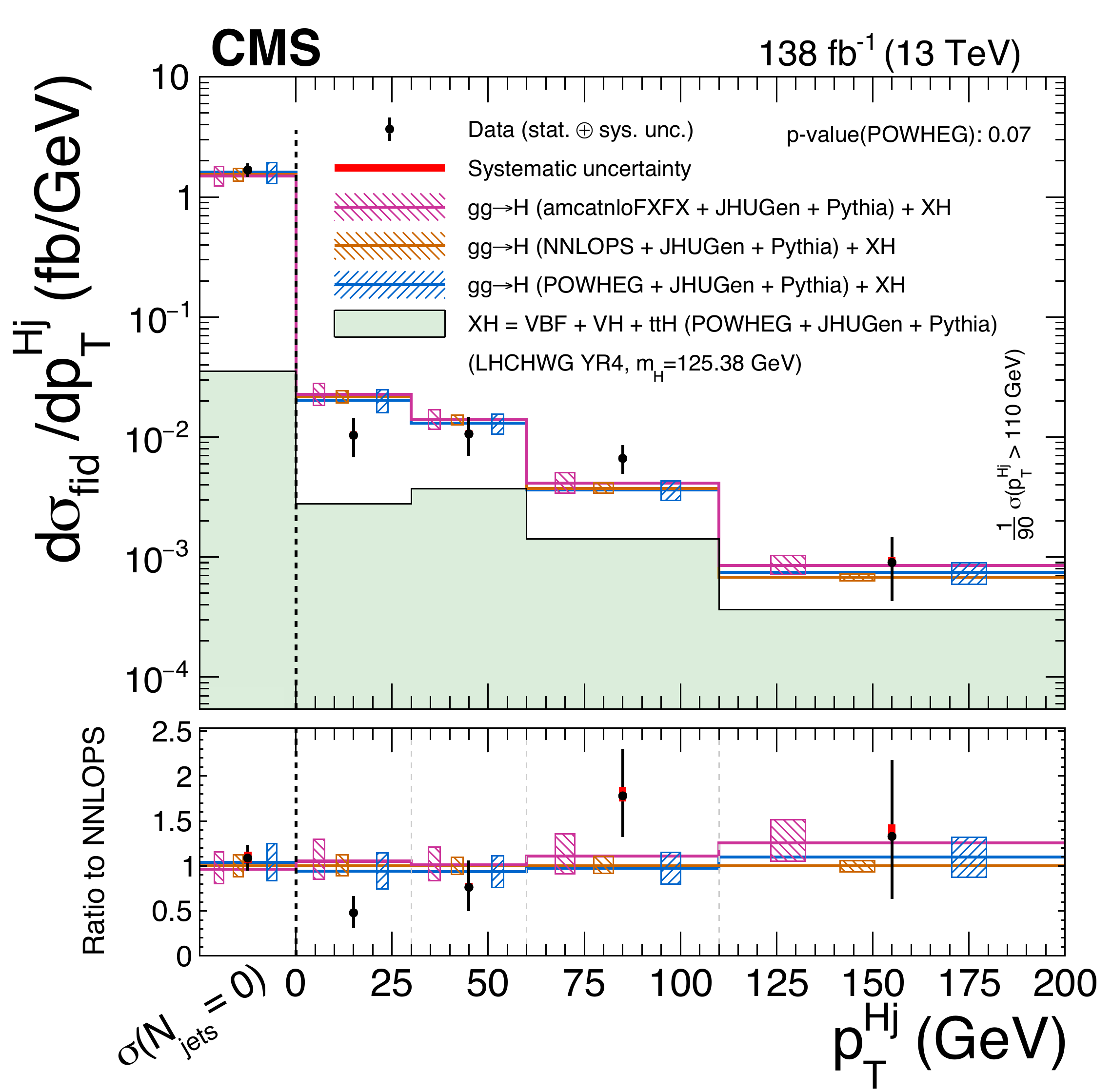}\\
		\includegraphics[width=0.48\textwidth]{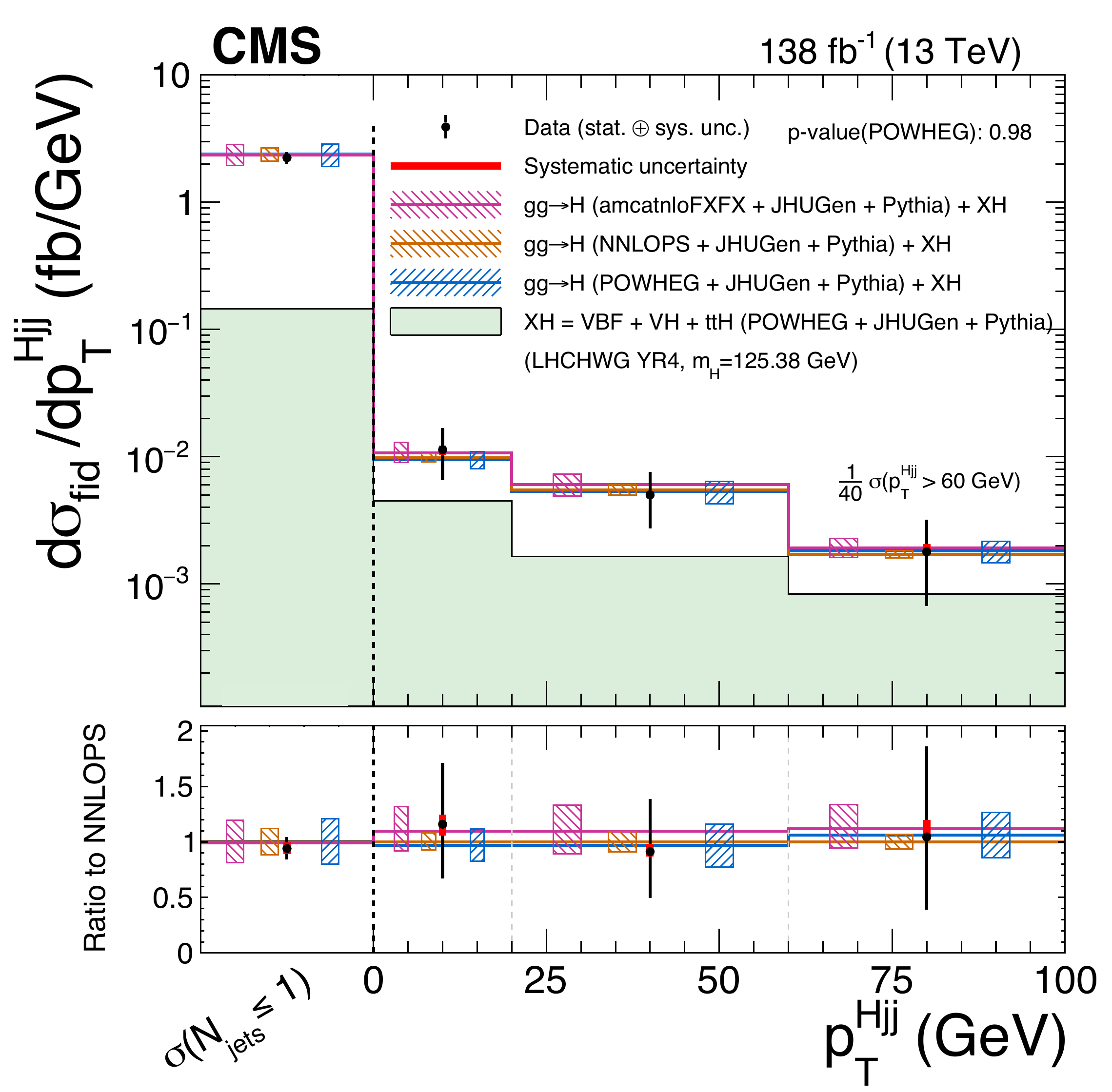}
		\caption{
			Upper left: differential cross sections as functions of the invariant mass of the $\PH+j$ system $m_{\PH j}$, where j is the leading jet in the event.
			The fiducial cross section in the last bin is measured for events with $m_{\PH j}>600\GeV$ and normalized to a bin width of $280\GeV$.
			The first bin comprises all events with less than one jet, for which $m_{\PH j}$  is undefined.
			Upper right: differential cross sections as functions of the transverse momentum of the $\PH+j$ system $\pt^{\PH j}$.
			The fiducial cross section in the last bin is measured for events with $\pt^{\PH j}>110\GeV$ and normalized to a bin width of $90\GeV$.
			The first bin comprises all events with less than one jet, for which $\pt^{\PH j}$  is undefined.
			Lower: differential cross sections as functions of the transverse momentum of the $\PH+jj$ system $\pt^{\PH jj}$.
			The fiducial cross section in the last bin is measured for events with $\pt^{\PH jj}>60\GeV$ and normalized to a bin width of $40\GeV$.
			The first bin comprises all events with less than two jet, for which $\pt^{\PH jj}$  is undefined.
			The acceptance and theoretical uncertainties in the differential bins are calculated using the $\ggH$ predictions from three different generators normalized to next-to-next-to-next-to-leading order ($\mathrm{N^3LO}$)~\cite{deFlorian:2016spz}.
			The subdominant component of the signal ($\VBF + \VH + \ttH$) is denoted as XH and is fixed to the SM prediction.
			The measured cross sections are compared with the  $\ggH$ predictions from \POWHEG (blue), \textsc{NNLOPS} (orange), and \MGvATNLO (pink).
			The hatched areas correspond to the systematic uncertainties in the theoretical predictions.
			Black points represent the measured fiducial cross sections in each bin, black error bars the total uncertainty in each measurement, red boxes the systematic uncertainties.
			The lower panels display the ratios of the measured cross sections and of the predictions from \POWHEG and \MGvATNLO to the \textsc{NNLOPS} theoretical predictions.
			\label{fig:fidHJ}}
	\end{figure}
\end{center}

Cross sections in bins of observables of the $\PH+j$ and $\PH+jj$ systems are also measured.
The results in differential bins of the invariant mass and \pt of the $\PH+j$ system are presented in Fig.~\ref{fig:fidHJ} together with the results in differential bins of the \pt of the $\PH+jj$ system.

\begin{center}
	\begin{figure}[!htb]
		\centering
		\includegraphics[width=0.48\textwidth]{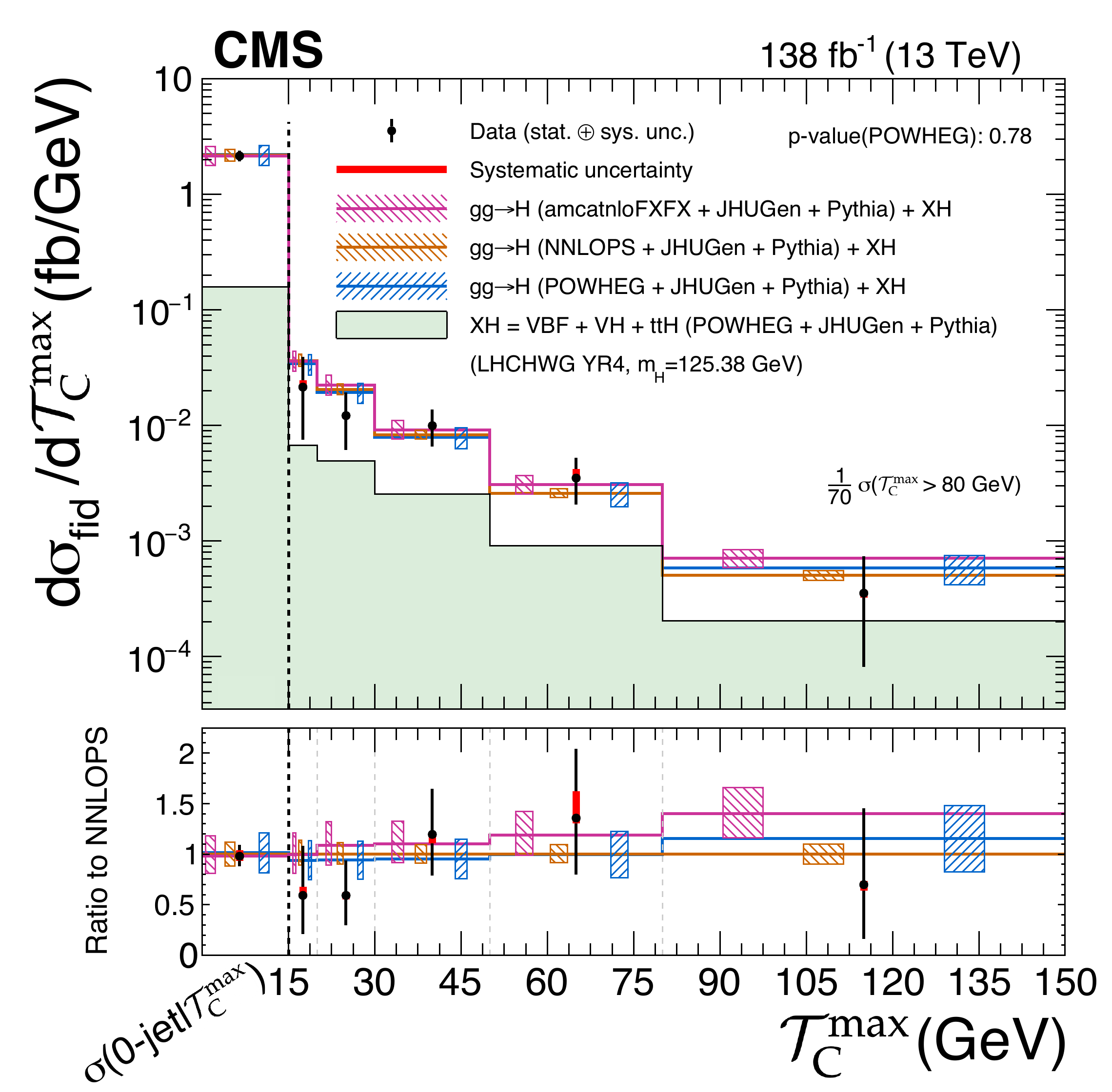}
		\includegraphics[width=0.48\textwidth]{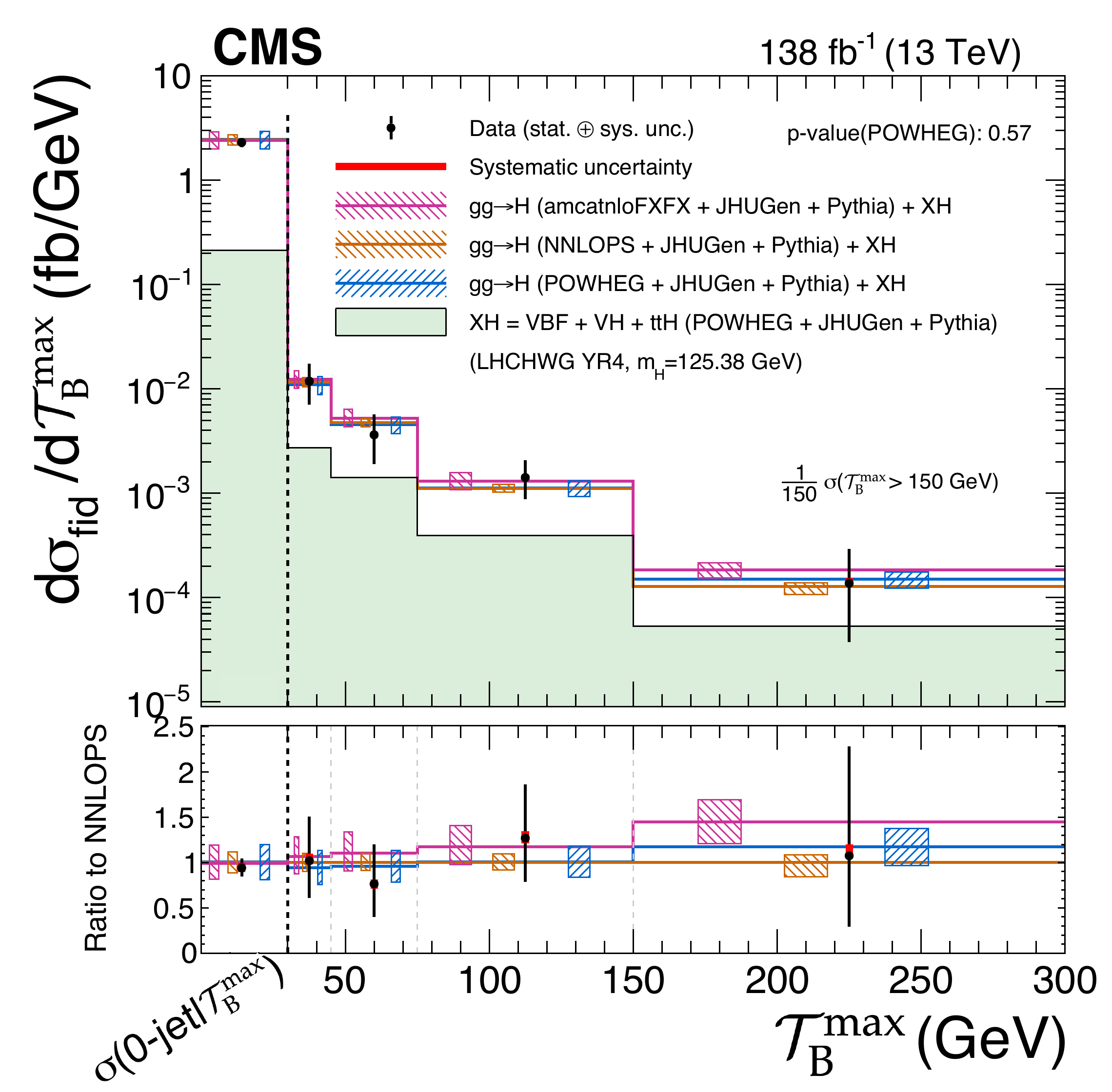}
		
		\caption{
			Left: differential cross sections as functions of the rapidity-weighed jet veto $\mathcal{T}_{\text{C}}^{\text{max}}$.
			The fiducial cross section in the last bin is measured for events with $\mathcal{T}_{\text{C}}^{\text{max}}>80\GeV$ and normalized to a bin width of $70\GeV$.
			The first bin comprises all events in the 0-jet phase space region redefined as a function of  $\mathcal{T}_{\text{C}}^{\text{max}}$, \ie, events with less than one jet, for which $\mathcal{T}_{\text{C}}^{\text{max}}$ is undefined, and events with $\mathcal{T}_{\text{C}}^{\text{max}}<15\GeV$.
			Right: differential cross sections as functions of the rapidity-weighed jet veto $\mathcal{T}_{\text{B}}^{\text{max}}$.
			The fiducial cross section in the last bin is measured for events with $\mathcal{T}_{\text{B}}^{\text{max}}>150\GeV$ and normalized to a bin width of $150\GeV$.
			The first bin comprises all events in the 0-jet phase space region redefined as a function of  $\mathcal{T}_{\text{B}}^{\text{max}}$, \ie, events with less than one jet, for which $\mathcal{T}_{\text{B}}^{\text{max}}$ is undefined, and events with $\mathcal{T}_{\text{B}}^{\text{max}}<30\GeV$.
			The acceptance and theoretical uncertainties in the differential bins are calculated using the $\ggH$ predictions from three different generators normalized to next-to-next-to-next-to-leading order ($\mathrm{N^3LO}$)~\cite{deFlorian:2016spz}.
			The subdominant component of the signal ($\VBF + \VH + \ttH$) is denoted as XH and is fixed to the SM prediction.
			The measured cross sections are compared with the  $\ggH$ predictions from \POWHEG (blue), \textsc{NNLOPS} (orange), and \MGvATNLO (pink).
			The hatched areas correspond to the systematic uncertainties in the theoretical predictions.
			Black points represent the measured fiducial cross sections in each bin, black error bars the total uncertainty in each measurement, red boxes the systematic uncertainties.
			The lower panels display the ratios of the measured cross sections and of the predictions from \POWHEG and \MGvATNLO to the \textsc{NNLOPS} theoretical predictions.
			\label{fig:fidTC_TB}}
	\end{figure}
\end{center}

Cross sections are also measured in differential bins of the rapidity-weighted jet vetoes introduced in Section~\ref{sec:observables}, to enhance the sensitivity to phase space regions that probe directly the departure from LO kinematics and the QCD emission pattern. Figure~\ref{fig:fidTC_TB} presents the results for  $\mathcal{T}_{\text{C}}^{\text{max}}$ and $\mathcal{T}_{\text{B}}^{\text{max}}$.

\clearpage

\subsection{Differential cross sections: decay}

In this section the measurements of fiducial cross sections in differential bins of observables sensitive to the $\HZZfl$ decay are presented.
Since the final state is sensitive to interference effects in the case of identical particles, the results for decay observables are also presented separately for same- and different-flavor final states.
This ensures a complete coverage of the whole phase space and a more model-independent set of results.

\begin{center}
	\begin{figure}[!htb]
		\centering
		\includegraphics[width=0.48\textwidth]{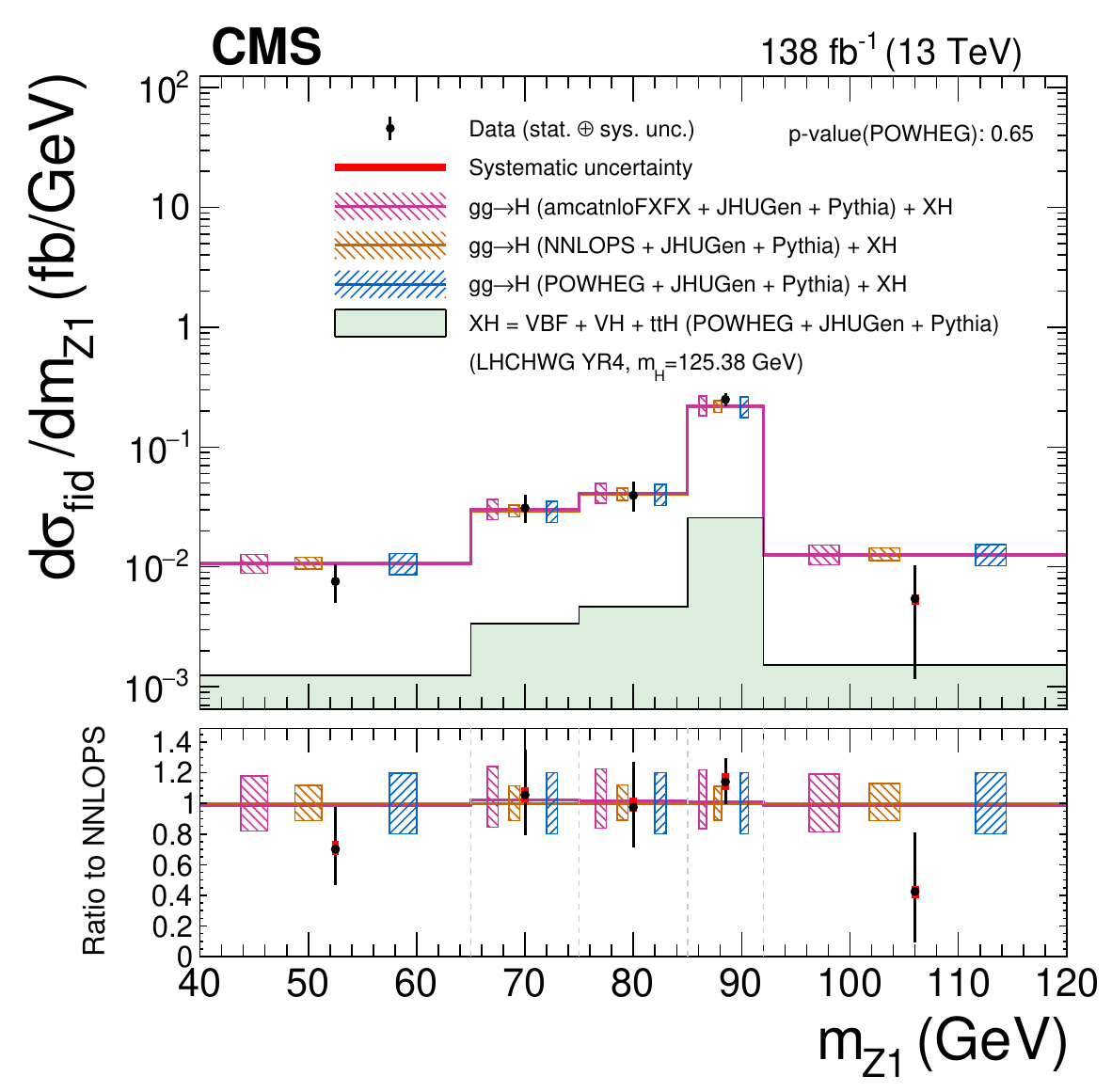}\\
		\includegraphics[width=0.48\textwidth]{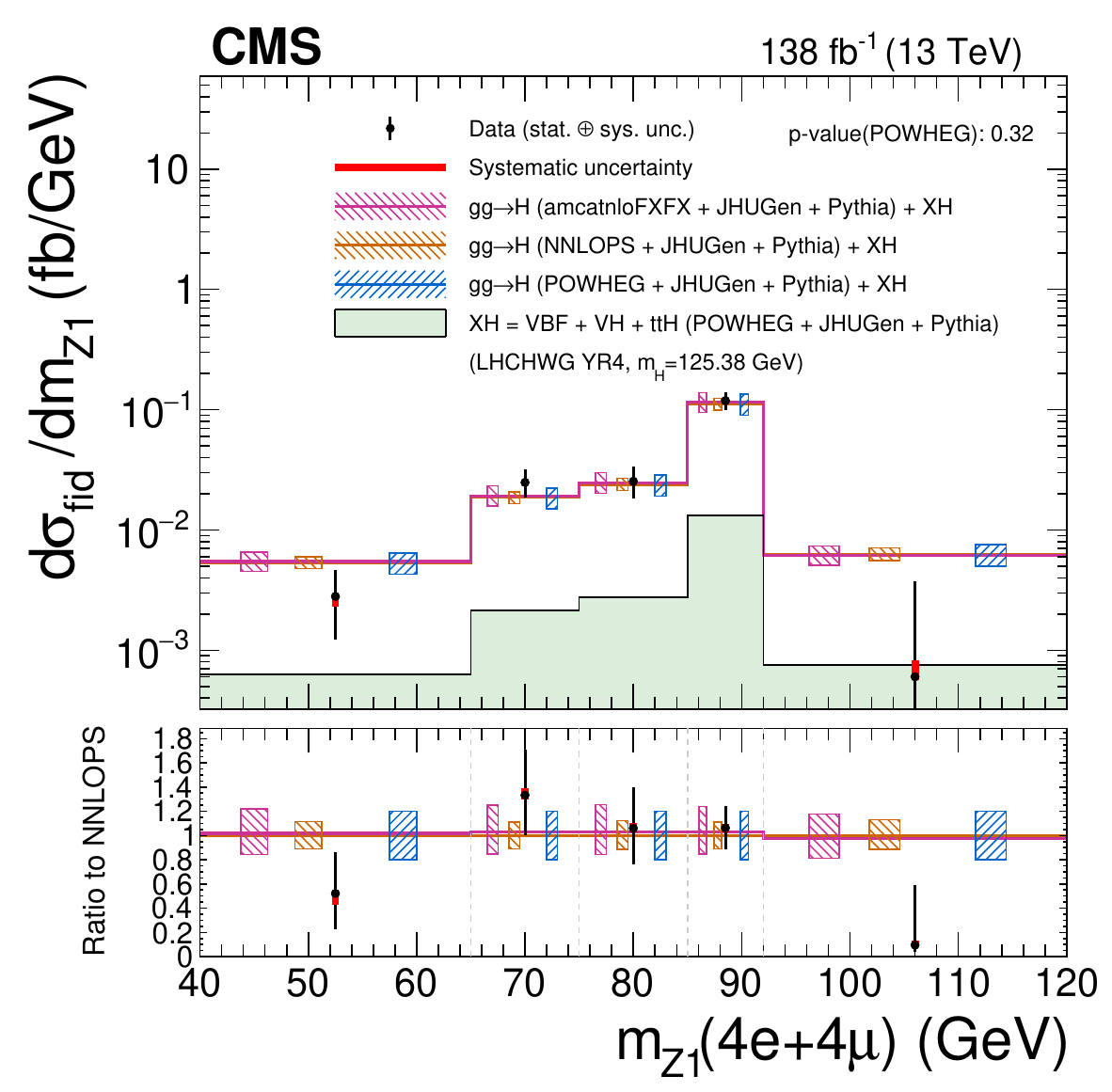}
		\includegraphics[width=0.48\textwidth]{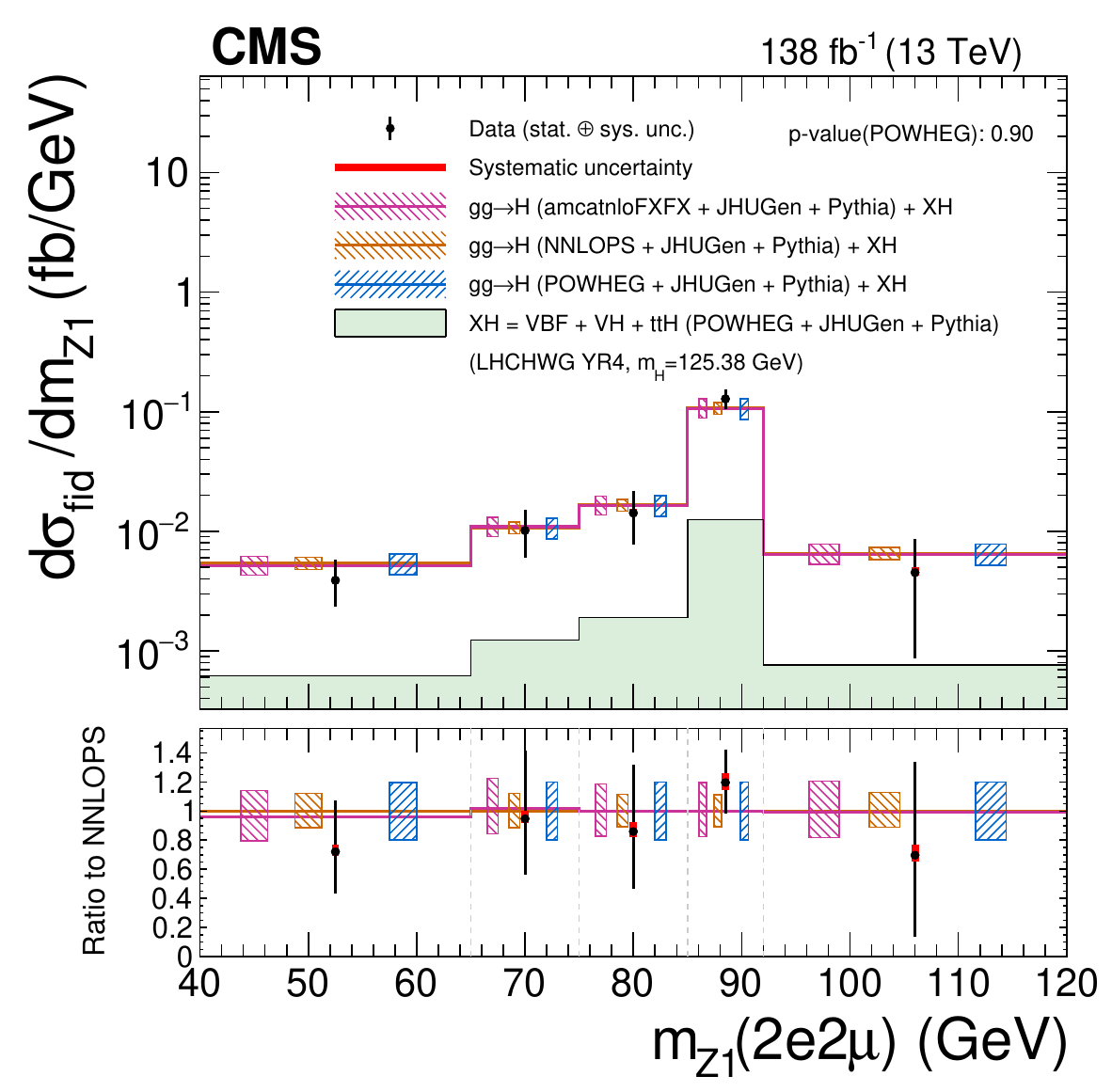}
		\caption{
			Differential cross sections as functions of the invariant mass of the leading dilepton pair $m_{\PZ_{1}}$ in the $4\ell$ (upper) and in the same-flavor (lower left) and different-flavor (lower right) final states.
			The acceptance and theoretical uncertainties in the differential bins are calculated using the $\ggH$ predictions from three different generators normalized to next-to-next-to-next-to-leading order ($\mathrm{N^3LO}$)~\cite{deFlorian:2016spz}.
			The subdominant component of the signal ($\VBF + \VH + \ttH$) is denoted as XH and is fixed to the SM prediction.
			The measured cross sections are compared with the  $\ggH$ predictions from \POWHEG (blue), \textsc{NNLOPS} (orange), and \MGvATNLO (pink).
			The hatched areas correspond to the systematic uncertainties in the theoretical predictions.
			Black points represent the measured fiducial cross sections in each bin, black error bars the total uncertainty in each measurement, red boxes the systematic uncertainties.
			The lower panels display the ratios of the measured cross sections and of the predictions from \POWHEG and \MGvATNLO to the \textsc{NNLOPS} theoretical predictions.
			\label{fig:fidMZ1}}
	\end{figure}
\end{center}

\begin{center}
	\begin{figure}[!htb]
		\centering
		\includegraphics[width=0.48\textwidth]{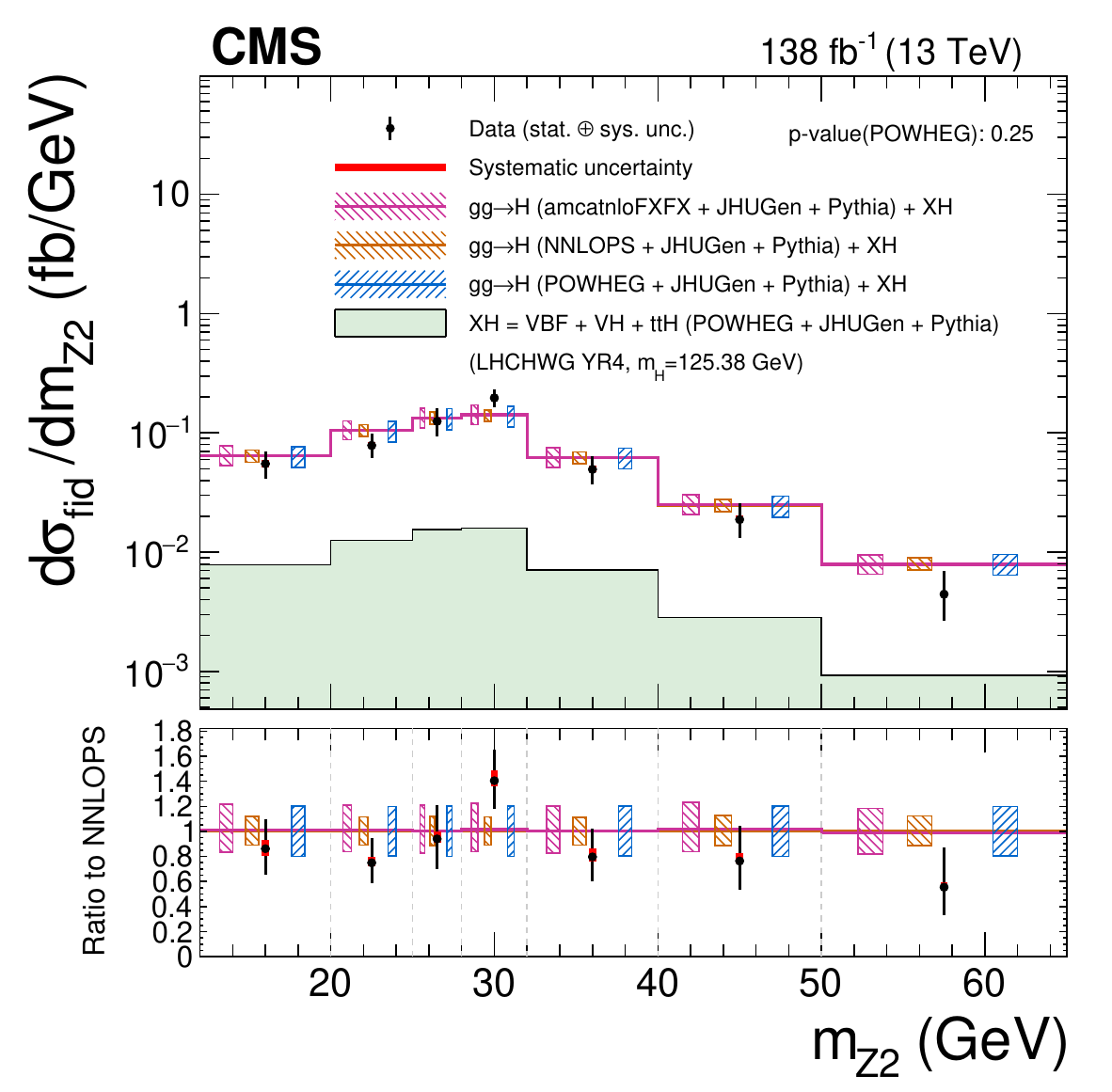}\\
		\includegraphics[width=0.48\textwidth]{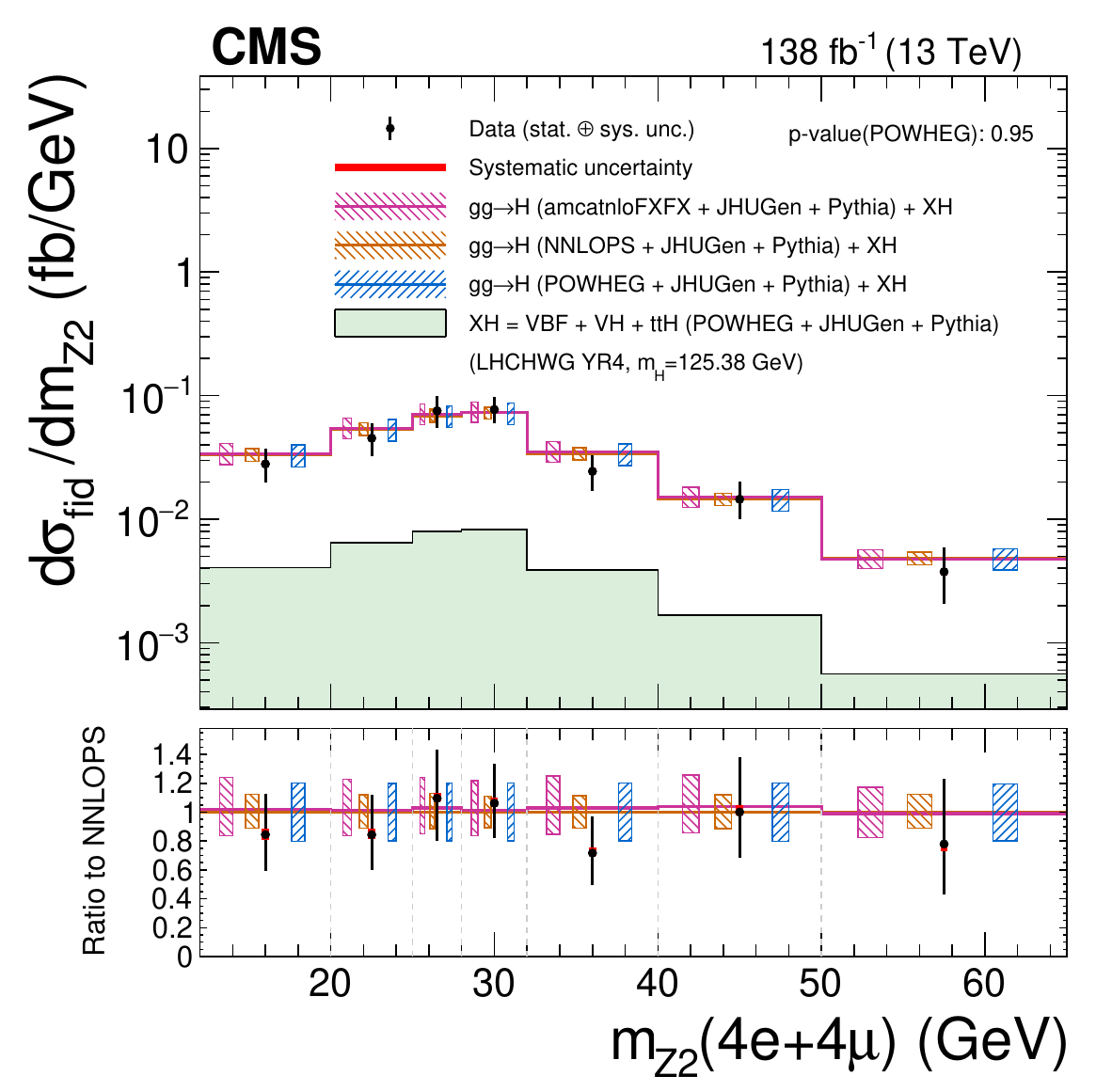}
		\includegraphics[width=0.48\textwidth]{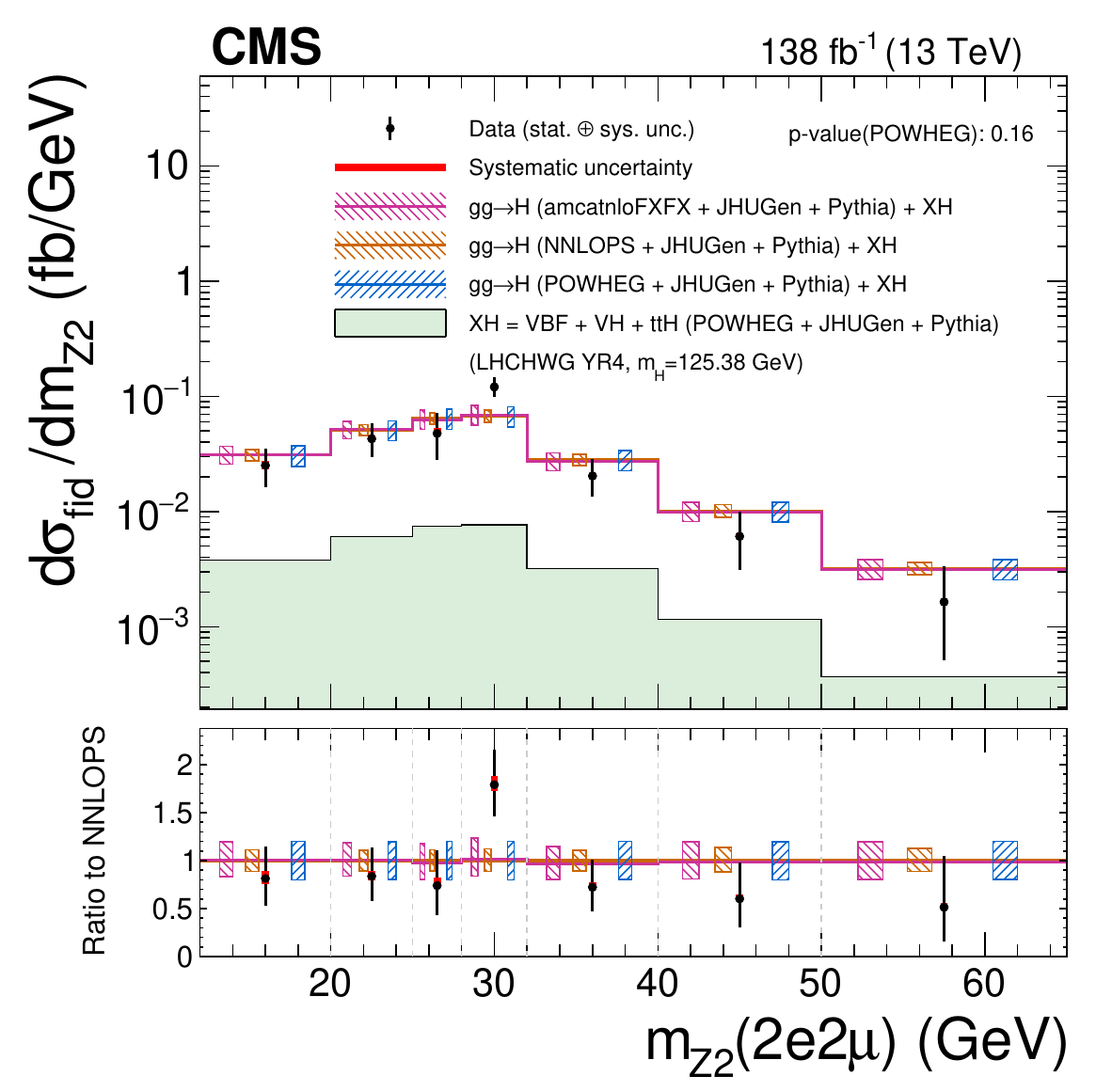}
		\caption{
			Differential cross sections as functions of the invariant mass of the subleading dilepton pair $m_{\PZ_{2}}$ in the $4\ell$ (upper) and in the same-flavor (lower left) and different-flavor (lower right) final states.
			The acceptance and theoretical uncertainties in the differential bins are calculated using the $\ggH$ predictions from three different generators normalized to next-to-next-to-next-to-leading order ($\mathrm{N^3LO}$)~\cite{deFlorian:2016spz}.
			The subdominant component of the signal ($\VBF + \VH + \ttH$) is denoted as XH and is fixed to the SM prediction.
			The measured cross sections are compared with the  $\ggH$ predictions from \POWHEG (blue), \textsc{NNLOPS} (orange), and \MGvATNLO (pink).
			The hatched areas correspond to the systematic uncertainties in the theoretical predictions.
			Black points represent the measured fiducial cross sections in each bin, black error bars the total uncertainty in each measurement, red boxes the systematic uncertainties.
			The lower panels display the ratios of the measured cross sections and of the predictions from \POWHEG and \MGvATNLO to the \textsc{NNLOPS} theoretical predictions.
			\label{fig:fidMZ2}}
	\end{figure}
\end{center}

\begin{center}
	\begin{figure}[!htb]
		\centering
		\includegraphics[width=0.48\textwidth]{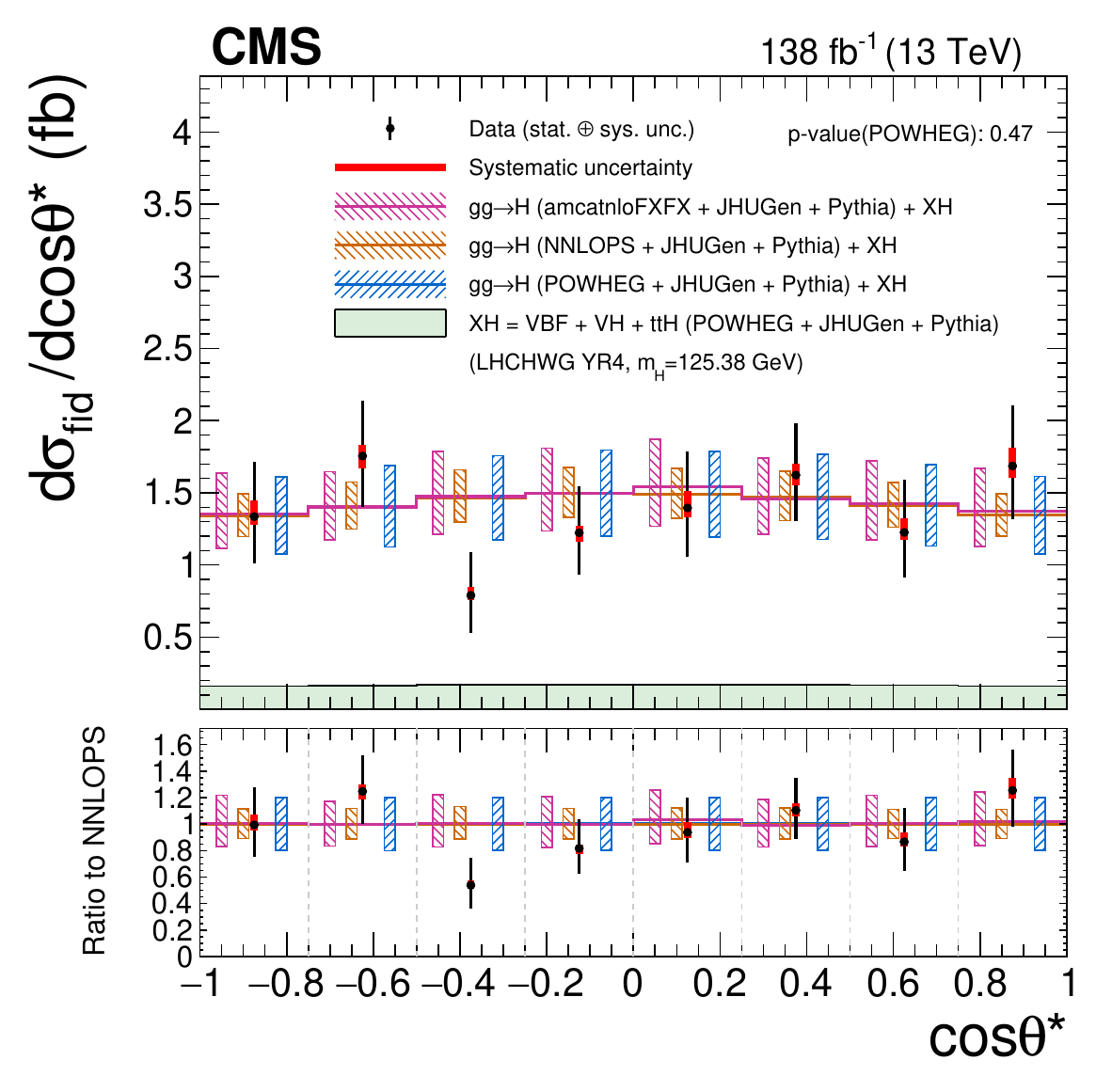}\\
		\includegraphics[width=0.48\textwidth]{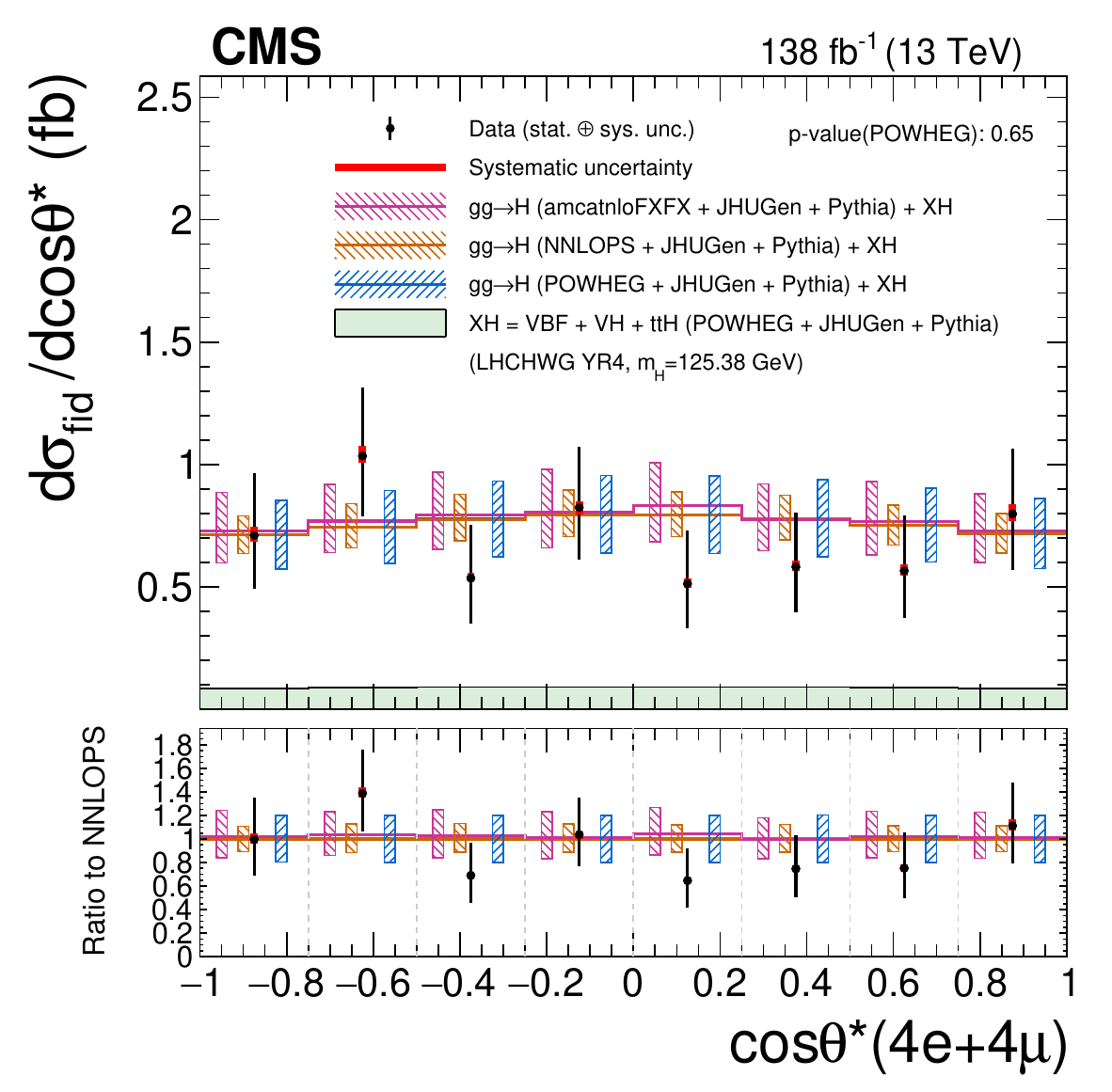}
		\includegraphics[width=0.48\textwidth]{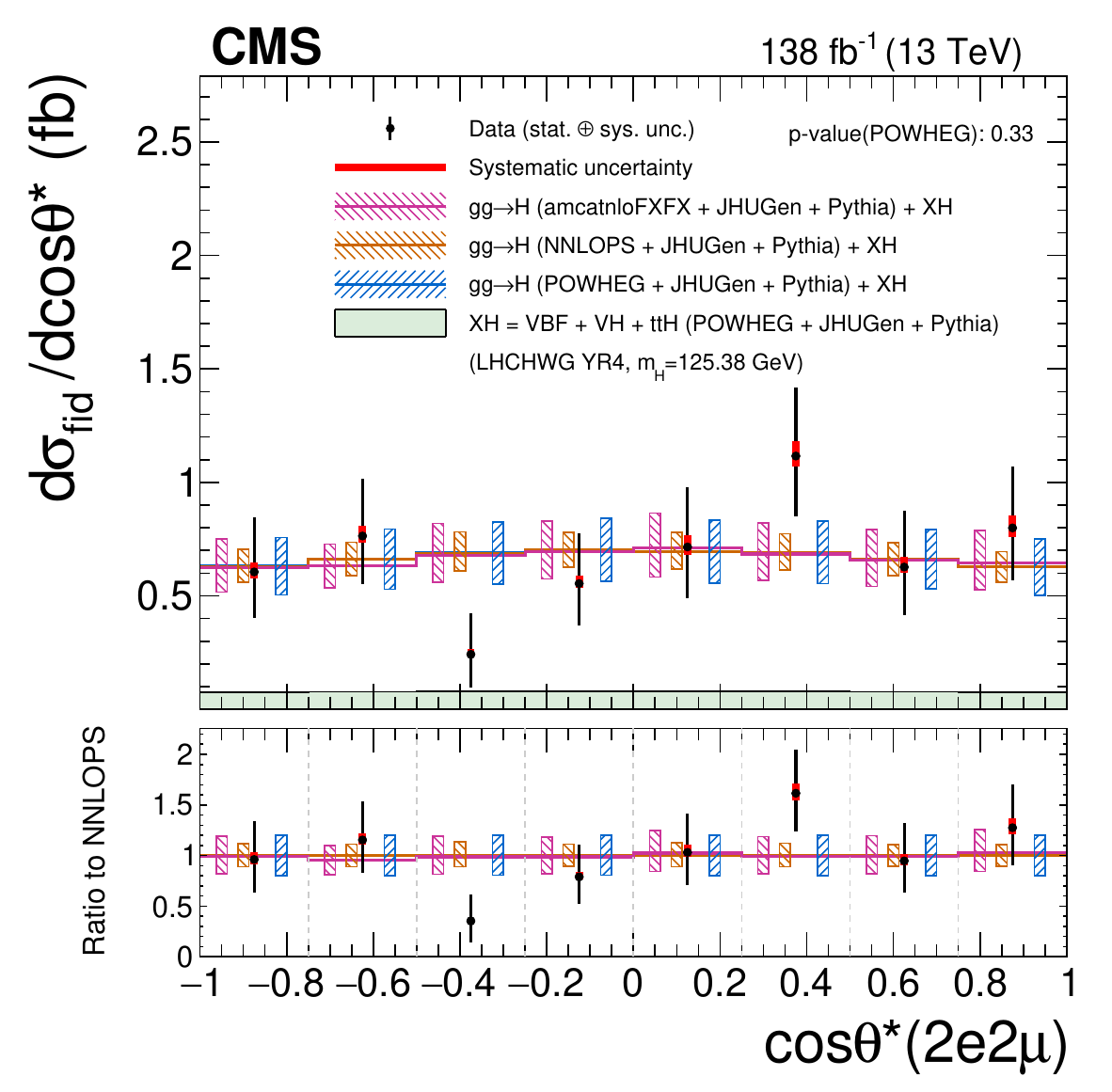}	
		\caption{
			Differential cross sections as functions of  $\cos \theta^*$ in the $4\ell$ (upper) and in the same-flavor (lower left) and different-flavor (lower right) final states.
			The acceptance and theoretical uncertainties in the differential bins are calculated using the $\ggH$ predictions from three different generators normalized to next-to-next-to-next-to-leading order ($\mathrm{N^3LO}$)~\cite{deFlorian:2016spz}.
			The subdominant component of the signal ($\VBF + \VH + \ttH$) is denoted as XH and is fixed to the SM prediction.
			The measured cross sections are compared with the  $\ggH$ predictions from \POWHEG (blue), \textsc{NNLOPS} (orange), and \MGvATNLO (pink).
			The hatched areas correspond to the systematic uncertainties in the theoretical predictions.
			Black points represent the measured fiducial cross sections in each bin, black error bars the total uncertainty in each measurement, red boxes the systematic uncertainties.
			The lower panels display the ratios of the measured cross sections and of the predictions from \POWHEG and \MGvATNLO to the \textsc{NNLOPS} theoretical predictions.
			\label{fig:fidCOSTS}}
	\end{figure}
\end{center}

\begin{center}
	\begin{figure}[!htb]
		\centering
		\includegraphics[width=0.48\textwidth]{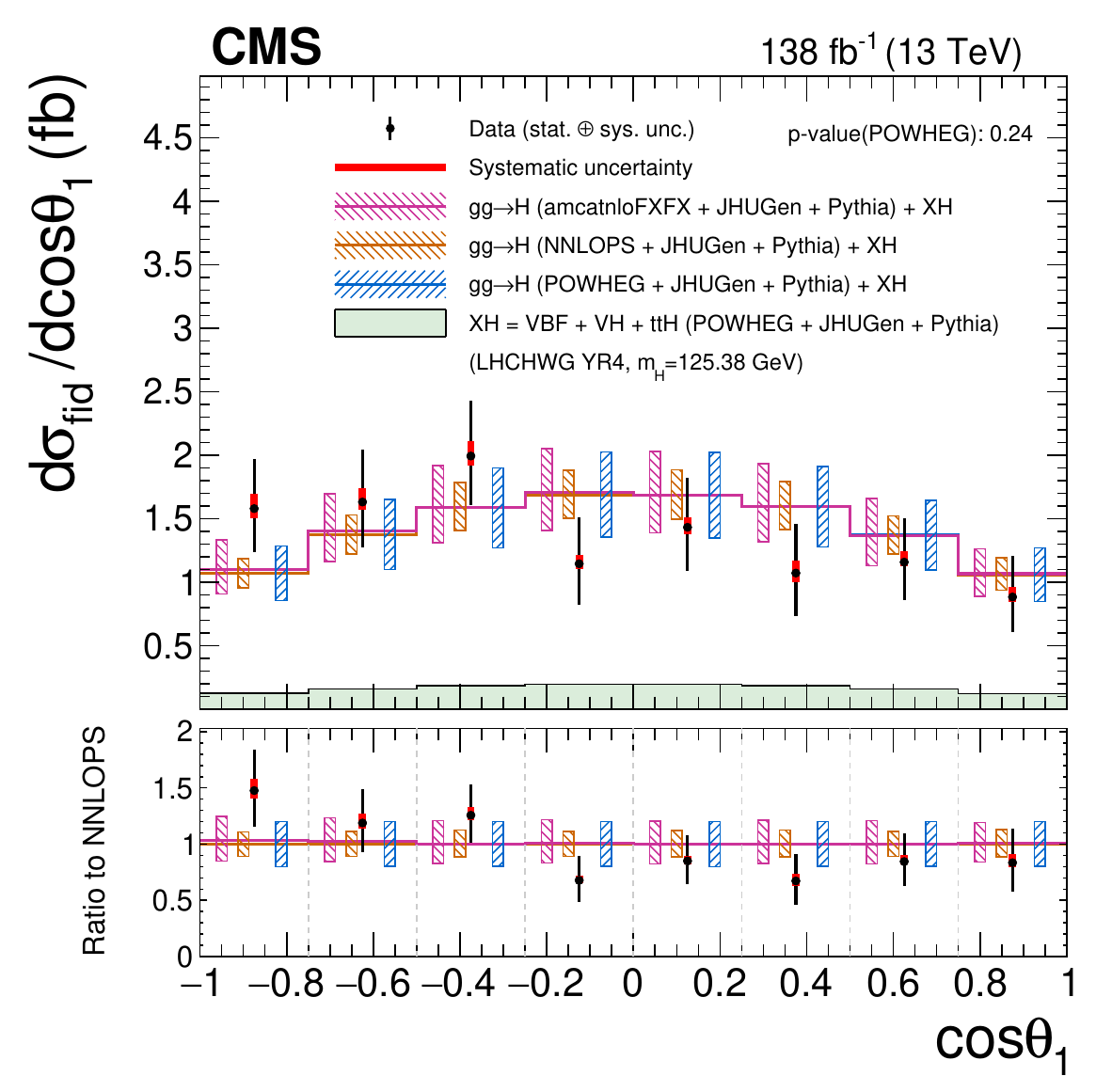}\\
		\includegraphics[width=0.48\textwidth]{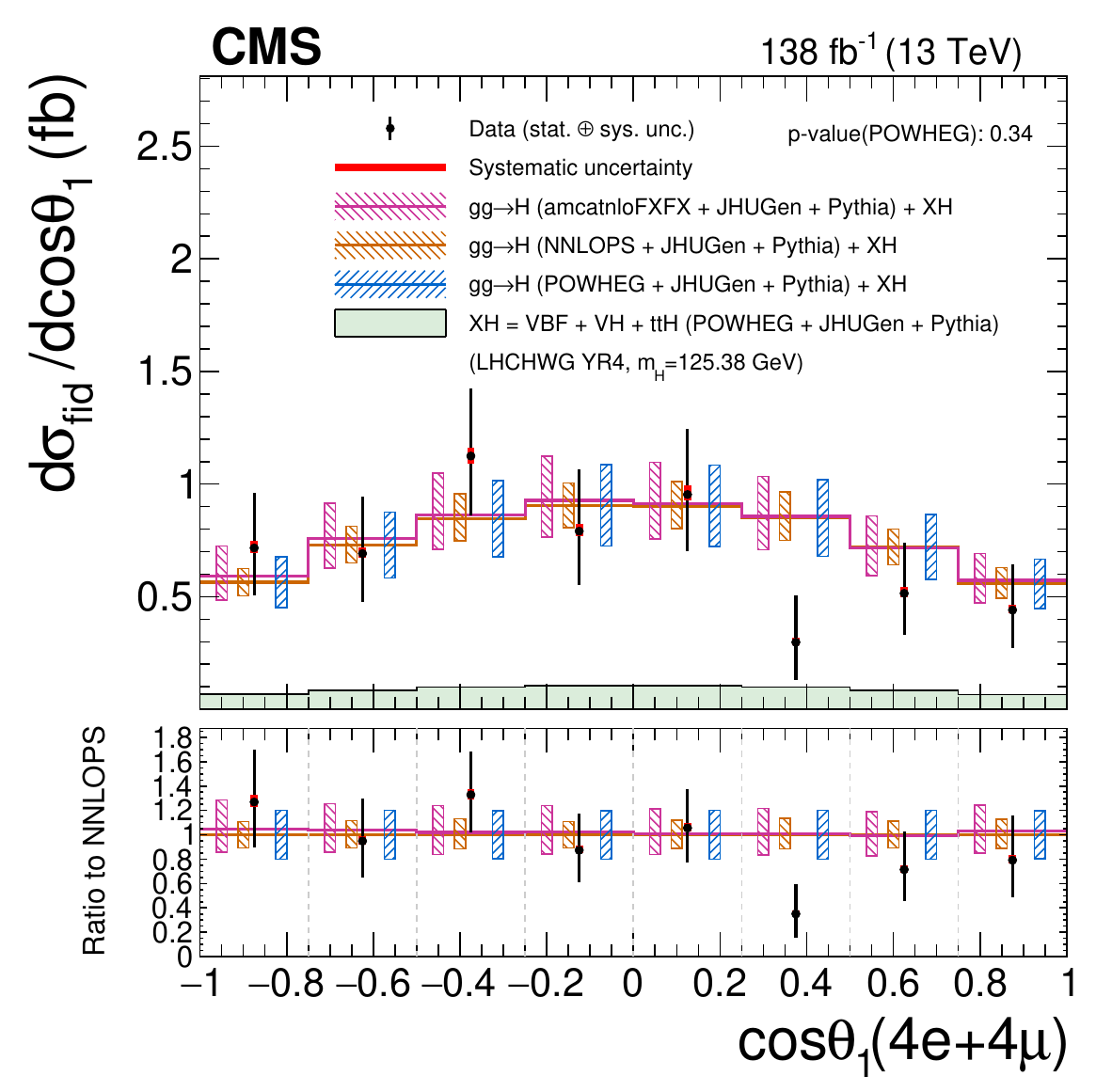}
		\includegraphics[width=0.48\textwidth]{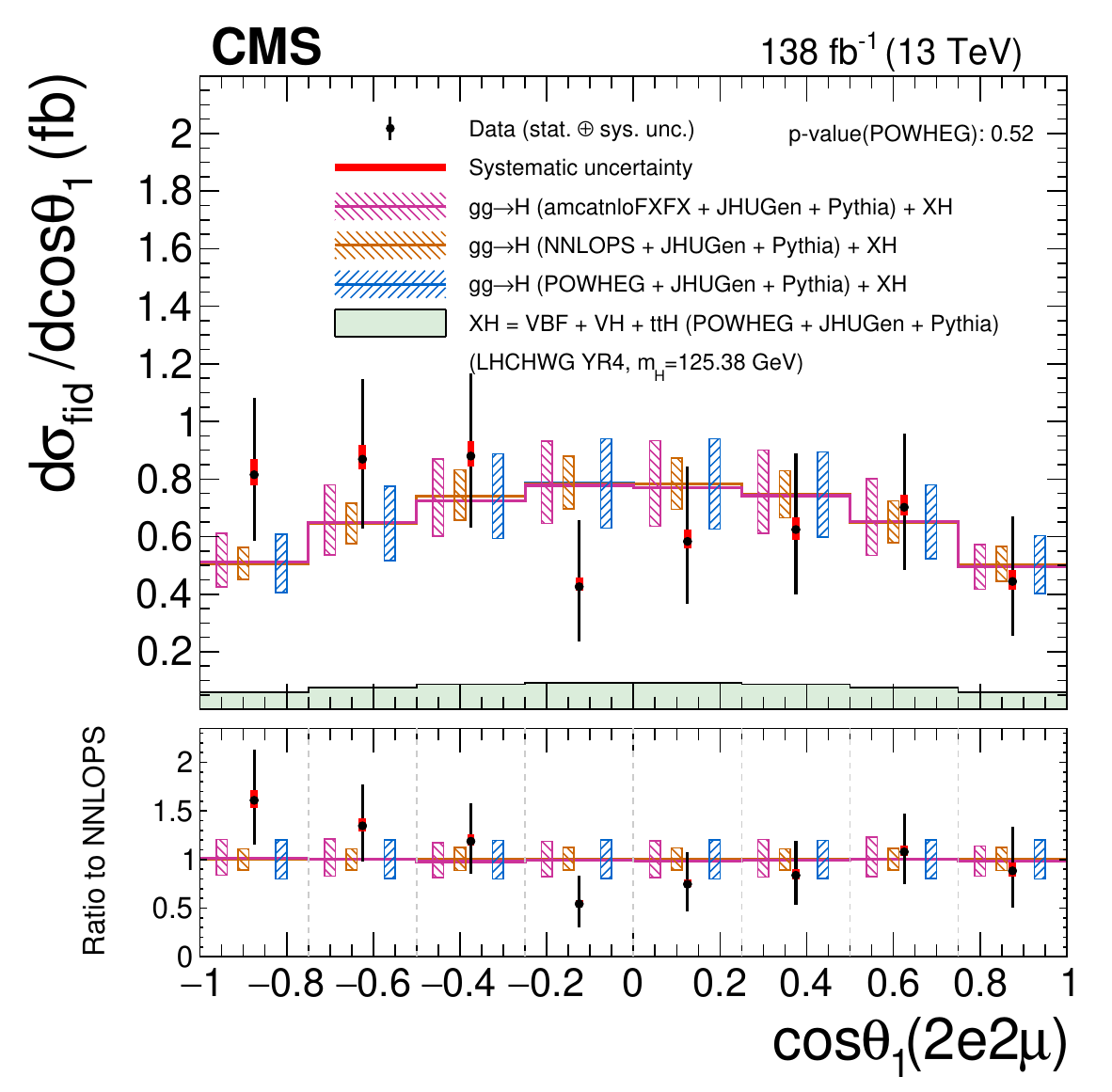}
		\caption{
			Differential cross sections as functions of $\cos \theta_\text{1}$ in the $4\ell$ (upper) and in the same-flavor (lower left) and different-flavor (lower right) final states.
			The acceptance and theoretical uncertainties in the differential bins are calculated using the $\ggH$ predictions from three different generators normalized to next-to-next-to-next-to-leading order ($\mathrm{N^3LO}$)~\cite{deFlorian:2016spz}.
			The subdominant component of the signal ($\VBF + \VH + \ttH$) is denoted as XH and is fixed to the SM prediction.
			The measured cross sections are compared with the  $\ggH$ predictions from \POWHEG (blue), \textsc{NNLOPS} (orange), and \MGvATNLO (pink).
			The hatched areas correspond to the systematic uncertainties in the theoretical predictions.
			Black points represent the measured fiducial cross sections in each bin, black error bars the total uncertainty in each measurement, red boxes the systematic uncertainties.
			The lower panels display the ratios of the measured cross sections and of the predictions from \POWHEG and \MGvATNLO to the \textsc{NNLOPS} theoretical predictions.
			\label{fig:fidCOSZ1}}
	\end{figure}
\end{center}

\begin{center}
	\begin{figure}[!htb]
		\centering
		\includegraphics[width=0.48\textwidth]{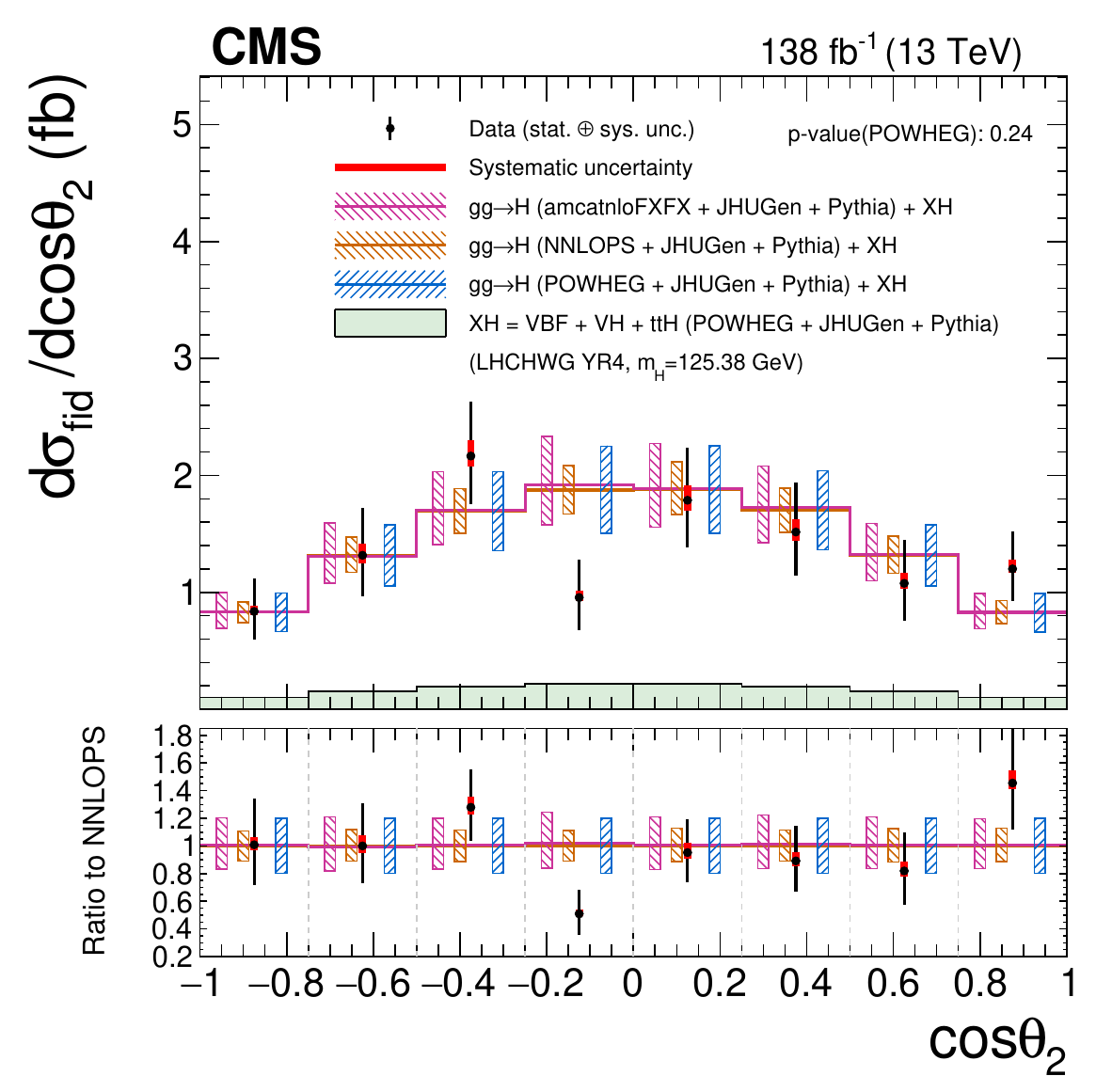}\\
		\includegraphics[width=0.48\textwidth]{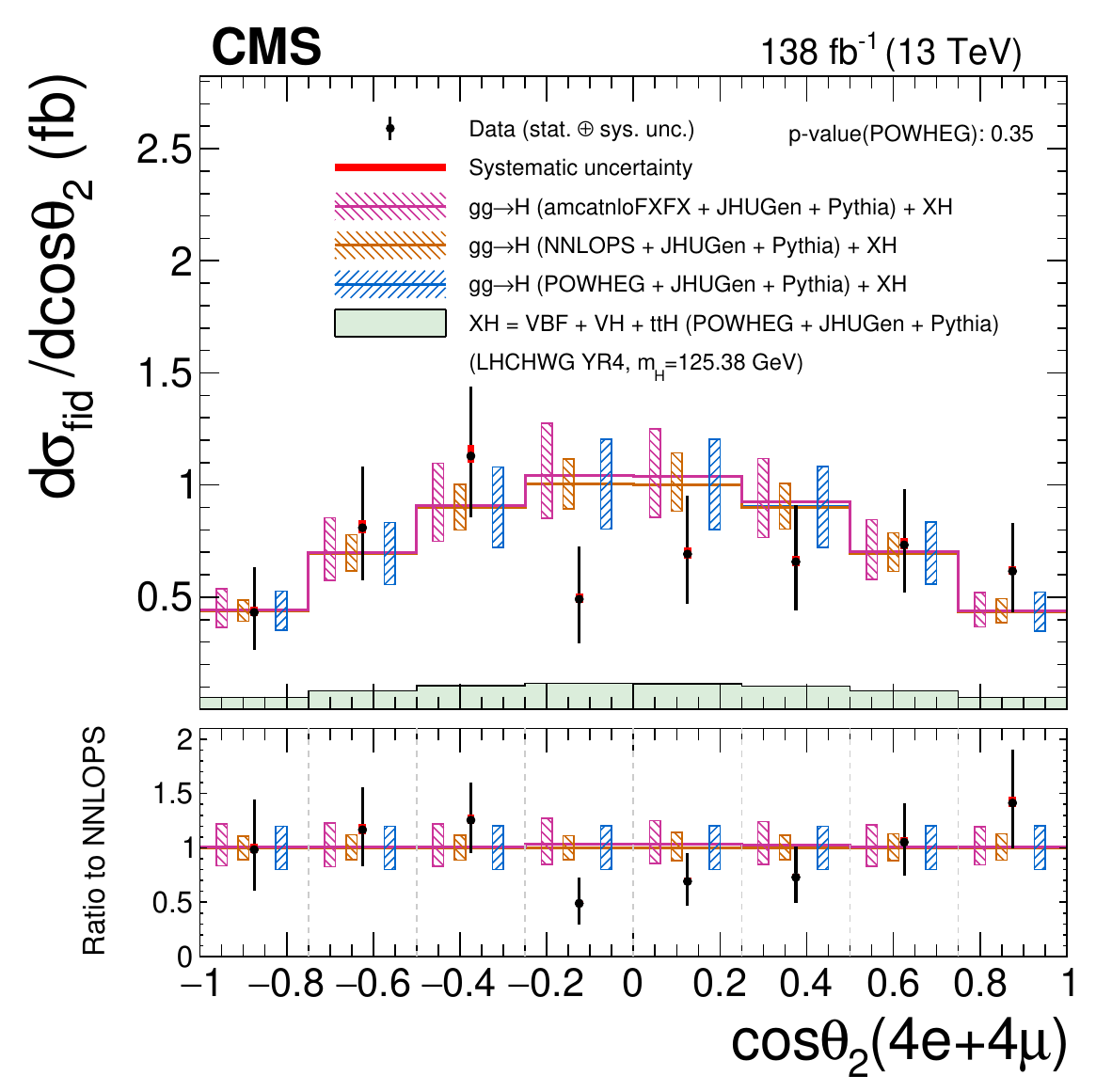}
		\includegraphics[width=0.48\textwidth]{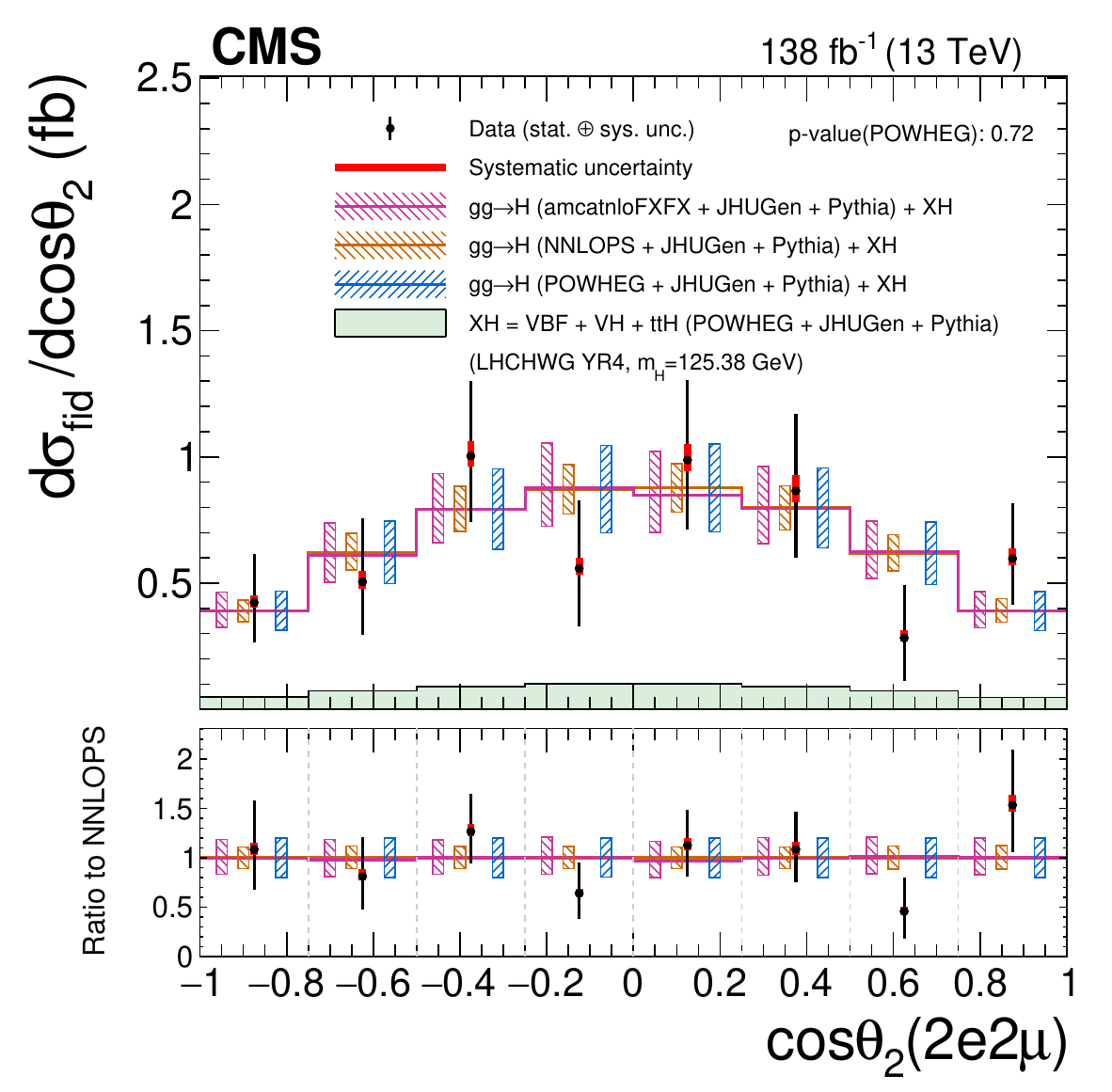}
		\caption{
			Differential cross sections as functions of $\cos \theta_\text{2}$ in the $4\ell$ (upper) and in the same-flavor (lower left) and different-flavor (lower right) final states.
			The acceptance and theoretical uncertainties in the differential bins are calculated using the $\ggH$ predictions from three different generators normalized to next-to-next-to-next-to-leading order ($\mathrm{N^3LO}$)~\cite{deFlorian:2016spz}.
			The subdominant component of the signal ($\VBF + \VH + \ttH$) is denoted as XH and is fixed to the SM prediction.
			The measured cross sections are compared with the  $\ggH$ predictions from \POWHEG (blue), \textsc{NNLOPS} (orange), and \MGvATNLO (pink).
			The hatched areas correspond to the systematic uncertainties in the theoretical predictions.
			Black points represent the measured fiducial cross sections in each bin, black error bars the total uncertainty in each measurement, red boxes the systematic uncertainties.
			The lower panels display the ratios of the measured cross sections and of the predictions from \POWHEG and \MGvATNLO to the \textsc{NNLOPS} theoretical predictions.
			\label{fig:fidCOSZ2}}
	\end{figure}
\end{center}

\begin{center}
	\begin{figure}[!htb]
		\centering
		\includegraphics[width=0.48\textwidth]{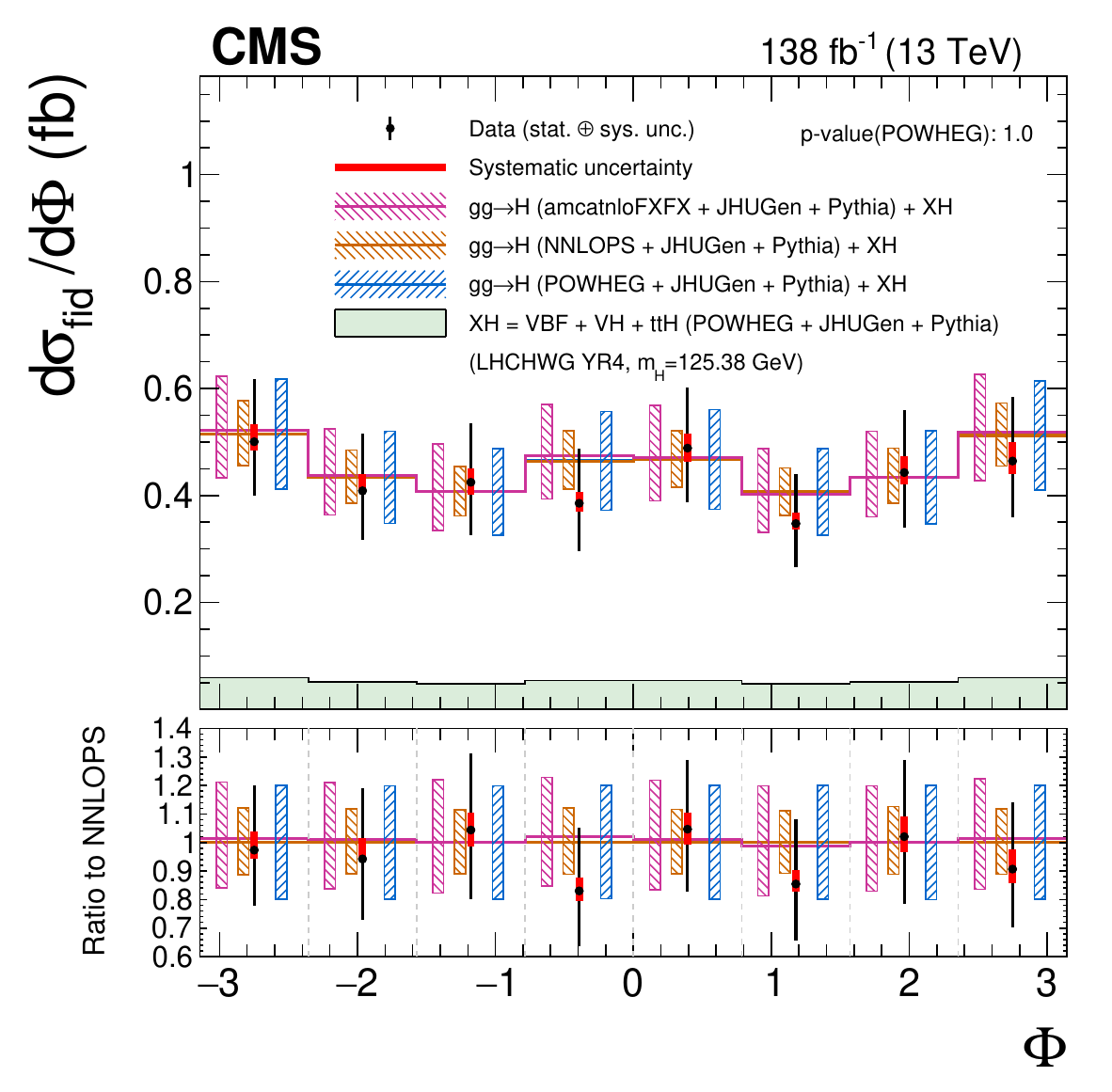}\\
		\includegraphics[width=0.48\textwidth]{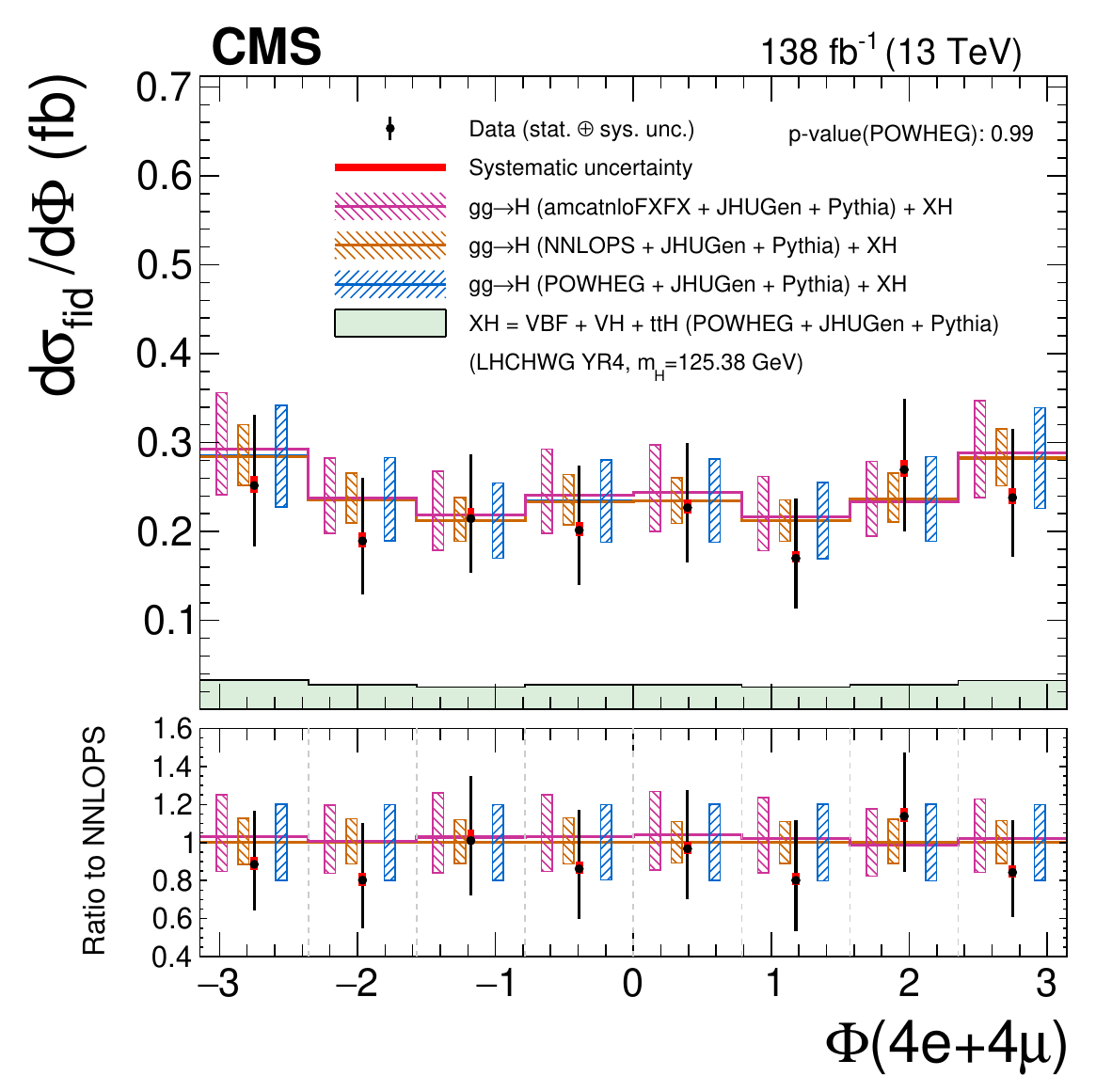}
		\includegraphics[width=0.48\textwidth]{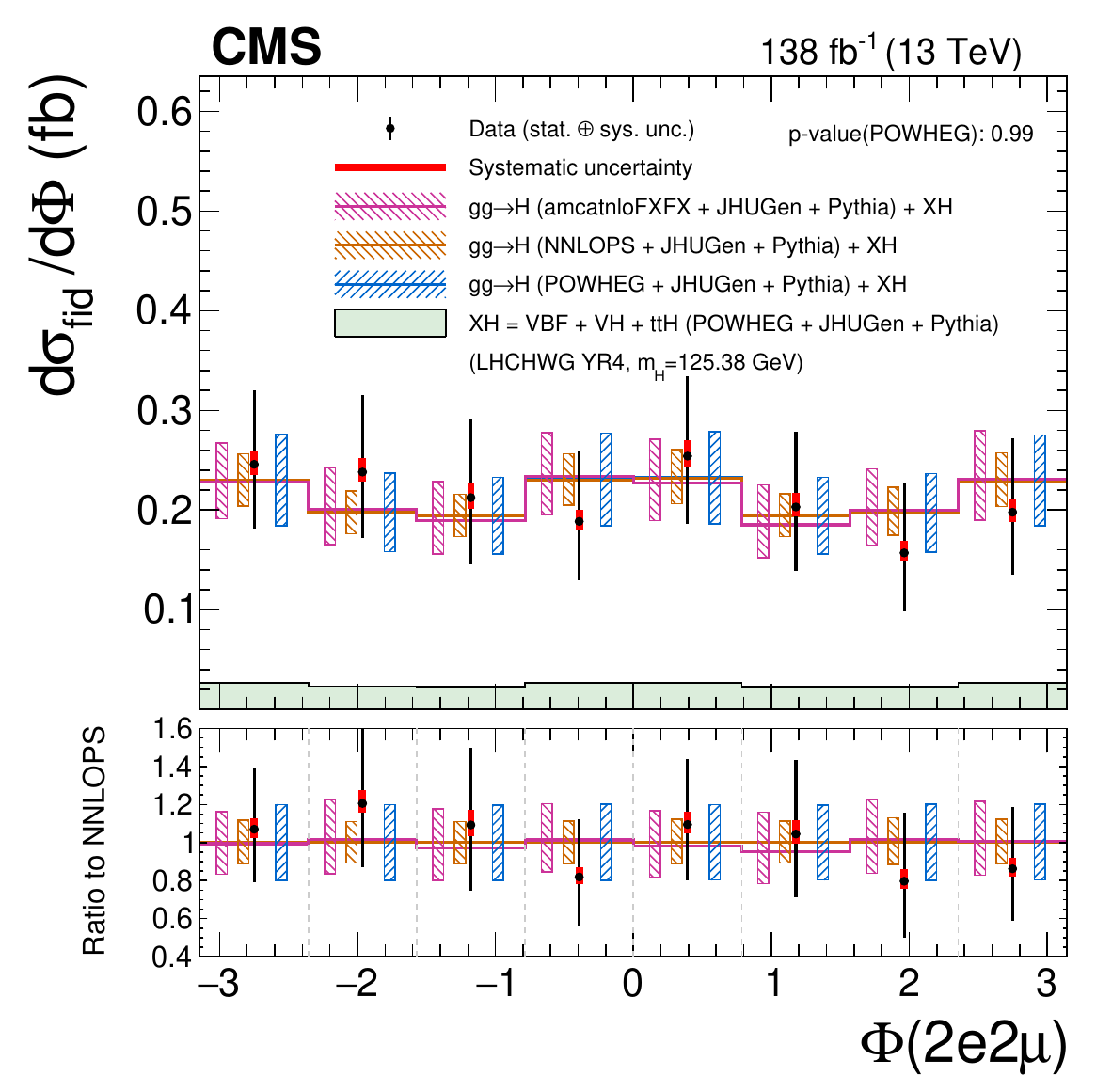}
		\caption{
			Differential cross sections as functions of the $\Phi$ angle in the $4\ell$ (upper) and in the same-flavor (lower left) and different-flavor (lower right) final states.
			The acceptance and theoretical uncertainties in the differential bins are calculated using the $\ggH$ predictions from three different generators normalized to next-to-next-to-next-to-leading order ($\mathrm{N^3LO}$)~\cite{deFlorian:2016spz}.
			The subdominant component of the signal ($\VBF + \VH + \ttH$) is denoted as XH and is fixed to the SM prediction.
			The measured cross sections are compared with the  $\ggH$ predictions from \POWHEG (blue), \textsc{NNLOPS} (orange), and \MGvATNLO (pink).
			The hatched areas correspond to the systematic uncertainties in the theoretical predictions.
			Black points represent the measured fiducial cross sections in each bin, black error bars the total uncertainty in each measurement, red boxes the systematic uncertainties.
			The lower panels display the ratios of the measured cross sections and of the predictions from \POWHEG and \MGvATNLO to the \textsc{NNLOPS} theoretical predictions.
			\label{fig:fidPHI}}
	\end{figure}
\end{center}

\begin{center}
	\begin{figure}[!htb]
		\centering
		\includegraphics[width=0.48\textwidth]{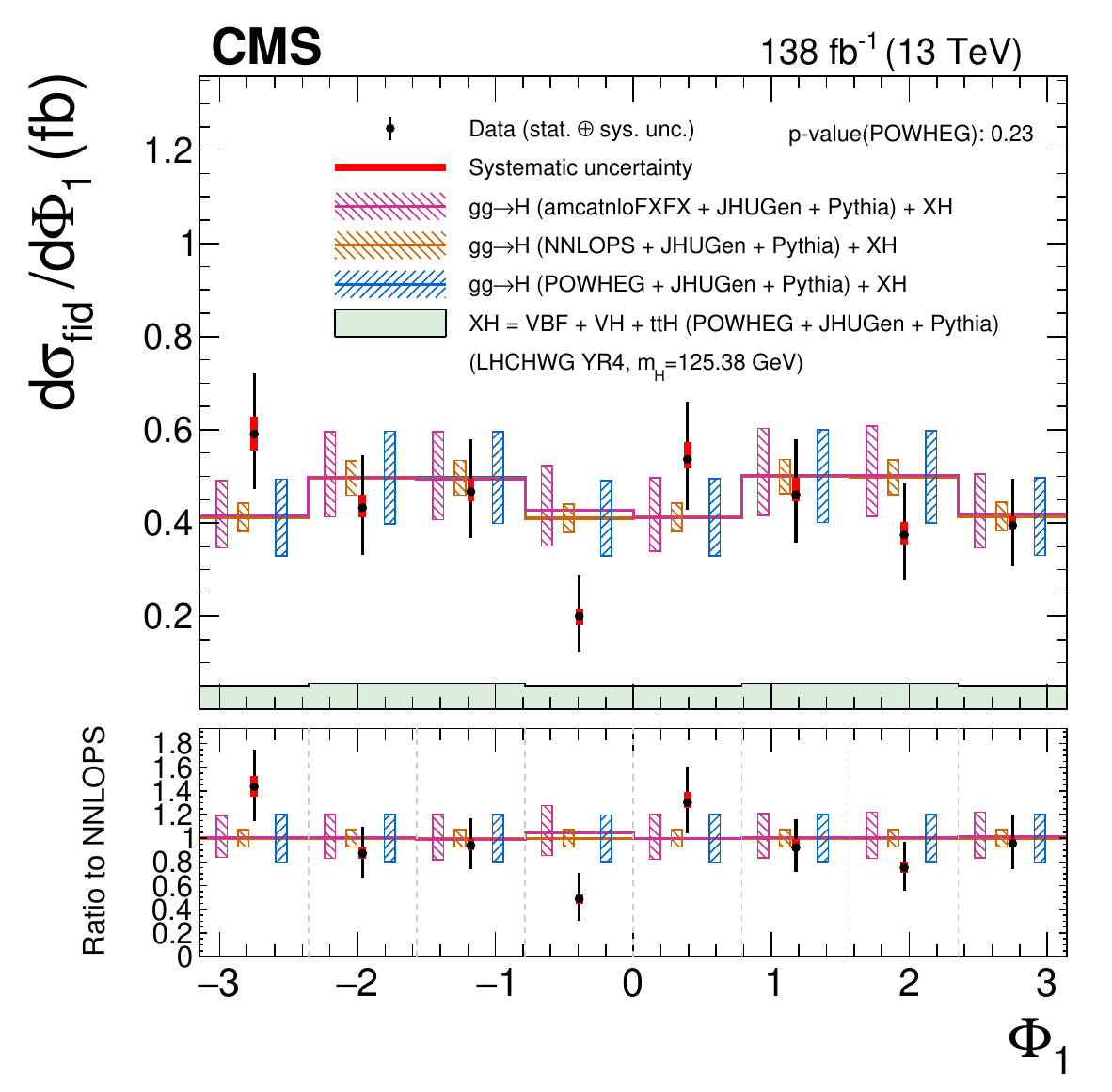}\\
		\includegraphics[width=0.48\textwidth]{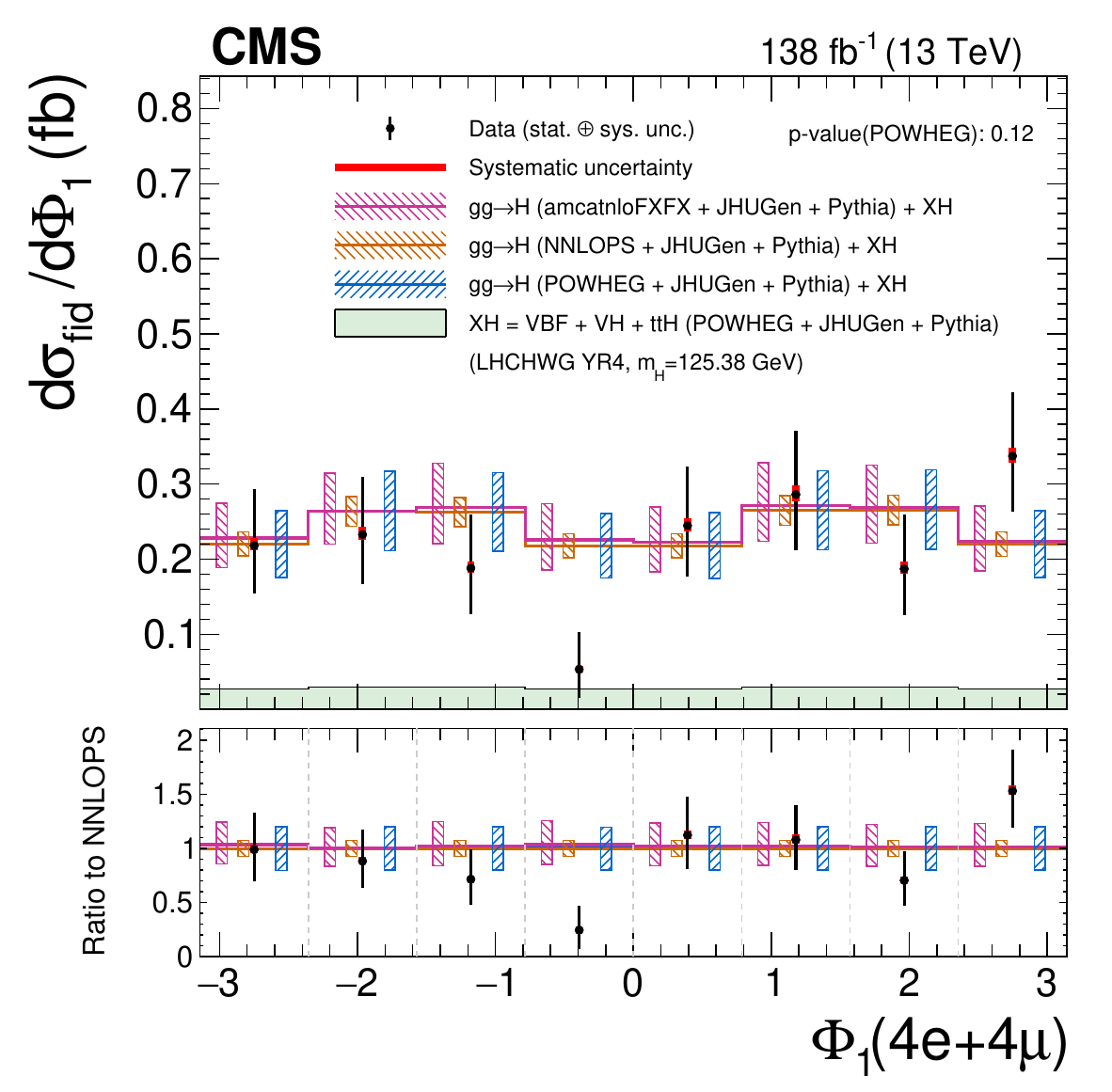}
		\includegraphics[width=0.48\textwidth]{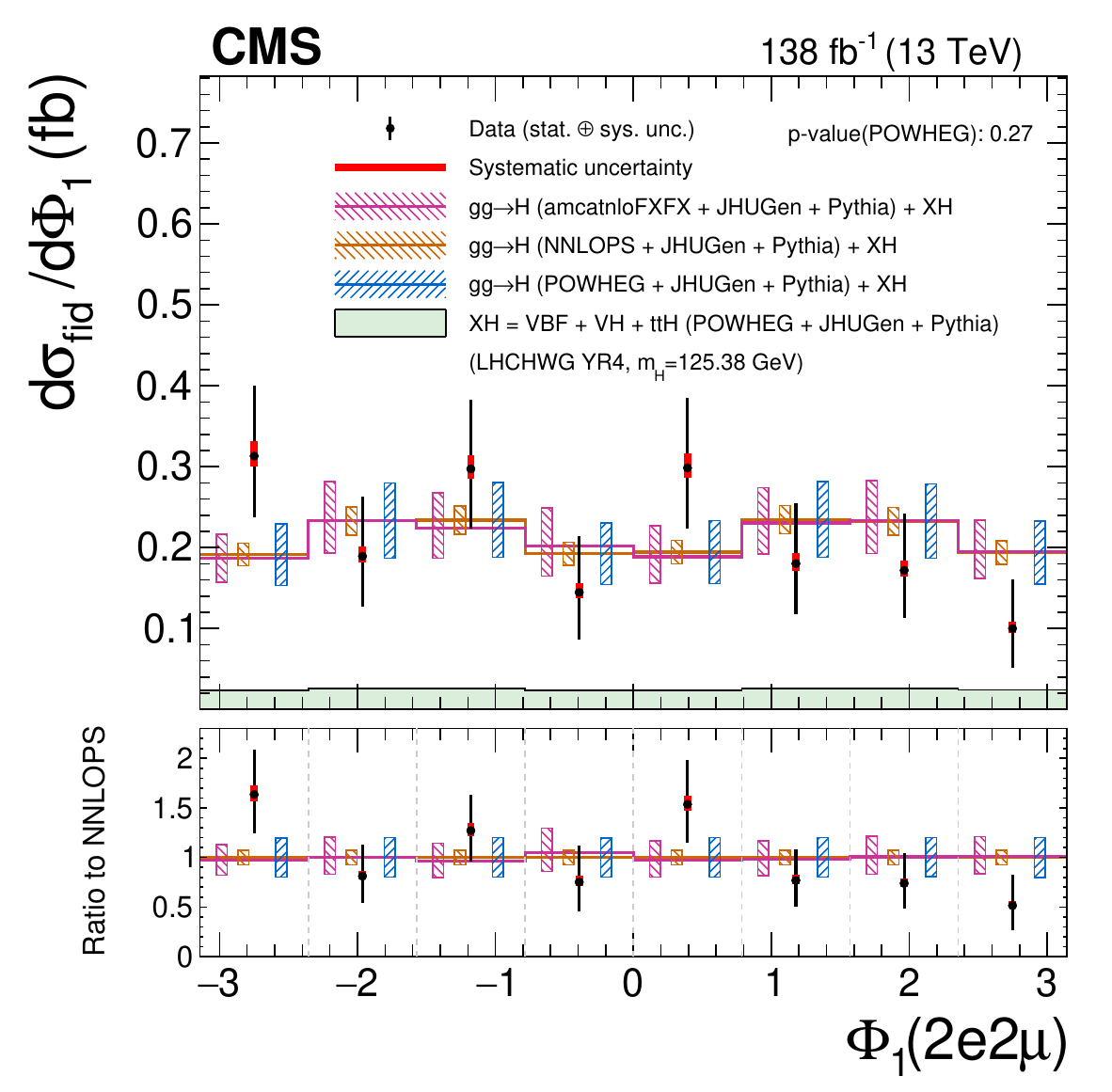}
		\caption{
			Differential cross sections as functions of the $\Phi_\text{1}$ angle in the $4\ell$ (upper) and in the same-flavor (lower left) and different-flavor (lower right) final states.
			The acceptance and theoretical uncertainties in the differential bins are calculated using the $\ggH$ predictions from three different generators normalized to next-to-next-to-next-to-leading order ($\mathrm{N^3LO}$)~\cite{deFlorian:2016spz}.
			The subdominant component of the signal ($\VBF + \VH + \ttH$) is denoted as XH and is fixed to the SM prediction.
			The measured cross sections are compared with the  $\ggH$ predictions from \POWHEG (blue), \textsc{NNLOPS} (orange), and \MGvATNLO (pink).
			The hatched areas correspond to the systematic uncertainties in the theoretical predictions.
			Black points represent the measured fiducial cross sections in each bin, black error bars the total uncertainty in each measurement, red boxes the systematic uncertainties.
			The lower panels display the ratios of the measured cross sections and of the predictions from \POWHEG and \MGvATNLO to the \textsc{NNLOPS} theoretical predictions.
			\label{fig:fidPHISTAR}}
	\end{figure}
\end{center}

The cross sections measured in bins of the invariant mass of the two \PZ boson candidates are shown in Figs.~\ref{fig:fidMZ1} and~\ref{fig:fidMZ2}.
The additional degrees of freedom that characterize the $\HZZfl$ decay are the five angles introduced in Section~\ref{sec:observables}. 
The cross sections in differential bins of the cosine of the $\theta$ angles are presented in Figs.~\ref{fig:fidCOSTS},~\ref{fig:fidCOSZ1}, and~\ref{fig:fidCOSZ2}, respectively.
Figures~\ref{fig:fidPHI} and~\ref{fig:fidPHISTAR} 
show the results for the measurements in bins of the $\Phi$ and $\Phi_1$ angles.

\clearpage

\clearpage

\clearpage

\clearpage

\clearpage

\clearpage

These observables can be used to compute matrix element discriminants sensitive to the presence of possible BSM physics effects as described in Section~\ref{sec:discriminants} and Ref.~\cite{CMSHIG19009}.

\begin{center}
	\begin{figure}[!htb]
		\centering
		\includegraphics[width=0.48\textwidth]{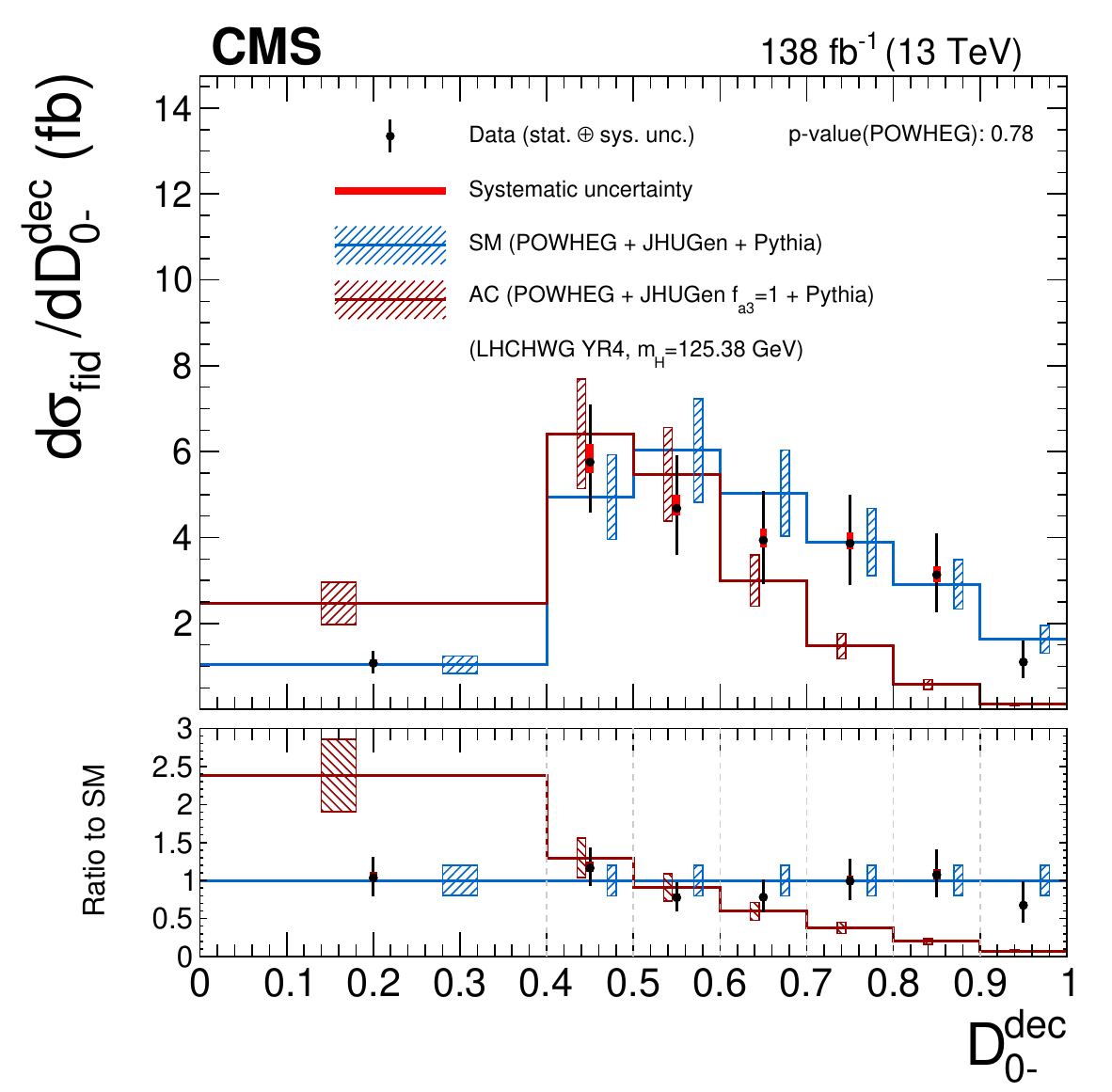}\\
		\includegraphics[width=0.48\textwidth]{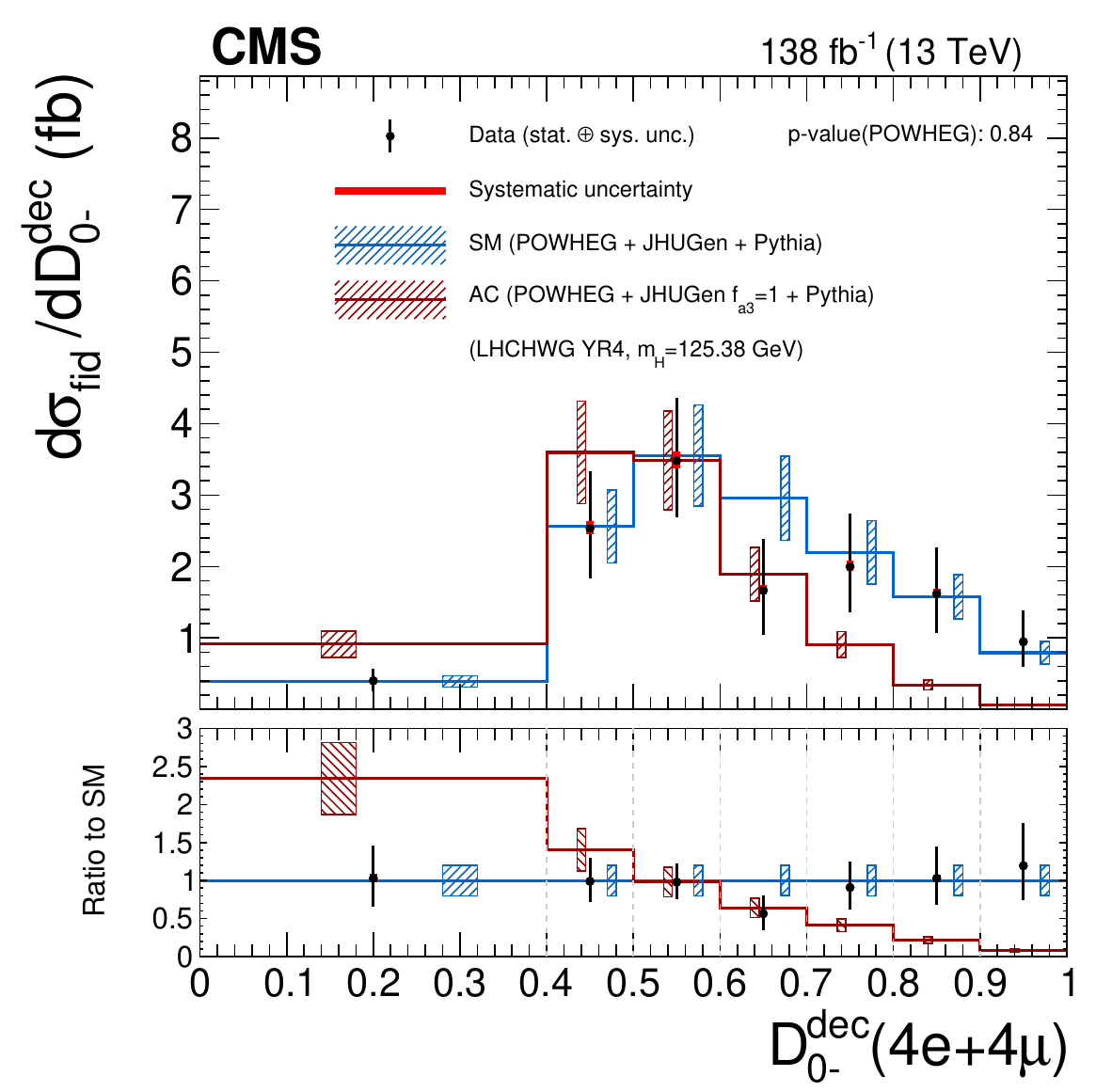}
		\includegraphics[width=0.48\textwidth]{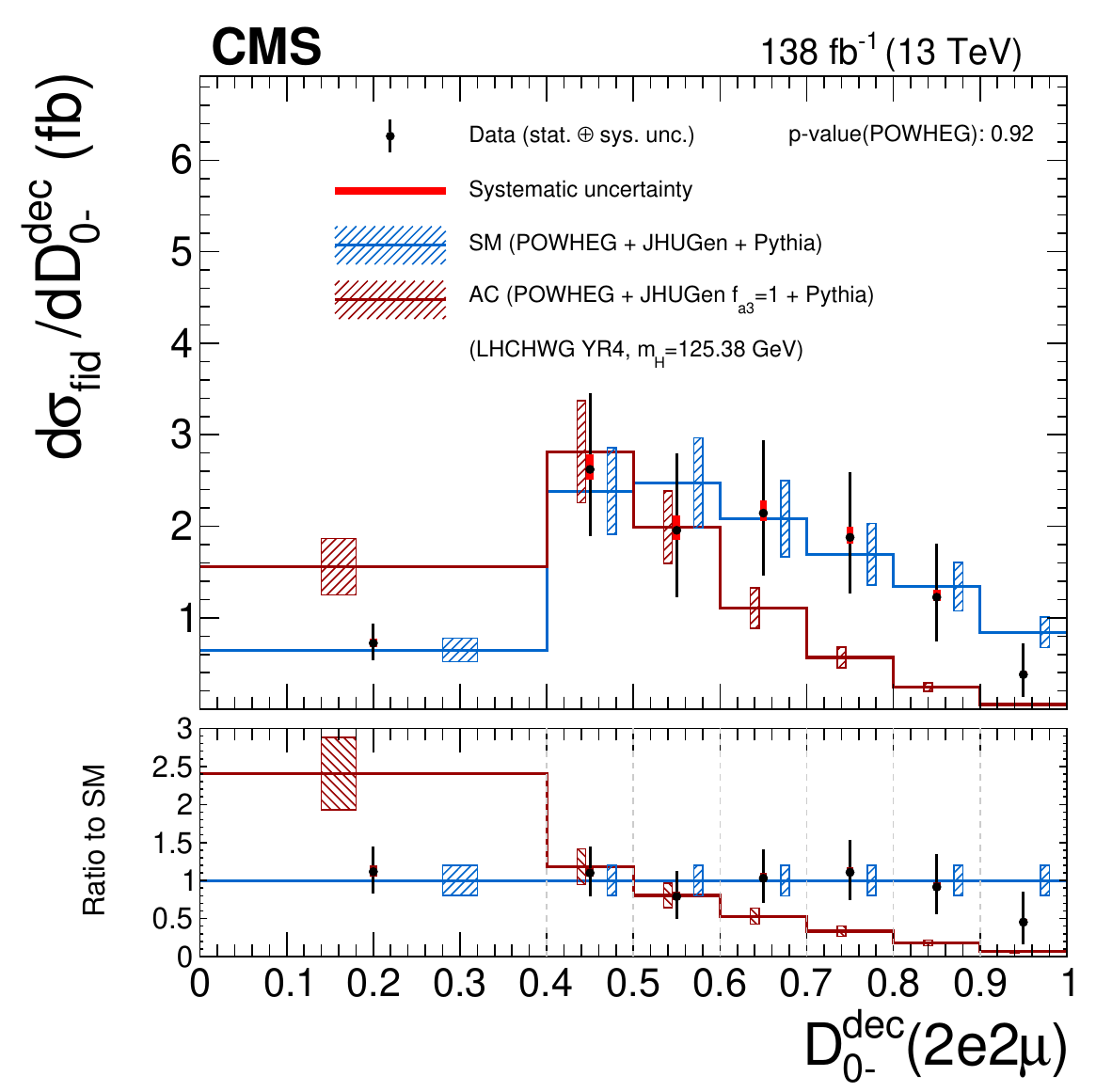}
		\caption{
			Differential cross sections as functions of the matrix element kinematic discriminant $\Dzm$ in the $4\ell$ (upper) and in the same-flavor (lower left) and different-flavor (lower right) final states.
			The brown histograms show the distribution of the matrix element discriminant for the HVV anomalous coupling scenario corresponding to $f_{a3} = 1$.
			The subdominant component of the signal ($\VBF + \VH + \ttH$) is fixed to the SM prediction.
			The hatched areas correspond to the systematic uncertainties in the theoretical predictions.
			Black points represent the measured fiducial cross sections in each bin, black error bars the total uncertainty in each measurement, red boxes the systematic uncertainties.
			The lower panels display the ratios of the measured cross sections and of the predictions from \POWHEG and \MGvATNLO to the \textsc{NNLOPS} theoretical predictions.
			\label{fig:fidDZM}}
	\end{figure}
\end{center}

\begin{center}
	\begin{figure}[!htb]
		\centering
		\includegraphics[width=0.48\textwidth]{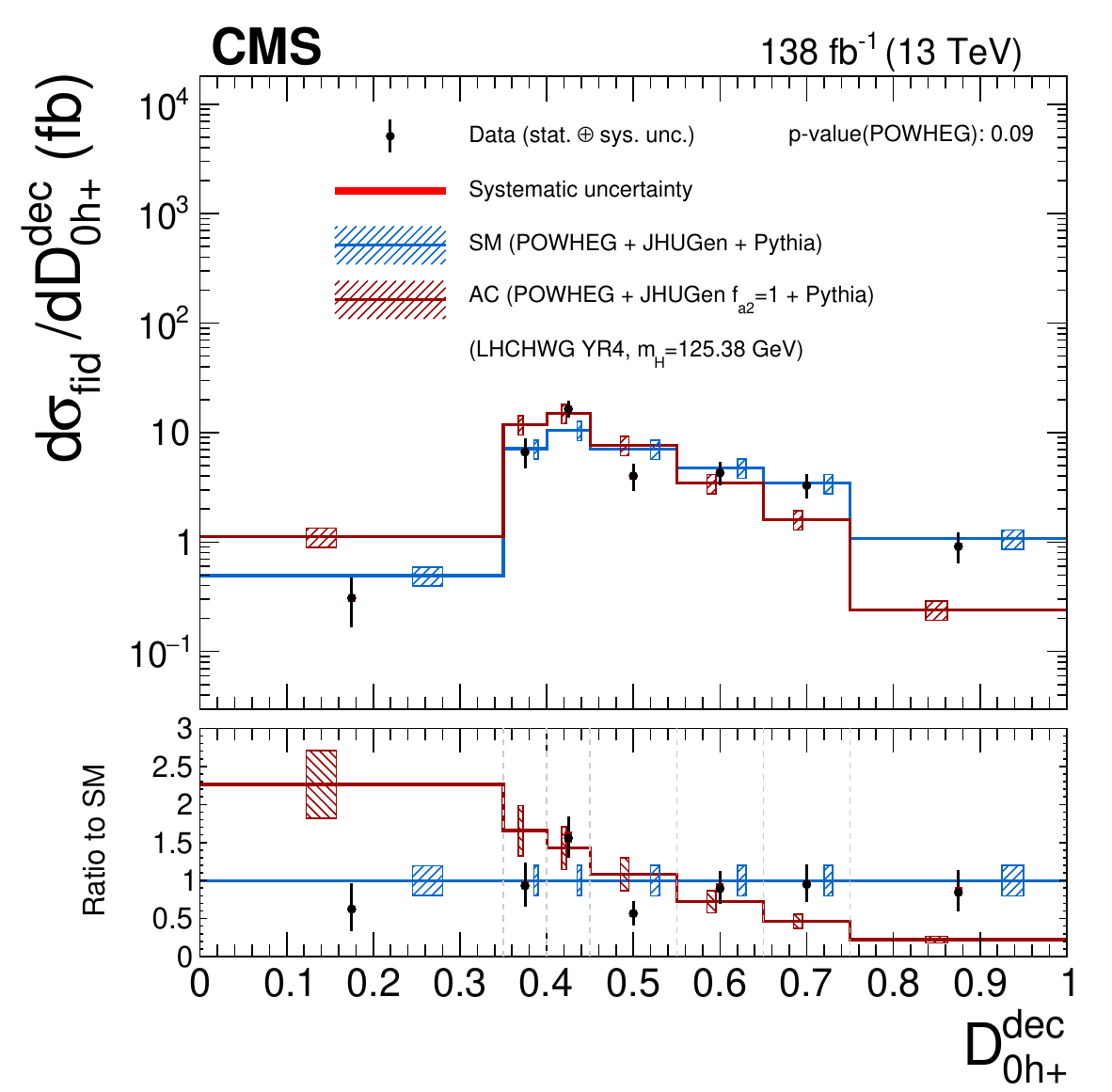}\\
		\includegraphics[width=0.48\textwidth]{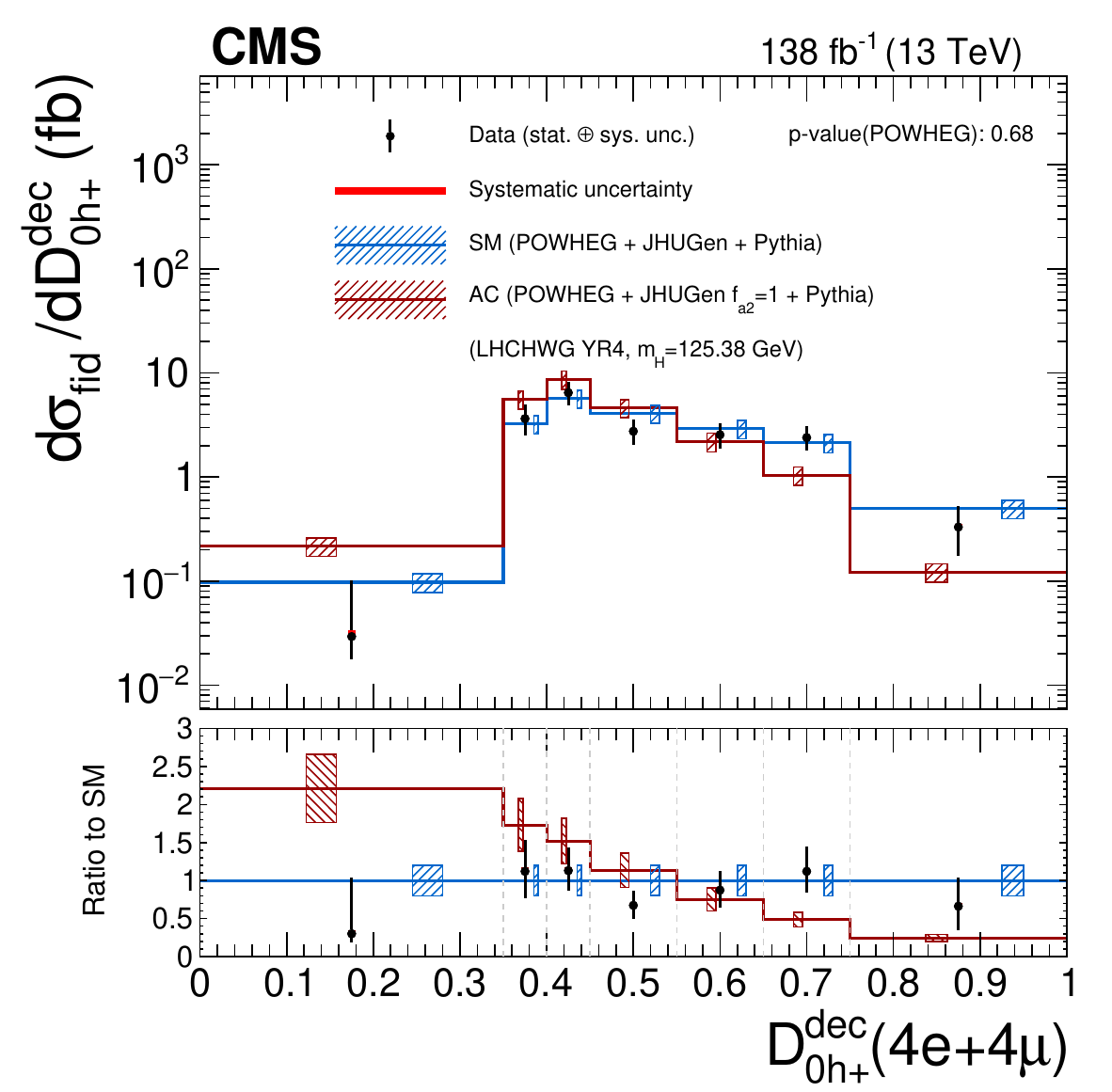}
		\includegraphics[width=0.48\textwidth]{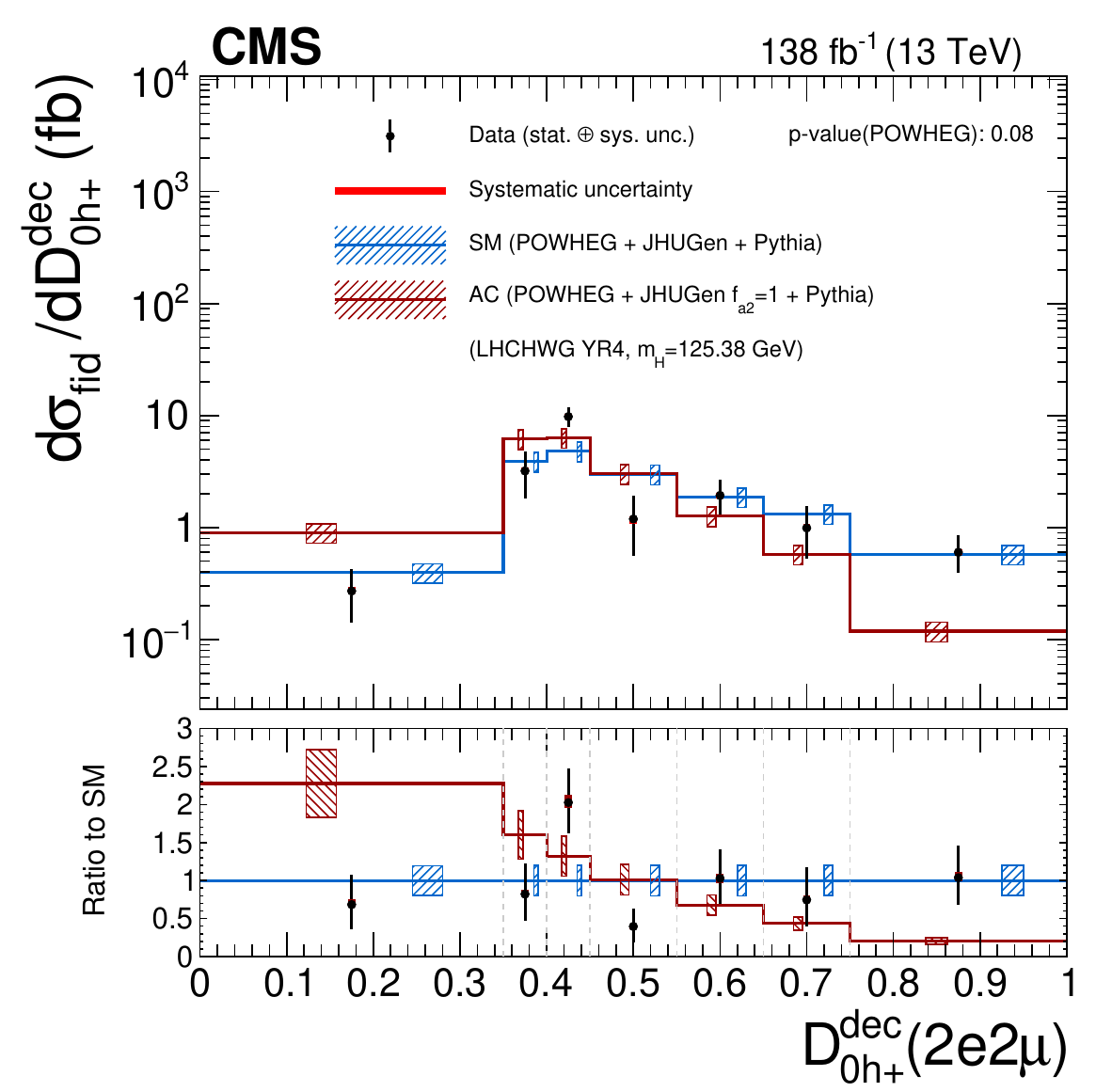}
		\caption{
			Differential cross sections as functions of the matrix element kinematic discriminant $\Dzhp$ in the $4\ell$ (upper) and in the same-flavor (lower left) and different-flavor (lower right) final states.
			The brown histograms show the distribution of the matrix element discriminant for the HVV anomalous coupling scenario corresponding to $f_{a2} = 1$.
			The subdominant component of the signal ($\VBF + \VH + \ttH$) is fixed to the SM prediction.
			The hatched areas correspond to the systematic uncertainties in the theoretical predictions.
			Black points represent the measured fiducial cross sections in each bin, black error bars the total uncertainty in each measurement, red boxes the systematic uncertainties.
			The lower panels display the ratios of the measured cross sections and of the predictions from \POWHEG and \MGvATNLO to the \textsc{NNLOPS} theoretical predictions.
			\label{fig:fidDOHP}}
	\end{figure}
\end{center}

\begin{center}
	\begin{figure}[!htb]
		\centering
		\includegraphics[width=0.48\textwidth]{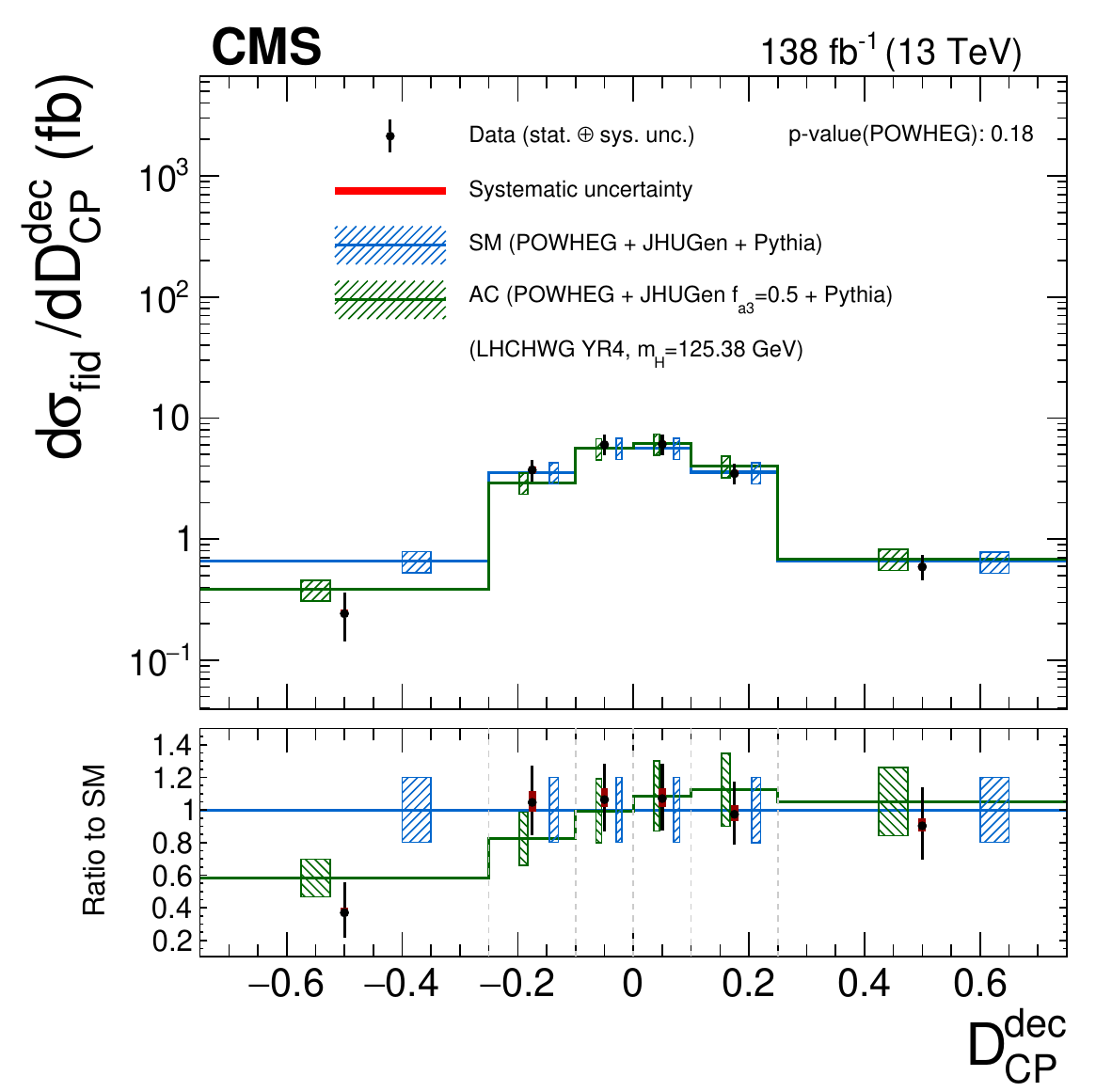}\\
		\includegraphics[width=0.48\textwidth]{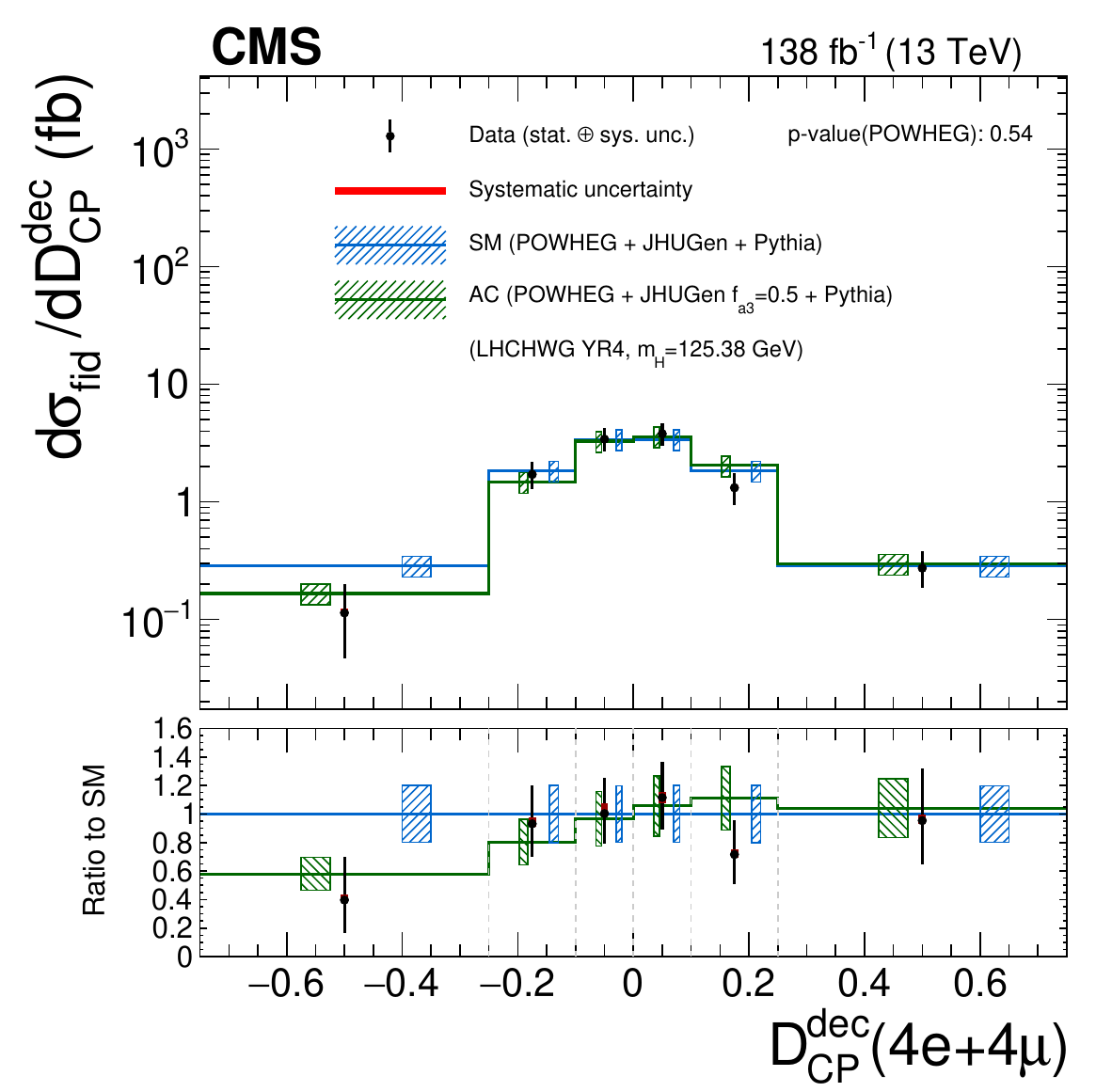}
		\includegraphics[width=0.48\textwidth]{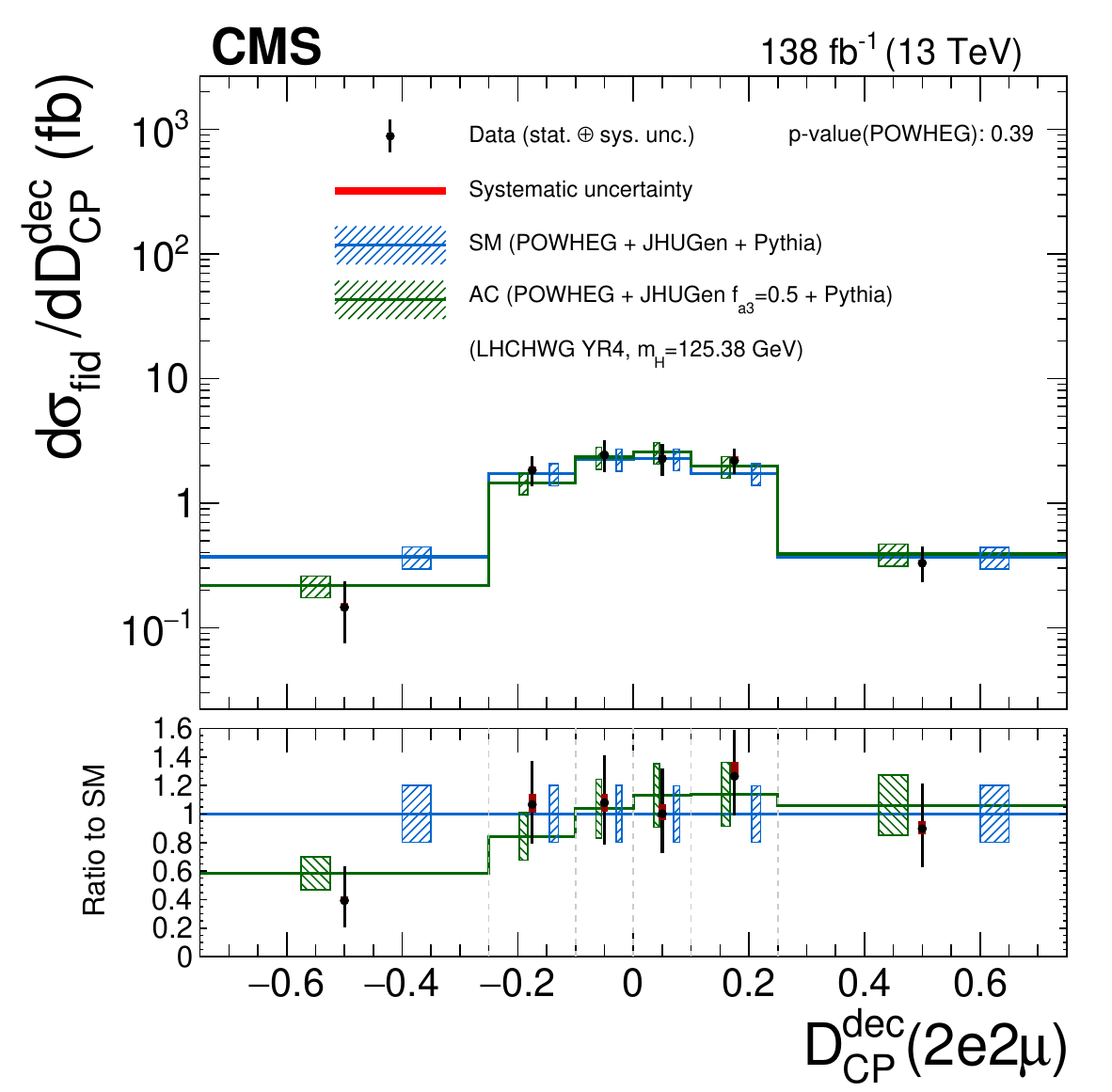}
		\caption{
			Differential cross sections as functions of the matrix element kinematic discriminant $\DCP$  in the $4\ell$ (upper) and in the same-flavor (lower left) and different-flavor (lower right) final states.
			The green histogram shows the distribution of the discriminant for the HVV anomalous coupling scenario corresponding to $f_{a3} = 0.5$.
			The subdominant component of the signal ($\VBF + \VH + \ttH$) is fixed to the SM prediction.
			The hatched areas correspond to the systematic uncertainties in the theoretical predictions.
			Black points represent the measured fiducial cross sections in each bin, black error bars the total uncertainty in each measurement, red boxes the systematic uncertainties.
			The lower panels display the ratios of the measured cross sections and of the predictions from \POWHEG and \MGvATNLO to the \textsc{NNLOPS} theoretical predictions.
			\label{fig:fidDCP}}
	\end{figure}
\end{center}

\begin{center}
	\begin{figure}[!htb]
		\centering
		\includegraphics[width=0.48\textwidth]{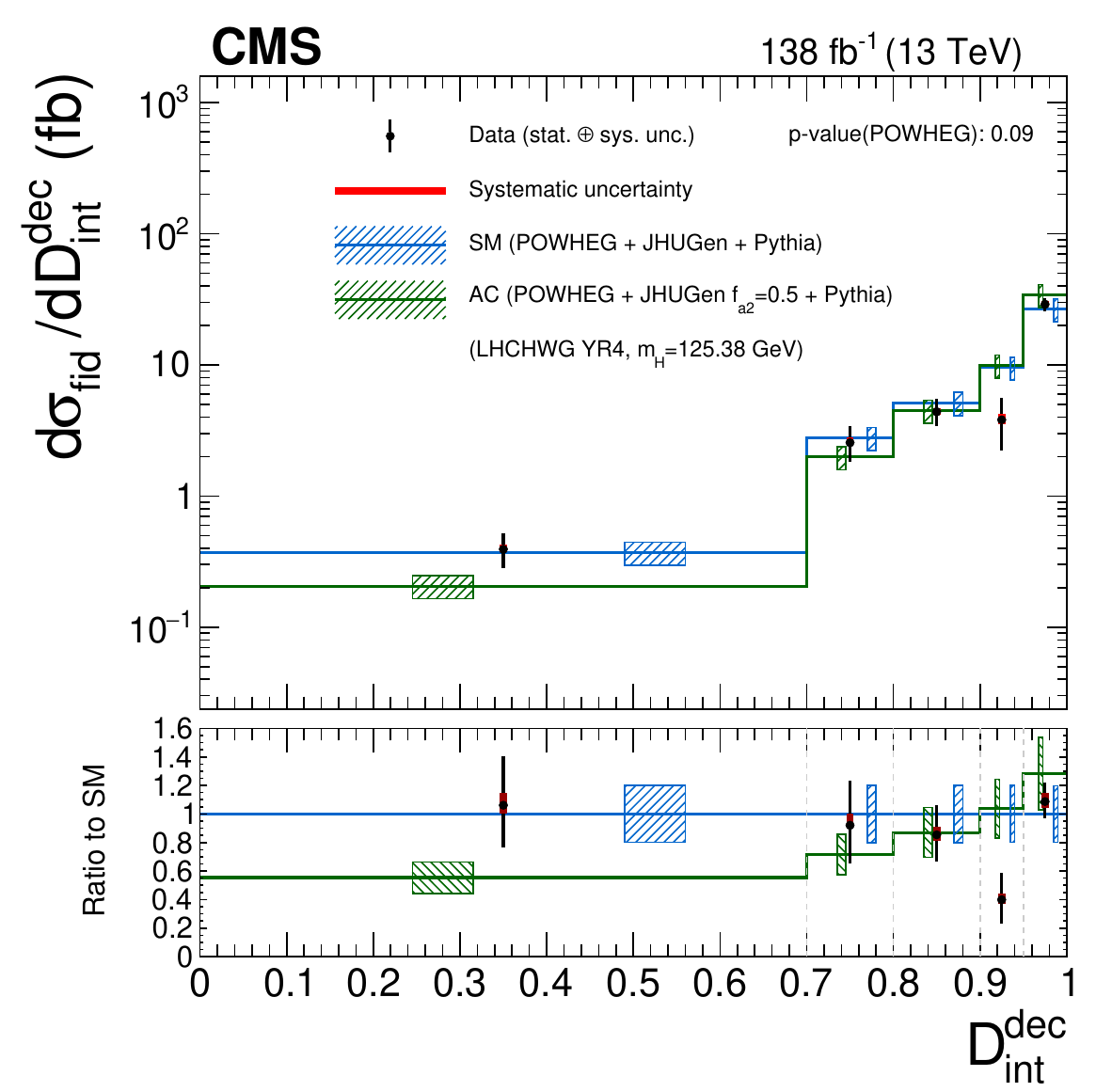}\\
		\includegraphics[width=0.48\textwidth]{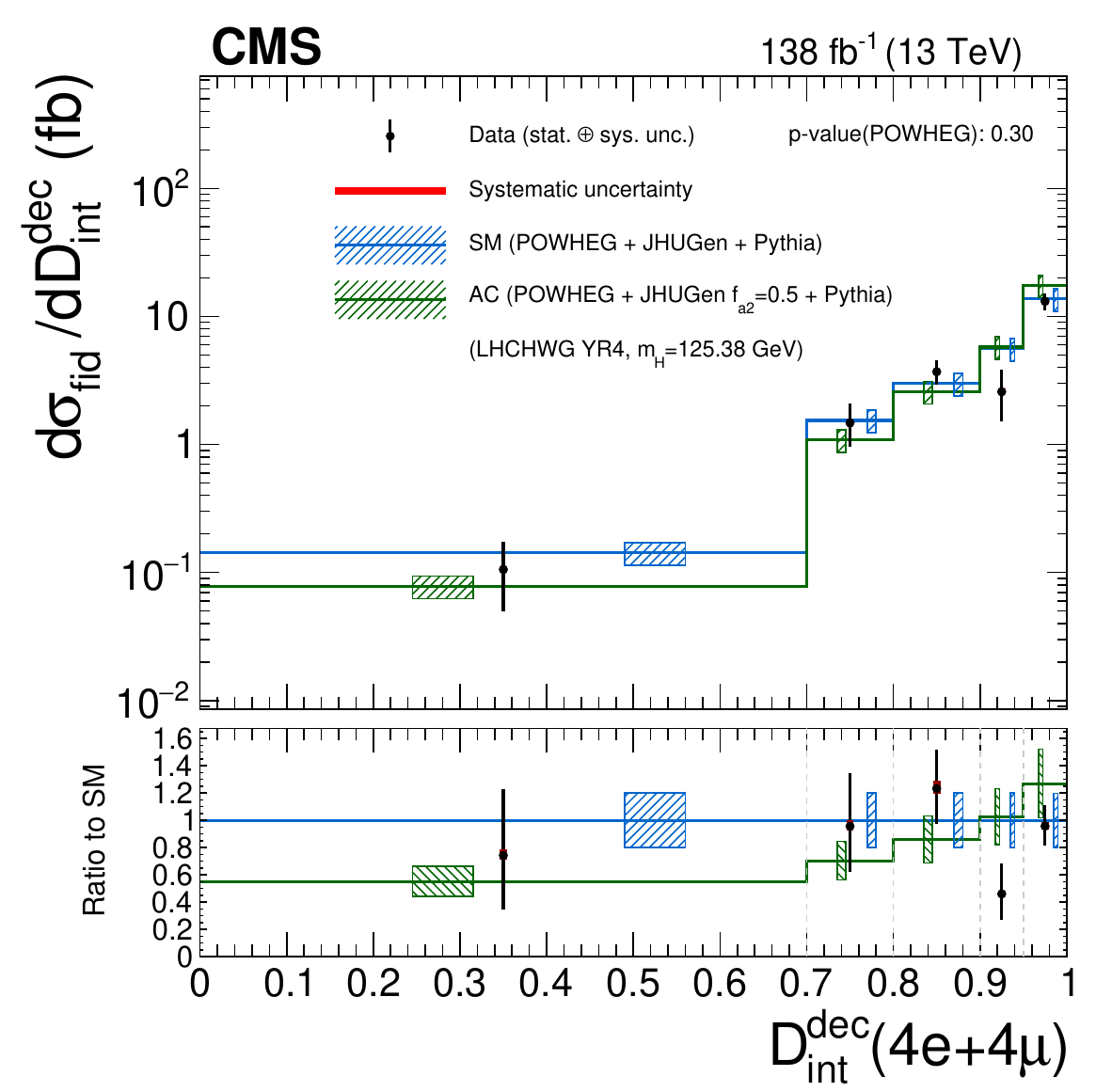}
		\includegraphics[width=0.48\textwidth]{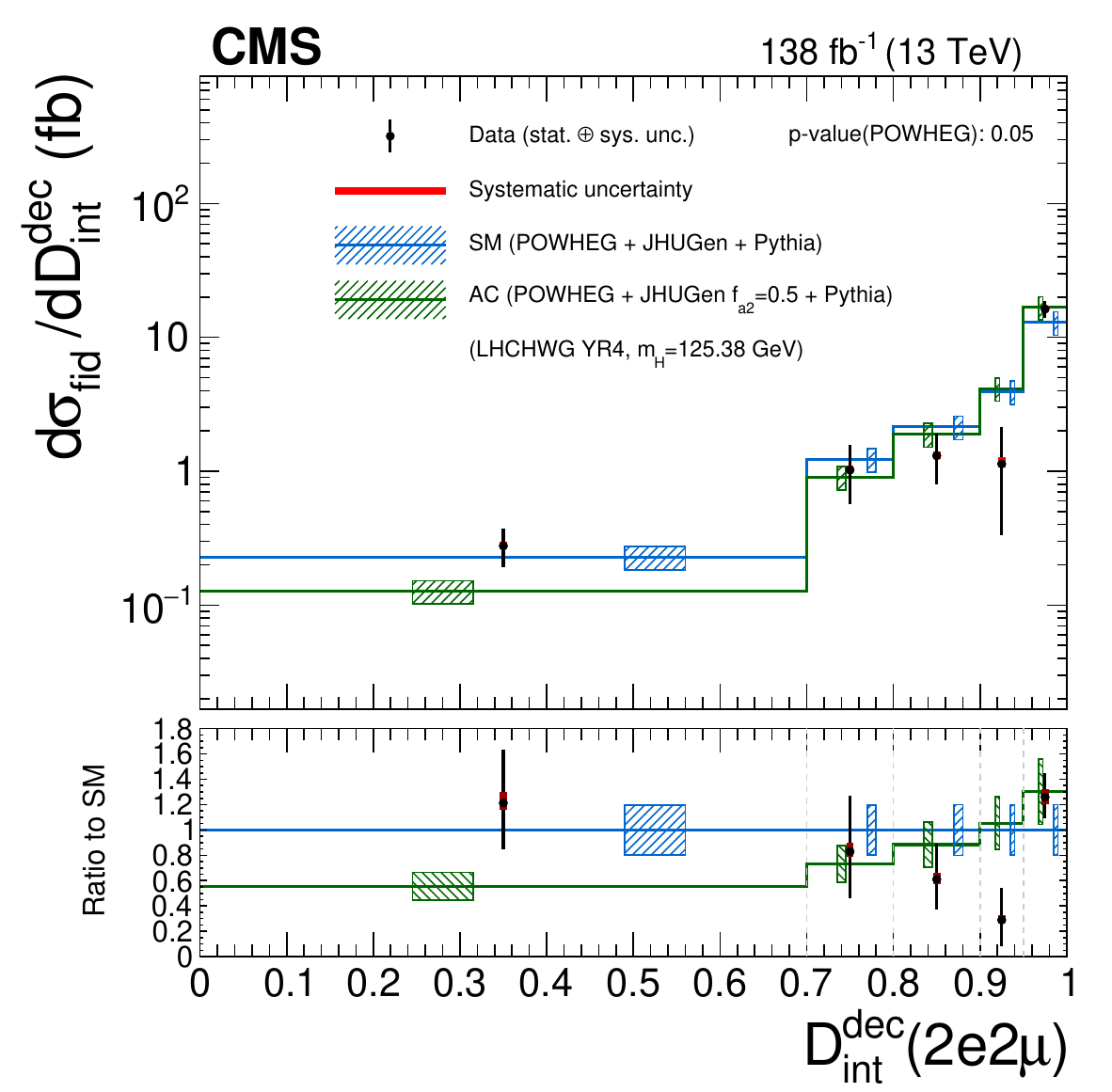}
		\caption{
			Differential cross sections as functions of the matrix element kinematic discriminant $\Dint$ in the $4\ell$ (upper) and in the same-flavor (lower left) and different-flavor (lower right) final states.
			The green histogram shows the distribution of the discriminant for the HVV anomalous coupling scenario corresponding to $f_{a2} = 0.5$.
			The subdominant component of the signal ($\VBF + \VH + \ttH$) is fixed to the SM prediction.
			The hatched areas correspond to the systematic uncertainties in the theoretical predictions.
			Black points represent the measured fiducial cross sections in each bin, black error bars the total uncertainty in each measurement, red boxes the systematic uncertainties.
			The lower panels display the ratios of the measured cross sections and of the predictions from \POWHEG and \MGvATNLO to the \textsc{NNLOPS} theoretical predictions.
			\label{fig:fidDINT}}
	\end{figure}
\end{center}

\begin{center}
	\begin{figure}[!htb]
		\centering
		\includegraphics[width=0.48\textwidth]{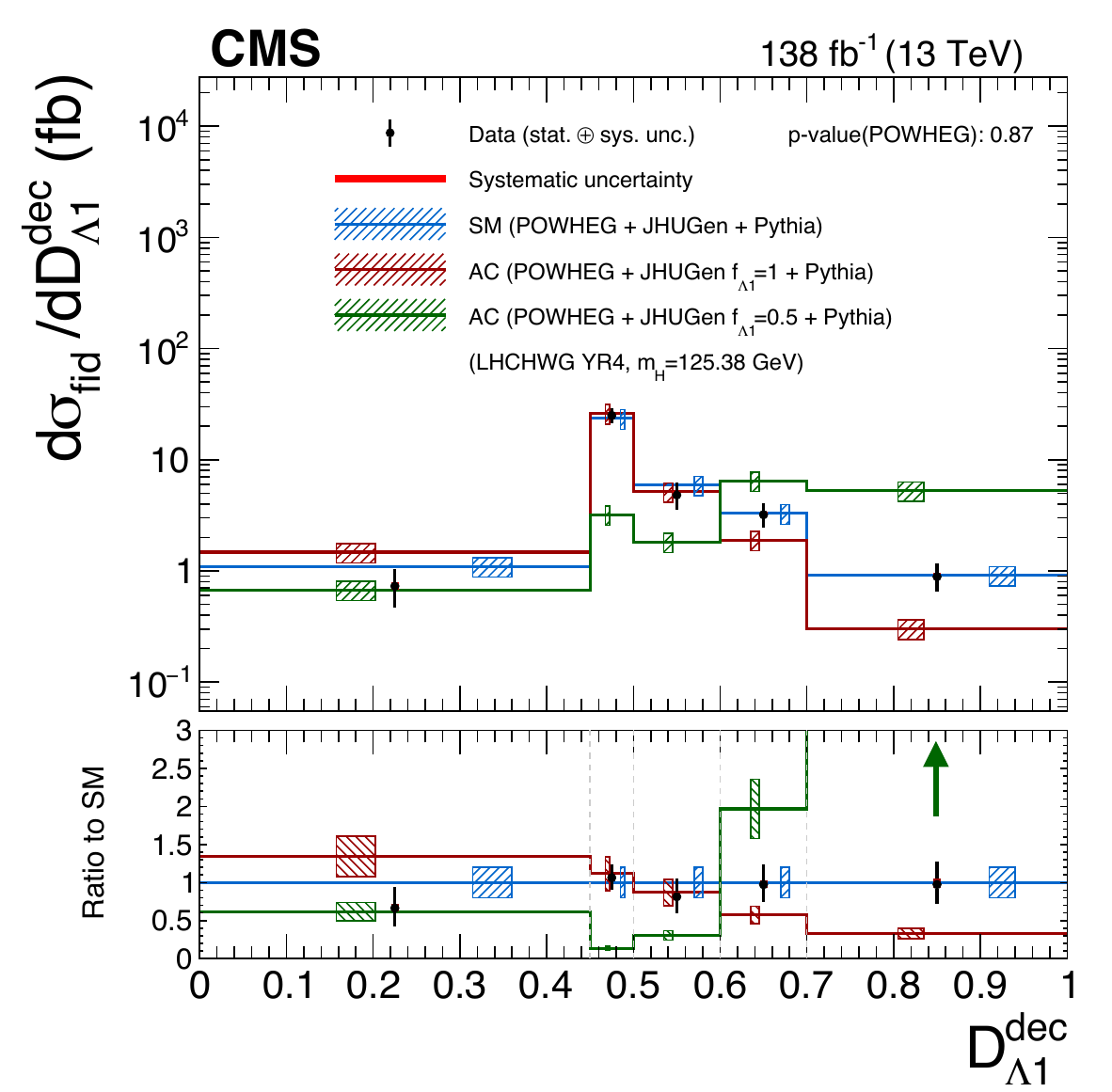}\\
		\includegraphics[width=0.48\textwidth]{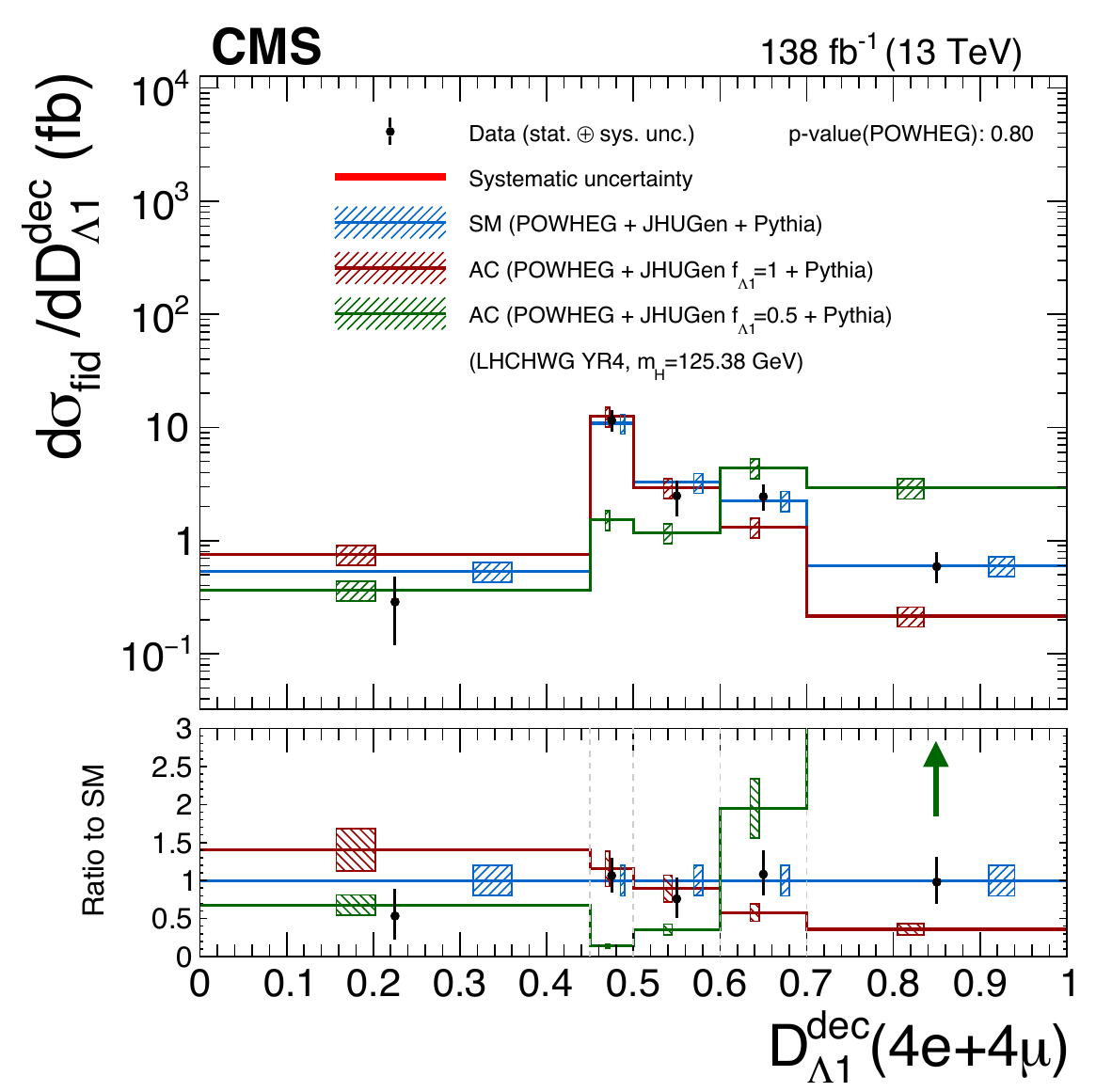}
		\includegraphics[width=0.48\textwidth]{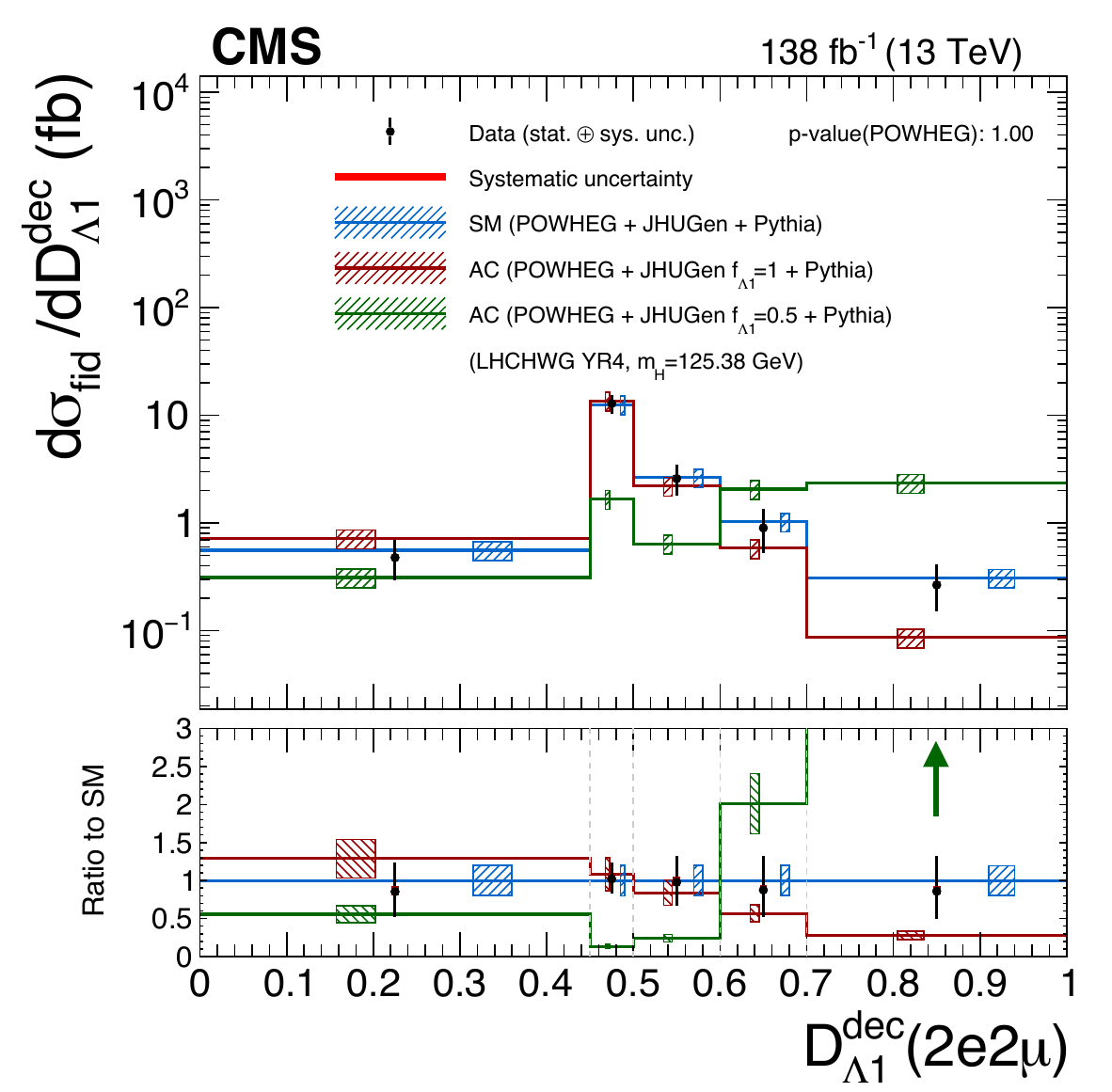}
		\caption{
			Differential cross sections as functions of the matrix element kinematic discriminant $\DLone$  in the $4\ell$ (upper) and in the same-flavor (lower left) and different-flavor (lower right) final states.
			The brown and green histograms show the distributions of the discriminant for the HVV anomalous coupling scenarios corresponding to $f_{\Lambda 1} = 1$ and $f_{\Lambda 1} = 0.5$.
			The subdominant component of the signal ($\VBF + \VH + \ttH$) is fixed to the SM prediction.
			The hatched areas correspond to the systematic uncertainties in the theoretical predictions.
			Black points represent the measured fiducial cross sections in each bin, black error bars the total uncertainty in each measurement, red boxes the systematic uncertainties.
			The lower panels display the ratio of the measured cross section and of the predictions from \POWHEG and \MGvATNLO to the \textsc{NNLOPS} theoretical expectation.
			\label{fig:fidDL1}}
	\end{figure}
\end{center}

\begin{center}
	\begin{figure}[!htb]
		\centering
		\includegraphics[width=0.48\textwidth]{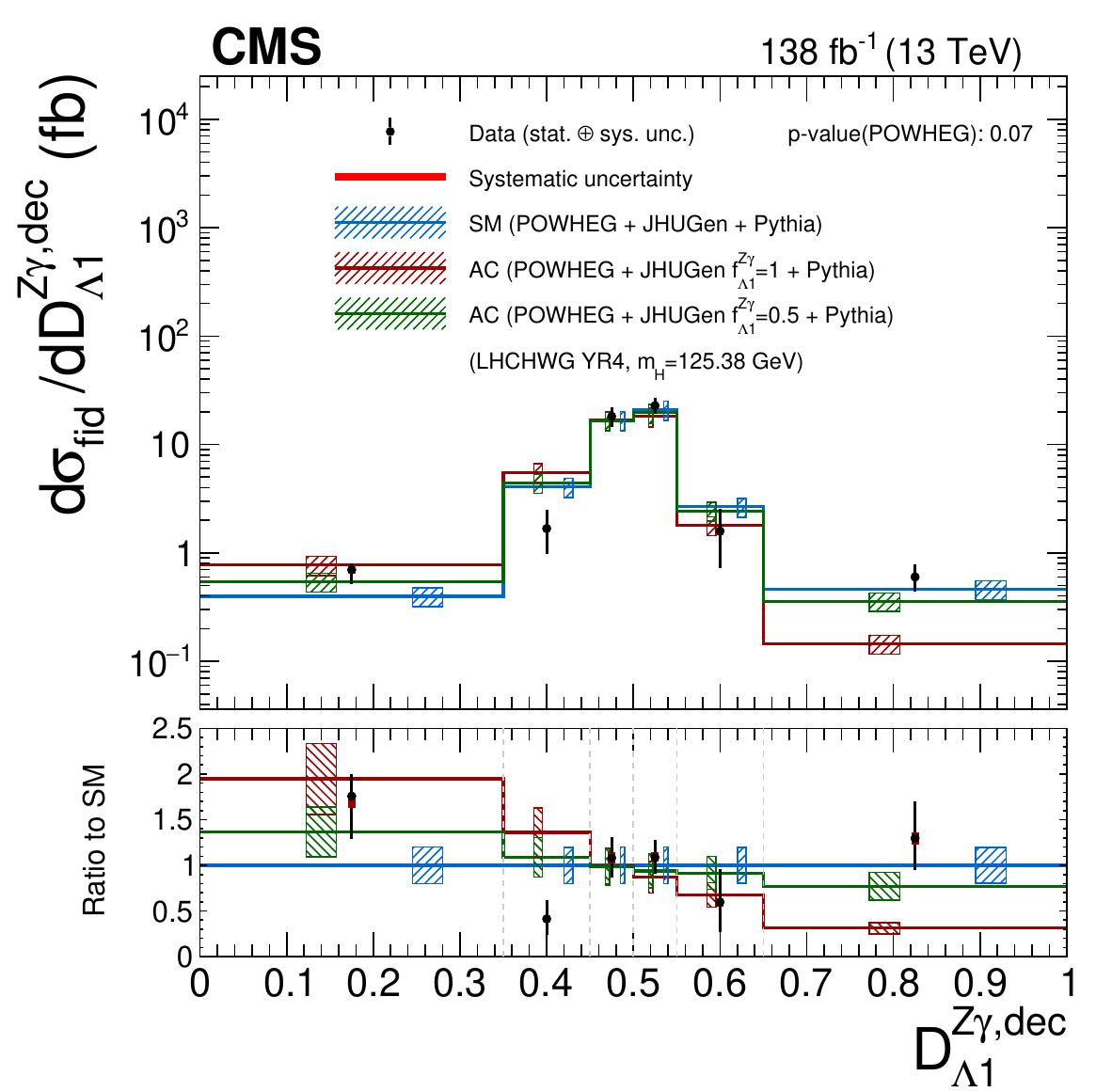}\\
		\includegraphics[width=0.48\textwidth]{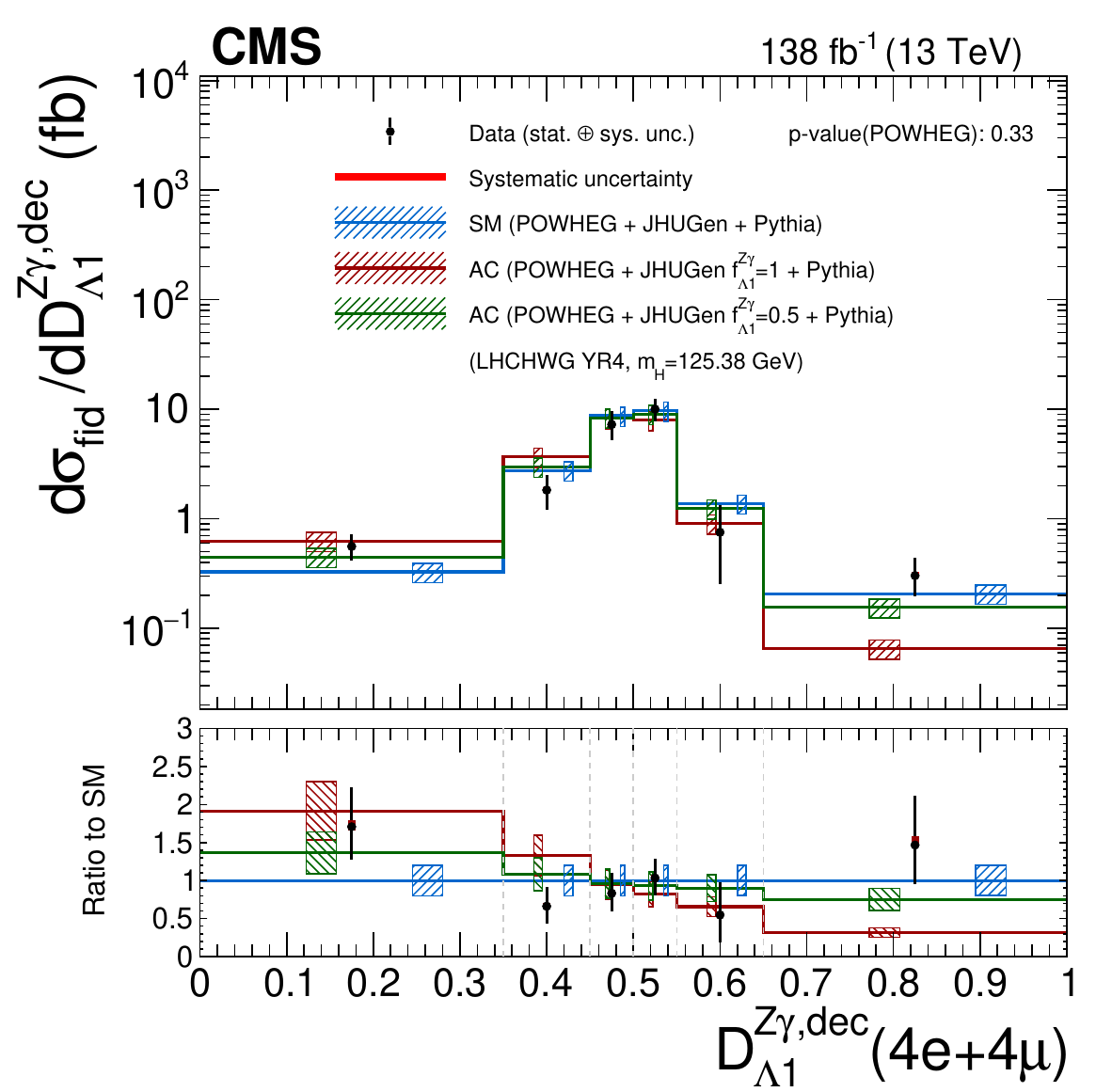}
		\includegraphics[width=0.48\textwidth]{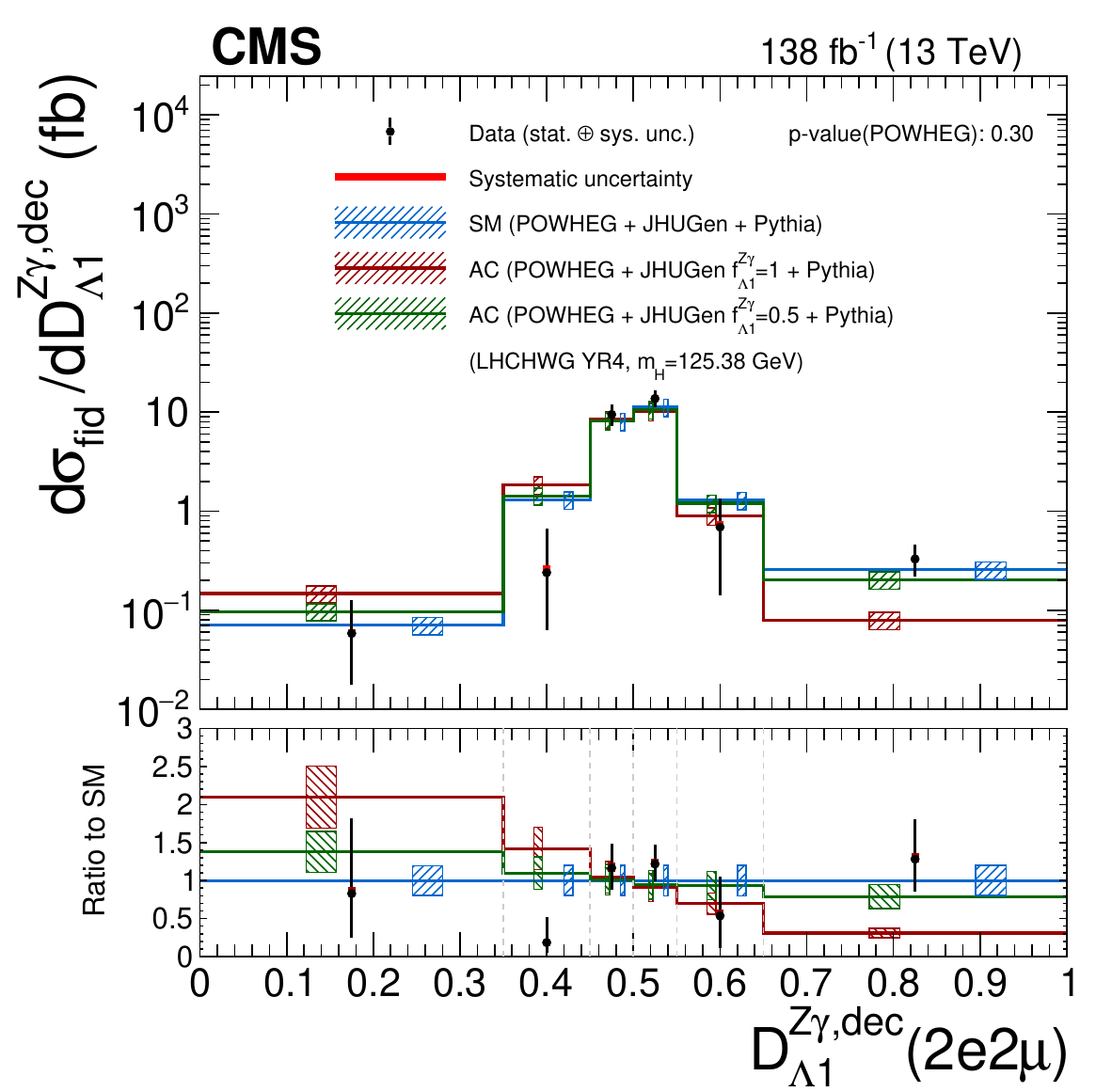}
		\caption{
			Differential cross sections as functions of the matrix element kinematic discriminant $\DLoneZg$  in the $4\ell$ (upper) and in the same-flavor (lower left) and different-flavor (lower right) final states.
			The brown and green histograms show the distributions of the discriminant for the HVV anomalous coupling scenarios corresponding to $f_{\Lambda 1}^{\PZ\gamma} = 1$ and $f_{\Lambda 1}^{\PZ\gamma} = 0.5$.
			The subdominant component of the signal ($\VBF + \VH + \ttH$) is fixed to the SM prediction.
			The hatched areas correspond to the systematic uncertainties in the theoretical predictions.
			Black points represent the measured fiducial cross sections in each bin, black error bars the total uncertainty in each measurement, red boxes the systematic uncertainties.
			The lower panels display the ratios of the measured cross sections and of the predictions from \POWHEG and \MGvATNLO to the \textsc{NNLOPS} theoretical predictions.
			\label{fig:fidDL1ZG}}
	\end{figure}
\end{center}

Cross sections are measured in differential bins of six kinematic discriminants sensitive to various HVV anomalous couplings and the interference between two model contributions (SM and a BSM scenario): $\Dzm$, $\Dzhp$, $\DCP$, $\Dint$, $\DLone$, and $\DLoneZg$.
The results of these measurements, shown in Figs.~\ref{fig:fidDZM}--\ref{fig:fidDL1ZG}, are compared to distributions for the matrix element discriminants corresponding to various anomalous coupling hypotheses.
Following the conventions adopted in Ref.~\cite{CMSHIG19009}, rather than using the value of the coupling to identify the type of the anomalous coupling sample, the cross sections fractions $f_{ai}$ are used:
\begin{equation}
	f_{ai} = \frac{\abs{a_i}^2\sigma_i}{\sum_{j}\abs{a_j}^2\sigma_j}\,\text{sign}\biggl(\frac{a_i}{a_1}\biggr),
\end{equation}
where $\sigma_i$ is the cross section for the process corresponding to $a_i=1,a_{j\neq i}=0$ in Eq.~(\ref{eq:formfact-fullampl-spin0}). 
The term for $\Lambda_1$ is $\tilde{\sigma}_{\Lambda1}/(\Lambda_1)^4$ instead of $\abs{a_i}^2\sigma_i$, where $\tilde{\sigma}_{\Lambda1}$ is the effective cross section for the process corresponding to $\Lambda_1=1\TeV$, given in units of $\text{fb}{\cdot}\TeVns^{4}$. 
To study the $a_2$ and $a_3$ couplings, discriminants of the form $\Dalt$ and $\mathcal{D}_{\text{int}}$ are built.
The former are compared to the distributions obtained for pure anomalous coupling scenarios corresponding to $f_{a3}=1$ and $f_{a2}=1$, while the latter are compared to the interference scenario where $f_{a3}=0.5$ and $f_{a2}=0.5$. A value of $f_{ai}=0.5$ corresponds to a maximal mixing between the SM and the BSM scenarios.
To inspect the couplings $\kappa_1$ and $\kappa_2^{\PZ\gamma}$, the interference discriminant is not built since it does not provide additional information and the corresponding $\Dalt$ can also be used to study the interference.
For this reason, the measurements of  $\DLone$ and $\DLoneZg$ are compared to the pure anomalous couplings scenario $f_{\Lambda 1} = 1$ and $f_{\Lambda 1}^{\PZ\gamma} = 1$, as well as to the interference hypotheses $f_{\Lambda 1} = 0.5$ and $f_{\Lambda 1}^{\PZ\gamma} = 0.5$.
These values of $f_{ai},\,f_{\Lambda 1}^{\PZ\gamma}$, and $f_{\Lambda 1}$ correspond to illustrative extreme scenarios chosen for a qualitative representation of the corresponding kinematic discriminants. The best constraints on these parameters are much stricter, as reported in Ref.~\cite{CMSHIG19009}

\clearpage

\clearpage

\clearpage

\clearpage

\clearpage

\subsection{Double-differential cross sections}
\begin{center}
	\begin{figure}[!htb]
		\centering
		\includegraphics[width=0.48\textwidth]{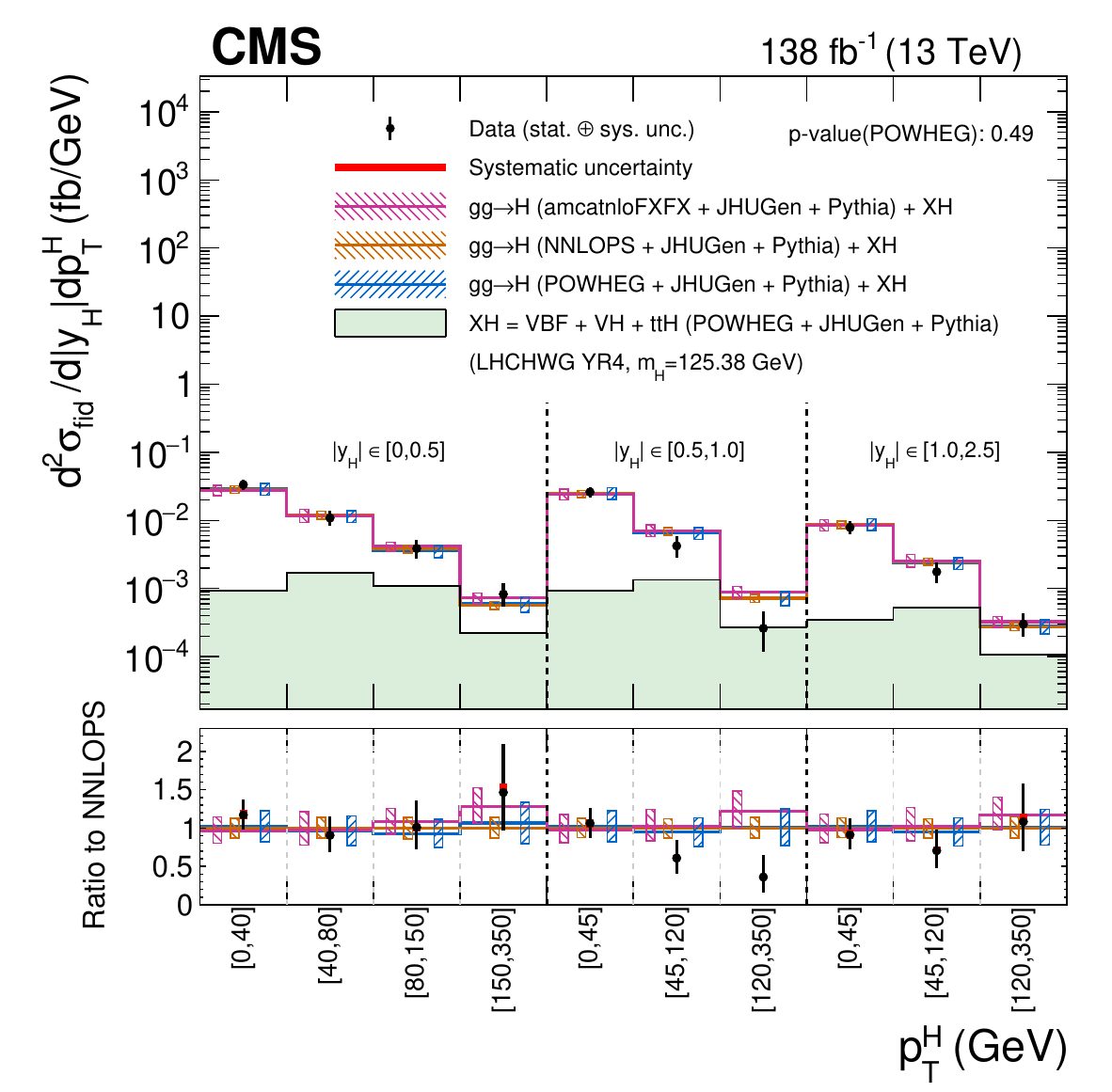}
		\includegraphics[width=0.48\textwidth]{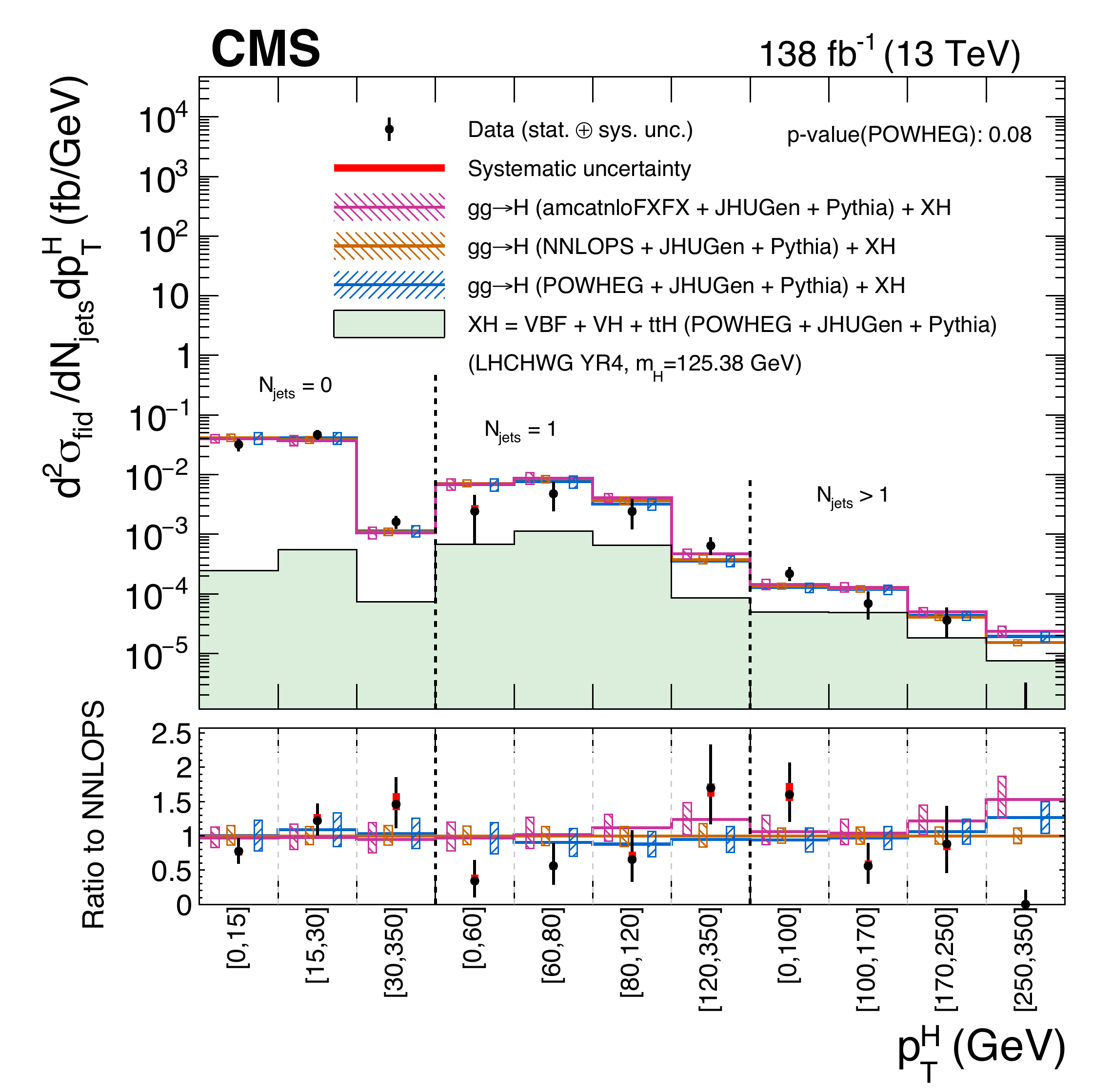}\\
		\includegraphics[width=0.48\textwidth]{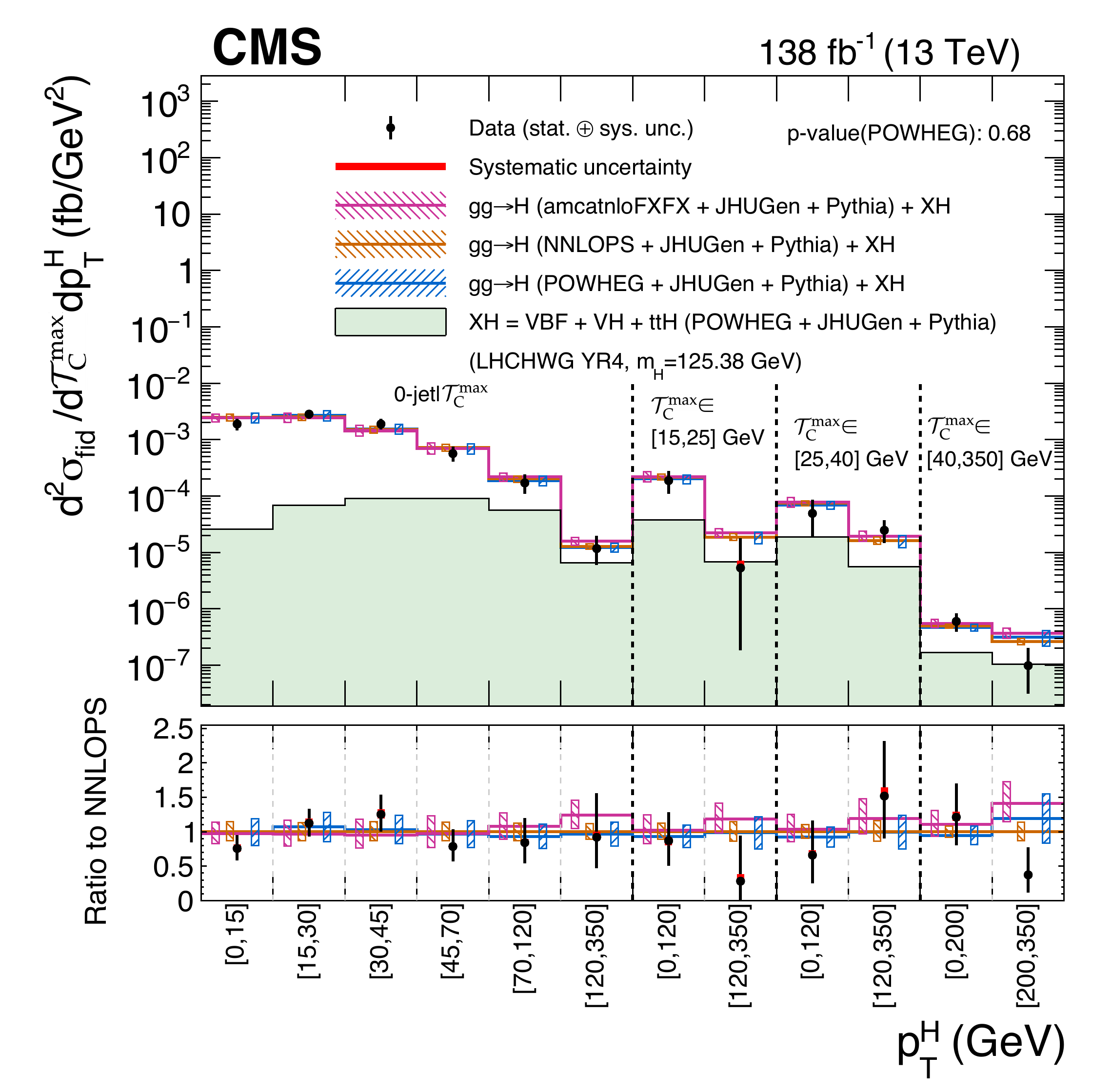}
		\caption{
			Double differential cross sections in bins of $\abs{y_{\PH}}$ vs. $\pt^{\PH}$ (upper \cmsLeft), number of associated jets vs.  $\pt^{\PH}$ (upper \cmsRight), and $\mathcal{T}_{\text{C}}^{\text{max}}$ vs. $\pt^{\PH}$ (lower).
			The binnings of the various measurements are reported in Table~\ref{tab:binBoundaries2D}.
			The content of each plot is described in the caption of Fig.~\ref{fig:fidPTH_YH}.
			\label{fig:fid2D_a}}
	\end{figure}
\end{center}

\begin{center}
	\begin{figure}[!htb]
		\centering
		\includegraphics[width=0.48\textwidth]{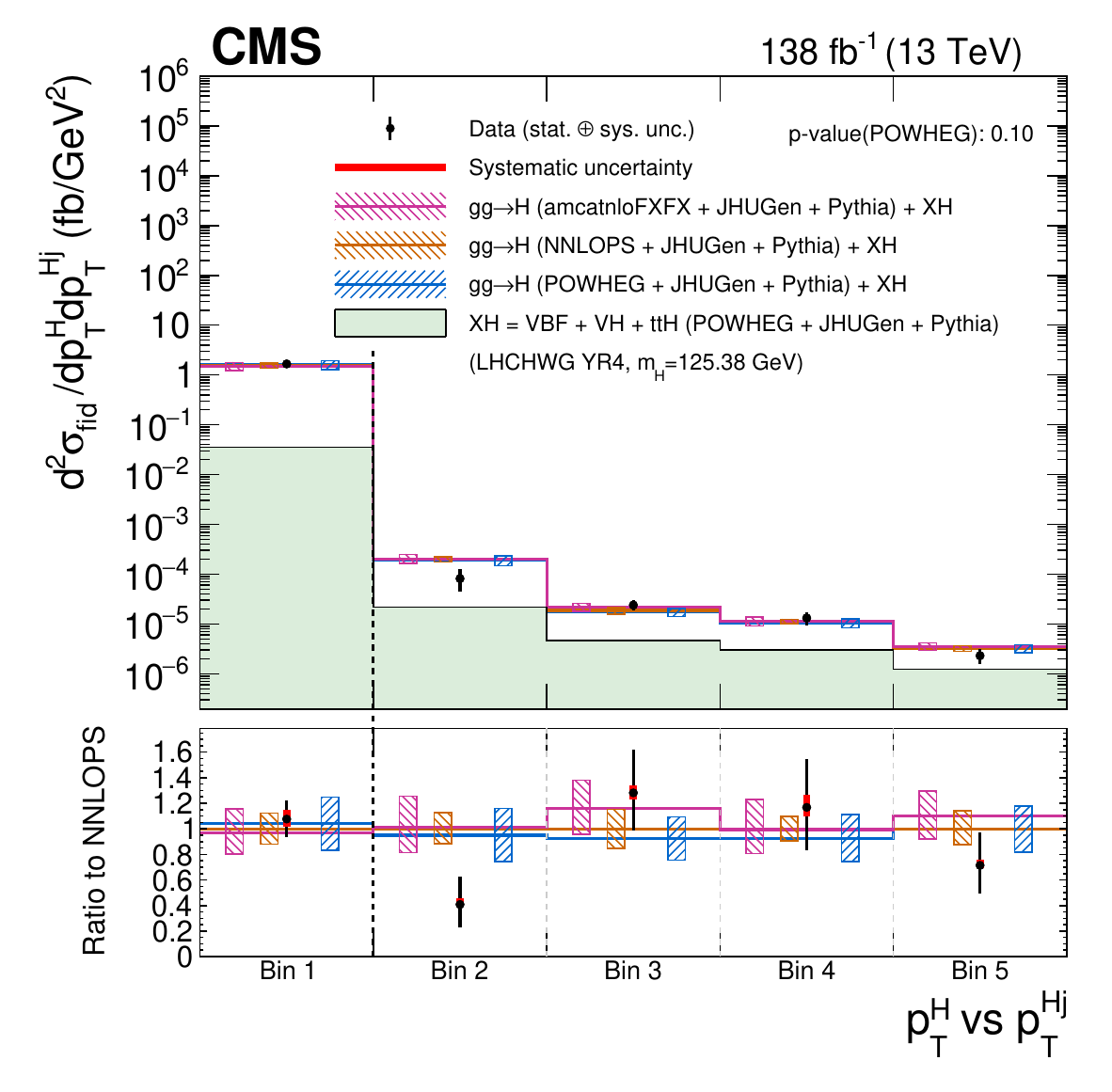}
		\includegraphics[width=0.48\textwidth]{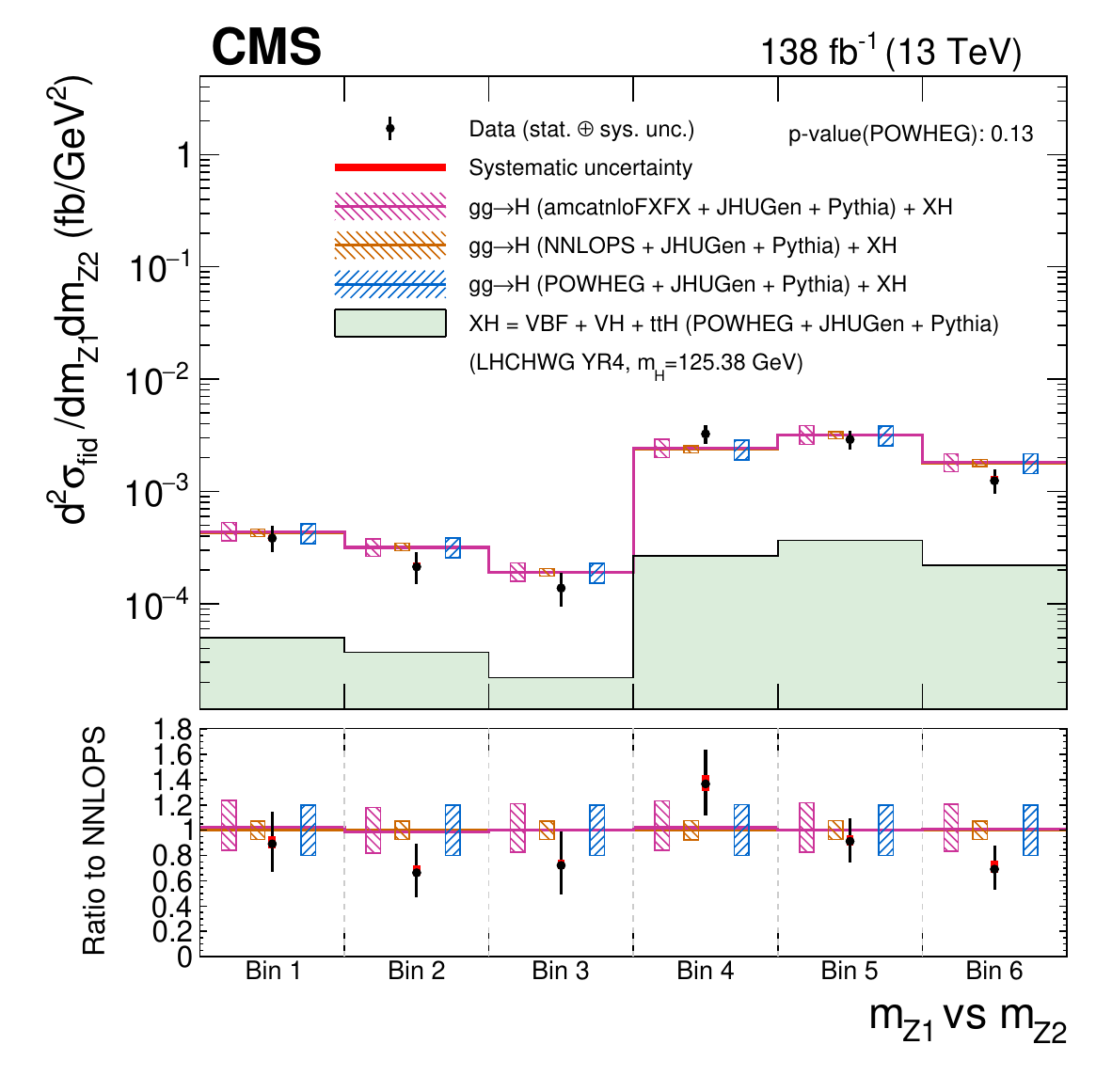}\\
		\includegraphics[width=0.48\textwidth]{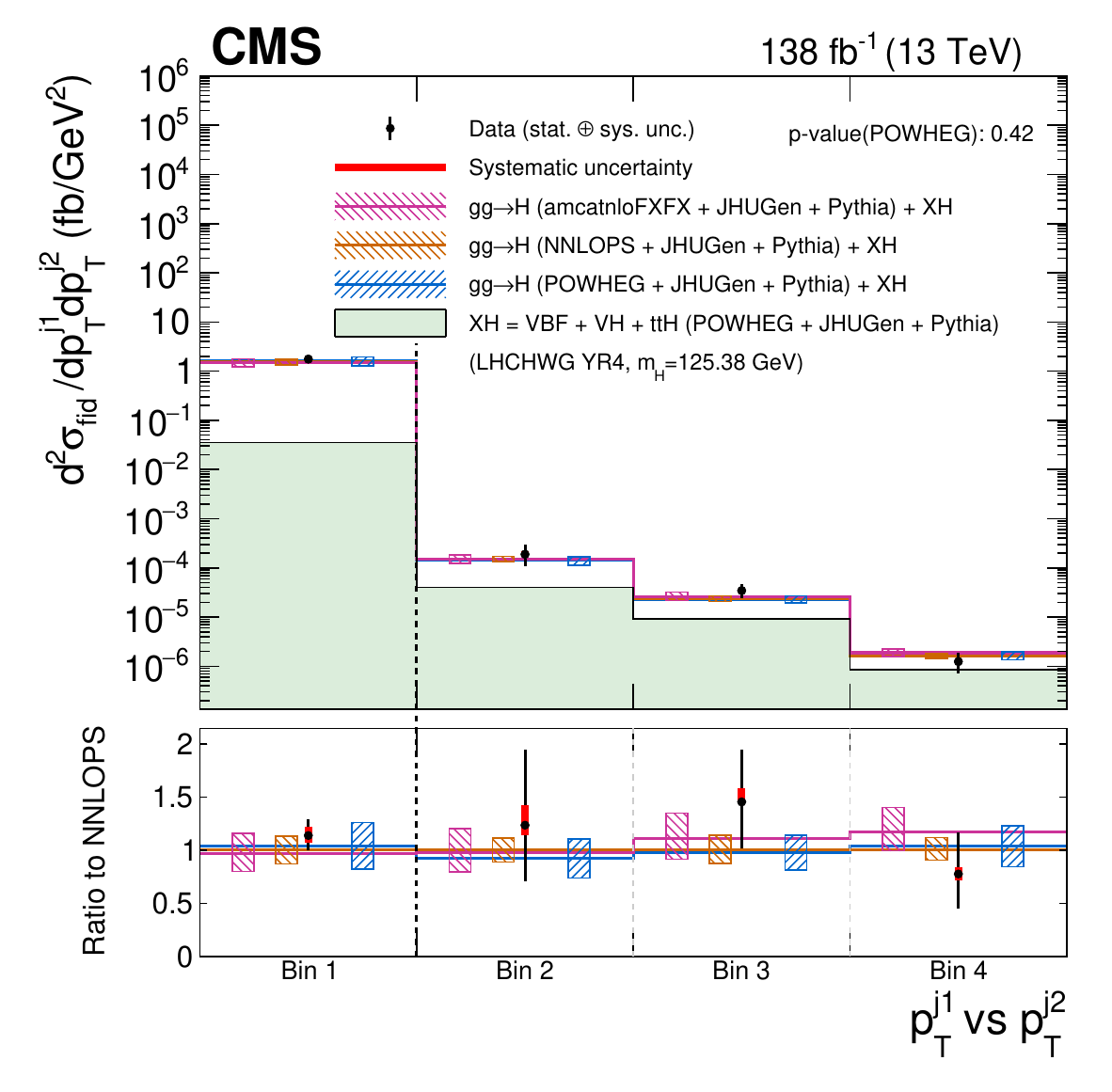}
		\caption{
			Double differential cross sections in bins of $\pt^{\PH \text{j}}$ vs. $\pt^{\PH}$ (upper \cmsLeft), $m_{\PZ_{1}}$ vs. $m_{\PZ_{2}}$ (upper \cmsRight), and \pt of the leading  vs. subleading jet (lower).
			The binnings of the various measurements are reported in Table~\ref{tab:binBoundaries2D}.
			The content of each plot is described in the caption of Fig.~\ref{fig:fidPTH_YH}.
			\label{fig:fid2D_b}}
	\end{figure}
\end{center}

The differential cross section measurements presented so far ensure a good coverage of the production and decay phase spaces in the $\HZZfl$ channel, together with a separation of possible interference effects present in the same- and different-flavor final states.
To improve the characterization of this decay channel and to maximize the coverage and separation of the different phase space regions, a set of double-differential measurements is also performed.
The results are shown in Fig.~\ref{fig:fid2D_a} for $\abs{y_{\PH}}$ vs. $\pt^{\PH}$ (upper left), the number of associated jets vs.  $\pt^{\PH}$ (upper right), and $\mathcal{T}_{\text{C}}^{\text{max}}$ vs. $\pt^{\PH}$ (lower) and in Fig.~\ref{fig:fid2D_b} for $\pt^{\PH \text{j}}$ vs. $\pt^{\PH}$ (upper left), , $m_{\PZ_{1}}$ vs. $m_{\PZ_{2}}$ (upper right), and \pt of the leading  vs. subleading jet (lower).
The results are consistent with the SM expectations, with the largest difference observed in the $\pt^{\PH}$ bins in the $N_\text{jets}=1$ phase space region.
The deficit in the low-$\pt^{\PH}$ bins for $N_\text{jets}=1$ is explained by large correlations with the high-$\pt^{\PH}$ bin of $N_\text{jets}=1$ and the first $\pt^{\PH}$ bin in the $N_\text{jets}>1$ phase space regions, where the fit to the data shows an excess with respect to the SM prediction.

\section{Interpretations}
\label{sec:interpretations}

\subsection{Constraints on the H boson self-coupling}
The differential cross section for the \PH boson production as a function of $\pt^\PH$ can be used to extract limits on the \PH boson self-coupling, following the approach described in Refs.~\cite{Degrassi:2016wml,Maltoni:2017ims,DiVita:2017eyz}.
At NLO in pQCD the \PH boson production includes processes sensitive to the trilinear  self-coupling ($\lambda_3$). 
The production modes $\ttH$ and $\VH$ introduce sizeable contributions to the \PH boson self-coupling due to the large vector boson and the top quark masses, whereas $\ggH$ and $\VBF$ production lead to much smaller contributions to the loop correction and are therefore less sensitive to possible modifications of $\lambda_3$.

The cross sections for the various production mechanisms of the \PH boson are parametrized as functions of a coupling modifier $\kappa_\lambda = \lambda_3/\lambda_3^\text{SM}$ in order to account for NLO terms arising from the \PH boson trilinear self-coupling.
The signal model defined in Section~\ref{sec:measurement} is modified by fixing the cross sections and branching fractions to their SM expectation values and by introducing scaling functions $\mu_{i,j}(\kappa_\lambda)$ in each bin $i$ of $\pt^\PH$, for each production mechanism $j$.
The dominant production mechanism is $\ggH$, for which a differential parametrization of the cross section as a function of  $\kappa_\lambda$ is not available yet, as discussed in Refs.~\cite{Degrassi:2016wml,Maltoni:2017ims,DiVita:2017eyz}.
The inclusive value is used for the parametrization of the \PH boson cross section for this production mechanism, taking into account an inclusive $\mathcal{O}(\lambda_3)$ correction factor.

In order to compute the scaling functions $\mu_{i,j}(\kappa_\lambda)$ for the other production modes, LO parton-level events are generated using \MGvATNLO~2.5.5 and are reweighted on an event-by-event basis using a dedicated EW reweighting tool, which computes the corresponding NLO $\lambda_3$-corrections ($\mathcal{O}(\lambda_3)$).
The ratio of the $\mathcal{O}(\lambda_3)$ to the LO distributions in bins of $\pt^\PH$ is used to derive the scaling functions $\mu_{i,j}(\kappa_\lambda)$ as detailed in Ref.~\cite{Maltoni:2017ims}.

\begin{figure}[!htb]
	\centering
	\includegraphics[width=0.6\textwidth]{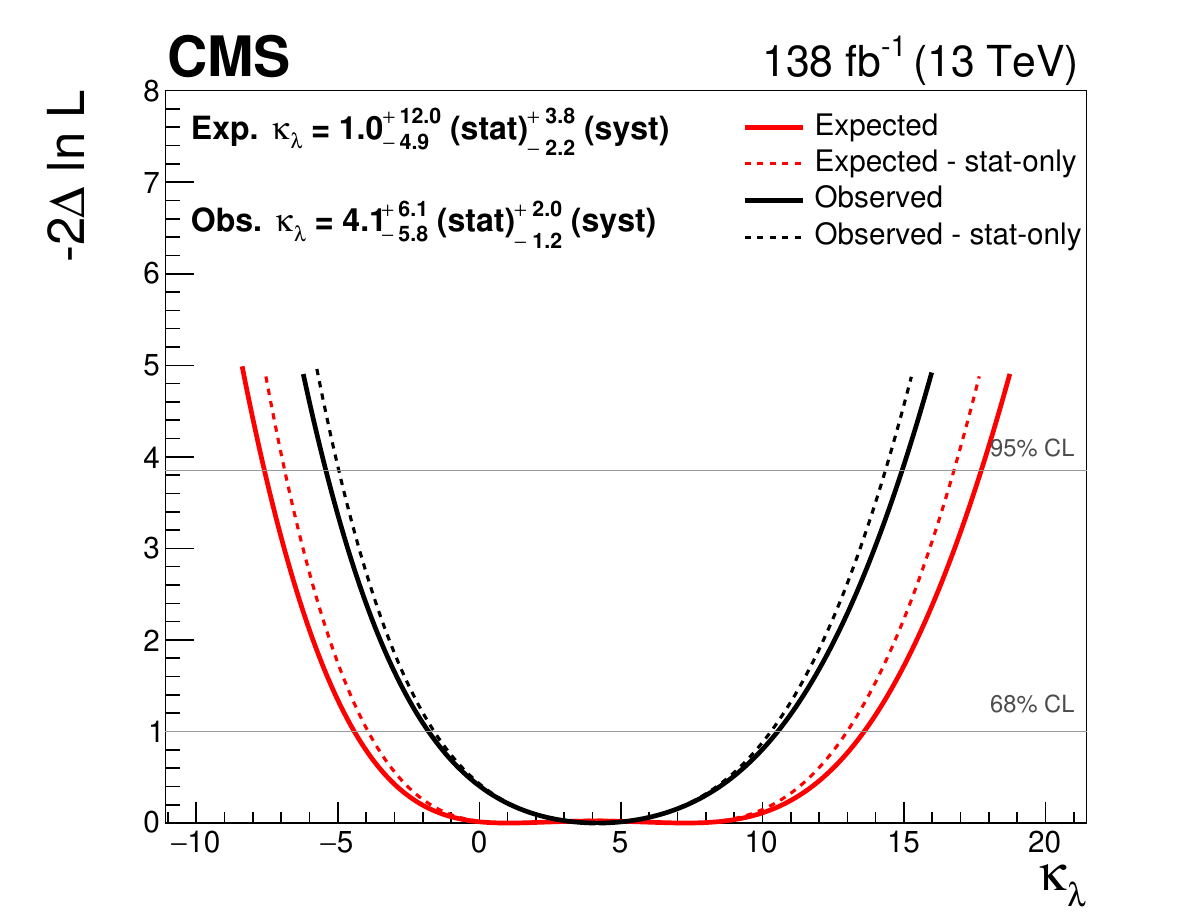}
	\caption{
		Likelihood scan as a function of $\kappa_\lambda$.
		The scan is shown with (solid line) and without (dashed line) systematic uncertainties profiled in the fit.
		\label{fig:klambda_Scan}}
\end{figure}

Constraints on $\kappa_\lambda$ are extracted from the maximum likelihood scan in the range $-10<\kappa_\lambda<20$, outside which the model is no longer valid as NLO effects start to dominate, while the other \PH couplings are fixed to their SM value.
The likelihood scan as a function of $\kappa_\lambda$ is shown in Fig.~\ref{fig:klambda_Scan}. 

The minimum of the negative log-likelihood ratio corresponds to a measured value of: 
\begin{linenomath*}
\begin{equation}
	\kappa_\lambda = 4.1^{+6.4}_{-5.9} = 4.1^{+6.1}_{-5.8}\stat^{+2.0}_{-1.2}\syst
\end{equation}
\end{linenomath*}
for an expected value of:
\begin{linenomath*}
\begin{equation}
	\kappa_\lambda = 1.0^{+12.6}_{-5.4} = 1.0^{+12.0}_{-4.9}\stat^{+3.8}_{-2.2}\syst.
\end{equation}
\end{linenomath*}

The corresponding observed (expected) excluded $\kappa_\lambda$ range at the 95\% confidence level (CL) is: $$-5.4\,(-7.6)\,<\kappa_\lambda<\,14.9\,(17.7).$$
The current best available constraints on $\kappa_\lambda$ are obtained from the combination of measurements of \PH boson pair production performed with the Run 2 data-set.
The limits set by the ATLAS and CMS Collaborations correspond to observed limits at the 95~\% CL of $-0.6<\kappa_\lambda<6.6$~\cite{ATLAS:2022jtk} 
and $-1.24<\kappa_\lambda<6.49$~\cite{CMS:2022dwd}, respectively.

\subsection{Constraints on the charm and bottom quark couplings}
The $\pt^\PH$ differential cross section of the $\ggH$ production mechanism is used to set constraints on the \PH boson coupling modifiers to $\PQb$ and $\PQc$ quarks in the context of the $\kappa$-framework~\cite{Heinemeyer:2013tqa}.
In fact, because of the presence of $\PQb$ and $\PQc$ quarks in the $\ggH$ loop~\cite{Bishara:2016jga}, anomalous values of these couplings can result in modifications of the $\ggH$ cross section.
The other production mechanisms are set to their SM expectation and no dependence from these couplings is assumed. The \PH boson coupling to the top quark is fixed to the SM value. The effects of the associated production with $\PQb$ quarks, whose contribution increases with increasing values of the coupling of the \PH boson to the $\PQb$ quark, are taken into account in the theoretical inputs to compute the parametrization.

The results are extracted from a maximum likelihood fit where the approach described in Section~\ref{sec:measurement} is modified by separating the $\ggH$ production from the other mechanisms, which are considered as background and constrained to the SM predictions with their respective uncertainties.
The combined effect of the \PH boson couplings to $\PQb$ ($\kappa_\PQb$) and $\PQc$ quarks ($\kappa_\PQc$) is modeled independently in each bin of the  $\pt^\PH$ spectrum by means of a quadratic polynomial, following the strategy of Ref.~\cite{CMS:2018gwt}.

\begin{figure}[!htb]
	\centering
	\includegraphics[width=0.45\textwidth]{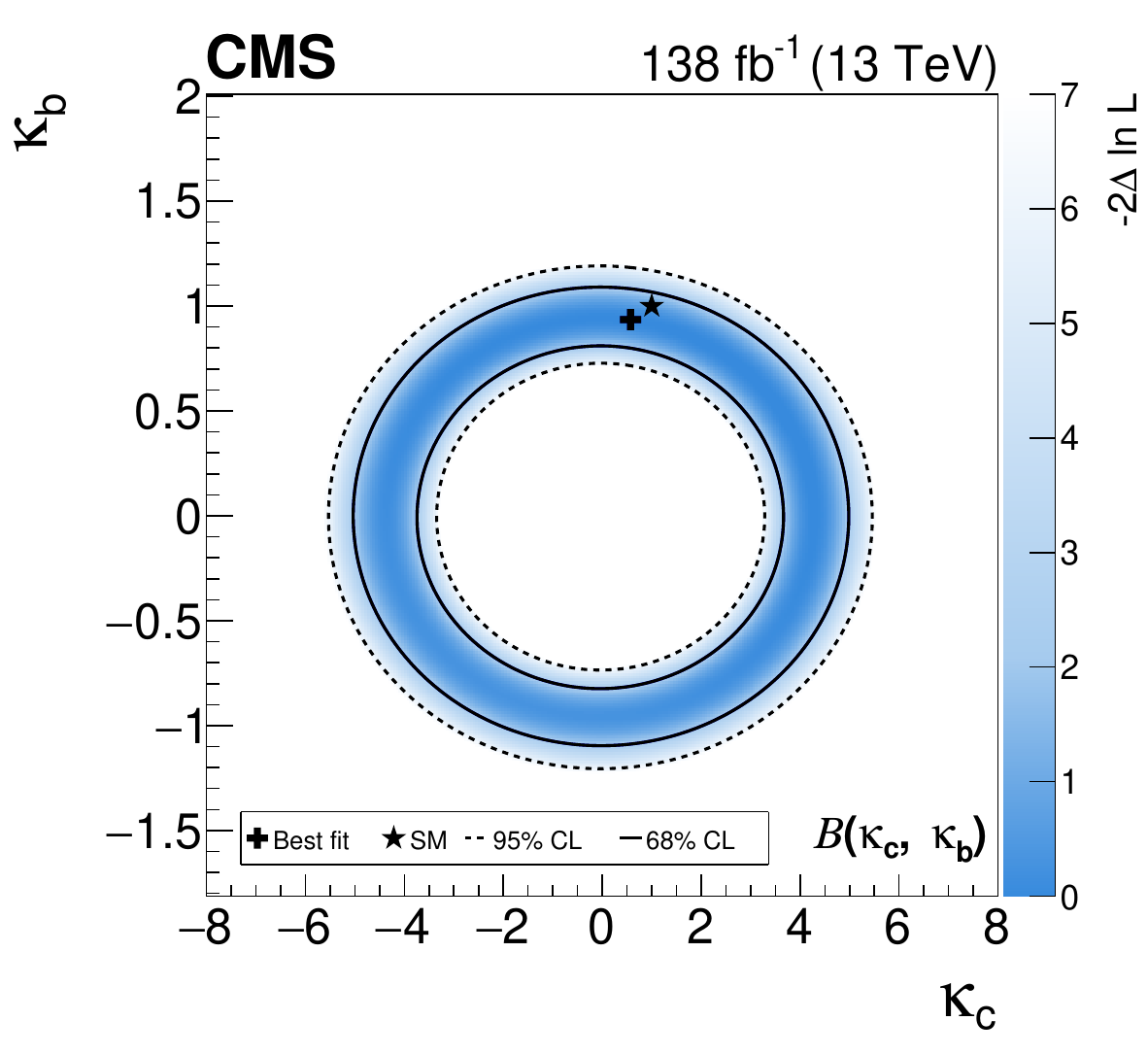}
	\includegraphics[width=0.45\textwidth]{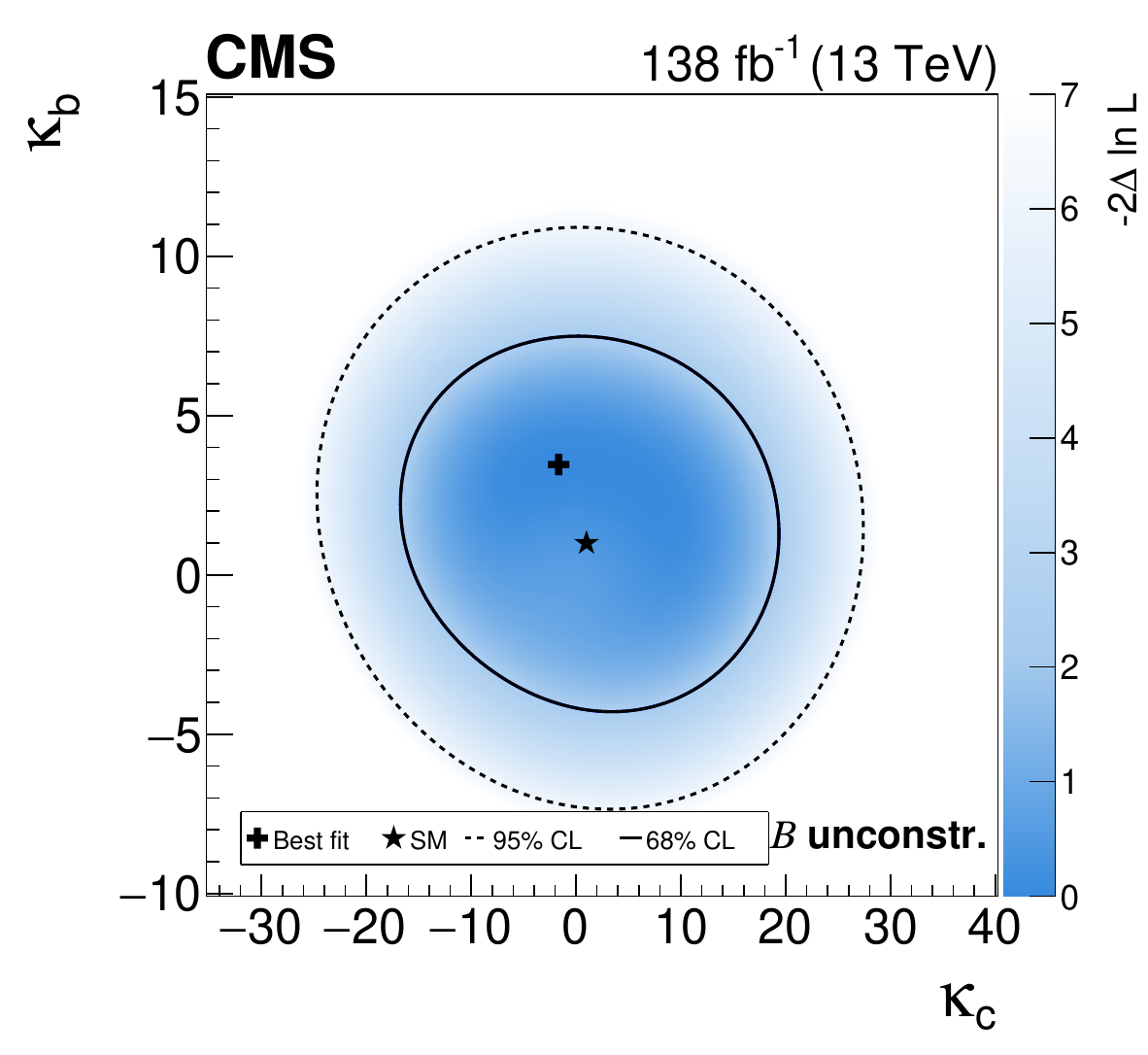}
	\caption{
		Simultaneous fit of $\kappa_\PQb$ and $\kappa_\PQc$, assuming a coupling dependence of the branching fraction (\cmsLeft) and treating it as an unconstrained parameter in the fit (\cmsRight).
		\label{fig:kbkc}}
\end{figure}

Figure~\ref{fig:kbkc} (\cmsLeft) shows the 2D likelihood scan of $\kappa_\PQb$ and $\kappa_\PQc$ under the assumption that the $\HZZ$ branching fraction is dependent on the \PH boson couplings $\mathcal{B} = \mathcal{B}(\kappa_\PQb,\kappa_\PQc)$, with all the other couplings fixed to their SM value, and assuming no beyond-the-SM contributions. As expected, the result is constrained by the saturation of the total width.

A simultaneous constraint on $\kappa_\PQb$ and $\kappa_\PQc$ is also derived by treating the $\HZZ$ branching fraction as an unconstrained parameter in the fit. The constraint from the total width and the overall normalization is removed in this way, and what remains is purely the constraint obtained from the shape of the $\pt^\PH$ spectrum. The result is shown in Fig.~\ref{fig:kbkc} (\cmsRight).

Confidence intervals on $\kappa_\PQb$ and $\kappa_\PQc$ are obtained from a maximum likelihood fit leaving one of the two parameters unconstrained in the fit and scanning the other.
The observed (expected) exclusion limits at the 95\% CL are:
\begin{equation}
	\begin{aligned}
	-1.1\, (-1.3) &<\kappa_\PQb<& 1.1\, (1.2)
	\\
	-5.3 \,(-5.7) &<\kappa_\PQc<&5.2\, (5.7),
	\label{eq:kbkc_scan_brfix}
\end{aligned}
\end{equation}
assuming a dependence of the branching fraction on $\kappa_\PQb$ and $\kappa_\PQc$, and:
\begin{equation}
\begin{aligned}
	-5.6 \,(-5.5) &<\kappa_\PQb<& 8.9\, (7.4)
	\\
	-20\, (-19) &<\kappa_\PQc<& 23\, (20),
	\label{eq:kbkc_scan_brfloat}
\end{aligned}
\end{equation}
treating the branching fraction as an unconstrained parameter in the fit.

\section{Summary}
\label{sec:summary}
This paper presents a comprehensive characterization of the $\HZZfl$ decay channel via the measurement of fiducial differential cross sections as functions of  several kinematic observables.
The \PH boson production is characterized via measurements of differential cross sections in bins of $\pt^\PH$ and $\abs{y_\text{\PH}}$, the \pt of the leading and subleading jets and observables of the dijet system, when associated with jets.
Fiducial cross sections are measured in bins of the seven kinematic observables that completely define the four-lepton decay: the invariant mass of the two $\PZ$ bosons and the five angles that describe the fermions kinematical properties and the production and decay planes.
Differential cross sections are also measured in bins of six matrix element kinematic discriminants sensitive to various anomalous couplings of the \PH boson to vector bosons.
The dynamical evolution of the renormalization and factorization scales, and resummation effects are probed by measuring cross sections in bins of rapidity-weighted jet vetoes, and in bins of observables of the \PH plus jets system.
An extensive set of double-differential measurements is presented, providing a complete coverage of the phase space under study.
The $\HZZfl$ inclusive fiducial cross section is $\sigma_{{\text{fid}}}=2.73\pm0.26\unit{fb}=2.73\pm0.22\stat\pm0.15\syst \unit{fb}$, in agreement with the SM expectation of $2.86\pm0.15\unit{fb}$.
The measurement of the fiducial cross section in differential bins of $\pt^\PH$ is used to set constraints on the trilinear self-coupling of the \PH boson, with an observed (expected) limit of $-5.4\,(-7.6)\,<\kappa_\lambda<\,14.9\,(17.7)$ at the 95\% CL.
Finally, constraints on the modifiers of \PH boson couplings to \PQb and \PQc quarks ($\kappa_\PQb$ and $\kappa_\PQc$) are also determined with an observed (expected) limit of $-1.1\,(-1.3)<\kappa_\PQb<1.1\,(1.2)$ and $-5.3\,(-5.7)<\kappa_\PQc<5.2\,(5.7)$ at the 95\% CL, obtained assuming a dependence of the branching fraction on $\kappa_\PQb$ and $\kappa_\PQc$.
All results are consistent with the SM predictions for the $\HZZfl$ decay channel in the considered fiducial phase space.

\begin{acknowledgments}
We congratulate our colleagues in the CERN accelerator departments for the excellent performance of the LHC and thank the technical and administrative staffs at CERN and at other CMS institutes for their contributions to the success of the CMS effort. In addition, we gratefully acknowledge the computing centres and personnel of the Worldwide LHC Computing Grid and other centres for delivering so effectively the computing infrastructure essential to our analyses. Finally, we acknowledge the enduring support for the construction and operation of the LHC, the CMS detector, and the supporting computing infrastructure provided by the following funding agencies: BMBWF and FWF (Austria); FNRS and FWO (Belgium); CNPq, CAPES, FAPERJ, FAPERGS, and FAPESP (Brazil); MES and BNSF (Bulgaria); CERN; CAS, MoST, and NSFC (China); MINCIENCIAS (Colombia); MSES and CSF (Croatia); RIF (Cyprus); SENESCYT (Ecuador); MoER, ERC PUT and ERDF (Estonia); Academy of Finland, MEC, and HIP (Finland); CEA and CNRS/IN2P3 (France); BMBF, DFG, and HGF (Germany); GSRI (Greece); NKFIH (Hungary); DAE and DST (India); IPM (Iran); SFI (Ireland); INFN (Italy); MSIP and NRF (Republic of Korea); MES (Latvia); LAS (Lithuania); MOE and UM (Malaysia); BUAP, CINVESTAV, CONACYT, LNS, SEP, and UASLP-FAI (Mexico); MOS (Montenegro); MBIE (New Zealand); PAEC (Pakistan); MES and NSC (Poland); FCT (Portugal); MESTD (Serbia); MCIN/AEI and PCTI (Spain); MOSTR (Sri Lanka); Swiss Funding Agencies (Switzerland); MST (Taipei); MHESI and NSTDA (Thailand); TUBITAK and TENMAK (Turkey); NASU (Ukraine); STFC (United Kingdom); DOE and NSF (USA).

\hyphenation{Rachada-pisek} Individuals have received support from the Marie-Curie programme and the European Research Council and Horizon 2020 Grant, contract Nos.\ 675440, 724704, 752730, 758316, 765710, 824093, 884104, 899987, and COST Action CA16108 (European Union); the Leventis Foundation; the Alfred P.\ Sloan Foundation; the Alexander von Humboldt Foundation; the Belgian Federal Science Policy Office; the Fonds pour la Formation \`a la Recherche dans l'Industrie et dans l'Agriculture (FRIA-Belgium); the Agentschap voor Innovatie door Wetenschap en Technologie (IWT-Belgium); the F.R.S.-FNRS and FWO (Belgium) under the ``Excellence of Science -- EOS" -- be.h project n.\ 30820817; the Beijing Municipal Science \& Technology Commission, No. Z191100007219010 and Fundamental Research Funds for the Central Universities (China); the Ministry of Education, Youth and Sports (MEYS) of the Czech Republic; the Hellenic Foundation for Research and Innovation (HFRI), Project Number 2288 (Greece); the Deutsche Forschungsgemeinschaft (DFG), under Germany's Excellence Strategy -- EXC 2121 ``Quantum Universe" -- 390833306, and under project number 400140256 - GRK2497; the Hungarian Academy of Sciences, the New National Excellence Program - \'UNKP, the NKFIH research grants K 124845, K 124850, K 128713, K 128786, K 129058, K 131991, K 133046, K 138136, K 143460, K 143477, 2020-2.2.1-ED-2021-00181, and TKP2021-NKTA-64 (Hungary); the Council of Science and Industrial Research, India; the Latvian Council of Science; the Ministry of Education and Science, project no. 2022/WK/14, and the National Science Center, contracts Opus 2021/41/B/ST2/01369 and 2021/43/B/ST2/01552 (Poland); the Funda\c{c}\~ao para a Ci\^encia e a Tecnologia, grant CEECIND/01334/2018 (Portugal); the National Priorities Research Program by Qatar National Research Fund; MCIN/AEI/10.13039/501100011033, ERDF ``a way of making Europe", and the Programa Estatal de Fomento de la Investigaci{\'o}n Cient{\'i}fica y T{\'e}cnica de Excelencia Mar\'{\i}a de Maeztu, grant MDM-2017-0765 and Programa Severo Ochoa del Principado de Asturias (Spain); the Chulalongkorn Academic into Its 2nd Century Project Advancement Project, and the National Science, Research and Innovation Fund via the Program Management Unit for Human Resources \& Institutional Development, Research and Innovation, grant B05F650021 (Thailand); the Kavli Foundation; the Nvidia Corporation; the SuperMicro Corporation; the Welch Foundation, contract C-1845; and the Weston Havens Foundation (USA).
\end{acknowledgments}\section*{Data availability} Release and preservation of data used by the CMS Collaboration as the basis for publications is guided by the  \href{https://doi.org/10.7483/OPENDATA.CMS.1BNU.8V1W}{CMS data preservation, re-use and open access policy}.

\bibliography{auto_generated}

@ARTICLE{Aad:2012tfa,
	AUTHOR=	"{ATLAS Collaboration}",
	TITLE=	"Observation of a new particle in the search for the Standard Model {Higgs} boson with the {ATLAS} detector at the {LHC}",
	JOURNAL=	"Phys. Lett. B",
	VOLUME=	"716",
	PAGES=	"1",
	DOI=	"10.1016/j.physletb.2012.08.020",
	YEAR=	"2012",
	EPRINT=	"1207.7214",
	ARCHIVEPREFIX=	"arXiv",
	PRIMARYCLASS=	"hep-ex",
	REPORTNUMBER=	"CERN-PH-EP-2012-218",
	SLACCITATION=	"%%CITATION = ARXIV:1207.7214;%%",
}

@ARTICLE{Chatrchyan:2012ufa,
	AUTHOR=	"{CMS Collaboration}",
	TITLE=	"Observation of a new boson at a mass of 125 {GeV} with the {CMS} experiment at the {LHC}",
	JOURNAL=	"Phys. Lett. B",
	VOLUME=	"716",
	PAGES=	"30",
	DOI=	"10.1016/j.physletb.2012.08.021",
	YEAR=	"2012",
	EPRINT=	"1207.7235",
	ARCHIVEPREFIX=	"arXiv",
	PRIMARYCLASS=	"hep-ex",
	REPORTNUMBER=	"CMS-HIG-12-028, CERN-PH-EP-2012-220",
	SLACCITATION=	"%%CITATION = ARXIV:1207.7235;%%",
}

@ARTICLE{Chatrchyan:2013lba,
	AUTHOR=	"{CMS Collaboration}",
	TITLE=	"Observation of a new boson with mass near 125 {GeV} in pp collisions at {$\sqrt{s}$ = 7 and 8 TeV}",
	JOURNAL=	"JHEP",
	VOLUME=	"06",
	PAGES=	"081",
	DOI=	"10.1007/JHEP06(2013)081",
	YEAR=	"2013",
	EPRINT=	"1303.4571",
	ARCHIVEPREFIX=	"arXiv",
	PRIMARYCLASS=	"hep-ex",
	REPORTNUMBER=	"CMS-HIG-12-036, CERN-PH-EP-2013-035",
	SLACCITATION=	"%%CITATION = ARXIV:1303.4571;%%",
}

@ARTICLE{Englert:1964et,
	AUTHOR=	"Englert, F. and Brout, R.",
	TITLE=	"Broken Symmetry and the Mass of Gauge Vector Mesons",
	JOURNAL=	"Phys. Rev. Lett.",
	VOLUME=	"13",
	PAGES=	"321",
	DOI=	"10.1103/PhysRevLett.13.321",
	YEAR=	"1964",
}

@ARTICLE{Higgs:1964ia,
	AUTHOR=	"Higgs, Peter W.",
	TITLE=	"Broken symmetries, massless particles and gauge fields",
	JOURNAL=	"Phys. Lett.",
	VOLUME=	"12",
	PAGES=	"132",
	DOI=	"10.1016/0031-9163(64)91136-9",
	YEAR=	"1964",
}

@ARTICLE{Higgs:1964pj,
	AUTHOR=	"Higgs, Peter W.",
	TITLE=	"Broken Symmetries and the Masses of Gauge Bosons",
	JOURNAL=	"Phys. Rev. Lett.",
	VOLUME=	"13",
	PAGES=	"508",
	DOI=	"10.1103/PhysRevLett.13.508",
	YEAR=	"1964",
}

@ARTICLE{Guralnik:1964eu,
	AUTHOR=	"Guralnik, G. S. and Hagen, C. R. and Kibble, T. W. B.",
	TITLE=	"Global Conservation Laws and Massless Particles",
	JOURNAL=	"Phys. Rev. Lett.",
	VOLUME=	"13",
	PAGES=	"585",
	DOI=	"10.1103/PhysRevLett.13.585",
	YEAR=	"1964",
}

@ARTICLE{Higgs:1966ev,
	AUTHOR=	"Higgs, Peter W.",
	TITLE=	"Spontaneous Symmetry Breakdown without Massless Bosons",
	JOURNAL=	"Phys. Rev.",
	VOLUME=	"145",
	PAGES=	"1156",
	DOI=	"10.1103/PhysRev.145.1156",
	YEAR=	"1966",
}

@ARTICLE{Kibble:1967sv,
	AUTHOR=	"Kibble, T. W. B.",
	TITLE=	"Symmetry breaking in nonAbelian gauge theories",
	JOURNAL=	"Phys. Rev.",
	VOLUME=	"155",
	PAGES=	"1554",
	DOI=	"10.1103/PhysRev.155.1554",
	YEAR=	"1967",
}

@ARTICLE{ATLASPropertiesRun1,
	AUTHOR=	"{ATLAS Collaboration}",
	TITLE=	"Measurements of the {Higgs} boson production and decay rates and coupling strengths using pp collision data at {$\sqrt{s}=7$ and 8 TeV} in the {ATLAS} experiment",
	JOURNAL=	"Eur. Phys. J. C",
	VOLUME=	"76",
	PAGES=	"6",
	YEAR=	"2016",
	DOI=	"10.1140/epjc/s10052-015-3769-y",
	EPRINT=	"1507.04548",
	ARCHIVEPREFIX=	"arXiv",
	PRIMARYCLASS=	"hep-ex",
	REPORTNUMBER=	"CERN-PH-EP-2015-125",
	SLACCITATION=	"%%CITATION = ARXIV:1507.04548;%%",
}

@ARTICLE{CMSPropertiesRun1,
	AUTHOR=	"{CMS Collaboration}",
	TITLE=	"Precise determination of the mass of the {Higgs} boson and tests of compatibility of its couplings with the standard model predictions using proton collisions at {7 and 8 $\,\text {TeV}$}",
	JOURNAL=	"Eur. Phys. J. C",
	VOLUME=	"75",
	YEAR=	"2015",
	PAGES=	"212",
	DOI=	"10.1140/epjc/s10052-015-3351-7",
	EPRINT=	"1412.8662",
	ARCHIVEPREFIX=	"arXiv",
	PRIMARYCLASS=	"hep-ex",
	REPORTNUMBER=	"CMS-HIG-14-009, CERN-PH-EP-2014-288",
	SLACCITATION=	"%%CITATION = ARXIV:1412.8662;%%",
}

@ARTICLE{ATLASCMSMassRun1,
	AUTHOR=	"{ATLAS and CMS Collaborations}",
	TITLE=	"Combined Measurement of the {Higgs} Boson Mass in pp Collisions at {$\sqrt{s}=7$ and 8 TeV} with the {ATLAS} and {CMS} Experiments",
	JOURNAL=	"Phys. Rev. Lett.",
	VOLUME=	"114",
	YEAR=	"2015",
	PAGES=	"191803",
	DOI=	"10.1103/PhysRevLett.114.191803",
	EPRINT=	"1503.07589",
	ARCHIVEPREFIX=	"arXiv",
	PRIMARYCLASS=	"hep-ex",
	REPORTNUMBER=	"ATLAS-HIGG-2014-14, CMS-HIG-14-042, CERN-PH-EP-2015-075",
	SLACCITATION=	"%%CITATION = ARXIV:1503.07589;%%",
}

@ARTICLE{ATLASCMSPropertiesRun1,
	AUTHOR=	"{ATLAS and CMS Collaborations}",
	TITLE=	"Measurements of the {Higgs} boson production and decay rates and constraints on its couplings from a combined {ATLAS} and {CMS} analysis of the {LHC} pp collision data at {$\sqrt{s}=7$ and 8 TeV}",
	JOURNAL=	"JHEP",
	YEAR=	"2016",
	VOLUME=	"08",
	PAGES=	"45",
	DOI=	"10.1007/JHEP08(2016)045",
	EPRINT=	"1606.02266",
	ARCHIVEPREFIX=	"arXiv",
	PRIMARYCLASS=	"hep-ex",
}

@ARTICLE{ATLASH4lLegacyRun1,
	AUTHOR=	"{ATLAS Collaboration}",
	TITLE=	"Measurements of {Higgs} boson production and couplings in the four-lepton channel in pp collisions at center-of-mass energies of 7 and 8 {TeV} with the {ATLAS} detector",
	JOURNAL=	"Phys. Rev. D",
	VOLUME=	"91",
	YEAR=	"2015",
	PAGES=	"012006",
	DOI=	"10.1103/PhysRevD.91.012006",
	EPRINT=	"1408.5191",
	ARCHIVEPREFIX=	"arXiv",
	PRIMARYCLASS=	"hep-ex",
	REPORTNUMBER=	"CERN-PH-EP-2014-170",
	SLACCITATION=	"%%CITATION = ARXIV:1408.5191;%%",
}

@ARTICLE{CMSH4lLegacyRun1,
	AUTHOR=	"{CMS Collaboration}",
	TITLE=	"Measurement of the properties of a {Higgs} boson in the four-lepton final state",
	JOURNAL=	"Phys. Rev. D",
	VOLUME=	"89",
	YEAR=	"2014",
	PAGES=	"092007",
	DOI=	"10.1103/PhysRevD.89.092007",
	EPRINT=	"1312.5353",
	ARCHIVEPREFIX=	"arXiv",
	PRIMARYCLASS=	"hep-ex",
	REPORTNUMBER=	"CMS-HIG-13-002, CERN-PH-EP-2013-220",
	SLACCITATION=	"%%CITATION = ARXIV:1312.5353;%%",
}

@ARTICLE{CMSH4lSpinParity,
	AUTHOR=	"{CMS Collaboration}",
	TITLE=	"Study of the Mass and Spin-Parity of the {Higgs} Boson Candidate via Its Decays to {$Z$} Boson Pairs",
	JOURNAL=	"Phys. Rev. Lett.",
	VOLUME=	"110",
	PAGES=	"081803",
	DOI=	"10.1103/PhysRevLett.110.081803",
	YEAR=	"2013",
	EPRINT=	"1212.6639",
	ARCHIVEPREFIX=	"arXiv",
	PRIMARYCLASS=	"hep-ex",
	REPORTNUMBER=	"CMS-HIG-12-041, CERN-PH-EP-2012-372",
	SLACCITATION=	"%%CITATION = ARXIV:1212.6639;%%",
}

@ARTICLE{CMSH4lAnomalousCouplings,
	AUTHOR=	"{CMS Collaboration}",
	TITLE=	"Constraints on the spin-parity and anomalous {HVV} couplings of the {Higgs} boson in proton collisions at {7 and 8\TeV}",
	JOURNAL=	"Phys. Rev. D",
	VOLUME=	"92",
	PAGES=	"012004",
	YEAR=	"2015",
	DOI=	"10.1103/PhysRevD.92.012004",
	EPRINT=	"1411.3441",
	ARCHIVEPREFIX=	"arXiv",
	PRIMARYCLASS=	"hep-ex",
	REPORTNUMBER=	"CMS-HIG-14-018, CERN-PH-EP-2014-265",
}

@ARTICLE{CMSH4l2016,
	AUTHOR=	"{CMS Collaboration}",
	TITLE=	"Measurements of properties of the {Higgs} boson decaying into the four-lepton final state in pp collisions at $ \sqrt{s}=13 $ {TeV}",
	JOURNAL=	"JHEP",
	VOLUME=	"11",
	YEAR=	"2017",
	PAGES=	"047",
	DOI=	"10.1007/JHEP11(2017)047",
	EPRINT=	"1706.09936",
	ARCHIVEPREFIX=	"arXiv",
	PRIMARYCLASS=	"hep-ex",
	REPORTNUMBER=	"CMS-HIG-16-041, CERN-EP-2017-123",
	SLACCITATION=	"%%CITATION = ARXIV:1706.09936;%%",
}

@ARTICLE{ATLASH4l2016,
	AUTHOR=	"{ATLAS Collaboration}",
	TITLE=	"Measurement of the {Higgs} boson coupling properties in the {$H\rightarrow ZZ^{*} \rightarrow 4\ell$} decay channel at {$\sqrt{s}$ = 13 TeV} with the {ATLAS} detector",
	JOURNAL=	"JHEP",
	VOLUME=	"03",
	YEAR=	"2018",
	PAGES=	"095",
	DOI=	"10.1007/JHEP03(2018)095",
	EPRINT=	"1712.02304",
	ARCHIVEPREFIX=	"arXiv",
	PRIMARYCLASS=	"hep-ex",
	REPORTNUMBER=	"CERN-EP-2017-206",
	SLACCITATION=	"%%CITATION = ARXIV:1712.02304;%%",
}

@ARTICLE{CMSH4lWidth,
	AUTHOR=	"{CMS Collaboration}",
	TITLE=	"Constraints on the {Higgs} boson width from off-shell production and decay to {$\cPZ$}-boson pairs",
	JOURNAL=	"Phys. Lett. B",
	VOLUME=	"736",
	PAGES=	"64",
	DOI=	"10.1016/j.physletb.2014.06.077",
	YEAR=	"2014",
	EPRINT=	"1405.3455",
	ARCHIVEPREFIX=	"arXiv",
	PRIMARYCLASS=	"hep-ex",
	REPORTNUMBER=	"CMS-HIG-14-002, CERN-PH-EP-2014-078",
	SLACCITATION=	"%%CITATION = ARXIV:1405.3455;%%",
}

@ARTICLE{CMSH4lLifetime,
	AUTHOR=	"{CMS Collaboration}",
	TITLE=	"Limits on the {Higgs} boson lifetime and width from its decay to four charged leptons",
	JOURNAL=	"Phys. Rev. D",
	VOLUME=	"92",
	PAGES=	"072010",
	DOI=	"10.1103/PhysRevD.92.072010",
	YEAR=	"2015",
	EPRINT=	"1507.06656",
	ARCHIVEPREFIX=	"arXiv",
	PRIMARYCLASS=	"hep-ex",
	REPORTNUMBER=	"CMS-HIG-14-036, CERN-PH-EP-2015-159",
	SLACCITATION=	"%%CITATION = ARXIV:1507.06656;%%",
}

@ARTICLE{ATLASH4lWidth,
	AUTHOR=	"{ATLAS Collaboration}",
	TITLE=	"Constraints on the off-shell {Higgs} boson signal strength in the high-mass {$ZZ$ and $WW$} final states with the {ATLAS detector}",
	JOURNAL=	"Eur. Phys. J. C",
	VOLUME=	"75",
	YEAR=	"2015",
	PAGES=	"335",
	DOI=	"10.1140/epjc/s10052-015-3542-2",
	EPRINT=	"1503.01060",
	ARCHIVEPREFIX=	"arXiv",
	PRIMARYCLASS=	"hep-ex",
	REPORTNUMBER=	"CERN-PH-EP-2015-026",
	SLACCITATION=	"%%CITATION = ARXIV:1503.01060;%%",
}

@ARTICLE{ATLASH4lWidth2016,
	AUTHOR=	"{ATLAS Collaboration}",
	TITLE=	"Constraints on off-shell {Higgs} boson production and the {Higgs} boson total width in {$ZZ\to4\ell$ and $ZZ\to2\ell2\nu$} final states with the {ATLAS detector}",
	JOURNAL=	"Phys. Lett. B",
	VOLUME=	"786",
	YEAR=	"2018",
	PAGES=	"223",
	DOI=	"10.1016/j.physletb.2018.09.048",
	EPRINT=	"1808.01191",
	ARCHIVEPREFIX=	"arXiv",
	PRIMARYCLASS=	"hep-ex",
	REPORTNUMBER=	"CERN-EP-2018-178",
	SLACCITATION=	"%%CITATION = ARXIV:1808.01191;%%",
}

@ARTICLE{ATLASH4lFiducial8TeV,
	AUTHOR=	"{ATLAS Collaboration}",
	TITLE=	"Fiducial and differential cross sections of {Higgs} boson production measured in the four-lepton decay channel in pp collisions at $\sqrt{s}=8$ {TeV} with the {ATLAS detector}",
	JOURNAL=	"Phys. Lett. B",
	VOLUME=	"738",
	YEAR=	"2014",
	PAGES=	"234",
	DOI=	"10.1016/j.physletb.2014.09.054",
	EPRINT=	"1408.3226",
	ARCHIVEPREFIX=	"arXiv",
	PRIMARYCLASS=	"hep-ex",
	REPORTNUMBER=	"CERN-PH-EP-2014-186",
	SLACCITATION=	"%%CITATION = ARXIV:1408.3226;%%",
}

@ARTICLE{CMSH4lFiducial8TeV,
	AUTHOR=	"{CMS Collaboration}",
	TITLE=	"Measurement of differential and integrated fiducial cross sections for {Higgs} boson production in the four-lepton decay channel in pp collisions at $ \sqrt{s}=7$ and 8 {TeV}",
	JOURNAL=	"JHEP",
	VOLUME=	"04",
	YEAR=	"2016",
	PAGES=	"005",
	DOI=	"10.1007/JHEP04(2016)005",
	EPRINT=	"1512.08377",
	ARCHIVEPREFIX=	"arXiv",
	PRIMARYCLASS=	"hep-ex",
}

@ARTICLE{ATLASH4lFiducial2016,
	AUTHOR=	"{ATLAS Collaboration}",
	TITLE=	"Measurement of inclusive and differential cross sections in the {$H \rightarrow ZZ^* \rightarrow 4\ell$} decay channel in pp collisions at $\sqrt{s}=13$ {TeV} with the {ATLAS detector}",
	JOURNAL=	"JHEP",
	VOLUME=	"10",
	YEAR=	"2017",
	PAGES=	"132",
	DOI=	"10.1007/JHEP10(2017)132",
	EPRINT=	"1708.02810",
	ARCHIVEPREFIX=	"arXiv",
	PRIMARYCLASS=	"hep-ex",
	REPORTNUMBER=	"CERN-EP-2017-139",
	SLACCITATION=	"%%CITATION = ARXIV:1708.02810;%%",
}

@ARTICLE{ATLASH4lLegacyRun2,
	AUTHOR=	"{ATLAS Collaboration}",
	TITLE=	"Higgs boson production cross-section measurements and their {EFT} interpretation in the $4\ell$ decay channel at $\sqrt{s}=13$ {TeV} with the {ATLAS detector}",
	JOURNAL=	"Eur. Phys. J. C",
	VOLUME=	"80",
	YEAR=	"2020",
	PAGES=	"957",
	DOI=	"10.1140/epjc/s10052-020-8227-9",
	EPRINT=	"2004.03447v2",
	ARCHIVEPREFIX=	"arXiv",
	PRIMARYCLASS=	"hep-ex",
}

@ARTICLE{ATLASH4lFiducialRun2,
	AUTHOR=	"{ATLAS Collaboration}",
	TITLE=	"Measurements of the {Higgs} boson inclusive and differential fiducial cross sections in the 4$\ell$ decay channel at $\sqrt{s}=13$ {TeV}",
	JOURNAL=	"Eur. Phys. J. C",
	VOLUME=	"80",
	YEAR=	"2020",
	PAGES=	"941",
	DOI=	"10.1140/epjc/s10052-020-8223-0",
	EPRINT=	"arXiv:2004.03969v3",
	ARCHIVEPREFIX=	"arXiv",
	PRIMARYCLASS=	"hep-ex",
}

@ARTICLE{CMSH4lAnomalousCouplings2016,
	AUTHOR=	"{CMS Collaboration}",
	TITLE=	"Constraints on anomalous {Higgs} boson couplings using production and decay information in the four-lepton final state",
	JOURNAL=	"Phys. Lett. B",
	VOLUME=	"775",
	YEAR=	"2017",
	PAGES=	"1",
	DOI=	"10.1016/j.physletb.2017.10.021",
	EPRINT=	"1707.00541",
	ARCHIVEPREFIX=	"arXiv",
	PRIMARYCLASS=	"hep-ex",
	REPORTNUMBER=	"CMS-HIG-17-011, CERN-EP-2017-143",
	SLACCITATION=	"%%CITATION = ARXIV:1707.00541;%%",
}

@ARTICLE{CMSHVVAnomalousCouplings2016,
	AUTHOR=	"{CMS Collaboration}",
	TITLE=	"Measurements of the {Higgs} boson width and anomalous {HVV} couplings from on-shell and off-shell production in the four-lepton final state",
	JOURNAL=	"Phys. Rev. D",
	EPRINT=	"1901.00174",
	ARCHIVEPREFIX=	"arXiv",
	PRIMARYCLASS=	"hep-ex",
	REPORTNUMBER=	"CMS-HIG-18-002, CERN-EP-2018-329",
	DOI=	"10.1103/PhysRevD.99.112003",
	VOLUME=	"99",
	PAGES=	"112003",
	YEAR=	"2019",
}

@ARTICLE{CMSHIG19009,
	AUTHOR=	"{CMS Collaboration}",
	TITLE=	"{Constraints on anomalous Higgs boson couplings to vector bosons and fermions in its production and decay using the four-lepton final state}",
	EPRINT=	"2104.12152",
	ARCHIVEPREFIX=	"arXiv",
	PRIMARYCLASS=	"hep-ex",
	REPORTNUMBER=	"CERN-EP-2021-054",
	DOI=	"10.1103/PhysRevD.104.052004",
	JOURNAL=	"Phys. Rev. D",
	VOLUME=	"104",
	PAGES=	"052004",
	YEAR=	"2021",
}

@ARTICLE{CMS:2020xrn,
	AUTHOR=	"{CMS Collaboration}",
	TITLE=	"{A measurement of the Higgs boson mass in the diphoton decay channel}",
	EPRINT=	"2002.06398",
	ARCHIVEPREFIX=	"arXiv",
	PRIMARYCLASS=	"hep-ex",
	REPORTNUMBER=	"CMS-HIG-19-004, CERN-EP-2020-004",
	DOI=	"10.1016/j.physletb.2020.135425",
	JOURNAL=	"Phys. Lett. B",
	VOLUME=	"805",
	PAGES=	"135425",
	YEAR=	"2020",
}

@ARTICLE{CMS:2022ley,
	AUTHOR=	"{CMS Collaboration}",
	TITLE=	"{Measurement of the Higgs boson width and evidence of its off-shell contributions to ZZ production}",
	EPRINT=	"2202.06923",
	ARCHIVEPREFIX=	"arXiv",
	PRIMARYCLASS=	"hep-ex",
	REPORTNUMBER=	"CMS-HIG-21-013, CERN-EP-2021-272",
	DOI=	"10.1038/s41567-022-01682-0",
	YEAR=	"2022",
	JOURNAL=	"Nature Phys.",
	PAGES=	"1329",
	VOLUME=	"18",
}

@TECHREPORT{deFlorian:2016spz,
	AUTHOR=	"{LHC Higgs Cross Section Working Group}",
	EDITOR=	"de Florian, D. and C. Grojean and F. Maltoni and C. Mariotti and A. Nikitenko and M. Pieri and P. Savard and M. Schumacher and R. Tanaka",
	TITLE=	"{Handbook of LHC Higgs Cross Sections: 4. Deciphering the Nature of the Higgs Sector}",
	EPRINT=	"1610.07922",
	ARCHIVEPREFIX=	"arXiv",
	PRIMARYCLASS=	"hep-ph",
	REPORTNUMBER=	"CERN-2017-002-M, CERN-2017-002",
	DOI=	"10.23731/CYRM-2017-002",
	VOLUME=	"2/2017",
	YEAR=	"2016",
}

@ARTICLE{CMSHIG19001,
	AUTHOR=	"{CMS Collaboration}",
	TITLE=	"{Measurements of production cross sections of the Higgs boson in the four-lepton final state in proton\textendash{}proton collisions at $\sqrt{s} = 13\,\text {Te}\text {V} $}",
	EPRINT=	"2103.04956",
	ARCHIVEPREFIX=	"arXiv",
	PRIMARYCLASS=	"hep-ex",
	REPORTNUMBER=	"CMS-HIG-19-001, CERN-EP-2021-016",
	DOI=	"10.1140/epjc/s10052-021-09200-x",
	JOURNAL=	"Eur. Phys. J. C",
	VOLUME=	"81",
	PAGES=	"488",
	YEAR=	"2021",
}

@UNPUBLISHED{ATLAS:2022tnm,
	TITLE=	"{Measurement of the properties of Higgs boson production at $\sqrt{s} = 13$ TeV in the $H\to\gamma\gamma$ channel using $139$ fb$^{-1}$ of pp collision data with the ATLAS experiment}",
	EPRINT=	"2207.00348",
	ARCHIVEPREFIX=	"arXiv",
	PRIMARYCLASS=	"hep-ex",
	REPORTNUMBER=	"CERN-EP-2022-094",
	NOTE=	"Submitted to \textit{JHEP}",
	YEAR=	"2022",
	AUTHOR=	"{ATLAS Collaboration}",
}

@ARTICLE{CMS:2021kom,
	AUTHOR=	"{CMS Collaboration}",
	TITLE=	"{Measurements of Higgs boson production cross sections and couplings in the diphoton decay channel at $ \sqrt{\mathrm{s}} $ = 13 TeV}",
	EPRINT=	"2103.06956",
	ARCHIVEPREFIX=	"arXiv",
	PRIMARYCLASS=	"hep-ex",
	REPORTNUMBER=	"CMS-HIG-19-015, CERN-EP-2021-038",
	DOI=	"10.1007/JHEP07(2021)027",
	JOURNAL=	"JHEP",
	VOLUME=	"07",
	PAGES=	"027",
	YEAR=	"2021",
}

@UNPUBLISHED{ATLAS:2022ooq,
	TITLE=	"{Measurements of Higgs boson production by gluon$-$gluon fusion and vector-boson fusion using $H\rightarrow W W^* \rightarrow e\nu \mu\nu$ decays in pp collisions at $\sqrt{s}=13$ TeV with the ATLAS detector}",
	EPRINT=	"2207.00338",
	ARCHIVEPREFIX=	"arXiv",
	PRIMARYCLASS=	"hep-ex",
	REPORTNUMBER=	"CERN-EP-2022-078",
	NOTE=	"Submitted to \textit{Phys. Rev. D.}",
	YEAR=	"2022",
	AUTHOR=	"{ATLAS Collaboration}",
}

@UNPUBLISHED{CMS:2022uhn,
	TITLE=	"{Measurements of the Higgs boson production cross section and couplings in the W boson pair decay channel in proton-proton collisions at $\sqrt{s}$ = 13 TeV}",
	EPRINT=	"2206.09466",
	ARCHIVEPREFIX=	"arXiv",
	PRIMARYCLASS=	"hep-ex",
	REPORTNUMBER=	"CMS-HIG-20-013, CERN-EP-2022-120",
	NOTE=	"Submitted to \textit{Eur. Phys. J. C}",
	YEAR=	"2022",
	AUTHOR=	"{CMS Collaboration}",
}

@ARTICLE{ATLAS:2020fcp,
	AUTHOR=	"{ATLAS Collaboration}",
	TITLE=	"{Measurements of WH and ZH production in the $H \rightarrow b\bar{b}$ decay channel in pp collisions at 13 TeV with the ATLAS detector}",
	EPRINT=	"2007.02873",
	ARCHIVEPREFIX=	"arXiv",
	PRIMARYCLASS=	"hep-ex",
	REPORTNUMBER=	"CERN-EP-2020-087",
	DOI=	"10.1140/epjc/s10052-020-08677-2",
	JOURNAL=	"Eur. Phys. J. C",
	VOLUME=	"81",
	PAGES=	"178",
	YEAR=	"2021",
}

@ARTICLE{ATLAS:2022yrq,
	AUTHOR=	"{ATLAS Collaboration}",
	TITLE=	"{Measurements of Higgs boson production cross-sections in the~$H\to\tau^{+}\tau^{-}$ decay channel in pp collisions at $ \sqrt{s}=13 $ TeV with the ATLAS detector}",
	EPRINT=	"2201.08269",
	ARCHIVEPREFIX=	"arXiv",
	PRIMARYCLASS=	"hep-ex",
	REPORTNUMBER=	"CERN-EP-2021-217",
	DOI=	"10.1007/JHEP08(2022)175",
	JOURNAL=	"JHEP",
	VOLUME=	"08",
	PAGES=	"175",
	YEAR=	"2022",
}

@UNPUBLISHED{CMS:2022kdi,
	TITLE=	"{Measurements of Higgs boson production in the decay channel with a pair of $\tau$ leptons in proton-proton collisions at $\sqrt{s}= 13$ TeV}",
	EPRINT=	"2204.12957",
	ARCHIVEPREFIX=	"arXiv",
	PRIMARYCLASS=	"hep-ex",
	REPORTNUMBER=	"CMS-HIG-19-010, CERN-EP-2022-027",
	NOTE=	"Submitted to\textit{Eur. Phys. J. C}",
	YEAR=	"2022",
	AUTHOR=	"{CMS Collaboration}",
}

@ARTICLE{ATLAS:2022vkf,
	TITLE=	"{A detailed map of Higgs boson interactions by the ATLAS experiment ten years after the discovery}",
	EPRINT=	"2207.00092",
	ARCHIVEPREFIX=	"arXiv",
	PRIMARYCLASS=	"hep-ex",
	REPORTNUMBER=	"CERN-EP-2022-057",
	DOI=	"10.1038/s41586-022-04893-w",
	JOURNAL=	"Nature Phys.",
	VOLUME=	"607",
	PAGES=	"52",
	YEAR=	"2022",
	NOTE=	"[Erratum: Nature 612, E24 (2022)]",
	AUTHOR=	"{ATLAS Collaboration}",
}

@ARTICLE{CMS:2022dwd,
	TITLE=	"{A portrait of the Higgs boson by the CMS experiment ten years after the discovery}",
	EPRINT=	"2207.00043",
	ARCHIVEPREFIX=	"arXiv",
	PRIMARYCLASS=	"hep-ex",
	REPORTNUMBER=	"CMS-HIG-22-001, CERN-EP-2022-039",
	DOI=	"10.1038/s41586-022-04892-x",
	JOURNAL=	"Nature",
	VOLUME=	"607",
	PAGES=	"10",
	YEAR=	"2022",
	AUTHOR=	"{CMS Collaboration}",
}

@ARTICLE{ATLAS:2022fnp,
	AUTHOR=	"{ATLAS Collaboration}",
	TITLE=	"{Measurements of the Higgs boson inclusive and differential fiducial cross-sections in the diphoton decay channel with pp collisions at $ \sqrt{s} $ = 13 TeV with the ATLAS detector}",
	EPRINT=	"2202.00487",
	ARCHIVEPREFIX=	"arXiv",
	PRIMARYCLASS=	"hep-ex",
	REPORTNUMBER=	"CERN-EP-2021-227",
	DOI=	"10.1007/JHEP08(2022)027",
	JOURNAL=	"JHEP",
	VOLUME=	"08",
	PAGES=	"027",
	YEAR=	"2022",
}

@UNPUBLISHED{CMS:2022hyj,
	TITLE=	"{Measurement of the Higgs boson inclusive and differential fiducial production cross sections in the diphoton decay channel with pp collisions at $\sqrt{s}= 13$ TeV}",
	EPRINT=	"2208.12279",
	ARCHIVEPREFIX=	"arXiv",
	PRIMARYCLASS=	"hep-ex",
	REPORTNUMBER=	"CMS-HIG-19-016, CERN-EP-2022-142",
	NOTE=	"Submitted to \textit{JHEP}",
	YEAR=	"2022",
	AUTHOR=	"{CMS Collaboration}",
}

@UNPUBLISHED{ATLAS:2023hyd,
	TITLE=	"{Measurements of differential cross sections of Higgs boson production through gluon fusion in the $H\to WW^{\ast}\to e\nu\mu\nu$ final state at $\sqrt{s} = 13$ TeV with the ATLAS detector}",
	EPRINT=	"2301.06822",
	ARCHIVEPREFIX=	"arXiv",
	PRIMARYCLASS=	"hep-ex",
	REPORTNUMBER=	"CERN-EP-2022-228",
	YEAR=	"2023",
	NOTE=	"Submitted to\textit{Eur. Phys. J. C}",
	AUTHOR=	"{ATLAS Collaboration}",
}

@UNPUBLISHED{ATLAS:2023pwa,
	AUTHOR=	"{ATLAS Collaboration}",
	TITLE=	"{Fiducial and differential cross-section measurements for the vector-boson-fusion production of the Higgs boson in the $H \rightarrow WW^{\ast} \rightarrow e\nu\mu\nu$ decay channel at 13 $\text{TeV}$ with the ATLAS detector}",
	EPRINT=	"2304.03053",
	ARCHIVEPREFIX=	"arXiv",
	PRIMARYCLASS=	"hep-ex",
	REPORTNUMBER=	"CERN-EP-2023-025",
	YEAR=	"2023",
	NOTE=	"Submitted to \textit{Phys. Rev. D.}",
}

@ARTICLE{CMS:2020dvg,
	AUTHOR=	"{CMS Collaboration}",
	TITLE=	"{Measurement of the inclusive and differential Higgs boson production cross sections in the leptonic WW decay mode at $\sqrt{s} =13$ TeV}",
	EPRINT=	"2007.01984",
	ARCHIVEPREFIX=	"arXiv",
	PRIMARYCLASS=	"hep-ex",
	REPORTNUMBER=	"CMS-HIG-19-002, CERN-EP-2020-106",
	DOI=	"10.1007/JHEP03(2021)003",
	JOURNAL=	"JHEP",
	VOLUME=	"03",
	PAGES=	"003",
	YEAR=	"2021",
}

@ARTICLE{CMS:2021gxc,
	AUTHOR=	"{CMS Collaboration}",
	TITLE=	"{Measurement of the inclusive and differential Higgs boson production cross sections in the decay mode to a pair of $\tau$ leptons in pp collisions at $\sqrt{s} = 13$ TeV}",
	EPRINT=	"2107.11486",
	ARCHIVEPREFIX=	"arXiv",
	PRIMARYCLASS=	"hep-ex",
	REPORTNUMBER=	"CMS-HIG-20-015, CERN-EP-2021-134",
	DOI=	"10.1103/PhysRevLett.128.081805",
	JOURNAL=	"Phys. Rev. Lett.",
	VOLUME=	"128",
	PAGES=	"081805",
	YEAR=	"2022",
}

@ARTICLE{ATLAS:2022qef,
	AUTHOR=	"{ATLAS Collaboration}",
	TITLE=	"{Measurement of the total and differential Higgs boson production cross-sections at $\sqrt{s} = 13$ TeV with the ATLAS detector by combining the $H \rightarrow ZZ^* \rightarrow 4\ell$ and $H \rightarrow \gamma \gamma$ decay channels}",
	EPRINT=	"2207.08615",
	ARCHIVEPREFIX=	"arXiv",
	PRIMARYCLASS=	"hep-ex",
	REPORTNUMBER=	"CERN-EP-2022-143",
	DOI=	"10.1007/JHEP05(2023)028",
	JOURNAL=	"JHEP",
	VOLUME=	"05",
	PAGES=	"028",
	YEAR=	"2023",
}

@ARTICLE{CMS:2018gwt,
	AUTHOR=	"{CMS Collaboration}",
	TITLE=	"{Measurement and interpretation of differential cross sections for Higgs boson production at $\sqrt{s} =$ 13 TeV}",
	EPRINT=	"1812.06504",
	ARCHIVEPREFIX=	"arXiv",
	PAGES=	"369",
	PRIMARYCLASS=	"hep-ex",
	REPORTNUMBER=	"CMS-HIG-17-028, CERN-EP-2018-304",
	DOI=	"10.1016/j.physletb.2019.03.059",
	JOURNAL=	"Phys. Lett. B",
	VOLUME=	"792",
	YEAR=	"2019",
}

@ARTICLE{Sirunyan:2020zal,
	AUTHOR=	"{CMS Collaboration}",
	TITLE=	"{Performance of the CMS Level-1 trigger in proton-proton collisions at $\sqrt{s} = 13$\,TeV}",
	JOURNAL=	"JINST",
	VOLUME=	"15",
	PAGES=	"P10017",
	YEAR=	"2020",
	DOI=	"10.1088/1748-0221/15/10/P10017",
	EPRINT=	"2006.10165",
	ARCHIVEPREFIX=	"arXiv",
	PRIMARYCLASS=	"hep-ex",
	REPORTNUMBER=	"CMS-TRG-17-001, CERN-EP-2020-065",
}

@ARTICLE{Khachatryan:2016bia,
	AUTHOR=	"{CMS Collaboration}",
	TITLE=	"{The CMS trigger system}",
	JOURNAL=	"JINST",
	VOLUME=	"12",
	PAGES=	"P01020",
	DOI=	"10.1088/1748-0221/12/01/P01020",
	YEAR=	"2017",
	EPRINT=	"1609.02366",
	ARCHIVEPREFIX=	"arXiv",
	PRIMARYCLASS=	"physics.ins-det",
	REPORTNUMBER=	"CMS-TRG-12-001, CERN-EP-2016-160",
	SLACCITATION=	"%%CITATION = ARXIV:1609.02366;%%",
}

@TECHREPORT{CMS-TDR-15-02,
	AUTHOR=	"{CMS Collaboration}",
	TITLE=	"Technical proposal for the Phase-{II} upgrade of the Compact Muon Solenoid",
	URL=	"http://cds.cern.ch/record/2020886",
	TYPE=	"CMS Technical proposal",
	NUMBER=	"CERN-LHCC-2015-010, CMS-TDR-15-02",
	YEAR=	"2015",
}

@ARTICLE{CMS:2020uim,
	AUTHOR=	"{CMS Collaboration}",
	TITLE=	"Electron and photon reconstruction and identification with the {CMS} experiment at the {CERN} {LHC}",
	EPRINT=	"2012.06888",
	JOURNAL=	"JINST",
	VOLUME=	"16",
	PAGES=	"P05014",
	YEAR=	"2021",
	ARCHIVEPREFIX=	"arXiv",
	PRIMARYCLASS=	"hep-ex",
	REPORTNUMBER=	"CMS-EGM-17-001, CERN-EP-2020-219",
	DOI=	"10.1088/1748-0221/16/05/P05014",
}

@TECHREPORT{CMS-DP-2020-021,
	TITLE=	"{ECAL} 2016 refined calibration and {Run 2} summary plots",
	AUTHOR=	"{CMS Collaboration}",
	URL=	"https://cds.cern.ch/record/2717925",
	TYPE=	"CMS Detector Performance Summary",
	NUMBER=	"CMS-DP-2020-021",
	YEAR=	"2020",
}

@ARTICLE{Sirunyan:2018,
	AUTHOR=	"{CMS Collaboration}",
	TITLE=	"{Performance of the CMS muon detector and muon reconstruction with proton-proton collisions at $\sqrt{s}=$ 13 TeV}",
	EPRINT=	"1804.04528",
	ARCHIVEPREFIX=	"arXiv",
	PRIMARYCLASS=	"physics.ins-det",
	REPORTNUMBER=	"CMS-MUO-16-001, CERN-EP-2018-058",
	DOI=	"10.1088/1748-0221/13/06/P06015",
	JOURNAL=	"JINST",
	VOLUME=	"13",
	PAGES=	"P06015",
	YEAR=	"2018",
}

@ARTICLE{Chatrchyan:2008zzk,
	AUTHOR=	"{CMS Collaboration}",
	TITLE=	"The {CMS} experiment at the {CERN} {LHC}",
	JOURNAL=	"JINST",
	VOLUME=	"3",
	YEAR=	"2008",
	PAGES=	"S08004",
	DOI=	"10.1088/1748-0221/3/08/S08004",
	SLACCITATION=	"%%CITATION = JINST,3,S08004;%%",
}

@ARTICLE{CMS-LUM-17-003,
	AUTHOR=	"{CMS Collaboration}",
	TITLE=	"Precision luminosity measurement in proton-proton collisions at $\sqrt{s} =$ 13 {TeV} in 2015 and 2016 at {CMS}",
	EPRINT=	"2104.01927",
	ARCHIVEPREFIX=	"arXiv",
	PRIMARYCLASS=	"hep-ex",
	REPORTNUMBER=	"CMS-LUM-17-003, CERN-EP-2021-033",
	YEAR=	"2021",
	DOI=	"10.1140/epjc/s10052-021-09538-2",
	JOURNAL=	"Eur. Phys. J. C",
	VOLUME=	"81",
	PAGES=	"800",
}

@TECHREPORT{CMS-PAS-LUM-17-004,
	TITLE=	"{CMS luminosity measurement for the 2017 data-taking period at $\sqrt{s} = 13~\mathrm{TeV}$}",
	REPORTNUMBER=	"CMS-PAS-LUM-17-004",
	NUMBER=	"CMS-PAS-LUM-17-004",
	YEAR=	"2018",
	URL=	"https://cds.cern.ch/record/2621960",
	TYPE=	"CMS Physics Analysis Summary",
	AUTHOR=	"{CMS Collaboration}",
}

@TECHREPORT{CMS-PAS-LUM-18-002,
	TITLE=	"{CMS luminosity measurement for the 2018 data-taking period at $\sqrt{s} = 13~\mathrm{TeV}$}",
	REPORTNUMBER=	"CMS-PAS-LUM-18-002",
	NUMBER=	"CMS-PAS-LUM-18-002",
	YEAR=	"2019",
	URL=	"https://cds.cern.ch/record/2676164",
	TYPE=	"CMS Physics Analysis Summary",
	AUTHOR=	"{CMS Collaboration}",
}

@ARTICLE{CMS:2011aa,
	AUTHOR=	"{CMS Collaboration}",
	ARCHIVEPREFIX=	"arXiv",
	DOI=	"10.1007/JHEP10(2011)132",
	EPRINT=	"1107.4789",
	JOURNAL=	"JHEP",
	PAGES=	"132",
	PRIMARYCLASS=	"hep-ex",
	REPORTNUMBER=	"CERN-PH-EP-2011-107, CMS-EWK-10-005",
	TITLE=	"{Measurement of the Inclusive {$W$ and $Z$} Production Cross Sections in pp Collisions at {$\sqrt{s}=7$ TeV}}",
	VOLUME=	"10",
	YEAR=	"2011",
}

@ARTICLE{Nason:2004rx,
	AUTHOR=	"Nason, Paolo",
	TITLE=	"A new method for combining {NLO QCD} with shower {Monte Carlo} algorithms",
	JOURNAL=	"JHEP",
	VOLUME=	"11",
	YEAR=	"2004",
	PAGES=	"040",
	EPRINT=	"hep-ph/0409146",
	ARCHIVEPREFIX=	"arXiv",
	DOI=	"10.1088/1126-6708/2004/11/040",
	SLACCITATION=	"%%CITATION = HEP-PH/0409146;%%",
}

@ARTICLE{Frixione:2007vw,
	AUTHOR=	"Frixione, Stefano and Nason, Paolo and Oleari, Carlo",
	TITLE=	"Matching {NLO QCD} computations with parton shower simulations: the {POWHEG method}",
	JOURNAL=	"JHEP",
	VOLUME=	"11",
	YEAR=	"2007",
	PAGES=	"070",
	EPRINT=	"0709.2092",
	ARCHIVEPREFIX=	"arXiv",
	PRIMARYCLASS=	"hep-ph",
	DOI=	"10.1088/1126-6708/2007/11/070",
	SLACCITATION=	"%%CITATION = 0709.2092;%%",
}

@ARTICLE{Alioli:2010xd,
	AUTHOR=	"Alioli, Simone and Nason, Paolo and Oleari, Carlo and Re, Emanuele",
	TITLE=	"A general framework for implementing {NLO} calculations in shower {Monte Carlo} programs: the {POWHEG BOX}",
	JOURNAL=	"JHEP",
	VOLUME=	"06",
	YEAR=	"2010",
	PAGES=	"043",
	DOI=	"10.1007/JHEP06(2010)043",
	EPRINT=	"1002.2581",
	ARCHIVEPREFIX=	"arXiv",
	PRIMARYCLASS=	"hep-ph",
	REPORTNUMBER=	"DESY-10-018, SFB-CPP-10-22, IPPP-10-11, DCPT-10-22",
	SLACCITATION=	"%%CITATION = ARXIV:1002.2581;%%",
}

@ARTICLE{Alioli:2008tz,
	AUTHOR=	"Alioli, Simone and Nason, Paolo and Oleari, Carlo and Re, Emanuele",
	TITLE=	"{NLO Higgs} boson production via gluon fusion matched with shower in {POWHEG}",
	JOURNAL=	"JHEP",
	VOLUME=	"04",
	YEAR=	"2009",
	PAGES=	"002",
	DOI=	"10.1088/1126-6708/2009/04/002",
	EPRINT=	"0812.0578",
	ARCHIVEPREFIX=	"arXiv",
	PRIMARYCLASS=	"hep-ph",
	SLACCITATION=	"%%CITATION = ARXIV:0812.0578;%%",
}

@ARTICLE{Nason:2009ai,
	AUTHOR=	"Nason, Paolo and Oleari, Carlo",
	TITLE=	"{NLO Higgs} boson production via vector-boson fusion matched with shower in {POWHEG}",
	JOURNAL=	"JHEP",
	VOLUME=	"02",
	YEAR=	"2010",
	PAGES=	"037",
	DOI=	"10.1007/JHEP02(2010)037",
	EPRINT=	"0911.5299",
	ARCHIVEPREFIX=	"arXiv",
	PRIMARYCLASS=	"hep-ph",
	SLACCITATION=	"%%CITATION = ARXIV:0911.5299;%%",
}

@ARTICLE{Luisoni2013,
	AUTHOR=	"Luisoni, Gionata and Nason, Paolo and Oleari, Carlo and Tramontano, Francesco",
	TITLE=	"{HW$^{\pm}$/HZ + 0 and 1 jet at NLO} with the {POWHEG BOX} interfaced to {GoSam} and their merging within {MiNLO}",
	JOURNAL=	"JHEP",
	YEAR=	"2013",
	VOLUME=	"10",
	PAGES=	"1",
	DOI=	"10.1007/JHEP10(2013)083",
	EPRINT=	"1306.2542",
	ARCHIVEPREFIX=	"arXiv",
	PRIMARYCLASS=	"hep-ph",
}

@ARTICLE{Hartanto:2015uka,
	AUTHOR=	"Hartanto, Heribertus B. and Jager, Barbara and Reina, Laura and Wackeroth, Doreen",
	TITLE=	"Higgs boson production in association with top quarks in the {POWHEG BOX}",
	JOURNAL=	"Phys. Rev. D",
	VOLUME=	"91",
	YEAR=	"2015",
	PAGES=	"094003",
	DOI=	"10.1103/PhysRevD.91.094003",
	EPRINT=	"1501.04498",
	ARCHIVEPREFIX=	"arXiv",
	PRIMARYCLASS=	"hep-ph",
	SLACCITATION=	"%%CITATION = ARXIV:1501.04498;%%",
}

@ARTICLE{Hamilton:2013fea,
	AUTHOR=	"Hamilton, Keith and Nason, Paolo and Re, Emanuele and Zanderighi, Giulia",
	TITLE=	"{NNLOPS} simulation of {H}iggs boson production",
	JOURNAL=	"JHEP",
	VOLUME=	"10",
	YEAR=	"2013",
	PAGES=	"222",
	DOI=	"10.1007/JHEP10(2013)222",
	EPRINT=	"1309.0017",
	ARCHIVEPREFIX=	"arXiv",
	PRIMARYCLASS=	"hep-ph",
	REPORTNUMBER=	"MCNET-13-11, CERN-PH-TH-2013-205, OUTP-13-18P",
	SLACCITATION=	"%%CITATION = ARXIV:1309.0017;%%",
}

@ARTICLE{Gao:2010qx,
	AUTHOR=	"Gao, Yanyan and Gritsan, Andrei V. and Guo, Zijin and Melnikov, Kirill and Schulze, Markus and Tran, Nhan V.",
	TITLE=	"Spin determination of single-produced resonances at hadron colliders",
	JOURNAL=	"Phys. Rev. D",
	VOLUME=	"81",
	PAGES=	"075022",
	DOI=	"10.1103/PhysRevD.81.075022",
	YEAR=	"2010",
	NOTE=	"[Erratum: \DOI{10.1103/PhysRevD.81.079905}]",
	EPRINT=	"1001.3396",
	ARCHIVEPREFIX=	"arXiv",
	PRIMARYCLASS=	"hep-ph",
	REPORTNUMBER=	"FERMILAB-PUB-10-011-E",
	SLACCITATION=	"%%CITATION = ARXIV:1001.3396;%%",
}

@ARTICLE{Bolognesi:2012mm,
	AUTHOR=	"Bolognesi, Sara and Gao, Yanyan and Gritsan, Andrei V. and Melnikov, Kirill and Schulze, Markus and Tran, Nhan V. and Whitbeck, Andrew",
	TITLE=	"On the spin and parity of a single-produced resonance at the {LHC}",
	JOURNAL=	"Phys. Rev. D",
	VOLUME=	"86",
	PAGES=	"095031",
	DOI=	"10.1103/PhysRevD.86.095031",
	YEAR=	"2012",
	EPRINT=	"1208.4018",
	ARCHIVEPREFIX=	"arXiv",
	PRIMARYCLASS=	"hep-ph",
	REPORTNUMBER=	"ANL-HEP-PR-12-62, FERMILAB-PUB-12-475-PPD",
	SLACCITATION=	"%%CITATION = ARXIV:1208.4018;%%",
}

@ARTICLE{Anderson:2013afp,
	AUTHOR=	"Anderson, Ian and Bolognesi, Sara and Caola, Fabrizio and Gao, Yanyan and Gritsan, Andrei V. and Martin, Christopher B. and Melnikov, Kirill and Schulze, Markus and Tran, Nhan V. and Whitbeck, Andrew and Zhou, Yaofu",
	TITLE=	"Constraining anomalous {$HVV$} interactions at proton and lepton colliders",
	JOURNAL=	"Phys. Rev. D",
	VOLUME=	"89",
	PAGES=	"035007",
	DOI=	"10.1103/PhysRevD.89.035007",
	YEAR=	"2014",
	EPRINT=	"1309.4819",
	ARCHIVEPREFIX=	"arXiv",
	PRIMARYCLASS=	"hep-ph",
	REPORTNUMBER=	"FERMILAB-PUB-13-386-PPD",
	SLACCITATION=	"%%CITATION = ARXIV:1309.4819;%%",
}

@ARTICLE{Gritsan:2016hjl,
	AUTHOR=	"Gritsan, Andrei V. and R{\"o}ntsch, Raoul and Schulze, Markus and Xiao, Meng",
	TITLE=	"{Constraining anomalous Higgs boson couplings to the heavy flavor fermions using matrix element techniques}",
	JOURNAL=	"Phys. Rev. D",
	VOLUME=	"94",
	YEAR=	"2016",
	PAGES=	"055023",
	DOI=	"10.1103/PhysRevD.94.055023",
	EPRINT=	"1606.03107",
	ARCHIVEPREFIX=	"arXiv",
	PRIMARYCLASS=	"hep-ph",
	REPORTNUMBER=	"TTP16-020, CERN-TH-2016-135",
	SLACCITATION=	"%%CITATION = ARXIV:1606.03107;%%",
}

@ARTICLE{Gritsan:2020pib,
	AUTHOR=	"Gritsan, Andrei V. and Roskes, Jeffrey and Sarica, Ulascan and Schulze, Markus and Xiao, Meng and Zhou, Yaofu",
	TITLE=	"{New features in the JHU generator framework: constraining Higgs boson properties from on-shell and off-shell production}",
	EPRINT=	"2002.09888",
	ARCHIVEPREFIX=	"arXiv",
	PRIMARYCLASS=	"hep-ph",
	REPORTNUMBER=	"HU-EP-20/02",
	DOI=	"10.1103/PhysRevD.102.056022",
	JOURNAL=	"Phys. Rev. D",
	VOLUME=	"102",
	PAGES=	"056022",
	YEAR=	"2020",
}

@ARTICLE{Wiesemann:2014ioa,
	AUTHOR=	"Wiesemann, M. and Frederix, R. and Frixione, S. and Hirschi, V. and Maltoni, F. and Torrielli, P.",
	TITLE=	"{Higgs production in association with bottom quarks}",
	EPRINT=	"1409.5301",
	ARCHIVEPREFIX=	"arXiv",
	PRIMARYCLASS=	"hep-ph",
	REPORTNUMBER=	"CERN-PH-TH-2014-182, CP3-14-64, LPN14-114, MCNET-14-20, ZU-TH-33-14",
	DOI=	"10.1007/JHEP02(2015)132",
	JOURNAL=	"JHEP",
	VOLUME=	"02",
	PAGES=	"132",
	YEAR=	"2015",
}

@ARTICLE{Anastasiou:2015ema,
	AUTHOR=	"Anastasiou, Charalampos and Duhr, Claude and Dulat, Falko and Herzog, Franz and Mistlberger, Bernhard",
	TITLE=	"Higgs boson gluon-fusion production in {QCD at three loops}",
	JOURNAL=	"Phys. Rev. Lett.",
	YEAR=	"2015",
	VOLUME=	"114",
	PAGES=	"212001",
	DOI=	"10.1103/PhysRevLett.114.212001",
	EPRINT=	"1503.06056",
	ARCHIVEPREFIX=	"arXiv",
	PRIMARYCLASS=	"hep-ph",
}

@ARTICLE{Anastasiou2016,
	AUTHOR=	"Anastasiou, Charalampos and Duhr, Claude and Dulat, Falko and Furlan, Elisabetta and Gehrmann, Thomas and Herzog, Franz and Lazopoulos, Achilleas and Mistlberger, Bernhard",
	TITLE=	"High precision determination of the gluon fusion {Higgs} boson cross-section at the {LHC}",
	JOURNAL=	"JHEP",
	YEAR=	"2016",
	VOLUME=	"05",
	PAGES=	"58",
	DOI=	"10.1007/JHEP05(2016)058",
	EPRINT=	"1602.00695",
	ARCHIVEPREFIX=	"arXiv",
	PRIMARYCLASS=	"hep-ph",
}

@ARTICLE{Ciccolini:2007jr,
	ARCHIVEPREFIX=	"arXiv",
	AUTHOR=	"Ciccolini, M. and Denner, Ansgar and Dittmaier, S.",
	DOI=	"10.1103/PhysRevLett.99.161803",
	EPRINT=	"0707.0381",
	JOURNAL=	"Phys. Rev. Lett.",
	PAGES=	"161803",
	PRIMARYCLASS=	"hep-ph",
	TITLE=	"Strong and electroweak corrections to the production of a {Higgs} boson+2 Jets via weak interactions at the {Large Hadron Collider}",
	VOLUME=	"99",
	YEAR=	"2007",
}

@ARTICLE{Ciccolini:2007ec,
	ARCHIVEPREFIX=	"arXiv",
	AUTHOR=	"Ciccolini, Mariano and Denner, Ansgar and Dittmaier, Stefan",
	DOI=	"10.1103/PhysRevD.77.013002",
	EPRINT=	"0710.4749",
	JOURNAL=	"Phys. Rev. D",
	PAGES=	"013002",
	PRIMARYCLASS=	"hep-ph",
	SLACCITATION=	"%%CITATION = 0710.4749;%%",
	TITLE=	"Electroweak and {QCD} corrections to {Higgs} production via vector-boson fusion at the {LHC}",
	VOLUME=	"77",
	YEAR=	"2008",
}

@ARTICLE{Bolzoni:2010xr,
	ARCHIVEPREFIX=	"arXiv",
	AUTHOR=	"Bolzoni, Paolo and Maltoni, Fabio and Moch, Sven-Olaf and Zaro, Marco",
	DOI=	"10.1103/PhysRevLett.105.011801",
	EPRINT=	"1003.4451",
	JOURNAL=	"Phys. Rev. Lett.",
	PAGES=	"011801",
	PRIMARYCLASS=	"hep-ph",
	SLACCITATION=	"%%CITATION = 1003.4451;%%",
	TITLE=	"Higgs production via vector-boson fusion at {NNLO in QCD}",
	VOLUME=	"105",
	YEAR=	"2010",
}

@ARTICLE{Bolzoni:2011cu,
	AUTHOR=	"Bolzoni, Paolo and Maltoni, Fabio and Moch, Sven-Olaf and Zaro, Marco",
	TITLE=	"Vector boson fusion at next-to-next-to-leading order in {QCD}: Standard model {Higgs} boson and beyond",
	JOURNAL=	"Phys. Rev. D",
	YEAR=	"2012",
	VOLUME=	"85",
	PAGES=	"035002",
	DOI=	"10.1103/PhysRevD.85.035002",
	EPRINT=	"1109.3717",
	ARCHIVEPREFIX=	"arXiv",
	PRIMARYCLASS=	"hep-ph",
}

@ARTICLE{Brein:2003wg,
	ARCHIVEPREFIX=	"arXiv",
	AUTHOR=	"Brein, Oliver and Djouadi, Abdelhak and Harlander, Robert",
	DOI=	"10.1016/j.physletb.2003.10.112",
	EPRINT=	"hep-ph/0307206",
	JOURNAL=	"Phys. Lett. B",
	PAGES=	"149",
	SLACCITATION=	"%%CITATION = HEP-PH/0307206;%%",
	TITLE=	"{NNLO QCD} corrections to the {Higgs-strahlung} processes at hadron colliders",
	VOLUME=	"579",
	YEAR=	"2004",
}

@ARTICLE{Ciccolini:2003jy,
	ARCHIVEPREFIX=	"arXiv",
	AUTHOR=	"Ciccolini, M. L. and Dittmaier, S. and Kr{\"a}mer, M.",
	DOI=	"10.1103/PhysRevD.68.073003",
	EPRINT=	"hep-ph/0306234",
	JOURNAL=	"Phys. Rev. D",
	PAGES=	"073003",
	SLACCITATION=	"%%CITATION = HEP-PH/0306234;%%",
	TITLE=	"Electroweak radiative corrections to associated {$WH$ and $ZH$} production at hadron colliders",
	VOLUME=	"68",
	YEAR=	"2003",
}

@ARTICLE{Beenakker:2001rj,
	ARCHIVEPREFIX=	"arXiv",
	AUTHOR=	"Beenakker, W. and Dittmaier, S. and Kramer, M. and Plumper, B. and Spira, M. and Zerwas, P. M.",
	DOI=	"10.1103/PhysRevLett.87.201805",
	EPRINT=	"hep-ph/0107081",
	JOURNAL=	"Phys. Rev. Lett.",
	PAGES=	"201805",
	SLACCITATION=	"%%CITATION = HEP-PH/0107081;%%",
	TITLE=	"Higgs radiation off top quarks at the {Tevatron and the LHC}",
	VOLUME=	"87",
	YEAR=	"2001",
}

@ARTICLE{Beenakker:2002nc,
	ARCHIVEPREFIX=	"arXiv",
	AUTHOR=	"Beenakker, W. and Dittmaier, S. and Kramer, M. and Plumper, B. and Spira, M. and Zerwas, P. M.",
	DOI=	"10.1016/S0550-3213(03)00044-0",
	EPRINT=	"hep-ph/0211352",
	JOURNAL=	"Nucl. Phys. B",
	PAGES=	"151",
	SLACCITATION=	"%%CITATION = HEP-PH/0211352;%%",
	TITLE=	"{NLO QCD} corrections to $\ttbar$ {H} production in hadron collisions.",
	VOLUME=	"653",
	YEAR=	"2003",
}

@ARTICLE{Dawson:2002tg,
	ARCHIVEPREFIX=	"arXiv",
	AUTHOR=	"Dawson, S. and Orr, L. H. and Reina, L. and Wackeroth, D.",
	DOI=	"10.1103/PhysRevD.67.071503",
	EPRINT=	"hep-ph/0211438",
	JOURNAL=	"Phys. Rev. D",
	PAGES=	"071503",
	SLACCITATION=	"%%CITATION = HEP-PH/0211438;%%",
	TITLE=	"Associated top quark {Higgs} boson production at the {LHC}",
	VOLUME=	"67",
	YEAR=	"2003",
}

@ARTICLE{Dawson:2003zu,
	ARCHIVEPREFIX=	"arXiv",
	AUTHOR=	"Dawson, S. and Jackson, C. and Orr, L. H. and Reina, L. and Wackeroth, D.",
	DOI=	"10.1103/PhysRevD.68.034022",
	EPRINT=	"hep-ph/0305087",
	JOURNAL=	"Phys. Rev. D",
	PAGES=	"034022",
	SLACCITATION=	"%%CITATION = HEP-PH/0305087;%%",
	TITLE=	"Associated {Higgs} production with top quarks at the {Large Hadron Collider: NLO QCD} corrections",
	VOLUME=	"68",
	YEAR=	"2003",
}

@ARTICLE{Yu:2014cka,
	AUTHOR=	"Yu, Zhang and Wen-Gan, Ma and Ren-You, Zhang and Chong, Chen and Lei, Guo",
	TITLE=	"{QCD NLO and EW NLO} corrections to {$t\bar{t}H$} production with top quark decays at hadron collider",
	JOURNAL=	"Phys. Lett. B",
	YEAR=	"2014",
	VOLUME=	"738",
	PAGES=	"1",
	DOI=	"10.1016/j.physletb.2014.09.022",
	EPRINT=	"1407.1110",
	ARCHIVEPREFIX=	"arXiv",
	PRIMARYCLASS=	"hep-ph",
}

@ARTICLE{Frixione:2014qaa,
	AUTHOR=	"S. Frixione, S. and Hirschi, V. and Pagani, D. and Shao, H. S. and Zaro, M.",
	TITLE=	"Weak corrections to {Higgs} hadroproduction in association with a top-quark pair",
	JOURNAL=	"JHEP",
	YEAR=	"2014",
	VOLUME=	"09",
	PAGES=	"65",
	DOI=	"10.1007/JHEP09(2014)065",
	EPRINT=	"1407.0823",
	ARCHIVEPREFIX=	"arXiv",
	PRIMARYCLASS=	"hep-ph",
}

@ARTICLE{Demartin:2015uha,
	AUTHOR=	"Demartin, F. and Maltoni, F. and Mawatari, K. and Zaro, M.",
	TITLE=	"Higgs production in association with a single top quark at the {LHC}",
	JOURNAL=	"Eur. Phys. J. C",
	YEAR=	"2015",
	VOLUME=	"75",
	PAGES=	"267",
	DOI=	"10.1140/epjc/s10052-015-3475-9",
	EPRINT=	"1504.0611",
	ARCHIVEPREFIX=	"arXiv",
	PRIMARYCLASS=	"hep-ph",
}

@ARTICLE{Demartin:2016axk,
	AUTHOR=	"Demartin, F. and Maier, B. and Maltoni, F. and Mawatari, K. and Zaro, M.",
	TITLE=	"{tWH associated production at the LHC}",
	JOURNAL=	"Eur. Phys. J. C",
	YEAR=	"2017",
	VOLUME=	"77",
	PAGES=	"34",
	DOI=	"10.1140/epjc/s10052-017-4601-7",
	EPRINT=	"1607.05862",
	ARCHIVEPREFIX=	"arXiv",
	PRIMARYCLASS=	"hep-ph",
}

@ARTICLE{Denner:2011mq,
	ARCHIVEPREFIX=	"arXiv",
	AUTHOR=	"Denner, A. and Heinemeyer, S. and Puljak, I. and Rebuzzi, D. and Spira, M.",
	DOI=	"10.1140/epjc/s10052-011-1753-8",
	EPRINT=	"1107.5909",
	JOURNAL=	"Eur. Phys. J. C",
	PAGES=	"1753",
	PRIMARYCLASS=	"hep-ph",
	SLACCITATION=	"%%CITATION = ARXIV:1107.5909;%%",
	TITLE=	"Standard model {Higgs}-boson branching ratios with uncertainties",
	VOLUME=	"71",
	YEAR=	"2011",
}

@ARTICLE{Djouadi:1997yw,
	AUTHOR=	"Djouadi, A. and Kalinowski, J. and Spira, M.",
	TITLE=	"{HDECAY}: {A} Program for {Higgs} boson decays in the standard model and its supersymmetric extension",
	JOURNAL=	"Comput. Phys. Commun.",
	VOLUME=	"108",
	PAGES=	"56",
	DOI=	"10.1016/S0010-4655(97)00123-9",
	YEAR=	"1998",
	EPRINT=	"hep-ph/9704448",
	ARCHIVEPREFIX=	"arXiv",
	PRIMARYCLASS=	"hep-ph",
	REPORTNUMBER=	"DESY-97-079, IFT-96-29, PM-97-04",
	SLACCITATION=	"%%CITATION = HEP-PH/9704448;%%",
}

@INPROCEEDINGS{hdecay2,
	ARCHIVEPREFIX=	"arXiv",
	AUTHOR=	"Djouadi, A. and Kalinowski, J. and Muhlleitner, M. and Spira , M.",
	BOOKTITLE=	"{The Les Houches 2009 workshop on TeV colliders: The tools and Monte Carlo working group summary report}",
	EPRINT=	"1003.1643",
	PRIMARYCLASS=	"hep-ph",
	TITLE=	"An update of the program {HDECAY}",
	YEAR=	"2010",
}

@ARTICLE{Bredenstein:2006rh,
	ARCHIVEPREFIX=	"arXiv",
	AUTHOR=	"Bredenstein, A. and Denner, Ansgar and Dittmaier, S. and Weber, M. M.",
	DOI=	"10.1103/PhysRevD.74.013004",
	EPRINT=	"hep-ph/0604011",
	JOURNAL=	"Phys. Rev. D",
	PAGES=	"013004",
	SLACCITATION=	"%%CITATION = HEP-PH/0604011;%%",
	TITLE=	"Precise predictions for the {Higgs}-boson decay {H $\rightarrow$ WW/ZZ $\rightarrow$} 4 leptons",
	VOLUME=	"74",
	YEAR=	"2006",
}

@ARTICLE{Bredenstein:2006ha,
	ARCHIVEPREFIX=	"arXiv",
	AUTHOR=	"Bredenstein, A. and Denner, Ansgar and Dittmaier, S. and Weber, M. M.",
	DOI=	"10.1088/1126-6708/2007/02/080",
	EPRINT=	"hep-ph/0611234",
	JOURNAL=	"JHEP",
	PAGES=	"80",
	PRIMARYCLASS=	"hep-ph",
	TITLE=	"Radiative corrections to the semileptonic and hadronic {Higgs}-boson decays {H $\rightarrow $WW/ZZ$\rightarrow$} 4 fermions",
	VOLUME=	"02",
	YEAR=	"2007",
}

@ARTICLE{Boselli:2015aha,
	AUTHOR=	"Boselli, S. and Carloni Calame, C. M. and Montagna, G. and Nicrosini, O. and Piccinini, F.",
	TITLE=	"Higgs boson decay into four leptons at {NLOPS} electroweak accuracy",
	JOURNAL=	"JHEP",
	YEAR=	"2015",
	VOLUME=	"06",
	PAGES=	"23",
	DOI=	"10.1007/JHEP06(2015)023",
	EPRINT=	"1503.07394",
	ARCHIVEPREFIX=	"arXiv",
	PRIMARYCLASS=	"hep-ph",
}

@ARTICLE{Actis:2008ts,
	ARCHIVEPREFIX=	"arXiv",
	AUTHOR=	"Actis, Stefano and Passarino, Giampiero and Sturm, Christian and Uccirati, Sandro",
	DOI=	"10.1016/j.nuclphysb.2008.11.024",
	EPRINT=	"0809.3667",
	JOURNAL=	"Nucl. Phys. B",
	PAGES=	"182",
	PRIMARYCLASS=	"hep-ph",
	SLACCITATION=	"%%CITATION = 0809.3667;%%",
	TITLE=	"{NNLO} computational techniques: the cases {$H \to \gamma \gamma$ and $H \to g g$}",
	VOLUME=	"811",
	YEAR=	"2009",
}

@ARTICLE{Melia:2011tj,
	AUTHOR=	"Melia, Tom and Nason, Paolo and Rontsch, Raoul and Zanderighi, Giulia",
	TITLE=	"{W$^{+}$W$^{-}$, WZ and ZZ production in the POWHEG BOX}",
	EPRINT=	"1107.5051",
	ARCHIVEPREFIX=	"arXiv",
	PRIMARYCLASS=	"hep-ph",
	DOI=	"10.1007/JHEP11(2011)078",
	JOURNAL=	"JHEP",
	VOLUME=	"11",
	PAGES=	"078",
	YEAR=	"2011",
}

@ARTICLE{MCFM,
	AUTHOR=	"Campbell, John M. and Ellis, R. K.",
	TITLE=	"{MCFM for the Tevatron and the LHC}",
	JOURNAL=	"Nucl. Phys. Proc. Suppl.",
	VOLUME=	"205--206",
	YEAR=	"2010",
	PAGES=	"10",
	EPRINT=	"1007.3492",
	ARCHIVEPREFIX=	"arXiv",
	PRIMARYCLASS=	"hep-ph",
	DOI=	"10.1016/j.nuclphysbps.2010.08.011",
	SLACCITATION=	"%%CITATION = 1007.3492;%%",
}

@ARTICLE{Campbell:2011bn,
	AUTHOR=	"Campbell, John M. and Ellis, R. Keith and Williams, Ciaran",
	TITLE=	"{Vector boson pair production at the {LHC}}",
	JOURNAL=	"JHEP",
	VOLUME=	"07",
	PAGES=	"018",
	DOI=	"10.1007/JHEP07(2011)018",
	YEAR=	"2011",
	EPRINT=	"1105.0020",
	ARCHIVEPREFIX=	"arXiv",
	PRIMARYCLASS=	"hep-ph",
	REPORTNUMBER=	"FERMILAB-PUB-11-182-T",
	SLACCITATION=	"%%CITATION = ARXIV:1105.0020;%%",
}

@ARTICLE{Campbell:2013una,
	AUTHOR=	"Campbell, John M. and Ellis, R. Keith and Williams, Ciaran",
	TITLE=	"{Bounding the Higgs width at the {LHC} using full analytic results for $\Pg\Pg\to \Pe^{-}\Pe^{+} \PGm^{-} \PGm^{+}$}",
	JOURNAL=	"JHEP",
	VOLUME=	"04",
	PAGES=	"060",
	DOI=	"10.1007/JHEP04(2014)060",
	YEAR=	"2014",
	EPRINT=	"1311.3589",
	ARCHIVEPREFIX=	"arXiv",
	PRIMARYCLASS=	"hep-ph",
	REPORTNUMBER=	"FERMILAB-PUB-13-508-T",
	SLACCITATION=	"%%CITATION = ARXIV:1311.3589;%%",
}

@ARTICLE{Campbell:2015vwa,
	AUTHOR=	"Campbell, John M. and Ellis, R. Keith",
	TITLE=	"{Higgs constraints from vector boson fusion and scattering}",
	JOURNAL=	"JHEP",
	VOLUME=	"04",
	YEAR=	"2015",
	PAGES=	"030",
	DOI=	"10.1007/JHEP04(2015)030",
	EPRINT=	"1502.02990",
	ARCHIVEPREFIX=	"arXiv",
	PRIMARYCLASS=	"hep-ph",
	REPORTNUMBER=	"FERMILAB-PUB-15-030-T",
	SLACCITATION=	"%%CITATION = ARXIV:1502.02990;%%",
}

@ARTICLE{Sjostrand:2014zea,
	AUTHOR=	"Sj{\"o}strand, Torbj{\"o}rn and Ask, Stefan and Christiansen, Jesper R. and Corke, Richard and Desai, Nishita and Ilten, Philip and Mrenna, Stephen and Prestel, Stefan and Rasmussen, Christine O. and Skands, Peter Z.",
	TITLE=	"An introduction to {PYTHIA} 8.2",
	JOURNAL=	"Comput. Phys. Commun.",
	VOLUME=	"191",
	YEAR=	"2015",
	PAGES=	"159",
	DOI=	"10.1016/j.cpc.2015.01.024",
	EPRINT=	"1410.3012",
	ARCHIVEPREFIX=	"arXiv",
	PRIMARYCLASS=	"hep-ph",
	REPORTNUMBER=	"LU-TP-14-36, MCNET-14-22, CERN-PH-TH-2014-190, FERMILAB-PUB-14-316-CD, DESY-14-178, SLAC-PUB-16122",
	SLACCITATION=	"%%CITATION = ARXIV:1410.3012;%%",
}

@ARTICLE{Khachatryan:2015pea,
	AUTHOR=	"{CMS Collaboration}",
	TITLE=	"{Event generator tunes obtained from underlying event and multiparton scattering measurements}",
	JOURNAL=	"Eur. Phys. J. C",
	VOLUME=	"76",
	YEAR=	"2016",
	PAGES=	"155",
	DOI=	"10.1140/epjc/s10052-016-3988-x",
	EPRINT=	"1512.00815",
	ARCHIVEPREFIX=	"arXiv",
	PRIMARYCLASS=	"hep-ex",
	REPORTNUMBER=	"CMS-GEN-14-001, CERN-PH-EP-2015-291",
	SLACCITATION=	"%%CITATION = ARXIV:1512.00815;%%",
}

@ARTICLE{Sirunyan:2019dfx,
	AUTHOR=	"{CMS Collaboration}",
	TITLE=	"{Extraction and validation of a new set of {CMS PYTHIA8} tunes from underlying-event measurements}",
	EPRINT=	"1903.12179",
	ARCHIVEPREFIX=	"arXiv",
	PRIMARYCLASS=	"hep-ex",
	REPORTNUMBER=	"CMS-GEN-17-001, CERN-EP-2019-007",
	DOI=	"10.1140/epjc/s10052-019-7499-4",
	JOURNAL=	"Eur. Phys. J. C",
	VOLUME=	"80",
	PAGES=	"4",
	YEAR=	"2020",
}

@ARTICLE{Ball:2014uwa,
	AUTHOR=	"Ball, Richard D. and others",
	COLLABORATION=	"NNPDF",
	TITLE=	"{Parton distributions for the {LHC Run II}}",
	EPRINT=	"1410.8849",
	ARCHIVEPREFIX=	"arXiv",
	PRIMARYCLASS=	"hep-ph",
	REPORTNUMBER=	"EDINBURGH-2014-15, IFUM-1034-FT, CERN-PH-TH-2013-253, OUTP-14-11P, CAVENDISH-HEP-14-11",
	DOI=	"10.1007/JHEP04(2015)040",
	JOURNAL=	"JHEP",
	VOLUME=	"04",
	PAGES=	"040",
	YEAR=	"2015",
}

@ARTICLE{Agostinelli:2002hh,
	AUTHOR=	"Agostinelli, S. and others",
	TITLE=	"{\GEANTfour: a simulation toolkit}",
	COLLABORATION=	"\GEANTfour",
	JOURNAL=	"Nucl. Instrum. Meth. A",
	VOLUME=	"506",
	PAGES=	"250",
	DOI=	"10.1016/S0168-9002(03)01368-8",
	YEAR=	"2003",
	REPORTNUMBER=	"SLAC-PUB-9350, FERMILAB-PUB-03-339",
	SLACCITATION=	"%%CITATION = NUIMA,A506,250;%%",
}

@ARTICLE{GEANT,
	AUTHOR=	"Allison, John and others",
	TITLE=	"{\GEANTfour developments and applications}",
	JOURNAL=	"IEEE Trans. Nucl. Sci.",
	VOLUME=	"53",
	PAGES=	"270",
	DOI=	"10.1109/TNS.2006.869826",
	YEAR=	"2006",
	REPORTNUMBER=	"SLAC-PUB-11870",
	SLACCITATION=	"%%CITATION = IETNA,53,270;%%",
}

@ARTICLE{CMS-PRF-14-001,
	AUTHOR=	"{CMS Collaboration}",
	TITLE=	"{Particle-flow reconstruction and global event description with the CMS detector}",
	JOURNAL=	"JINST",
	VOLUME=	"12",
	YEAR=	"2017",
	PAGES=	"P10003",
	DOI=	"10.1088/1748-0221/12/10/P10003",
	EPRINT=	"1706.04965",
	ARCHIVEPREFIX=	"arXiv",
	PRIMARYCLASS=	"physics.ins-det",
	REPORTNUMBER=	"CMS-PRF-14-001, CERN-EP-2017-110",
	SLACCITATION=	"%%CITATION = ARXIV:1706.04965;%%",
}

@TECHREPORT{Chen:2016btl,
	AUTHOR=	"Chen, Tianqi and Guestrin, Carlos",
	TITLE=	"{XGBoost: A} Scalable Tree Boosting System",
	DOI=	"10.1145/2939672.2939785",
	YEAR=	"2016",
	EPRINT=	"1603.02754",
	ARCHIVEPREFIX=	"arXiv",
	PRIMARYCLASS=	"cs.LG",
	SLACCITATION=	"%%CITATION = ARXIV:1603.02754;%%",
}

@ARTICLE{PUmitigationCMS,
	AUTHOR=	"{CMS Collaboration}",
	TITLE=	"{Pileup mitigation at CMS in 13 TeV data}",
	EPRINT=	"2003.00503",
	ARCHIVEPREFIX=	"arXiv",
	PRIMARYCLASS=	"hep-ex",
	REPORTNUMBER=	"CMS-JME-18-001, CERN-EP-2020-017",
	DOI=	"10.1088/1748-0221/15/09/P09018",
	JOURNAL=	"JINST",
	VOLUME=	"15",
	PAGES=	"P09018",
	YEAR=	"2020",
}

@ARTICLE{Cacciari:2008gp,
	AUTHOR=	"Cacciari, Matteo and Salam, Gavin P. and Soyez, Gregory",
	TITLE=	"{The anti-\kt jet clustering algorithm}",
	JOURNAL=	"JHEP",
	VOLUME=	"04",
	YEAR=	"2008",
	PAGES=	"063",
	DOI=	"10.1088/1126-6708/2008/04/063",
	EPRINT=	"0802.1189",
	ARCHIVEPREFIX=	"arXiv",
	PRIMARYCLASS=	"hep-ex",
}

@ARTICLE{Cacciari:2011ma,
	AUTHOR=	"Cacciari, Matteo and Salam, Gavin P. and Soyez, Gregory",
	TITLE=	"{FastJet user manual}",
	JOURNAL=	"Eur. Phys. J. C",
	VOLUME=	"72",
	PAGES=	"1896",
	DOI=	"10.1140/epjc/s10052-012-1896-2",
	YEAR=	"2012",
	EPRINT=	"1111.6097",
	ARCHIVEPREFIX=	"arXiv",
	PRIMARYCLASS=	"hep-ph",
	REPORTNUMBER=	"CERN-PH-TH-2011-297",
	SLACCITATION=	"%%CITATION = ARXIV:1111.6097;%%",
}

@ARTICLE{Khachatryan:2016kdb,
	AUTHOR=	"{CMS Collaboration}",
	TITLE=	"{Jet energy scale and resolution in the CMS experiment in pp collisions at 8 TeV}",
	JOURNAL=	"JINST",
	VOLUME=	"12",
	YEAR=	"2017",
	PAGES=	"P02014",
	DOI=	"10.1088/1748-0221/12/02/P02014",
	EPRINT=	"1607.03663",
	ARCHIVEPREFIX=	"arXiv",
	PRIMARYCLASS=	"hep-ex",
	REPORTNUMBER=	"CMS-JME-13-004, CERN-PH-EP-2015-305",
	SLACCITATION=	"%%CITATION = ARXIV:1607.03663;%%",
}

@ARTICLE{Zyla:2020zbs,
	AUTHOR=	"{Particle Data Group} and Zyla, P. A. and others",
	TITLE=	"Review of particle physics",
	DOI=	"10.1093/ptep/ptac097",
	JOURNAL=	"Prog. Theor. Exp. Phys.",
	VOLUME=	"2022",
	PAGES=	"083C01",
	YEAR=	"2022",
}

@ARTICLE{Gangal:2014qda,
	AUTHOR=	"Gangal, Shireen and Stahlhofen, Maximilian and Tackmann, Frank J.",
	TITLE=	"{Rapidity-Dependent Jet Vetoes}",
	EPRINT=	"1412.4792",
	ARCHIVEPREFIX=	"arXiv",
	PRIMARYCLASS=	"hep-ph",
	REPORTNUMBER=	"DESY-14-240",
	DOI=	"10.1103/PhysRevD.91.054023",
	JOURNAL=	"Phys. Rev. D",
	VOLUME=	"91",
	PAGES=	"054023",
	YEAR=	"2015",
}

@ARTICLE{Grazzini2015407,
	TITLE=	"{ZZ production at the LHC: Fiducial cross sections and distributions in NNLO QCD}",
	JOURNAL=	"Phys. Lett. B ",
	VOLUME=	"750",
	PAGES=	"407",
	YEAR=	"2015",
	ISSN=	"0370-2693",
	DOI=	"10.1016/j.physletb.2015.09.055",
	EPRINT=	"1507.06257",
	ARCHIVEPREFIX=	"arXiv",
	PRIMARYCLASS=	"hep-ph",
	AUTHOR=	"Massimiliano Grazzini and Stefan Kallweit and Dirk Rathlev",
}

@ARTICLE{Bierweiler:2013dja,
	AUTHOR=	"Bierweiler, Anastasiya and Kasprzik, Tobias and K{\"u}hn, Johann H.",
	TITLE=	"{Vector-boson pair production at the LHC to $\mathcal{O}(\alpha^3)$ accuracy}",
	JOURNAL=	"JHEP",
	VOLUME=	"12",
	YEAR=	"2013",
	PAGES=	"071",
	DOI=	"10.1007/JHEP12(2013)071",
	EPRINT=	"1305.5402",
	ARCHIVEPREFIX=	"arXiv",
	PRIMARYCLASS=	"hep-ph",
	SLACCITATION=	"%%CITATION = ARXIV:1305.5402;%%",
}

@ARTICLE{Bonvini:1304.3053,
	AUTHOR=	"Bonvini, Marco and Caola, Fabrizio and Forte, Stefano and Melnikov, Kirill and Ridolfi, Giovanni",
	TITLE=	"{Signal-background interference effects in $gg \to H \to WW$ beyond leading order}",
	JOURNAL=	"Phys. Rev. D",
	VOLUME=	"88",
	PAGES=	"034032",
	YEAR=	"2013",
	DOI=	"10.1103/PhysRevD.88.034032",
	EPRINT=	"1304.3053",
	ARCHIVEPREFIX=	"arXiv",
	PRIMARYCLASS=	"hep-ph",
	SLACCITATION=	"%%CITATION = ARXIV:1304.3053;%%",
}

@ARTICLE{Melnikov:2015laa,
	AUTHOR=	"Melnikov, Kirill and Dowling, Matthew",
	TITLE=	"{Production of two Z-bosons in gluon fusion in the heavy top quark approximation}",
	JOURNAL=	"Phys. Lett. B",
	VOLUME=	"744",
	PAGES=	"43",
	DOI=	"10.1016/j.physletb.2015.03.030",
	YEAR=	"2015",
	EPRINT=	"1503.01274",
	ARCHIVEPREFIX=	"arXiv",
	PRIMARYCLASS=	"hep-ph",
	REPORTNUMBER=	"TTP15-009",
	SLACCITATION=	"%%CITATION = ARXIV:1503.01274;%%",
}

@ARTICLE{Li:2015jva,
	AUTHOR=	"Li, Chong Sheng and Li, Hai Tao and Shao, Ding Yu and Wang, Jian",
	ARCHIVEPREFIX=	"arXiv",
	DOI=	"10.1007/JHEP08(2015)065",
	EPRINT=	"1504.02388",
	JOURNAL=	"JHEP",
	PAGES=	"065",
	PRIMARYCLASS=	"hep-ph",
	REPORTNUMBER=	"MITP-15-016",
	TITLE=	"{Soft gluon resummation in the signal-background interference process of gg($\rightarrow$ h$^{*}$) $\rightarrow$ ZZ}",
	VOLUME=	"08",
	YEAR=	"2015",
}

@ARTICLE{Passarino:1312.2397v1,
	AUTHOR=	"Passarino, Giampiero",
	TITLE=	"{Higgs CAT}",
	JOURNAL=	"Eur. Phys. J. C",
	VOLUME=	"74",
	PAGES=	"2866",
	YEAR=	"2014",
	DOI=	"10.1140/epjc/s10052-014-2866-7",
	EPRINT=	"1312.2397",
	ARCHIVEPREFIX=	"arXiv",
	PRIMARYCLASS=	"hep-ph",
	SLACCITATION=	"%%CITATION = ARXIV:1105.0020;%%",
}

@ARTICLE{Catani:2007vq,
	AUTHOR=	"Catani, Stefano and Grazzini, Massimiliano",
	TITLE=	"{An NNLO subtraction formalism in hadron collisions and its application to Higgs boson production at the LHC}",
	JOURNAL=	"Phys. Rev. Lett.",
	VOLUME=	"98",
	YEAR=	"2007",
	PAGES=	"222002",
	DOI=	"10.1103/PhysRevLett.98.222002",
	EPRINT=	"hep-ph/0703012",
	ARCHIVEPREFIX=	"arXiv",
	PRIMARYCLASS=	"hep-ph",
	SLACCITATION=	"%%CITATION = HEP-PH/0703012;%%",
}

@ARTICLE{Grazzini:2008tf,
	AUTHOR=	"Grazzini, Massimiliano",
	TITLE=	"{NNLO predictions for the Higgs boson signal in the H $\to$ WW $\to\ell\nu\ell\nu$ and H$\to$ ZZ $\to4\ell$ decay channels}",
	JOURNAL=	"JHEP",
	VOLUME=	"02",
	YEAR=	"2008",
	PAGES=	"043",
	DOI=	"10.1088/1126-6708/2008/02/043",
	EPRINT=	"0801.3232",
	ARCHIVEPREFIX=	"arXiv",
	PRIMARYCLASS=	"hep-ph",
	SLACCITATION=	"%%CITATION = ARXIV:0801.3232;%%",
}

@ARTICLE{Grazzini:2013mca,
	AUTHOR=	"Grazzini, Massimiliano and Sargsyan, Hayk",
	TITLE=	"{Heavy-quark mass effects in Higgs boson production at the LHC}",
	JOURNAL=	"JHEP",
	VOLUME=	"09",
	YEAR=	"2013",
	PAGES=	"129",
	DOI=	"10.1007/JHEP09(2013)129",
	EPRINT=	"1306.4581",
	ARCHIVEPREFIX=	"arXiv",
	PRIMARYCLASS=	"hep-ph",
	REPORTNUMBER=	"ZU-TH-10-13",
	SLACCITATION=	"%%CITATION = ARXIV:1306.4581;%%",
}

@PHDTHESIS{Skwarnicki:1986xj,
	AUTHOR=	"Skwarnicki, Tomasz",
	TITLE=	"{A study of the radiative CASCADE transitions between the Upsilon-Prime and Upsilon resonances}",
	REPORTNUMBER=	"DESY-F31-86-02, DESY-F-31-86-02",
	SCHOOL=	"Cracow, INP",
	YEAR=	"1986",
}

@TECHREPORT{LHC-HCG,
	TITLE=	"Procedure for the {LHC} {Higgs} boson search combination in {Summer} 2011",
	AUTHOR=	"{ATLAS Collaboration, CMS Collaboration, and LHC Higgs Combination Group}",
	NUMBER=	"CMS-NOTE-2011-005, ATL-PHYS-PUB-2011-11",
	YEAR=	"2011",
	REPORTNUMBER=	"CMS-NOTE-2011-005",
	URL=	"https://cds.cern.ch/record/1379837",
}

@ARTICLE{Cowan_2011,
	AUTHOR=	"Cowan, Glen and Cranmer, Kyle and Gross, Eilam and Vitells, Ofer",
	TITLE=	"Asymptotic formulae for likelihood-based tests of new physics",
	JOURNAL=	"Eur. Phys. J. C",
	VOLUME=	"71",
	YEAR=	"2011",
	PAGES=	"1554",
	DOI=	"10.1140/epjc/s10052-011-1554-0",
	NOTE=	"[Erratum: \DOI{10.1140/epjc/s10052-013-2501-z}]",
	EPRINT=	"1007.1727",
	ARCHIVEPREFIX=	"arXiv",
	PRIMARYCLASS=	"physics.data-an",
	SLACCITATION=	"%%CITATION = ARXIV:1007.1727;%%",
}

@ARTICLE{CMSHggFiducial8TeV,
	AUTHOR=	"{CMS Collaboration}",
	TITLE=	"{Measurement of differential cross sections for Higgs boson production in the diphoton decay channel in pp collisions at $\sqrt{s}=8\,\text {TeV} $}",
	EPRINT=	"1508.07819",
	ARCHIVEPREFIX=	"arXiv",
	PRIMARYCLASS=	"hep-ex",
	REPORTNUMBER=	"CMS-HIG-14-016, CERN-PH-EP-2015-195",
	DOI=	"10.1140/epjc/s10052-015-3853-3",
	JOURNAL=	"Eur. Phys. J. C",
	VOLUME=	"76",
	PAGES=	"13",
	YEAR=	"2016",
}

@ARTICLE{Butterworth:2015oua,
	AUTHOR=	"Butterworth, Jon and others",
	TITLE=	"{PDF4LHC recommendations for LHC Run II}",
	EPRINT=	"1510.03865",
	ARCHIVEPREFIX=	"arXiv",
	PRIMARYCLASS=	"hep-ph",
	REPORTNUMBER=	"OUTP-15-17P, SMU-HEP-15-12, TIF-UNIMI-2015-14, LCTS-2015-27, CERN-PH-TH-2015-249",
	DOI=	"10.1088/0954-3899/43/2/023001",
	JOURNAL=	"J. Phys. G",
	VOLUME=	"43",
	PAGES=	"023001",
	YEAR=	"2016",
}

@ARTICLE{ATLAS:2019jvq,
	AUTHOR=	"{ATLAS Collaboration}",
	TITLE=	"{Electron reconstruction and identification in the ATLAS experiment using the 2015 and 2016 LHC proton-proton collision data at $\sqrt{s}= 13$ TeV}",
	EPRINT=	"1902.04655",
	ARCHIVEPREFIX=	"arXiv",
	PRIMARYCLASS=	"physics.ins-det",
	REPORTNUMBER=	"CERN-EP-2018-273",
	DOI=	"10.1140/epjc/s10052-019-7140-6",
	JOURNAL=	"Eur. Phys. J. C",
	VOLUME=	"79",
	PAGES=	"639",
	YEAR=	"2019",
}

@MISC{hepdata,
	HOWPUBLISHED=	"{HEPD}ata record for this analysis",
	DOI=	"10.17182/hepdata.140341",
	YEAR=	"2023",
}

@ARTICLE{deFlorian:2012mx,
	AUTHOR=	"de Florian, D. and Ferrera, G. and Grazzini, M. and Tommasini, D.",
	TITLE=	"{Higgs boson production at the LHC: transverse momentum resummation effects in the $H \to \gamma \gamma$, $H \to WW \to \ell\nu\ell\nu$ and $H \to ZZ \to 4\ell$ decay modes}",
	JOURNAL=	"JHEP",
	VOLUME=	"06",
	PAGES=	"132",
	DOI=	"10.1007/JHEP06(2012)132",
	YEAR=	"2012",
	EPRINT=	"1203.6321",
	ARCHIVEPREFIX=	"arXiv",
	PRIMARYCLASS=	"hep-ph",
	REPORTNUMBER=	"ZU-TH-04-12, IFUM-993-FT",
	SLACCITATION=	"%%CITATION = ARXIV:1203.6321;%%",
}

@ARTICLE{Degrassi:2016wml,
	AUTHOR=	"Degrassi, Giuseppe and Giardino, Pier Paolo and Maltoni, Fabio and Pagani, Davide",
	TITLE=	"{Probing the Higgs self coupling via single Higgs production at the LHC}",
	EPRINT=	"1607.04251",
	ARCHIVEPREFIX=	"arXiv",
	PRIMARYCLASS=	"hep-ph",
	REPORTNUMBER=	"CP3-16-38, RM3-TH-16-8",
	DOI=	"10.1007/JHEP12(2016)080",
	JOURNAL=	"JHEP",
	VOLUME=	"12",
	PAGES=	"080",
	YEAR=	"2016",
}

@ARTICLE{Maltoni:2017ims,
	AUTHOR=	"Maltoni, Fabio and Pagani, Davide and Shivaji, Ambresh and Zhao, Xiaoran",
	TITLE=	"{Trilinear Higgs coupling determination via single-Higgs differential measurements at the LHC}",
	EPRINT=	"1709.08649",
	ARCHIVEPREFIX=	"arXiv",
	PRIMARYCLASS=	"hep-ph",
	REPORTNUMBER=	"CP3-17-37, TUM-HEP-1099-17, MCNET-17-18",
	DOI=	"10.1140/epjc/s10052-017-5410-8",
	JOURNAL=	"Eur. Phys. J. C",
	VOLUME=	"77",
	PAGES=	"887",
	YEAR=	"2017",
}

@ARTICLE{DiVita:2017eyz,
	AUTHOR=	"Di Vita, Stefano and Grojean, Christophe and Panico, Giuliano and Riembau, Marc and Vantalon, Thibaud",
	TITLE=	"{A global view on the Higgs self-coupling}",
	EPRINT=	"1704.01953",
	ARCHIVEPREFIX=	"arXiv",
	PRIMARYCLASS=	"hep-ph",
	REPORTNUMBER=	"DESY-17-044",
	DOI=	"10.1007/JHEP09(2017)069",
	JOURNAL=	"JHEP",
	VOLUME=	"09",
	PAGES=	"069",
	YEAR=	"2017",
}

@ARTICLE{ATLAS:2022jtk,
	TITLE=	"{Constraints on the Higgs boson self-coupling from single- and double-Higgs production with the ATLAS detector using pp collisions at s=13 TeV}",
	EPRINT=	"2211.01216",
	ARCHIVEPREFIX=	"arXiv",
	PRIMARYCLASS=	"hep-ex",
	REPORTNUMBER=	"CERN-EP-2022-149",
	DOI=	"10.1016/j.physletb.2023.137745",
	JOURNAL=	"Phys. Lett. B",
	VOLUME=	"843",
	PAGES=	"137745",
	YEAR=	"2023",
	AUTHOR=	"{ATLAS Collaboration}",
}

@TECHREPORT{Heinemeyer:2013tqa,
	AUTHOR=	"{LHC Higgs Cross Section Working Group}",
	EDITOR=	"Heinemeyer, S",
	TITLE=	"{Handbook of LHC Higgs Cross Sections: 3. Higgs Properties}",
	DOI=	"10.5170/CERN-2013-004",
	TYPE=	"CERN Report",
	YEAR=	"2013",
	EPRINT=	"1307.1347",
	ARCHIVEPREFIX=	"arXiv",
	PRIMARYCLASS=	"hep-ph",
	NUMBER=	"CERN-2013-004",
	SLACCITATION=	"%%CITATION = ARXIV:1307.1347;%%",
}

@ARTICLE{Bishara:2016jga,
	AUTHOR=	"Bishara, Fady and Haisch, Ulrich and Monni, Pier Francesco and Re, Emanuele",
	TITLE=	"{Constraining Light-Quark Yukawa Couplings from Higgs Distributions}",
	EPRINT=	"1606.09253",
	ARCHIVEPREFIX=	"arXiv",
	PRIMARYCLASS=	"hep-ph",
	REPORTNUMBER=	"OUTP-16-18P, CERN-TH-2016-136, LAPTH-026/16",
	DOI=	"10.1103/PhysRevLett.118.121801",
	JOURNAL=	"Phys. Rev. Lett.",
	VOLUME=	"118",
	PAGES=	"121801",
	YEAR=	"2017",
}
\cleardoublepage \appendix\section{The CMS Collaboration \label{app:collab}}\begin{sloppypar}\hyphenpenalty=5000\widowpenalty=500\clubpenalty=5000
\cmsinstitute{Yerevan Physics Institute, Yerevan, Armenia}
{\tolerance=6000
A.~Hayrapetyan, A.~Tumasyan\cmsAuthorMark{1}\cmsorcid{0009-0000-0684-6742}
\par}
\cmsinstitute{Institut f\"{u}r Hochenergiephysik, Vienna, Austria}
{\tolerance=6000
W.~Adam\cmsorcid{0000-0001-9099-4341}, J.W.~Andrejkovic, T.~Bergauer\cmsorcid{0000-0002-5786-0293}, S.~Chatterjee\cmsorcid{0000-0003-2660-0349}, K.~Damanakis\cmsorcid{0000-0001-5389-2872}, M.~Dragicevic\cmsorcid{0000-0003-1967-6783}, A.~Escalante~Del~Valle\cmsorcid{0000-0002-9702-6359}, P.S.~Hussain\cmsorcid{0000-0002-4825-5278}, M.~Jeitler\cmsAuthorMark{2}\cmsorcid{0000-0002-5141-9560}, N.~Krammer\cmsorcid{0000-0002-0548-0985}, L.~Lechner\cmsorcid{0000-0002-3065-1141}, D.~Liko\cmsorcid{0000-0002-3380-473X}, I.~Mikulec\cmsorcid{0000-0003-0385-2746}, J.~Schieck\cmsAuthorMark{2}\cmsorcid{0000-0002-1058-8093}, R.~Sch\"{o}fbeck\cmsorcid{0000-0002-2332-8784}, D.~Schwarz\cmsorcid{0000-0002-3821-7331}, M.~Sonawane\cmsorcid{0000-0003-0510-7010}, S.~Templ\cmsorcid{0000-0003-3137-5692}, W.~Waltenberger\cmsorcid{0000-0002-6215-7228}, C.-E.~Wulz\cmsAuthorMark{2}\cmsorcid{0000-0001-9226-5812}
% ...
\par}
\vskip\cmsinstskip
\dag:~Deceased\\
$^{1}$Also at Yerevan State University, Yerevan, Armenia\\
$^{2}$Also at TU Wien, Vienna, Austria\\
% ...
\end{sloppypar}
\end{document}